\pdfoutput=1 
\documentclass[aps,eqsecnum,showpacs,amsmath,amssymb,preprint,
floatfix]{revtex4}
\usepackage{graphics,epsfig}
\pdfoutput=1

\begin{document}

\count255=\time\divide\count255 by 60
\xdef\hourmin{\number\count255}
  \multiply\count255 by-60\advance\count255 by\time
 \xdef\hourmin{\hourmin:\ifnum\count255<10 0\fi\the\count255}

\newcommand{\xbf}[1]{\mbox{\boldmath $ #1 $}}

\newcommand{\sixj}[6]{\mbox{$\left\{ \begin{array}{ccc} {#1} & {#2} &
{#3} \\ {#4} & {#5} & {#6} \end{array} \right\}$}}

\newcommand{\threej}[6]{\mbox{$\left( \begin{array}{ccc} {#1} & {#2} &
{#3} \\ {#4} & {#5} & {#6} \end{array} \right)$}}

\title{Pion Electroproduction Amplitude Relations in the $1/N_c$
Expansion}

\author{Richard F. Lebed}
\email{Richard.Lebed@asu.edu}

\author{Lang Yu}
\email{langyu@asu.edu}

\affiliation{Department of Physics, Arizona State University, Tempe,
AZ 85287-1504}

\date{April 2009}

\begin{abstract}
We derive expressions for pion electroproduction amplitudes in the
$1/N_c$ expansion of QCD, and obtain from them linear relations
between the electromagnetic multipole amplitudes that hold at all
energies.  The leading-order relations in $1/N_c$ compare favorably
with available data (especially away from resonances), but the
next-to-leading-order relations tend to provide only small or no
improvement.
\end{abstract}


\maketitle

\section{Introduction}

The $1/N_c$ expansion of QCD~\cite{'tHooft:1973jz}, where $N_c$ is the
number of color charges, has emerged as one of the principal tools for
studying low-energy hadronic processes and hadron static observables.
In the simplest construction, baryons are assembled from a collection
of $N_c$ quarks, each one transforming under the SU$(N_c)$ fundamental
representation, such that the aggregate forms a color
singlet~\cite{Witten:1979kh}.  While analyzing baryons using the $N_c$
quarks' spin-flavor and combinatoric properties has led to a large
variety of interesting results far too numerous to review
here~\cite{predecessors}, the most compelling physical picture for
describing the dynamical properties of large $N_c$ baryons
(particularly scattering amplitudes) is the chiral soliton
approach~\cite{Adkins:1983ya} originally motivated by the Skyrme
model~\cite{Skyrme:1961vq}.  One of the most intriguing properties of
these studies is the emergence of model-independent linear relations
among meson-baryon scattering amplitudes~\cite{Hayashi:1984bc}, whose
origin gradually became understood as connected to the large $N_c$
limit~\cite{Mattis:1984ak,Mattis:1988hg}.  In fact, the existence of
such relations can be traced back to the contracted SU(4) spin-flavor
symmetry that emerges in the single-baryon sector as $N_c \! \to \!
\infty$, which in turn is obtained by demanding consistent
order-by-order unitarity in meson-baryon scattering
processes~\cite{Gervais:1983wq,Dashen:1993as,Dashen:1993jt}; the fact
that the former can be derived from the latter was first demonstrated
in Refs.~\cite{Cohen03}.

The large $N_c$ scattering method~\cite{Cohen03} explains the
existence of baryon resonance multiplets that share similarities (but
are not identical to) those appearing in large $N_c$ quark models; in
particular, one can study resonances with arbitrarily large
widths~\cite{Cohen:2003fv}, exotics~\cite{Cohen:2003nk}, and
three-flavor resonances~\cite{CL_3flavor}.  In addition, the means by
which the spurious $N_c \! > \! 3$ states may be
removed~\cite{Cohen:2006en} has been explored, as well as the nature
of the competing chiral and large $N_c$ limits~\cite{Cohen:2006up},
and results for multipion processes~\cite{Kwee:2006xf}.

An essential ingredient to obtaining results useful for $N_c \! = \!
3$ phenomenology is understanding how to include $1/N_c$ corrections.
The original large $N_c$ scattering amplitude relations were noted
long ago to satisfy the $t$-channel isospin-spin exchange constraint
$I_t \! = \!  J_t$~\cite{Mattis:1988hg}.  Using operator techniques,
Ref.~\cite{Dashen:1994qi} showed that static pion-baryon couplings
with $I \neq J$ are suppressed by a relative factor of $1/N_c^{|I \! -
\! J|}$, and the same techniques were used to show~\cite{NNItJt} that
nucleon-nucleon interactions with $|I_t \! - \! J_t| \!  = \! n$ are
suppressed by $1/N_c^n$.  Much more recently, the same techniques were
used to show~\cite{Cohen:2004qt} the analogous result for $t$-channel
pion-baryon scattering amplitudes.  The expression of these
pion-baryon constraints in terms of $s$-channel observables is
explored in Refs.~\cite{t_to_s}.

In this paper we modify the approach for describing meson-baryon
scattering amplitudes in the $1/N_c$ expansion, as it was applied to
the case of pion photoproduction $\gamma N \to \pi
N$~\cite{Cohen:2004bk}, to provide a model-independent expansion for
electromagnetic multipole amplitudes of the related pion
electroproduction process $e^- N \to e^- \pi N$, which at its essence
is the virtual photon process $\gamma^* N \to \pi N$.  Studies of
$\gamma N \! \to \! \pi N$ using the more traditional ``operator''
approach to large $N_c$ baryons also appear in the
literature~\cite{otherphoto}.  The fundamentals of hadronic
electroproduction are reviewed in Ref.~\cite{Drechsel:1992pn}.  The
photon squared four-momentum $q^2$ generalizes from zero in the
photoproduction case to a nonzero value for electroproduction, joining
the center-of-momentum (c.m.) energy $W$ of the $\gamma^* N$ system as
an independent kinematic variable for all amplitudes.  Furthermore,
virtual photons possess not only the familiar electric and magnetic
transverse multipoles, but scalar and longitudinal multipoles as well,
leading to a much richer set of experimental possibilities.  Even so,
the analysis of electroproduction amplitudes in the $1/N_c$ expansion
is almost identical to that for the case of photoproduction.  In this
paper we derive expressions for pion electroproduction amplitudes at
leading order (LO) and next-to-leading order (NLO) in the $1/N_c$
expansion and examine our findings using results of the MAID~2007
partial wave analysis from Universit\"{a}t
Mainz~\cite{Drechsel:2007if}.

Our purpose in this paper is to present only the results of a strict
$1/N_c$ analysis.  While the $1/N_c$ expansion relates combinations of
distinct amplitudes and predicts the magnitude of these differences at
each value of $q^2$ and $W$---an infinite number of testable
predictions---it does {\em not\/} predict the shapes of their $q^2$ or
$W$ dependences.  Obtaining such predictions would require imposing
dynamical assumptions that lie beyond the raw mandates of the $1/N_c$
expansion.  One may impose calculations using, for example, chiral
perturbation theory, specific quark models, or generalized parton
distributions on top of the amplitude predictions of this paper
(granted that they have been generalized to allow for $N_c$ to be
arbitrary) to obtain predictions for the detailed $q^2$ and $W$
amplitude shapes, but such modifications lie outside the intentionally
limited scope of this paper.  The only explicit $1/N_c$ dynamical
effects we discuss below arise due to the displaced pole positions of
baryon resonances in different channels.

This paper is organized as follows: In Sec.~\ref{relns} we derive the
linear electroproduction scattering amplitude relations.  In
Sec.~\ref{results} we confront the relations with the extensive
results of the MAID~2007 partial wave analysis, and comment upon the
quality of the comparisons.  We summarize briefly in Sec.~\ref{concl}.

\section{Derivation of Linear Relations} \label{relns}

Virtually all results stated in this section are a direct reprise of
those for the photoproduction case, Ref.~\cite{Cohen:2004bk}; the
differences are particularly noted.  The results for either process
are obtained by starting with general meson-baryon scattering
processes of the form $\Phi_1 + B_1 \to \Phi_2 + B_2$, where
$\Phi_{1,2}$ are mesons and $B_{1,2}$ are baryons, each pair carrying
fixed strangeness.  Since the amplitude
relations~\cite{Mattis:1988hg,Cohen03} for such processes depend upon
the mesons only through their quantum numbers, the same results may be
used for electroproduction, for which $\Phi_1$ is a virtual photon (or
technically, a meson interpolating field with the quantum numbers of a
virtual photon).

The master scattering formula for the observable scattering amplitudes
$S_{L_iL_fS_iS_fIJ}$ reads~\cite{Cohen03,Mattis:1988hg}
\begin{eqnarray}\label{master}
\lefteqn{S_{L_iL_fS_iS_fIJ}} \nonumber \\
  & = &\sum_{K, \tilde{K}_i, \tilde{K}_f}[K]([R_i][R_f][S_i][S_f]
  [\tilde{K}_i][\tilde{K}_f])^{1/2}
  \left\{
             \begin{array}{ccc}
               L_i & i_i & \tilde{K}_i \\
               S_i & R_i & s_i \\
               J & I & K \\
             \end{array}
           \right\}
  \left\{
             \begin{array}{ccc}
               L_f & i_f & \tilde{K}_f \\
               S_f & R_f & s_f \\
               J & I & K \\
             \end{array}
           \right\}
  \tau_{K\tilde{K}_i\tilde{K}_fL_iL_f} \, , \nonumber \\
\end{eqnarray}
where subscripts $i$ and $f$ label initial- and final-state
quantities, respectively.  Here, $R$ indicates the baryon spin (which
equals its isospin for both $N$ and $\Delta$), $s$ and $i$ label the
spin and isospin of the meson (or photon), respectively, $L$ labels
the orbital angular momentum of the meson (or photon) relative to the
baryon target, $S$ labels the total spin of the system, and
$\tilde{\textbf{K}} \equiv \textbf{i} + \textbf{L}$ is a hybrid
quantity that provides good quantum numbers $\tilde{K}_i$ and
$\tilde{K}_f$ in the large $N_c$ limit.  The overall conserved quantum
numbers $I$, $J$, and $K$ arise from the total isospin $\textbf{I}
\equiv \textbf{i}_i + \textbf{R}_i = \textbf{i}_f + \textbf{R}_f$, the
total angular momentum $\textbf{J} \equiv \textbf{L}_i + \textbf{S}_i
= \textbf{L}_f + \textbf{S}_f$, and the so-called grand spin
$\textbf{K} \equiv \textbf{I} + \textbf{J}$ (so that $K$ is also a
good quantum number in the large $N_c$ limit).  The braces are
conventional $9j$ symbols, and the multiplicity $2X \!  + \! 1$ of
each SU(2) representation $X$ is denoted by $[X]$.  The sums run over
all values of $K$, $\tilde{K}_i$, and $\tilde{K}_f$ consistent with
the nonvanishing of the $9j$ symbols, {\it i.e.}, each row and column
satisfies the triangle rule.  Beyond all the group-theoretical factors
lie the {\it reduced amplitudes\/} $\tau$, which are the undetermined
finite dynamical quantities in the large $N_c$ scattering amplitude
approach, analogous to reduced matrix elements in the Wigner-Eckart
theorem; their precise calculation would be tantamount to solving QCD
exactly at leading order in the $1/N_c$ expansion.

Virtual photons can carry either spin one or zero (the latter in
distinction to real photons); however, the angular momentum $\ell$
labeling each electromagnetic multipole is comprised of the
combination of the photon intrinsic spin with its orbital angular
momentum relative to the target~\cite{LL} (in this case, the initial
baryon).  In terms of Eq.~(\ref{master}), one effectively handles
photon angular momentum by setting $s_i \!  = \! 0$ and $L_i \! = \!
\ell$.

The photon isospin content is more complicated; photons carry both
isovector and isoscalar contributions, which couple to baryons through
operators carrying different $N_c$ counting.  Since this coupling
occurs through the photon polarization vector $\varepsilon_\mu$, and
since transverse, longitudinal, and scalar polarizations (the latter
two arising only for virtual photons) all have nonvanishing components
in spatial ($\mu \! = \! i$) directions, the relevant operators
representing photon couplings to baryons are
\begin{equation} \label{isov}
G^{ia} \equiv \sum_{\alpha=1}^{N_c} \, q^\dagger_\alpha \left(
\frac{\sigma^i}{2} \otimes \frac{\tau^a}{2} \right) q_\alpha \ ,
\end{equation}
and
\begin{equation}
J^i \equiv \sum_{\alpha=1}^{N_c} \, q^\dagger_\alpha \left(
\frac{\sigma^i}{2} \right) q_\alpha \ ,
\end{equation}
where $\sigma$ and $\tau$ are Pauli matrices in spin and isospin,
respectively, and $\alpha$ sums over the $N_c$ quark fields $q_\alpha$
in the baryon.  For the ground-state baryons ({\it e.g.}, $N$ and
$\Delta$), matrix elements of the isoscalar operator $J^i$ are of
course $O(N_c^0)$ (since they are just components of the total baryon
angular momentum), but those of the isovector operator $G^{ia}$ are
$O(N_c^1)$ due to the collective contribution of the $N_c$ quarks.
However, Eq.~(\ref{master}) does not incorporate this constraint, and
therefore it must be put in by hand: The full version of
Eq.~(\ref{master}) for electroproduction is the sum of a LO isovector
($i_i \! = \! 1$) piece and an isoscalar ($i_i \! = \! 0$) piece
carrying an explicit $1/N_c$ suppression factor, as expressed below in
Eqs.~(\ref{ivec_LO})--(\ref{ivec_NLO}).

The master amplitude expression Eq.~(\ref{master}) applied to pion
electroproduction off a nucleon target has $s_i \! = \! 0$ and $L_i
\! = \! \ell$ as mentioned above, as well as $i_i\equiv i_{\gamma} \in
\{0, 1\}$, $s_f \! = \! 0$, $i_f \! = \! 1$ (but retaining for the
moment the explicit symbol $i_f$), and $R_i \! = \! R_f \! = \! \frac
1 2$.  The triangle rules then force $S_i \! = S_f \! = \! \frac 1 2$,
and we relabel $L_f \! \to \! L$, which is the pion partial wave.  One
finds
\begin{equation}\label{S6}
    S_{\ell L\frac{1}{2}\frac{1}{2}IJ}=2(-1)^{L-\ell}\sum_{K}[K]
    \left\{ \begin{array}{ccc} J & \ell & \frac{1}{2} \\ i_{\gamma}& I
    & K \\ \end{array} \right\} \left\{ \begin{array}{ccc} J & L &
    \frac{1}{2} \\ i_f & I & K \\ \end{array} \right\}
    \tau^{\lambda}_{K\ell L} \, .
\end{equation}
Amplitudes for specific charge channels are obtained by attaching the
appropriate isospin Clebsch-Gordan coefficients to Eq.~(\ref{S6}).
Labeling the isospin third component of the incoming nucleon by $m_I$
and that of the outgoing pion by $\nu$, one has
\begin{eqnarray}\label{S7}
    M^{\lambda I i_{\gamma}}_{\ell LJm_I\nu} & = &
    2(-1)^{L-\ell}\sum_{K}[K]
        \left\{
             \begin{array}{ccc}
               J & \ell & \frac{1}{2} \\
               i_{\gamma}& I & K \\
             \end{array}
           \right\}
           \left\{
             \begin{array}{ccc}
              J   & L & \frac{1}{2} \\
              i_f & I & K \\
             \end{array}
           \right\}
           \tau^{\lambda}_{K\ell L}
\nonumber \\ & \times &
       \left(
             \begin{array}{cc|c}
               i_f   & \frac{1}{2} & I \\
               \nu & m_I-\nu & m_I \\
             \end{array}
       \right)
       \left(
             \begin{array}{cc|c}
               i_{\gamma} & \frac{1}{2} & I \\
               0 & m_I & m_I \\
             \end{array}
       \right) \, .
\end{eqnarray}
The label $\lambda$ indicates the type of multipole amplitude: $(\ell
- L)$ odd gives electric (e), longitudinal (l), and scalar (s)
multipoles (the latter two being absent from photoproduction), and
$(\ell - L)$ even gives magnetic (m) multipoles.  The l and s
multipoles are linearly dependent due to current
conservation~\cite{Drechsel:1992pn}, so we choose in this analysis to
eliminate l multipoles in favor of s multipoles.

We now exploit the result that amplitudes with $|I_t - J_t| = n$ are
suppressed by a relative factor $1/N_c^n$.  The first step is to
rewrite the $9j$ symbols in Eq.~(\ref{master}) in terms of $t$-channel
quantum numbers using the well-known SU(2) relation known as the
Biedenharn-Elliot sum rule~\cite{Edmonds}.  In terms of modified $6j$
symbols (called $[6j]$ symbols in Ref.~\cite{Cohen:2004qt}),
\begin{equation}\label{6j}
       \left\{
             \begin{array}{ccc}
               a & b & e \\
               c & d & f \\
             \end{array}
       \right\}
       \equiv\frac{(-1)^{-(b+d+e+f)}}{([a][b][c][d])^{1/4}}
       \left[
             \begin{array}{ccc}
               a & b & e \\
               c & d & f \\
             \end{array}
       \right] \, ,
\end{equation}
one obtains
\begin{eqnarray}
\left\{
\begin{array}{ccc}
  J & \ell & \frac{1}{2} \\
  i_\gamma & I & K
\end{array}
\right\} \left\{
\begin{array}{ccc}
  J & L & \frac{1}{2} \\
  i_f & I & K
\end{array}
\right\} & = & \sum_{{\cal J}} \frac{(-1)^{2{\cal
J}-i_f+i_\gamma}[{\cal J}]} {2\sqrt{[i_f][i_\gamma][L][\ell]}}
\left[
\begin{array}{ccc}
  i_f & L & K \\ \ell & i_\gamma & {\cal J}
\end{array}
\right] \left[
\begin{array}{ccc}
  i_f & \frac{1}{2} & I \\
  \frac{1}{2} & i_\gamma & {\cal J}
\end{array}
\right] \left[
\begin{array}{ccc}
  L & \frac{1}{2} & J \\
  \frac{1}{2} & \ell & {\cal J}
\end{array}
\right] \ , \label{BiedEll}
\nonumber \\ & &
\end{eqnarray}
where the quantum number ${\cal J}$ clearly adopts the role of both
$I_t$ (as seen from the second $[6j]$ symbol) and $J_t$ (as seen from
the third).  Setting at last $i_f \! = \! 1$ and defining the
$t$-channel reduced amplitudes by
\begin{equation}\label{t}
       \tau^{t\lambda i_{\gamma}}_{\mathcal {J}\ell L}
       \equiv\frac{(-1)^{2\mathcal {J}-1+i_{\gamma}}[\mathcal {J}]}
       {\sqrt{[1][i_{\gamma}][L][\ell]}}
       \sum_{K}[K]
       \left[
             \begin{array}{ccc}
               1 & L & K \\
               \ell & i_{\gamma} &\mathcal {J} \\
             \end{array}
       \right]\tau^{\lambda}_{K\ell L} \, ,
\end{equation}
one obtains the multipoles for the isovector case ($i_{\gamma}=1$):
\begin{equation}\label{ivec_LO}
    M^{\lambda I 1}_{\ell LJm_I\nu}=
       (-1)^{L-\ell}
       \left(
             \begin{array}{cc|c}
               1 & \frac{1}{2} & I \\
               \nu & m_I-\nu & m_I \\
             \end{array}
       \right)
       \left(
             \begin{array}{cc|c}
               1 & \frac{1}{2} & I \\
               0 & m_I & m_I \\
             \end{array}
       \right)
    \sum_{\mathcal {J}}
        \left[
             \begin{array}{ccc}
               1 & \frac{1}{2} & I \\
               \frac{1}{2}& 1 & \mathcal {J} \\
             \end{array}
           \right]
           \left[
             \begin{array}{ccc}
              L & \frac{1}{2} & J \\
              \frac{1}{2} &\ell & \mathcal {J} \\
             \end{array}
           \right]
           \tau^{t\lambda1}_{\mathcal {J}\ell L} \, ,
\end{equation}
and those for isoscalar case ($i_{\gamma}=0$):
\begin{equation}\label{iscalar}
    M^{\lambda I 0}_{\ell LJm_I\nu}=
       \frac{(-1)^{L-\ell}}{N_c}
       \left(
             \begin{array}{cc|c}
               1 & \frac{1}{2} & \frac{1}{2} \\
               \nu & m_I-\nu & m_I \\
             \end{array}
       \right)
    \frac{\delta_{I,\frac{1}{2}}}{[1]^{1/4}}
           \left[
             \begin{array}{ccc}
              L & \frac{1}{2} & J \\
              \frac{1}{2} &\ell & 1 \\
             \end{array}
           \right]
           \tau^{t\lambda0}_{1\ell L} \, .
\end{equation}
Note the explicit $1/N_c$ suppression factor in the isoscalar
expression.  In order to achieve relations at a consistent order in
the $1/N_c$ expansion, one must also include the independent NLO
isovector amplitudes, which have $|I_t - J_t|=1$.  Generalizing
Eq.~(\ref{ivec_LO}) gives
\begin{eqnarray}\label{ivec_NLO}
  M^{\lambda I 1(\rm NLO)}_{\ell LJm_I\nu} &=&
  \frac{(-1)^{L-\ell}}{N_c}
       \left(
             \begin{array}{cc|c}
               1 & \frac{1}{2} & I \\
               \nu & m_I-\nu & m_I \\
             \end{array}
       \right)
       \left(
             \begin{array}{cc|c}
               1 & \frac{1}{2} & I \\
               0 & m_I & m_I \\
             \end{array}
       \right)\nonumber\\
       & \times & \! \!
  \left\{ \sum_{x}
        \left[
             \begin{array}{ccc}
               1 & \frac{1}{2} & I \\
               \frac{1}{2}& 1 & x \\
             \end{array}
           \right] \!
           \left[
             \begin{array}{ccc}
              L & \frac{1}{2} & J \\
              \frac{1}{2} &\ell & x+1 \\
             \end{array}
           \right]
           \tau^{t\lambda(+)}_{x\ell L}
           \! +
        \sum_{y}
        \left[
             \begin{array}{ccc}
               1 & \frac{1}{2} & I \\
               \frac{1}{2}& 1 & y \\
             \end{array}
           \right] \!
           \left[
             \begin{array}{ccc}
              L & \frac{1}{2} & J \\
              \frac{1}{2} &\ell & y-1 \\
             \end{array}
           \right]
           \tau^{t\lambda(-)}_{y\ell L}
           \right\} ,
\nonumber \\ & &
\end{eqnarray}
where $x$ in the first sum and $y$ in the second represent $I_t$, and
the amplitudes $\tau^{t \lambda (\pm)}$ are independent of those at
leading order.  The total multipole amplitude, including LO and NLO
terms to consistent order in $1/N_c$, is therefore the sum of
Eqs.~(\ref{ivec_LO}), (\ref{iscalar}), and (\ref{ivec_NLO}):
\begin{equation}\label{S11}
    M^{\lambda I }_{\ell LJm_I\nu}=M^{\lambda I 1}_{\ell LJm_I\nu}+
    M^{\lambda I 0}_{\ell LJm_I\nu} +
    M^{\lambda I 1(\rm NLO)}_{\ell LJm_I\nu} \, .
\end{equation}
By including all values of $\mathcal {J}$, $x$, and $y$ allowed by the
triangle rules in the [6j] symbols and simplifying, one obtains the
expression
\begin{eqnarray}\label{full}
  M^{\lambda m_I \nu}_{\ell LJ} &=&
               \sum_{I}(-1)^{L-\ell}
       \left(\begin{array}{cc|c}
               1 & \frac{1}{2} & I \\
               \nu & m_I-\nu & m_I \\
             \end{array}\right)
       \left[
       \left(\begin{array}{cc|c}
               1 & \frac{1}{2} & I \\
               0 & m_I & m_I \\
             \end{array} \right) \right. \nonumber \\ & \times &
       \left\{ \delta_{\ell, L} \tau^{t\lambda 1}_{0LL}
               +\sqrt{\frac{2}{3}}
               \left( \delta_{I,\frac{1}{2}}- \frac 1 2
           \delta_{I,\frac{3}{2}} \right)
        \left[
             \begin{array}{ccc}
               L & \frac{1}{2} & J \\
               \frac{1}{2}& \ell & 1 \\
             \end{array}\right]\tau^{t\lambda 1}_{1\ell L}
\right. \nonumber \\ & & \left.
           +\frac{1}{N_c}
           \left(
           \left[\begin{array}{ccc}
              L & \frac{1}{2} & J \\
              \frac{1}{2} &\ell & 1 \\
             \end{array}\right]
           \tau^{t\lambda(+)}_{0\ell L}
             +\sqrt{\frac{2}{3}}
        \left( \delta_{I,\frac{1}{2}}- \frac 1 2
        \delta_{I,\frac{3}{2}} \right) \delta_{\ell, L}
        \tau^{t\lambda(-)}_{1 L L}
        \right)
        \right\}
\nonumber\\
        &&\left.+\frac{1}{N_c}\frac{\delta_{I,\frac{1}{2}}}{[1]^{1/4}}
           \left[
             \begin{array}{ccc}
              L & \frac{1}{2} & J \\
              \frac{1}{2} &\ell & 1 \\
             \end{array}
           \right]
           \tau^{t\lambda 0}_{1\ell L}
           \right] \, ,
\end{eqnarray}
which is identical in form to the expression found for
photoproduction~\cite{Cohen:2004bk}.

Pion electroproduction possesses four charged channels: $\gamma^* p
\rightarrow \pi^+ n$, $\gamma^* n \rightarrow \pi^- p$, $\gamma^* p
\rightarrow \pi^0 p$ and $\gamma^* n \rightarrow \pi^0 n$.  Assuming
isospin invariance, only three of these are
independent~\cite{Drechsel:1992pn}, and so we choose to eliminate the
channel $\gamma^* n \rightarrow \pi^0 n$ for which all the particles
are neutral.  Since the initial- and final-state nucleons carry only
spin 1/2, parity invariance constrains electric and scalar multipole
amplitudes to $\ell = L \pm 1$, while magnetic multipole amplitudes
satisfy $\ell = L$.  The spinless pion and spin-1/2 nucleon combine to
give possible angular momenta $J=L\pm\frac{1}{2}$; it is convenient to
express this information using the combination $2(J-L)$, which equals
$\pm 1$ for $J=L\pm\frac{1}{2}$, respectively, and write the
amplitudes as $M^\lambda_{\ell, L, 2(J-L)}$.  Using the
conventions~\cite{Drechsel:1992pn} employed by MAID, the multipole
amplitude $M^\lambda_{\ell, L, 2(J-L)}$ (where $\lambda$ = e, s, or m)
is proportional to $\Lambda_{L, 2(J-L)}$; for example, $M^{{\rm
s}}_{L-1, L, -} \propto S_{L-}$.  The proportionality factor depends
upon particle energies, $\lambda$, $L$, and $2(J-L)$, but is not
important for this analysis since in each relation presented below it
cancels out as a common factor.

Using Eq.~(\ref{full}), one obtains very similar relations for
electric and scalar multipoles:
\begin{eqnarray}\label{E1}
  M^{{\rm e/s}, \, p(\pi^{+})n}_{L-1, L, -} &=& M^{{\rm e/s},
  n(\pi^{-})p}_{L-1, L, -} + O(N^{-1}_c) \;
  (L\geq 2 \mbox{ for e, } L\geq 1 \mbox{ for s}) \, ,
\end{eqnarray}
\begin{eqnarray}\label{E2}
  M^{{\rm e/s}, \, p(\pi^{+})n}_{L+1, L, +} &=&
  M^{{\rm e/s}, \, n(\pi^{-})p}_{L+1, L, +} + O(N^{-1}_c)  \;
  (L\geq0) \, ,
\end{eqnarray}
and
\begin{eqnarray}\label{E3}
  M^{{\rm e/s}, \, p(\pi^{0})p}_{L\pm1, L, \pm} &=& O(N^{-1}_c) \, .
\end{eqnarray}
The presence of an $L \! = \! 1$ relation for s but not e amplitudes
in Eq.~(\ref{E1}) reflects the existence of C0 but not E0
electromagnetic multipoles.  Referring to Eq.~(\ref{full}), one
obtains four relations at LO because one has six observable
amplitudes, arising through three charged channels each with two
allowed values of $2(J-L)$, but only two reduced amplitudes, $\tau^{t
\, {\rm e/s} \, 1}_{1,L\pm1,L}$.  However, no relations survive at NLO
because four new amplitudes ($\tau^{t \, {\rm e/s}(+)}_{0,L\pm1,L}$
and $\tau^{t \, {\rm e/s} \, 0}_{1,L\pm1,L}$) appear at this order.
Equation~(\ref{E3}) implies the vanishing of the e and s multipole
amplitudes at LO for the process $\gamma^* p \rightarrow
\pi^{0}p$, which means that they are expected to be, on average, a
factor of about $N_c = 3$ smaller than the charged amplitudes; but
this is a rather qualitative statement [in particular, Eq.~(\ref{E3})
does not test the equality of measurable amplitudes], and we omit
numerical analysis of such amplitudes below.

Using Eq.~(\ref{full}) for the magnetic multipole amplitudes, one
again has six observable amplitudes expressed at LO in terms of only
two reduced amplitudes ($\tau^{t{\rm m}1}_{1LL}$ and $\tau^{t{\rm
m}1}_{0LL}$), leading to four relations:
\begin{eqnarray}\label{M1}
  M^{{\rm m}, \, p(\pi^{0})p}_{L, L, -} &=& M^{{\rm m},
  p(\pi^{0})p}_{L, L, +} + O(N^{-1}_c) \;
  (L\geq1) \, ,
\end{eqnarray}
\begin{eqnarray}\label{M2}
  M^{{\rm m}, \, p(\pi^{+})n}_{L, L, -} &=&
  M^{{\rm m}, \, n(\pi^{-})p}_{L, L, -}=
  -\frac{L+1}{L}M^{{\rm m}, \, p(\pi^{+})n}_{L, L, +}=
  -\frac{L+1}{L}M^{{\rm m}, \, n(\pi^{-})p}_{L, L, +}
+ O(N^{-1}_c) \; (L\geq1) \, . \nonumber \\
\end{eqnarray}
Only three new reduced amplitudes ($\tau^{t{\rm m}(+)}_{0LL}$,
$\tau^{t{\rm m}(-)}_{1LL}$, and $\tau^{t{\rm m}0}_{1LL}$) appear at
NLO, meaning that one special combination holds at both LO and NLO:
\begin{eqnarray}\label{M3}
  M^{{\rm m}, \, p(\pi^{+})n}_{L, L, -} &=&
  M^{{\rm m}, \, n(\pi^{-})p}_{L, L, -}
  -\left(\frac{L+1}{L}\right)
  \left[M^{{\rm m}, \, p(\pi^{+})n}_{L, L, +}-
        M^{{\rm m}, \, n(\pi^{-})p}_{L, L, +}
  \right]+O(N^{-2}_c)
\; (L\geq1) \, . \nonumber \\
\end{eqnarray}
One expects this relation to improve generically upon the predictions
of the LO relations by a factor of about $N_c \! = \! 3$.

Note that several of the LO relations, specifically those of the form
$M^{\lambda, p(\pi^+)n}_{\ell, L, \pm} = M^{\lambda, n(\pi^-)p}_{\ell,
L, \pm}$ [which are Eqs.~(\ref{E1})--(\ref{E2}) and the first and
third of relations in Eq.~(\ref{M2})] follow directly from isospin
symmetry and the LO dominance of the isovector amplitude, as may be
checked simply from the Clebsch-Gordan coefficients in
Eq.~(\ref{ivec_LO}).  All of the electroproduction multipole relations
presented here also appear for the photoproduction, of course
excepting the scalar multipole relations.

\section{Results} \label{results}

Our results consist of a comparison of the relations
Eqs.~(\ref{E1})--(\ref{E2}) and (\ref{M1})--(\ref{M3}) to the
experimental data as obtained from the MAID~2007 partial-wave
analysis~\cite{Drechsel:2007if}.  The exceptionally large volume of
available information is presented as concisely as possible: Each
allowed partial wave [$L$ can be arbitrarily large, and
Eqs.~(\ref{E1})--(\ref{M3}) provide relations between such amplitudes,
but data is typically compiled only up to $L = 5$] is a complex-valued
amplitude and depends upon two independent dynamical variables: the
photon virtuality $Q^2 \! \equiv \! -q^2$ and the $\gamma^* N$ c.m.\
energy $W$.  In each plot $W$ ranges from threshold to 2~GeV, and
$Q^2$ ranges from 0--5~GeV$^2$.  In all figures we present both real
and imaginary parts of the left-hand and right-hand sides (l.h.s.\ and
r.h.s.) of each multipole amplitude combination appearing in
Eqs.~(\ref{E1})--(\ref{E2}) and (\ref{M1})--(\ref{M3}) (except for $L
= 4$ and $5$ imaginary parts, given as zero by MAID), and then also
plot scale-independent amplitude ratios according to the prescription
\begin{equation}
\mbox{Ratio} = \frac{{\rm l.h.s.} - {\rm r.h.s.}}{\frac{1}{2}
\left( |\max ({\rm l.h.s.})| + |\max ({\rm r.h.s.}) | \right) } \, .
\end{equation}
For a LO relation, this quantity is dimensionless and predicted to be
of order $1/N_c$, while for NLO relations the prediction is
$O(1/N_c^2)$.  Absolute values appear in the denominator to avoid
physically uninteresting behavior when l.h.s.\ and r.h.s.\ are both
near zero and happen to be equal and opposite.  Maximally poor
relationships ($|{\rm l.h.s.}| \gg |{\rm r.h.s.}|$ or vice versa) are
manifested by Ratio~$\to \pm 2$.  On the other hand, relations that
truly hold to $O(1/N_c)$ might be expected to lie between $\pm 1/3$;
however, this conclusion neglects the order-unity coefficient implicit
in $O(1/N_c)$.  We choose as a useful metric to distinguish between
$O(1/N_c)$ and $O(N_c^0)$ quantities their geometric mean, $\pm
1/N_c^{1/2} \! \approx \! \pm 0.577$ for $N_c \! = \! 3$.  Similarly,
we take the largest $O(1/N_c^2)$ effects to be $\pm 1/N_c^{3/2} \!
\approx \! \pm 0.192$; nevertheless, the reader is free to choose
their own figure of merit upon viewing the plots.

\begin{figure}[htp]
\caption{Electric multipole data ($J = L - \frac 1 2$ amplitudes
$E_{L-}$) from MAID~2007.  The l.h.s., r.h.s., and ratio of
relation~(\ref{E1}) for $L \geq 2$ are presented in separate rows,
with separate columns for the real and imaginary parts (except for the
$L = 4$ and $5$ imaginary parts, given as zero by MAID).}
\label{E-plot}
\epsfxsize=0.45\textwidth\epsfbox{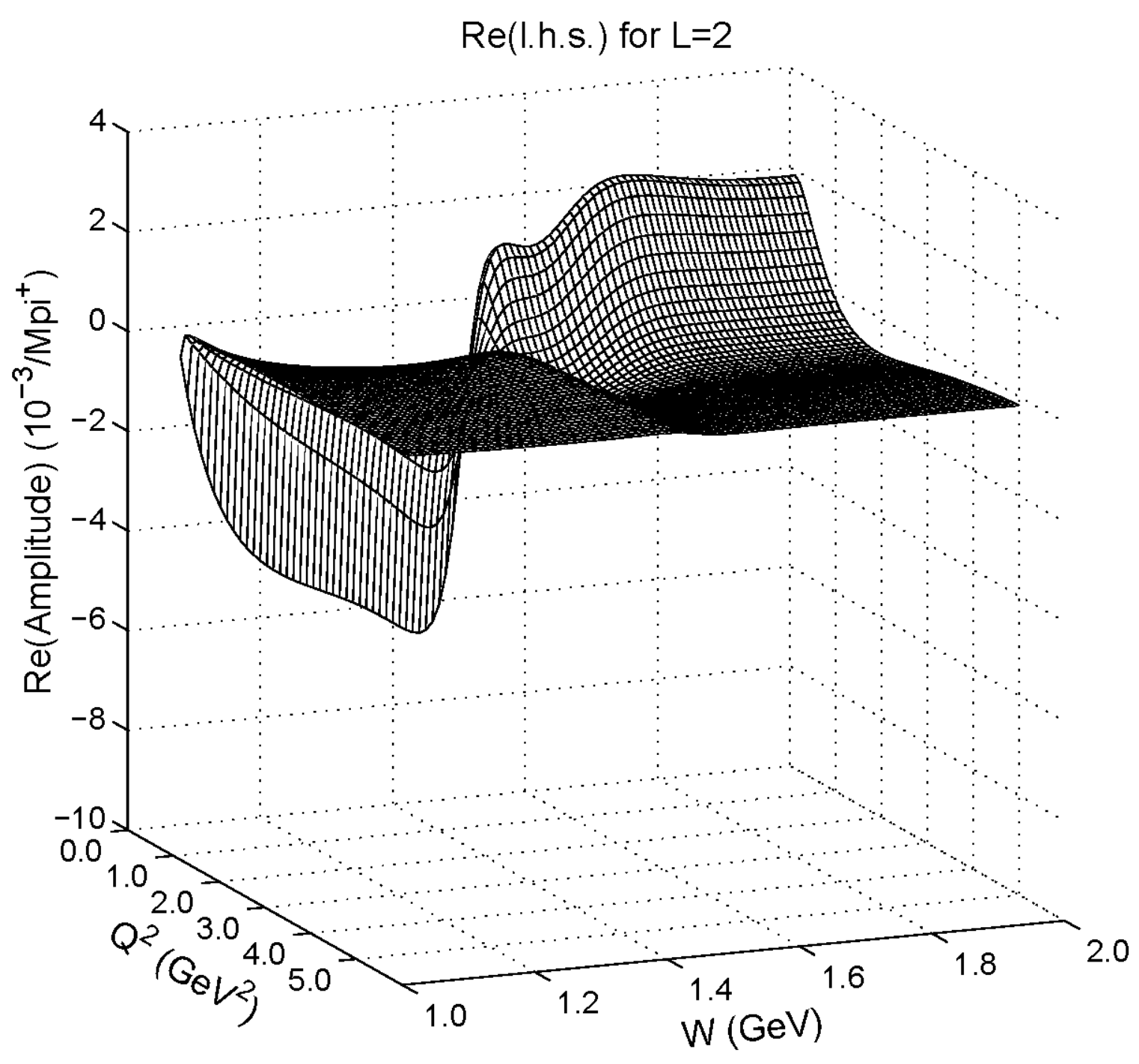}
\epsfxsize=0.45\textwidth\epsfbox{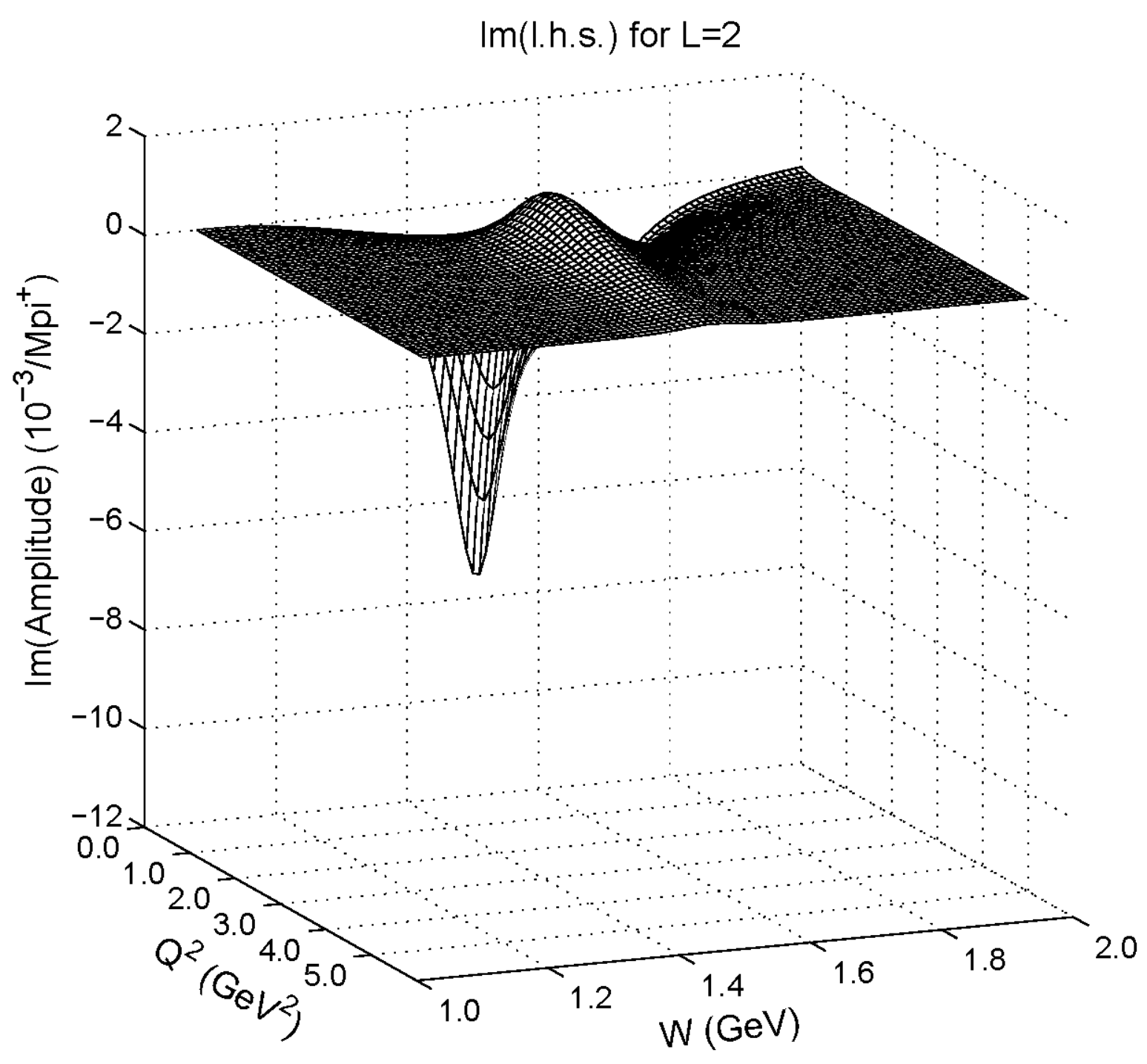}\\[1mm]
\epsfxsize=0.45\textwidth\epsfbox{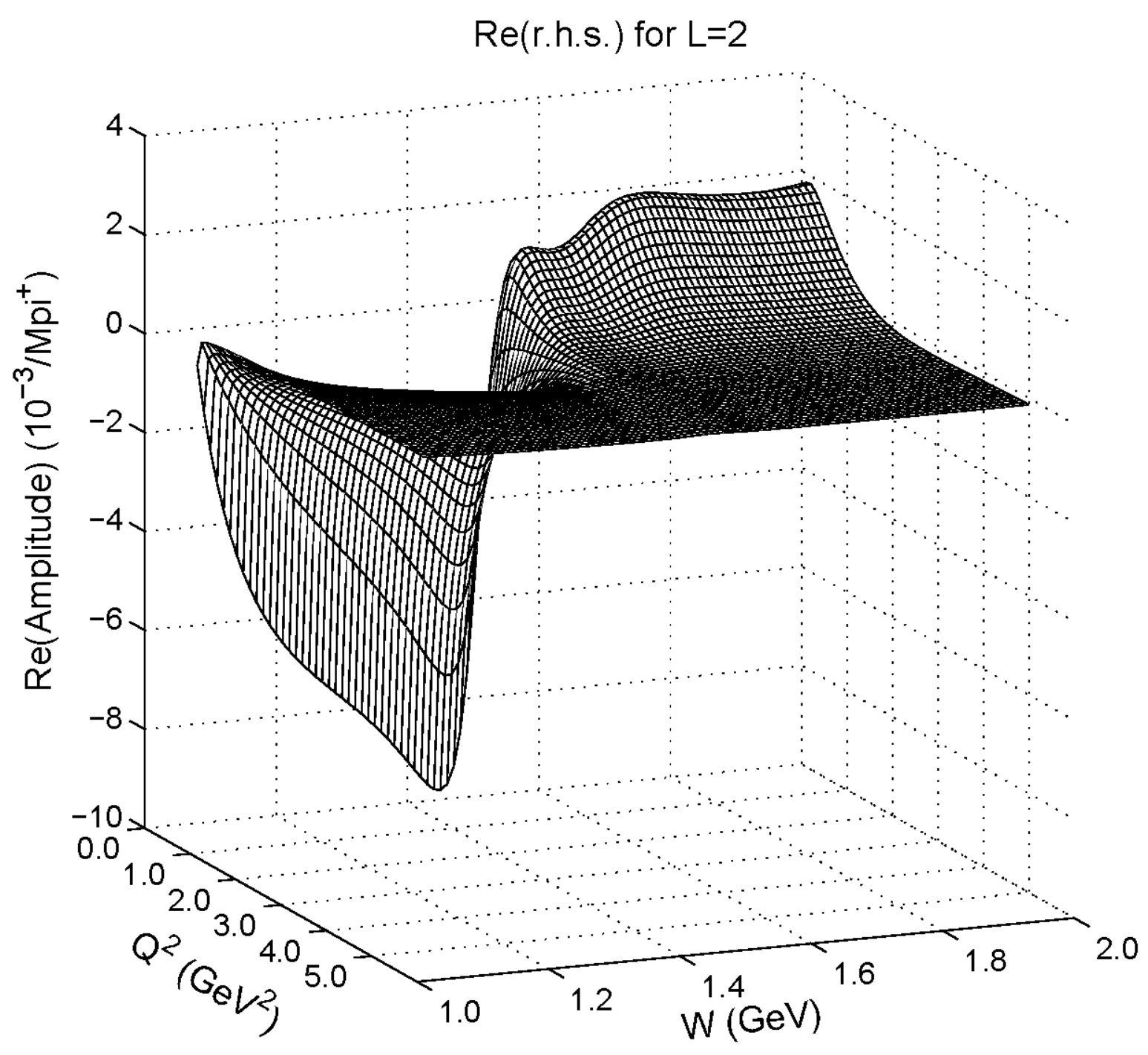}
\epsfxsize=0.45\textwidth\epsfbox{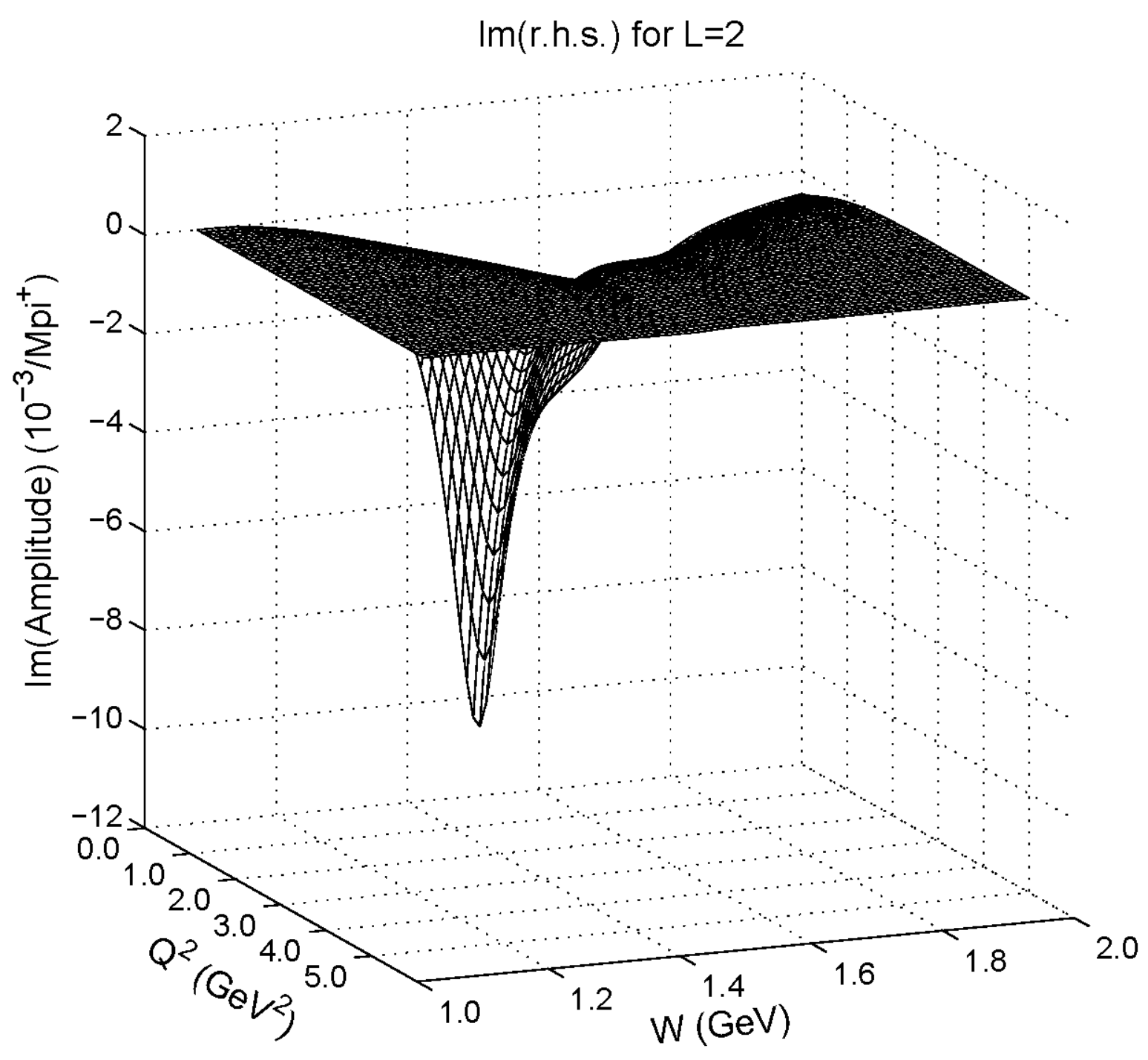}\\[1mm]
\epsfxsize=0.45\textwidth\epsfbox{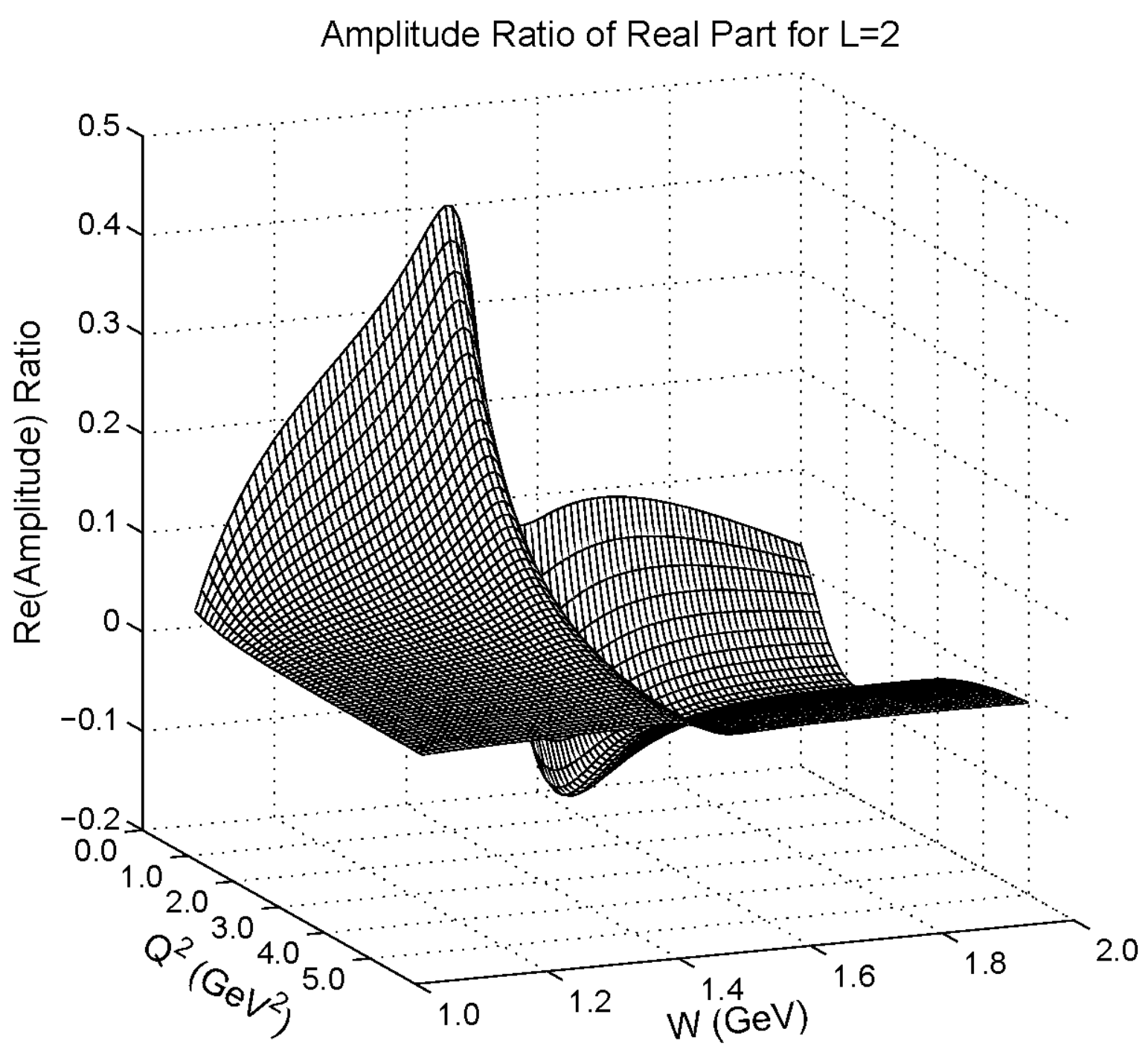}
\epsfxsize=0.45\textwidth\epsfbox{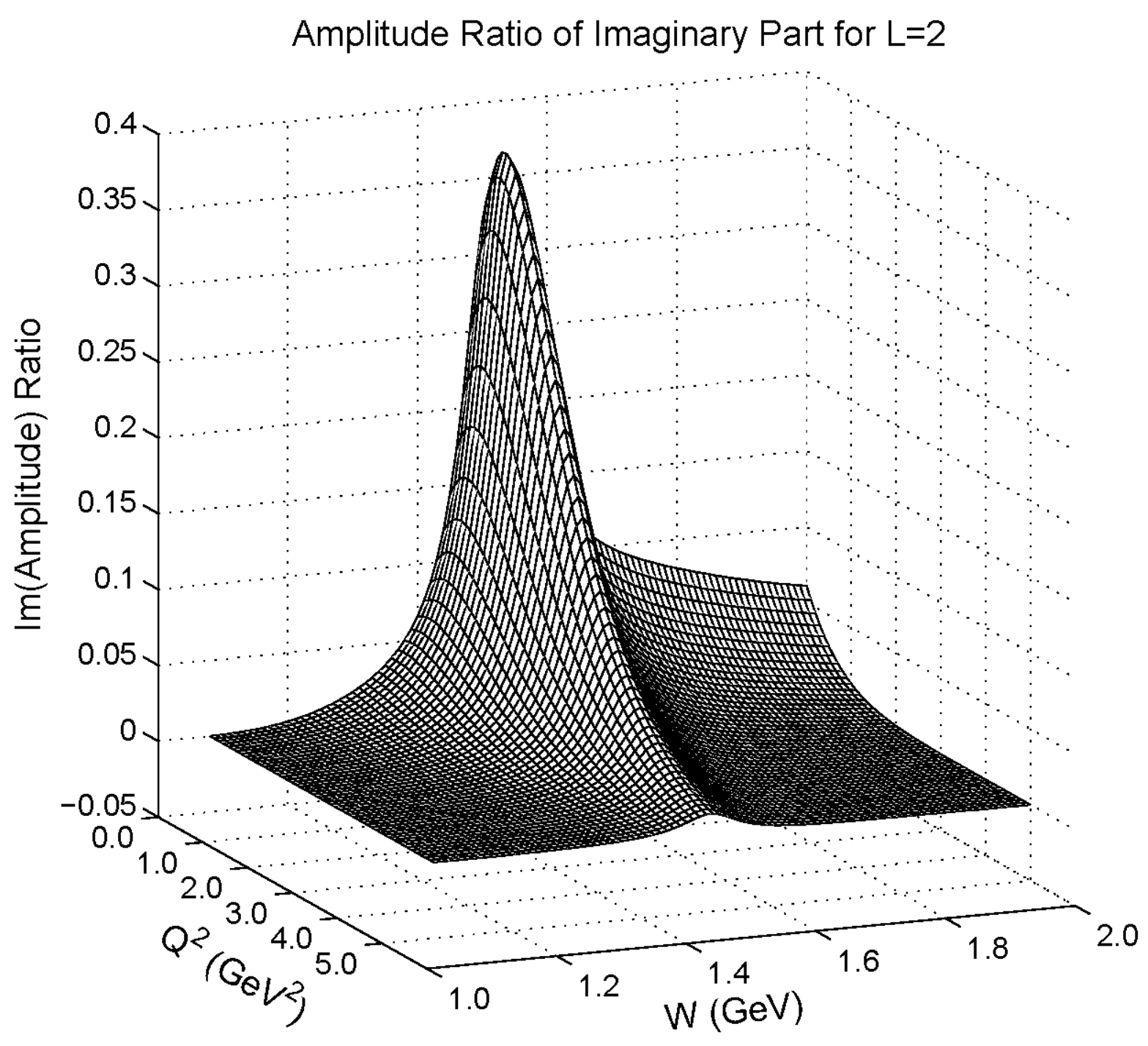}\\
\end{figure}

\begin{figure}[htp]
\caption{Scalar multipole data ($J = L - \frac 1 2$ amplitudes
$S_{L-}$) from MAID~2007.  The l.h.s., r.h.s., and ratio of
relation~(\ref{E1}) for $L \geq 1$ are presented in separate rows,
with separate columns for the real and imaginary parts (except for the
$L = 4$ and $5$ imaginary parts, given as zero by MAID).}
\label{S-plot}
\epsfxsize=0.44\textwidth\epsfbox{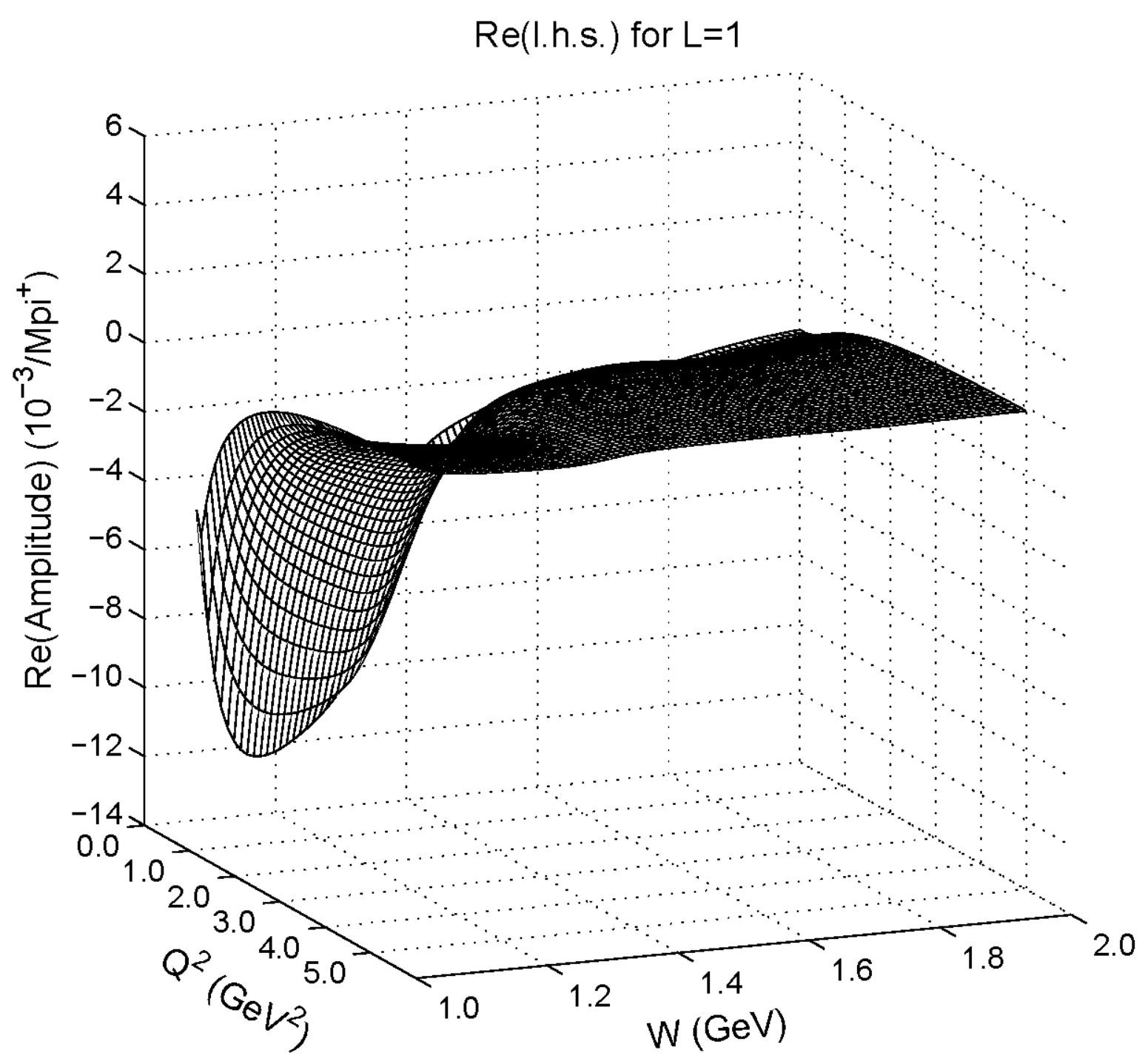}
\epsfxsize=0.44\textwidth\epsfbox{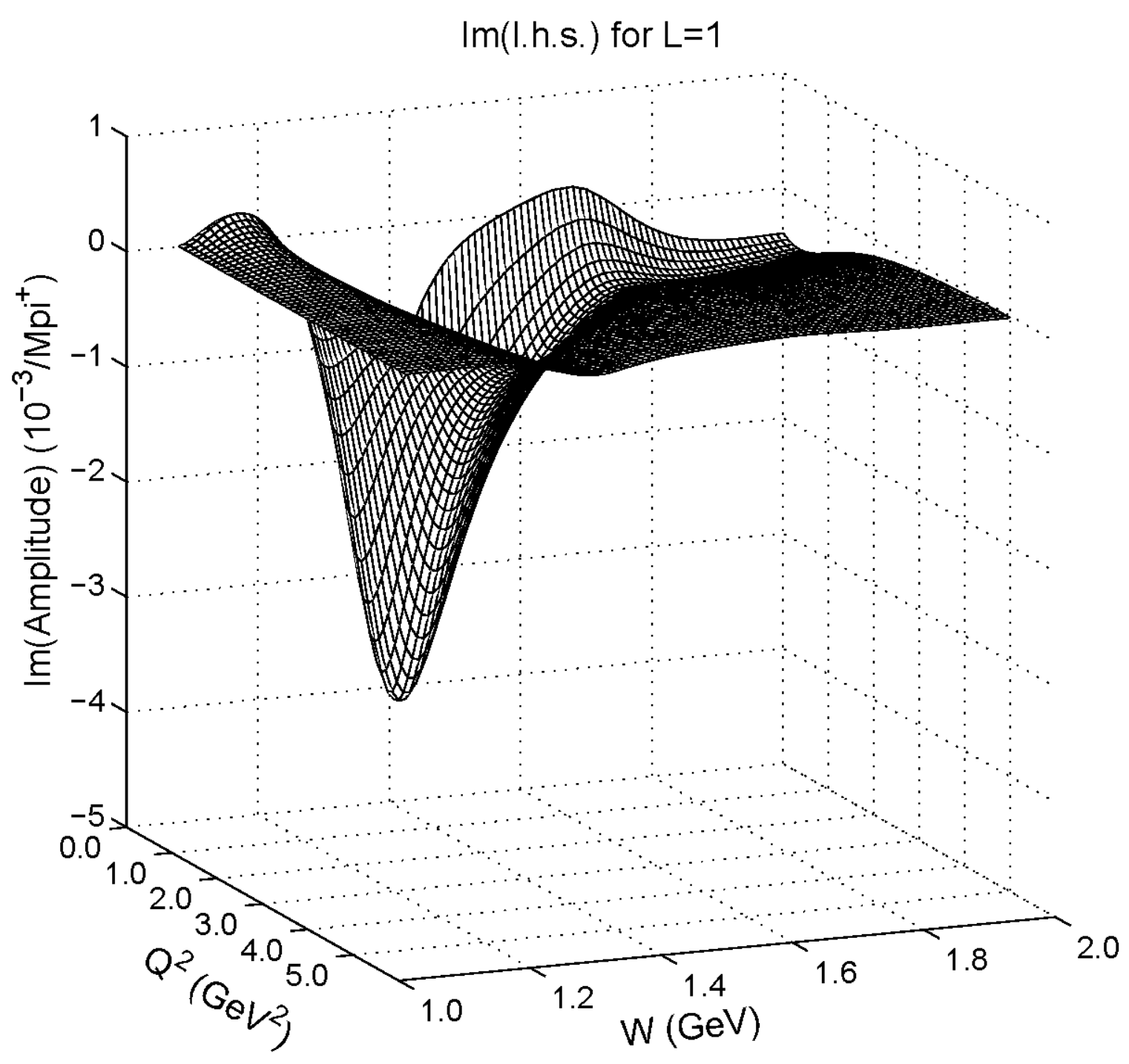}\\[1mm]
\epsfxsize=0.44\textwidth\epsfbox{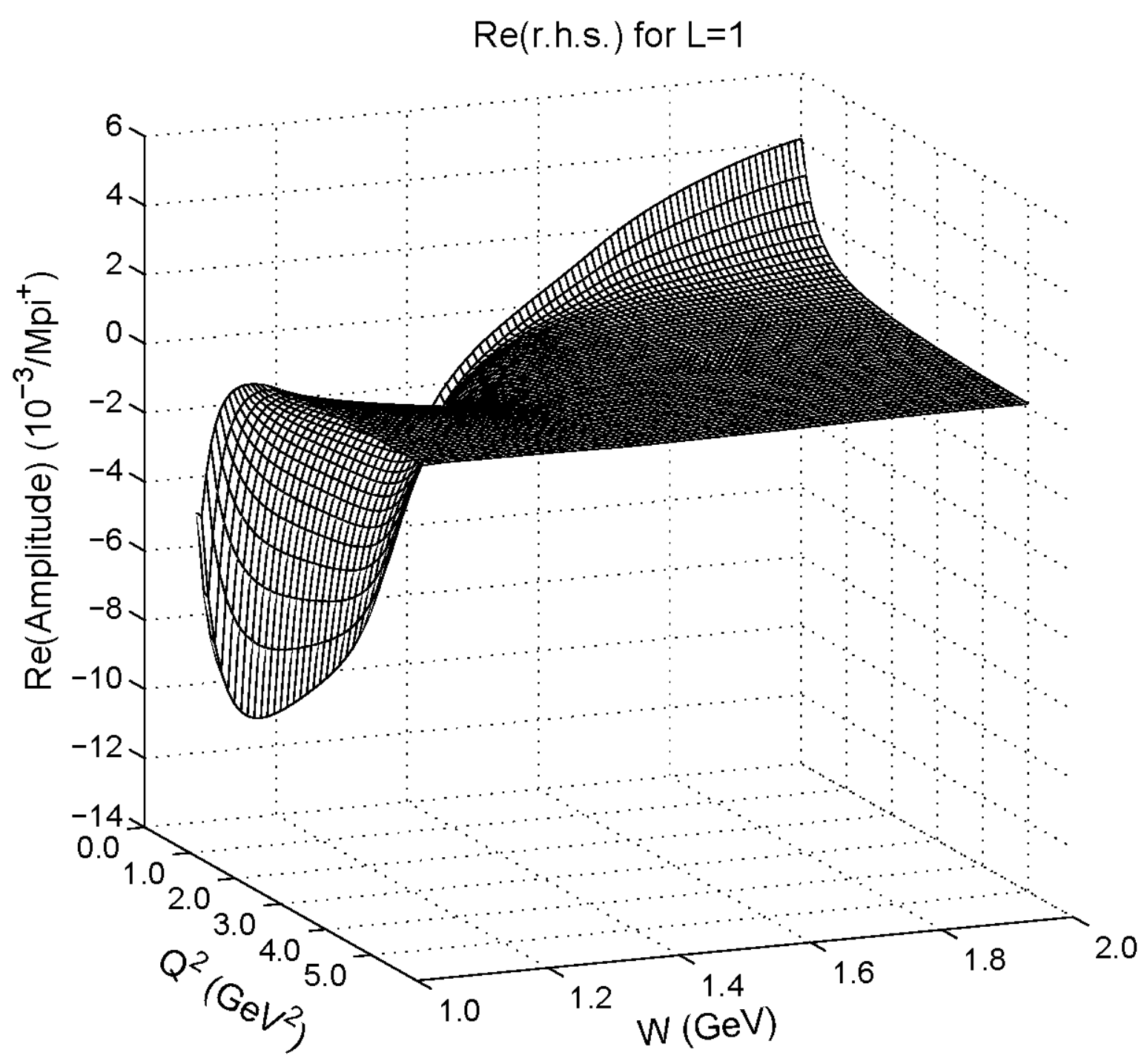}
\epsfxsize=0.44\textwidth\epsfbox{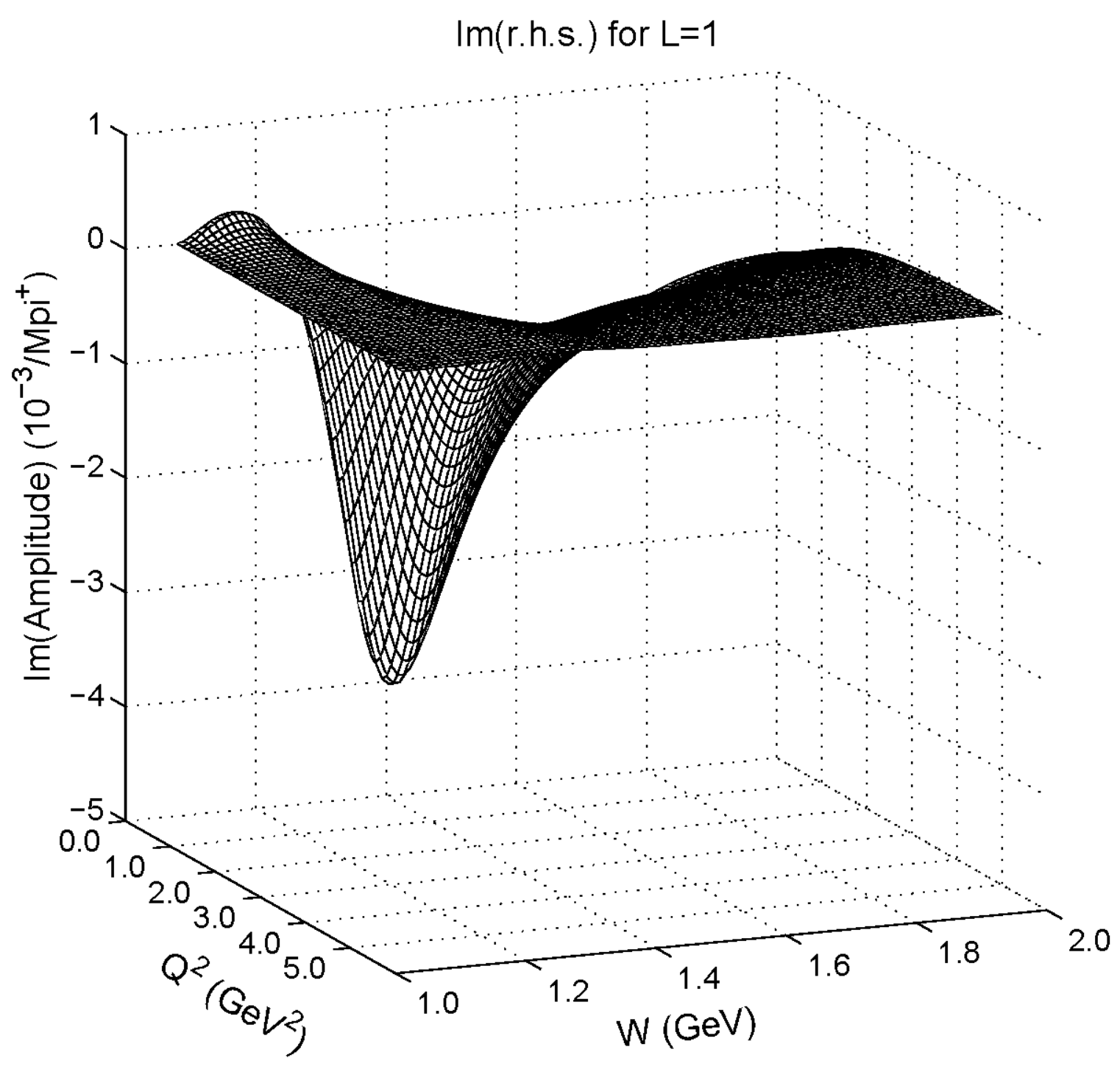}\\[1mm]
\epsfxsize=0.44\textwidth\epsfbox{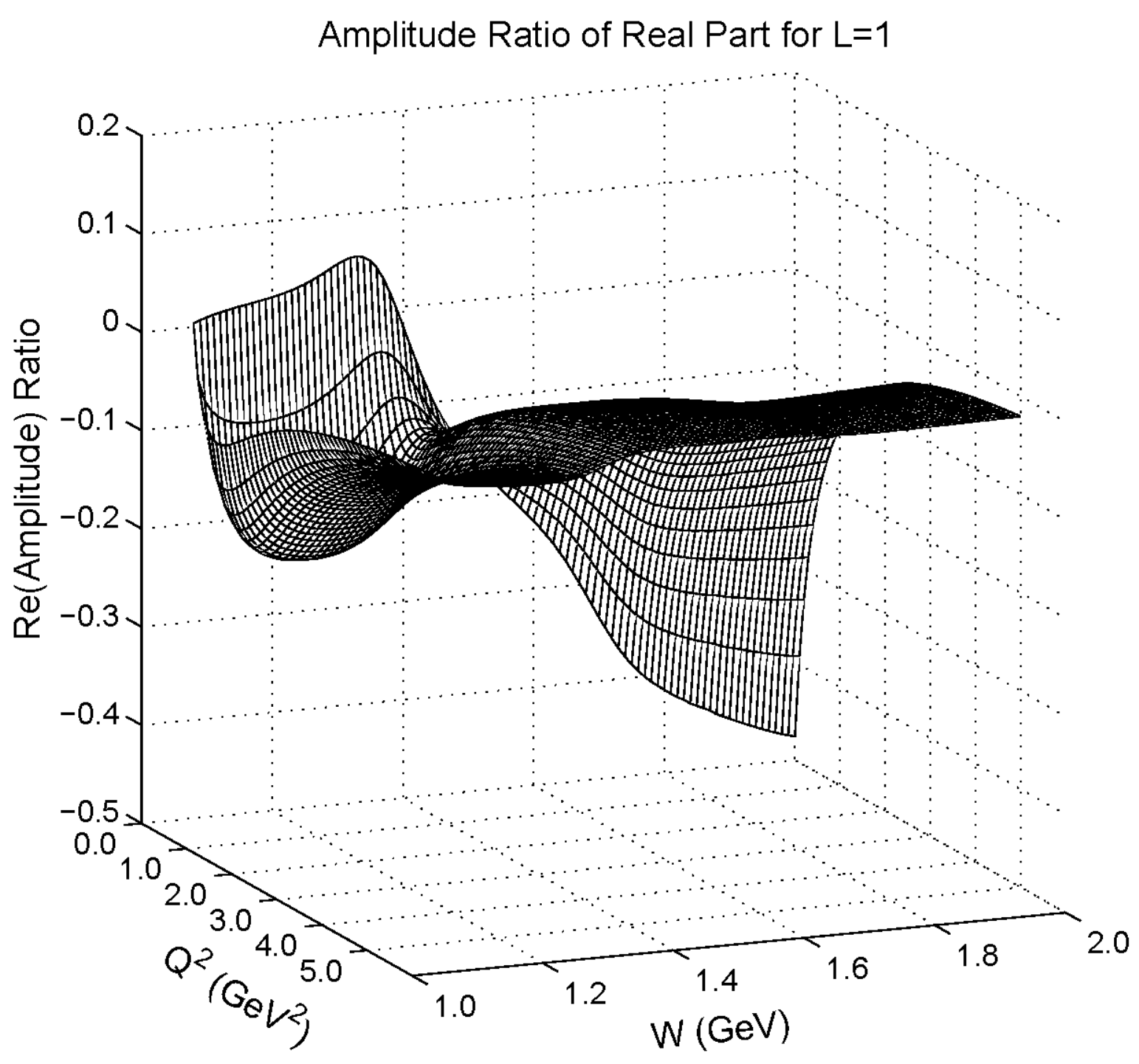}
\epsfxsize=0.44\textwidth\epsfbox{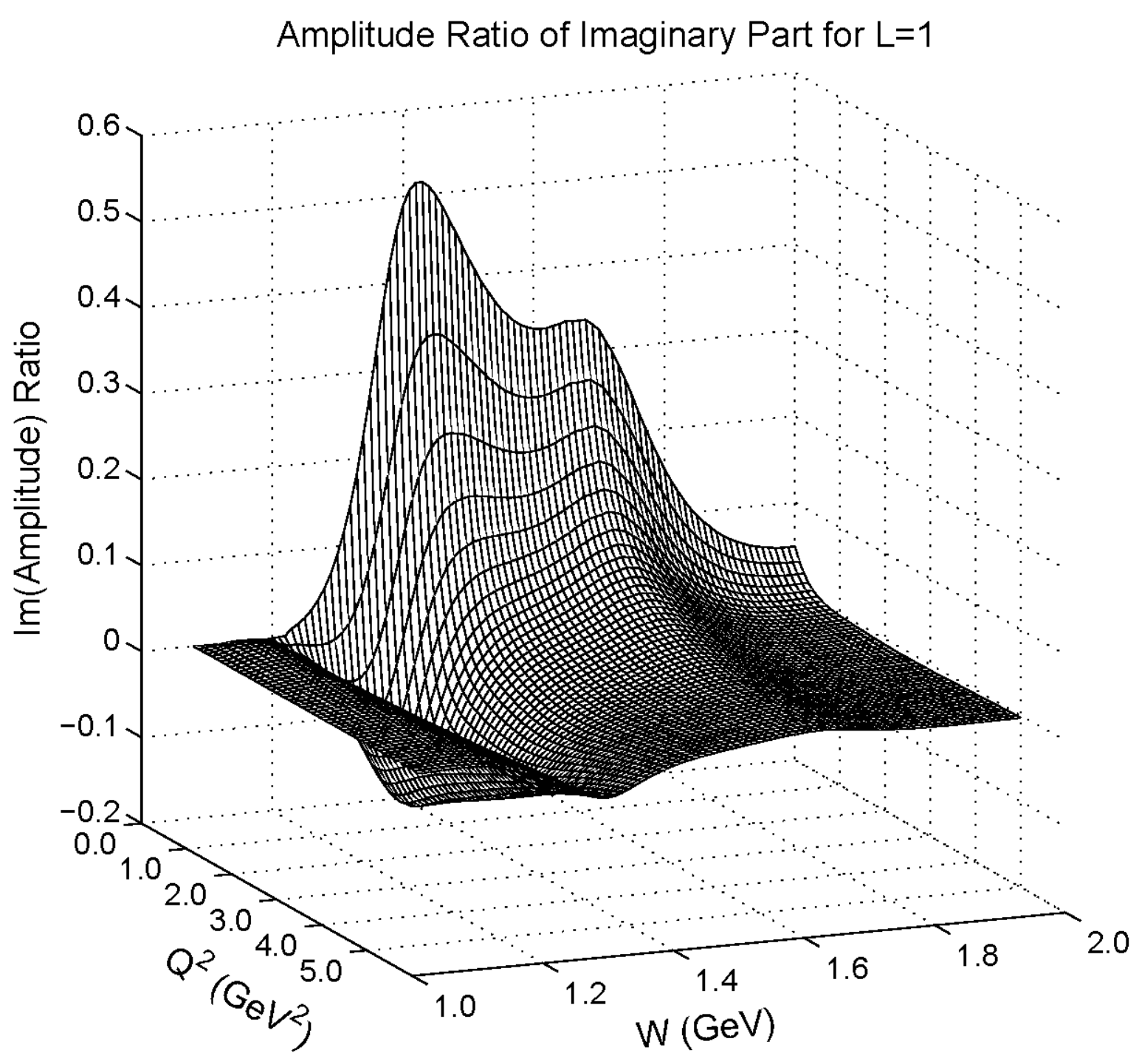}\\
\end{figure}

\begin{figure}[htp]
\caption{Electric multipole data ($J = L + \frac 1 2$ amplitudes
$E_{L+}$) from MAID~2007.  The l.h.s., r.h.s., and ratio of
relation~(\ref{E2}) for $L \geq 0$ are presented in separate rows,
with separate columns for the real and imaginary parts (except for the
$L = 4$ and $5$ imaginary parts, given as zero by MAID).}
\label{E+plot}
\epsfxsize=0.44\textwidth\epsfbox{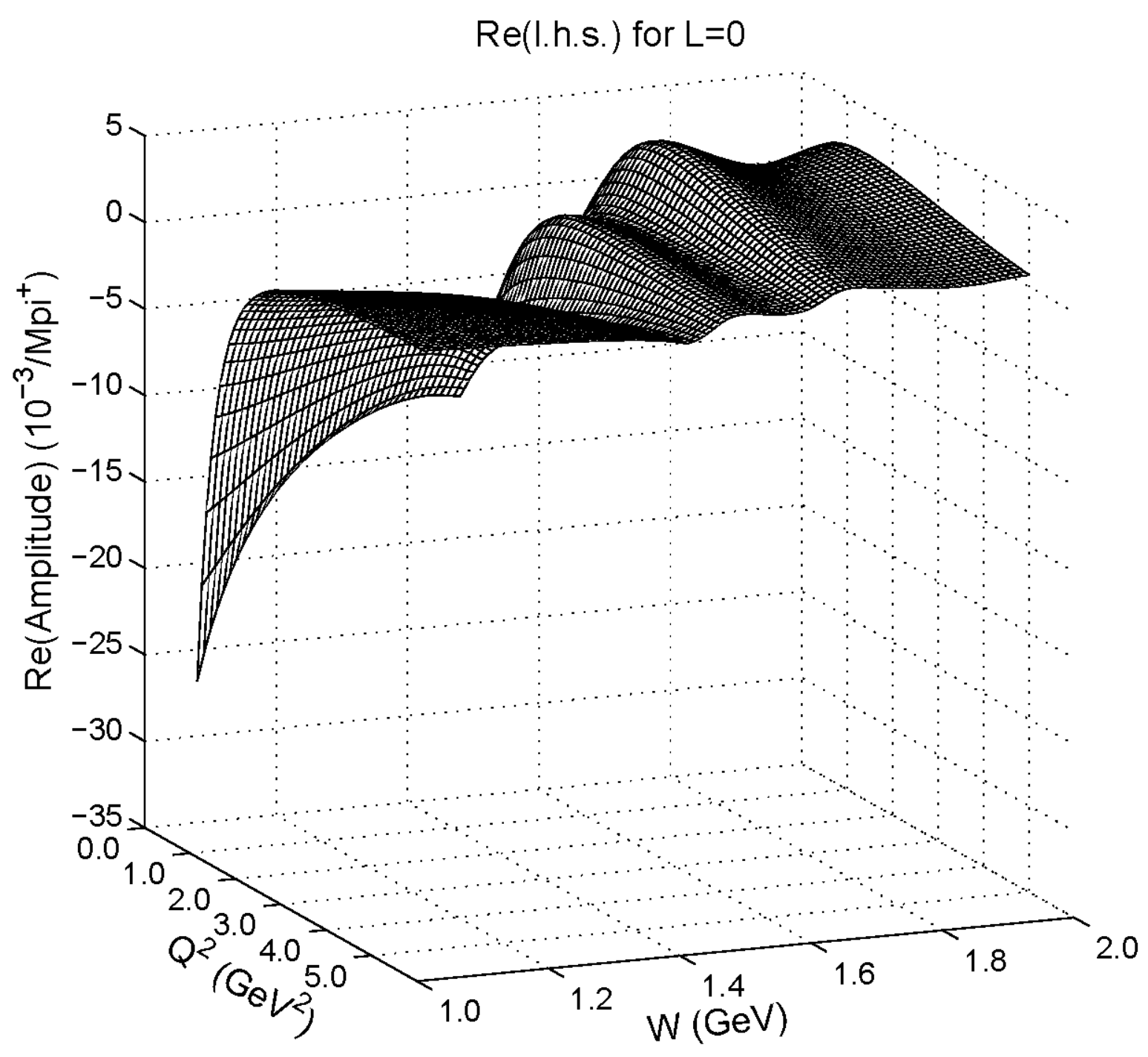}
\epsfxsize=0.44\textwidth\epsfbox{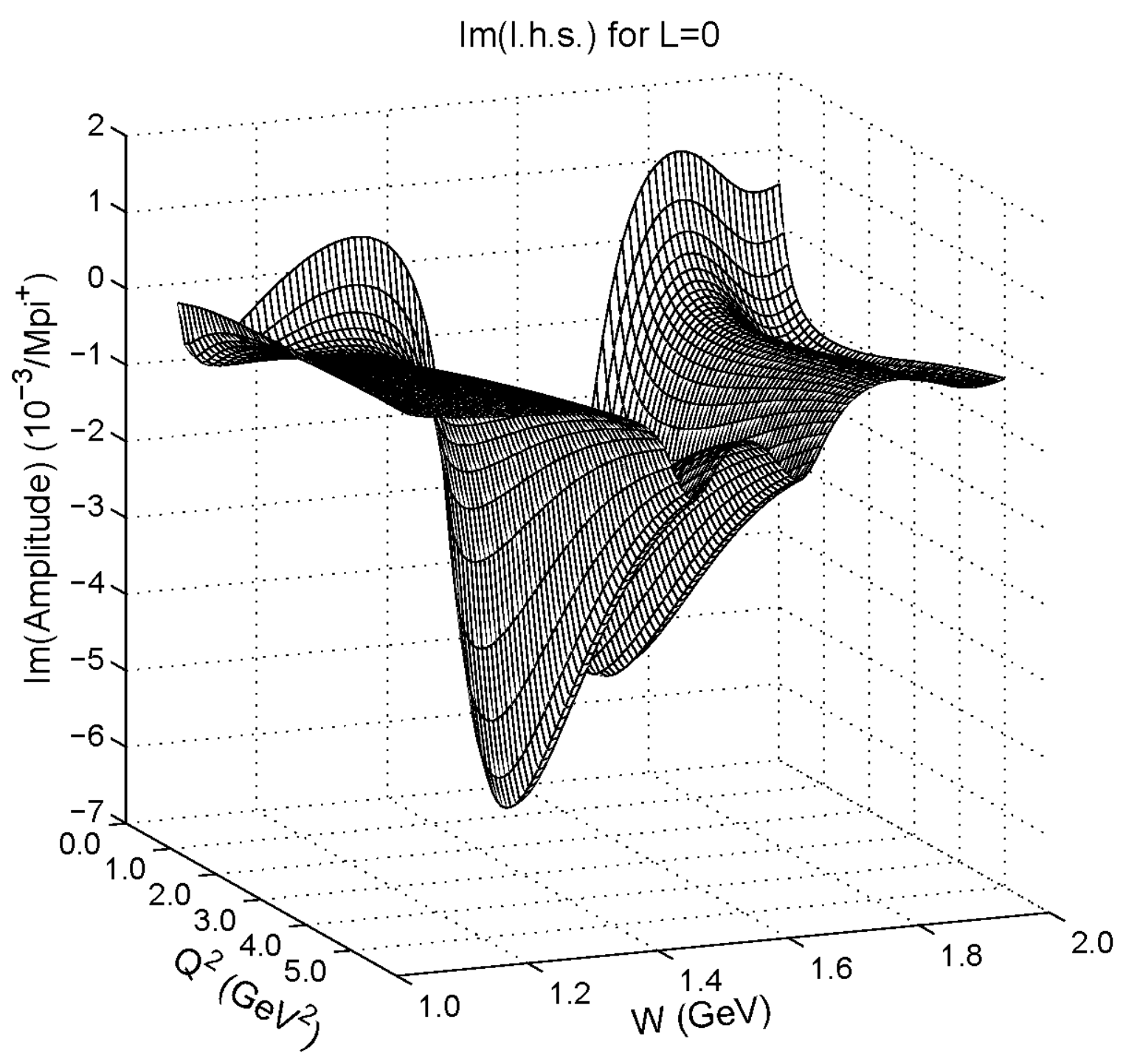}\\[1mm]
\epsfxsize=0.44\textwidth\epsfbox{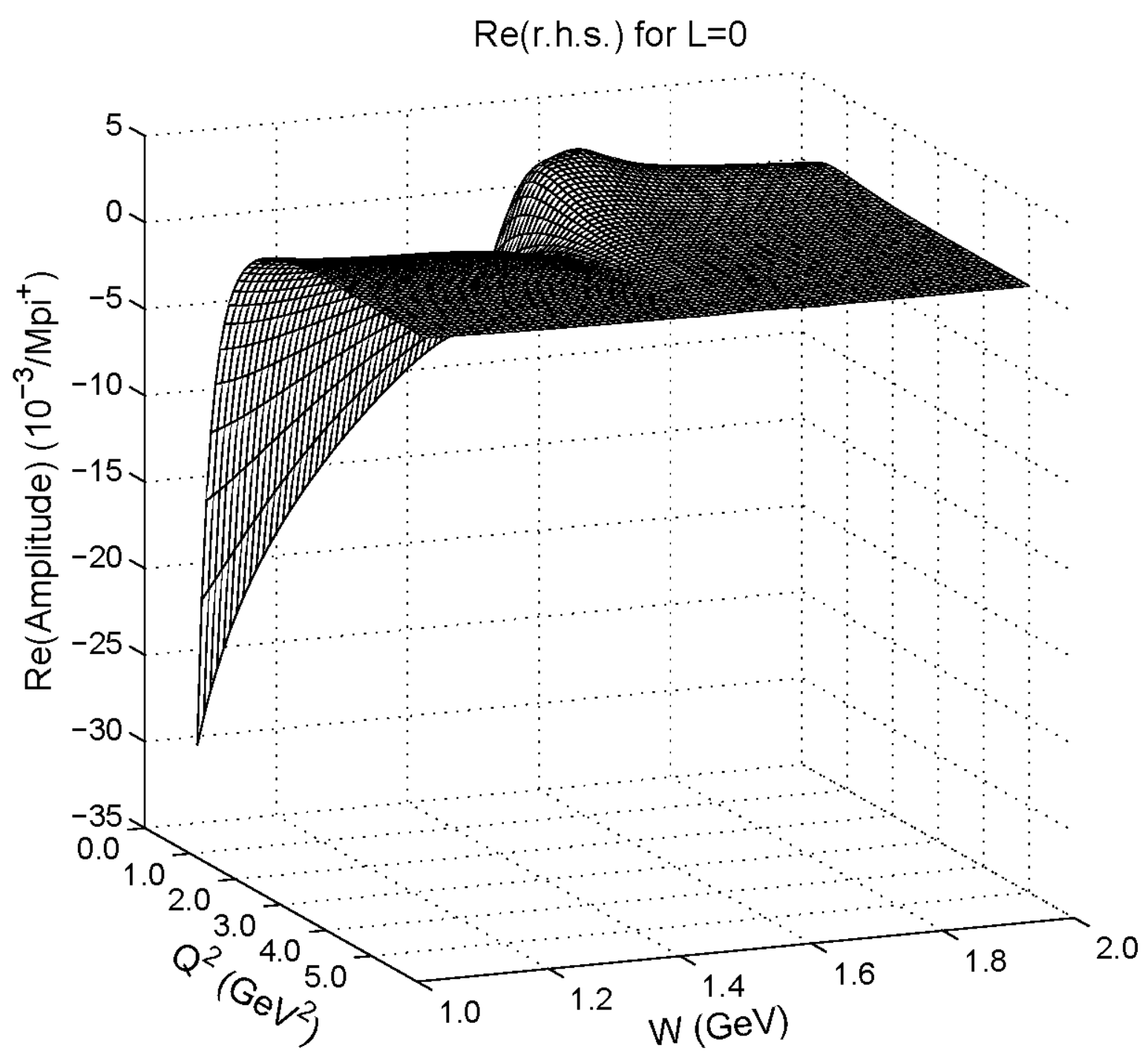}
\epsfxsize=0.44\textwidth\epsfbox{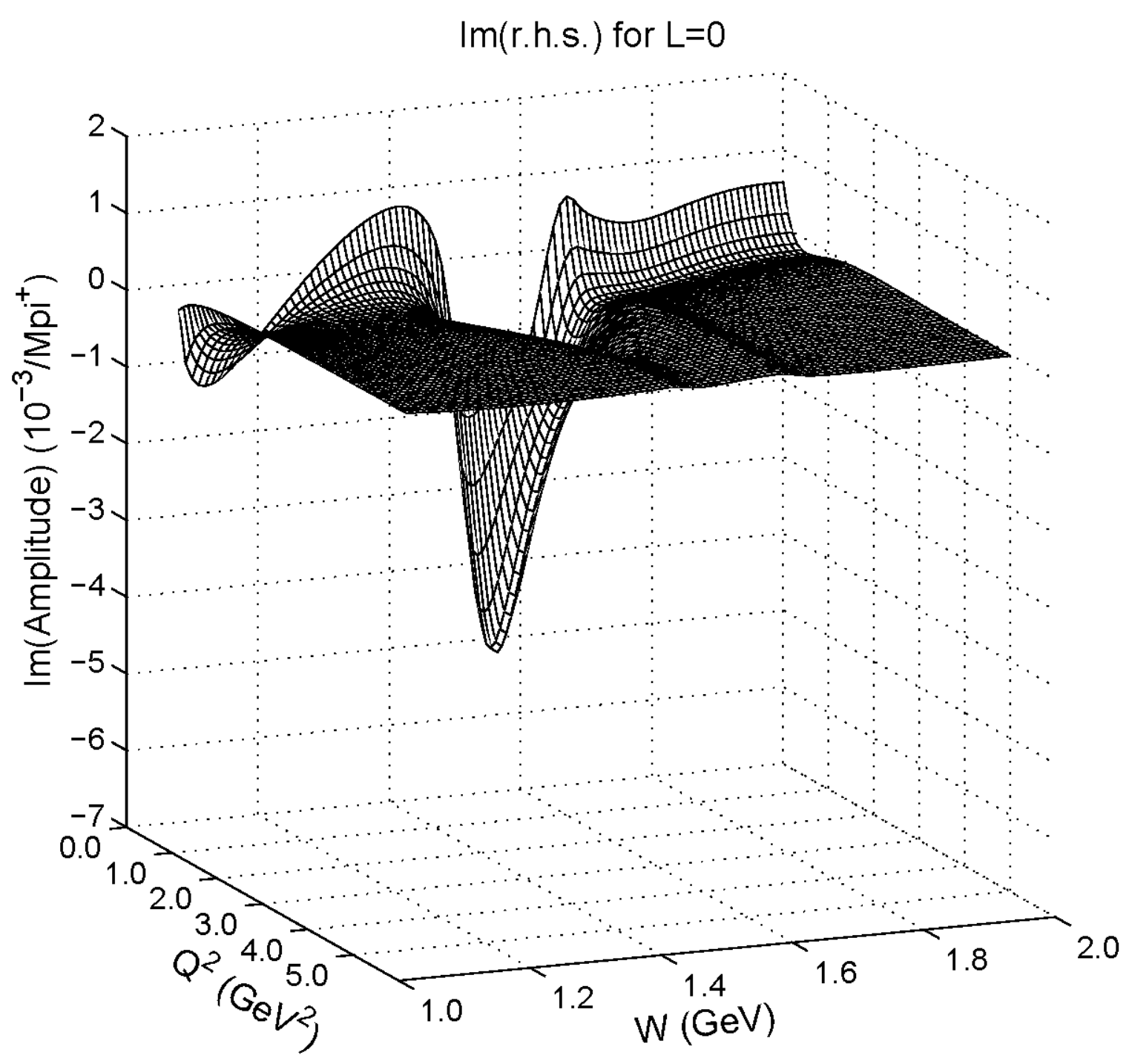}\\[1mm]
\epsfxsize=0.44\textwidth\epsfbox{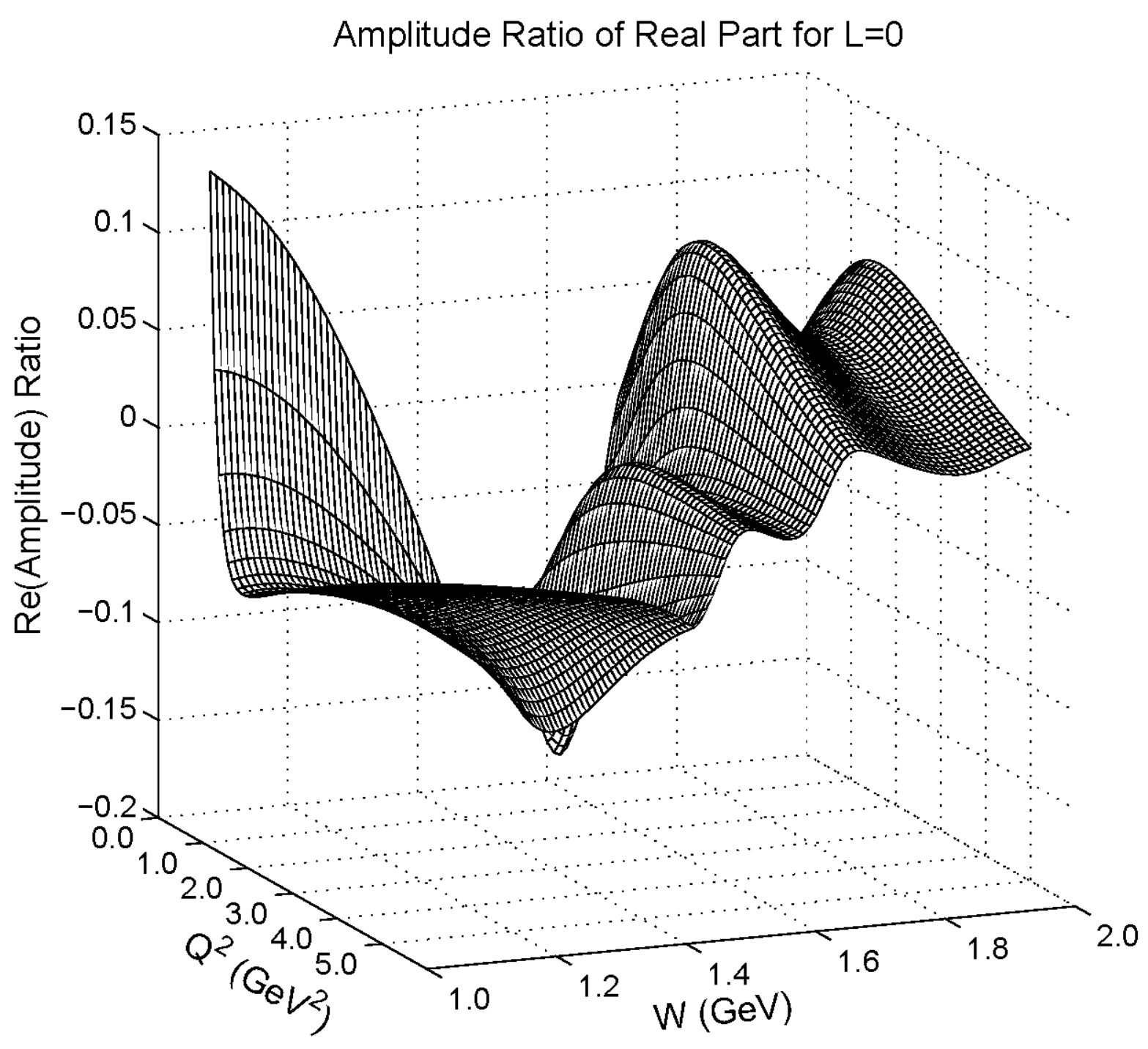}
\epsfxsize=0.44\textwidth\epsfbox{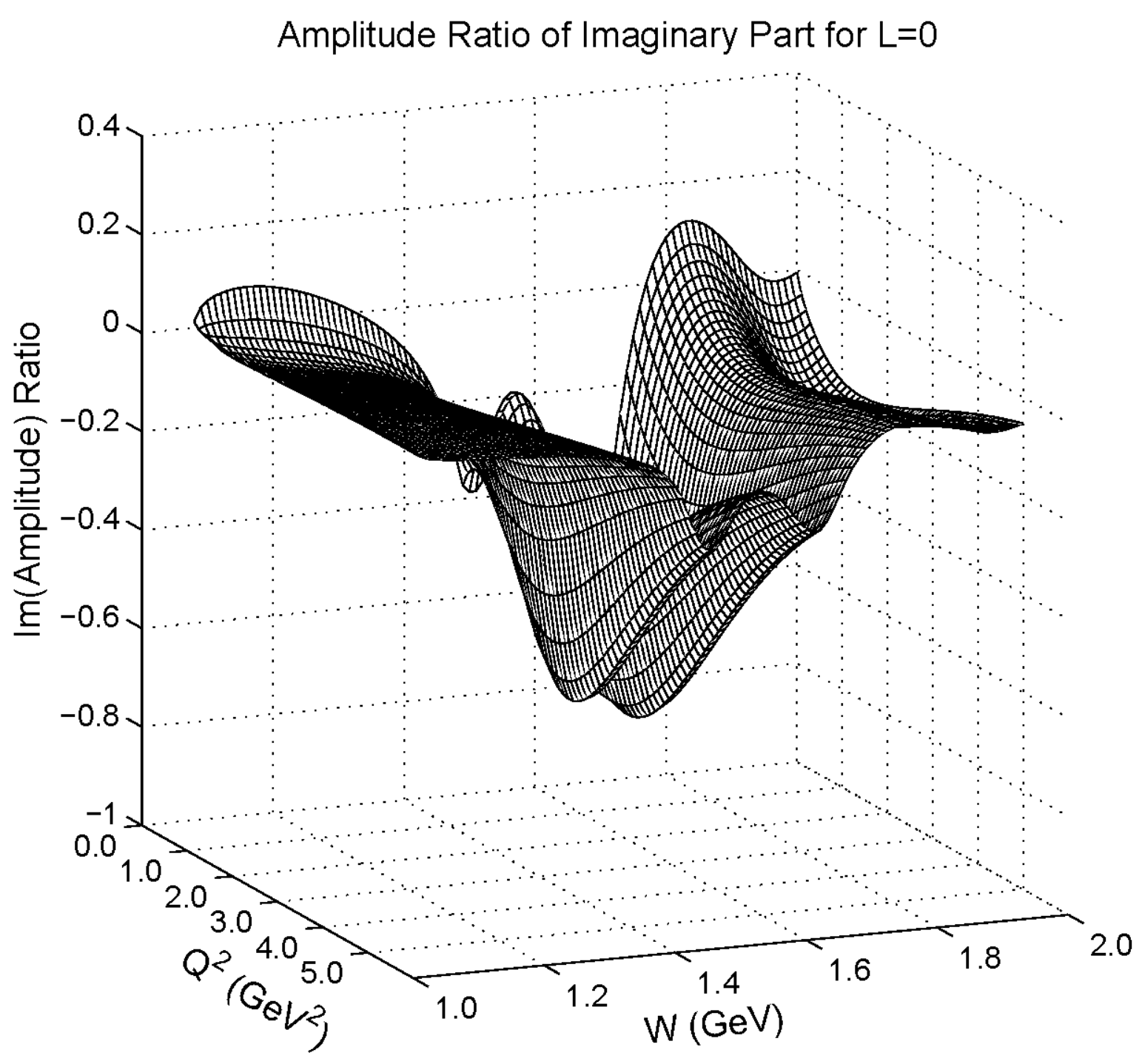}\\
\end{figure}

\begin{figure}[htp]
\caption{Scalar multipole data ($J = L + \frac 1 2$ amplitudes
$S_{L+}$) from MAID~2007.  The l.h.s., r.h.s., and ratio of
relation~(\ref{E2}) for $L \geq 0$ are presented in separate rows,
with separate columns for the real and imaginary parts (except for the
$L = 4$ and $5$ imaginary parts, given as zero by MAID).}
\label{S+plot}
\epsfxsize=0.44\textwidth\epsfbox{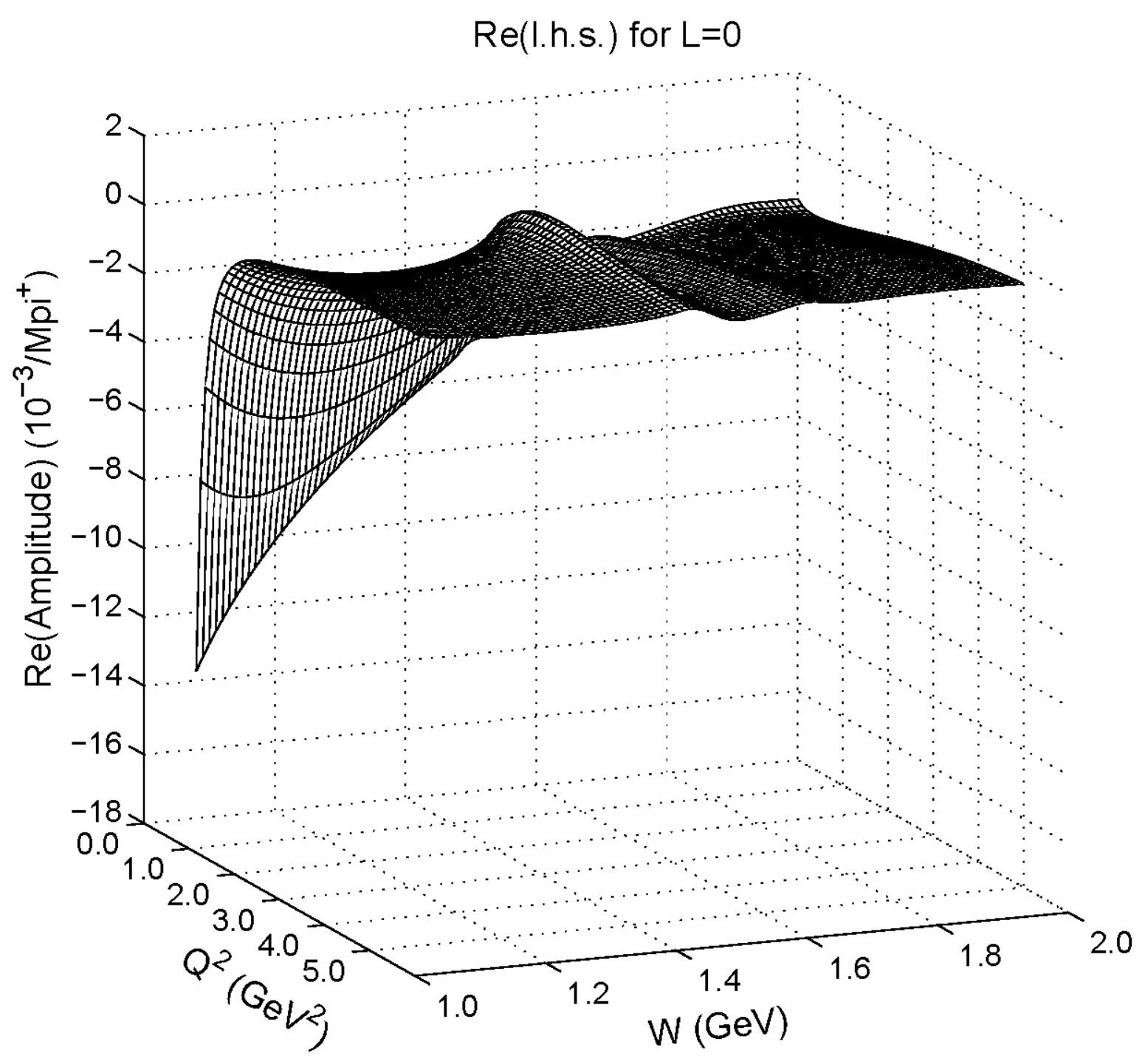}
\epsfxsize=0.44\textwidth\epsfbox{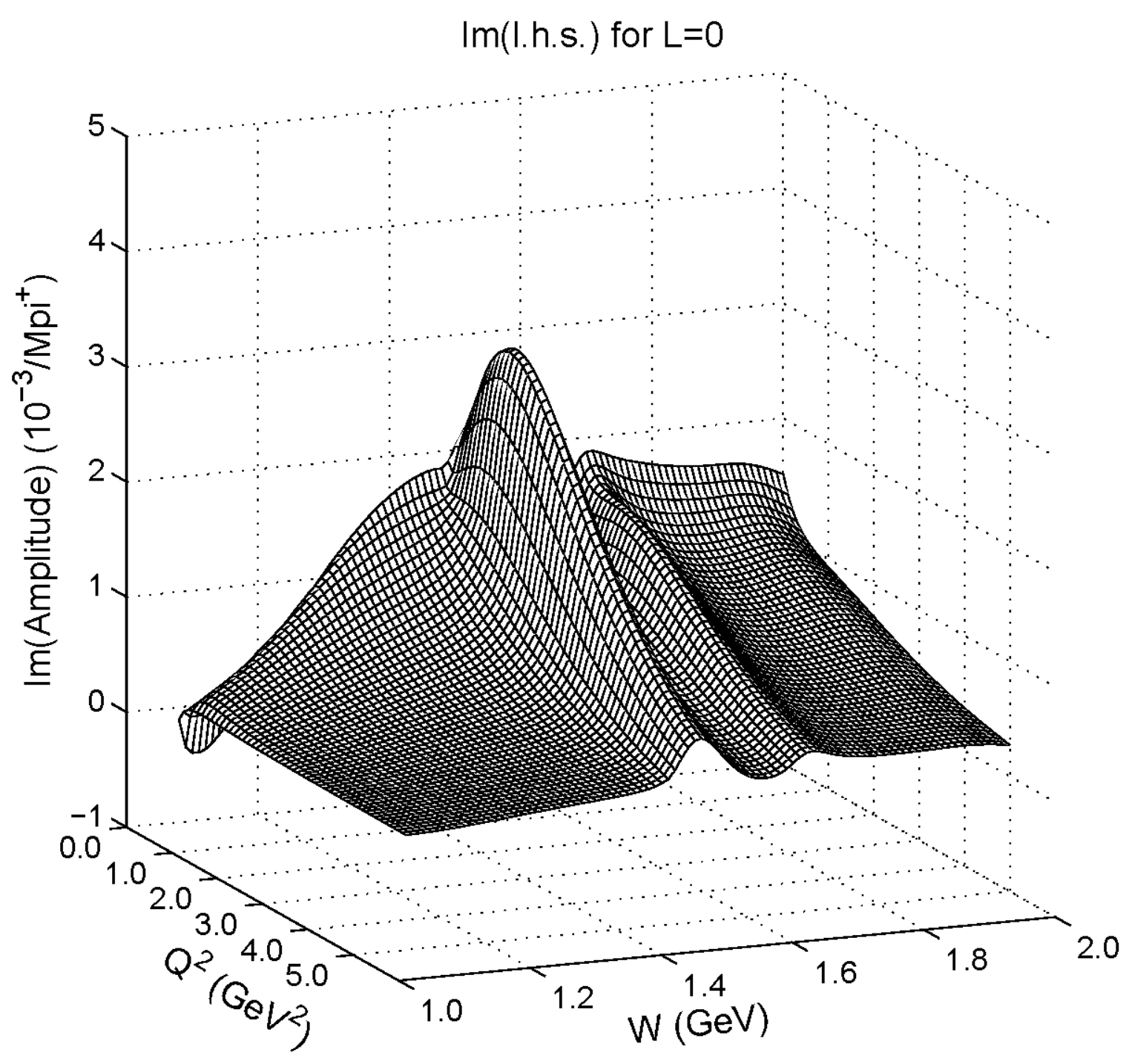}\\[1mm]
\epsfxsize=0.44\textwidth\epsfbox{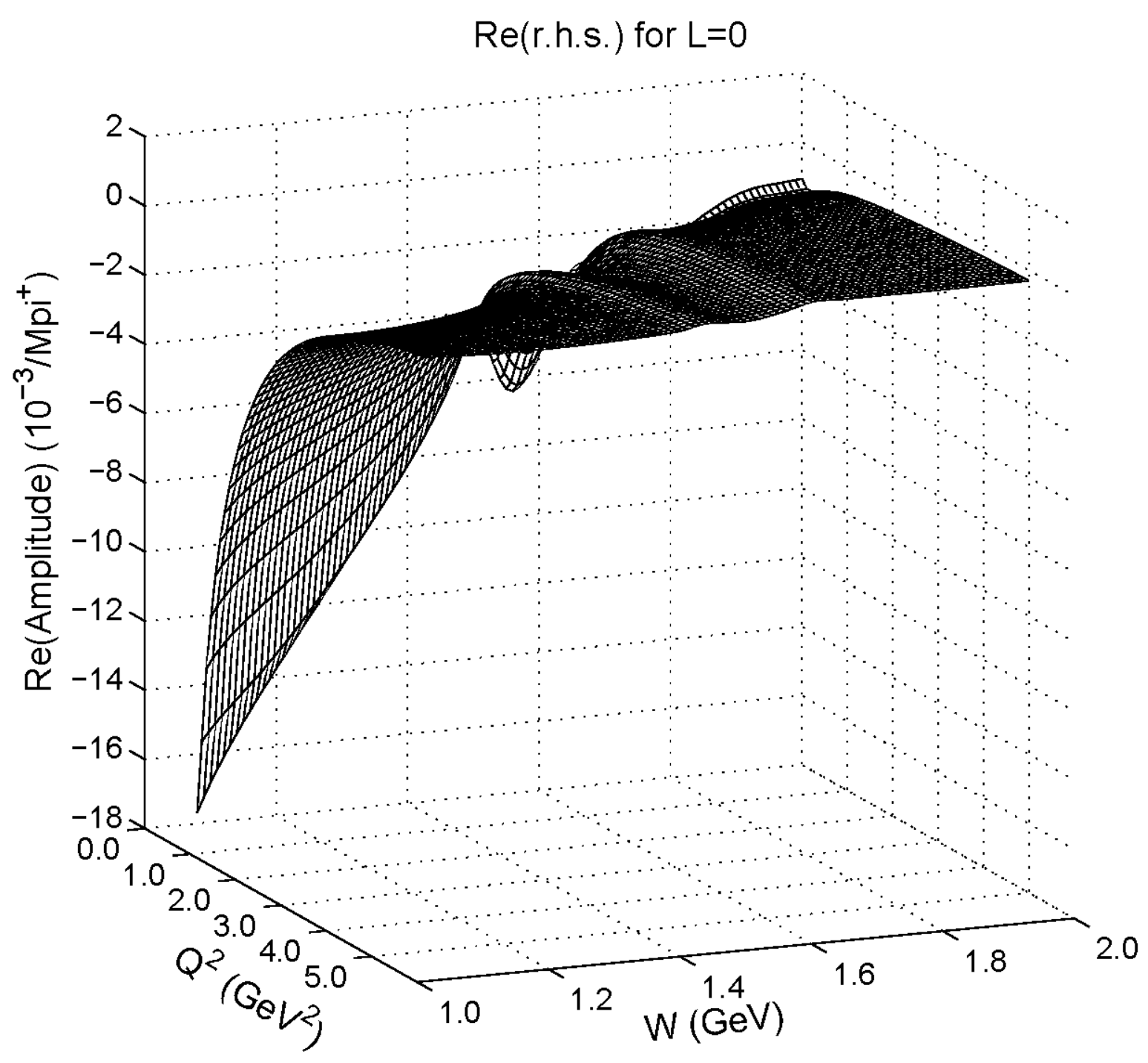}
\epsfxsize=0.44\textwidth\epsfbox{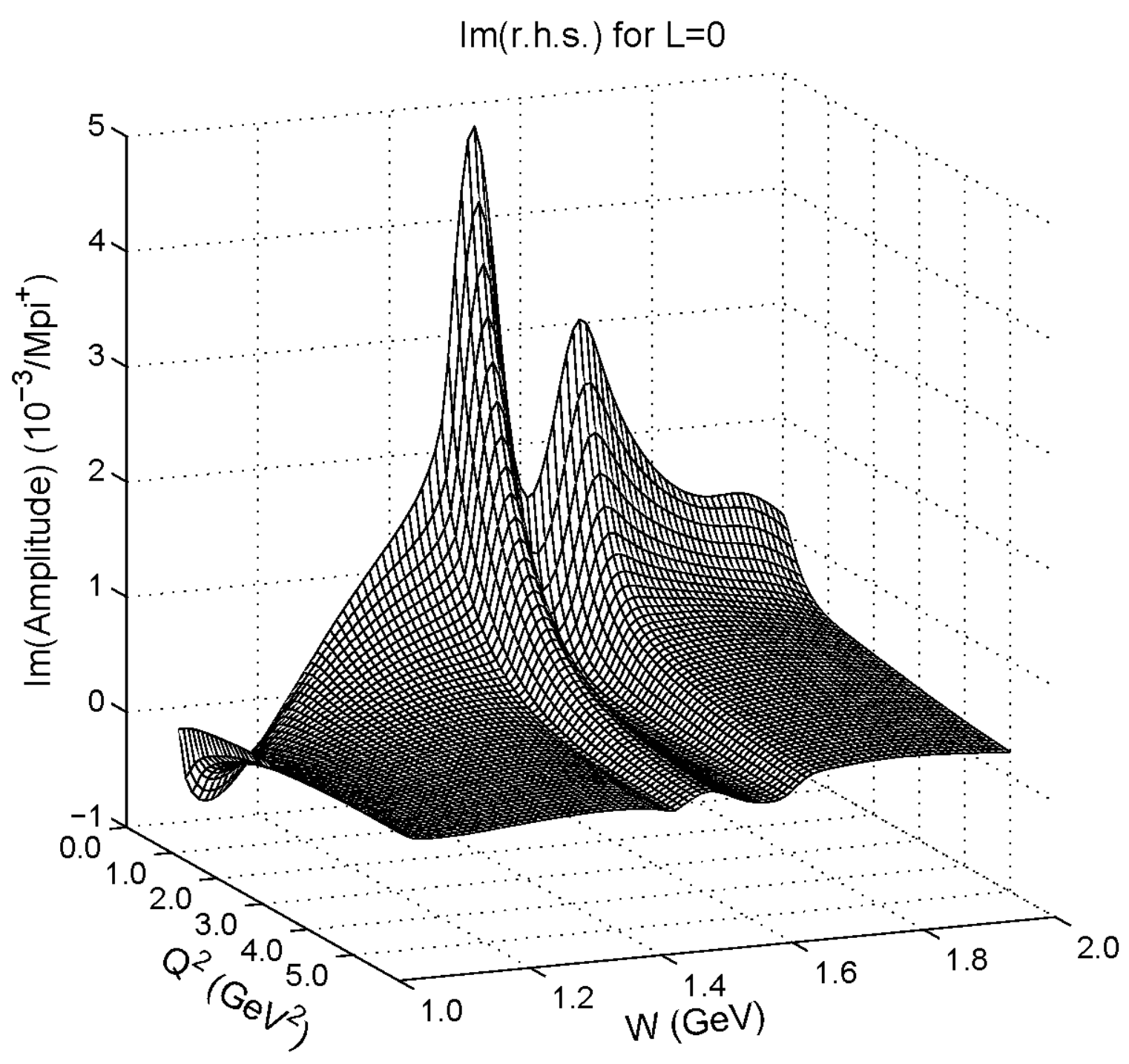}\\[1mm]
\epsfxsize=0.44\textwidth\epsfbox{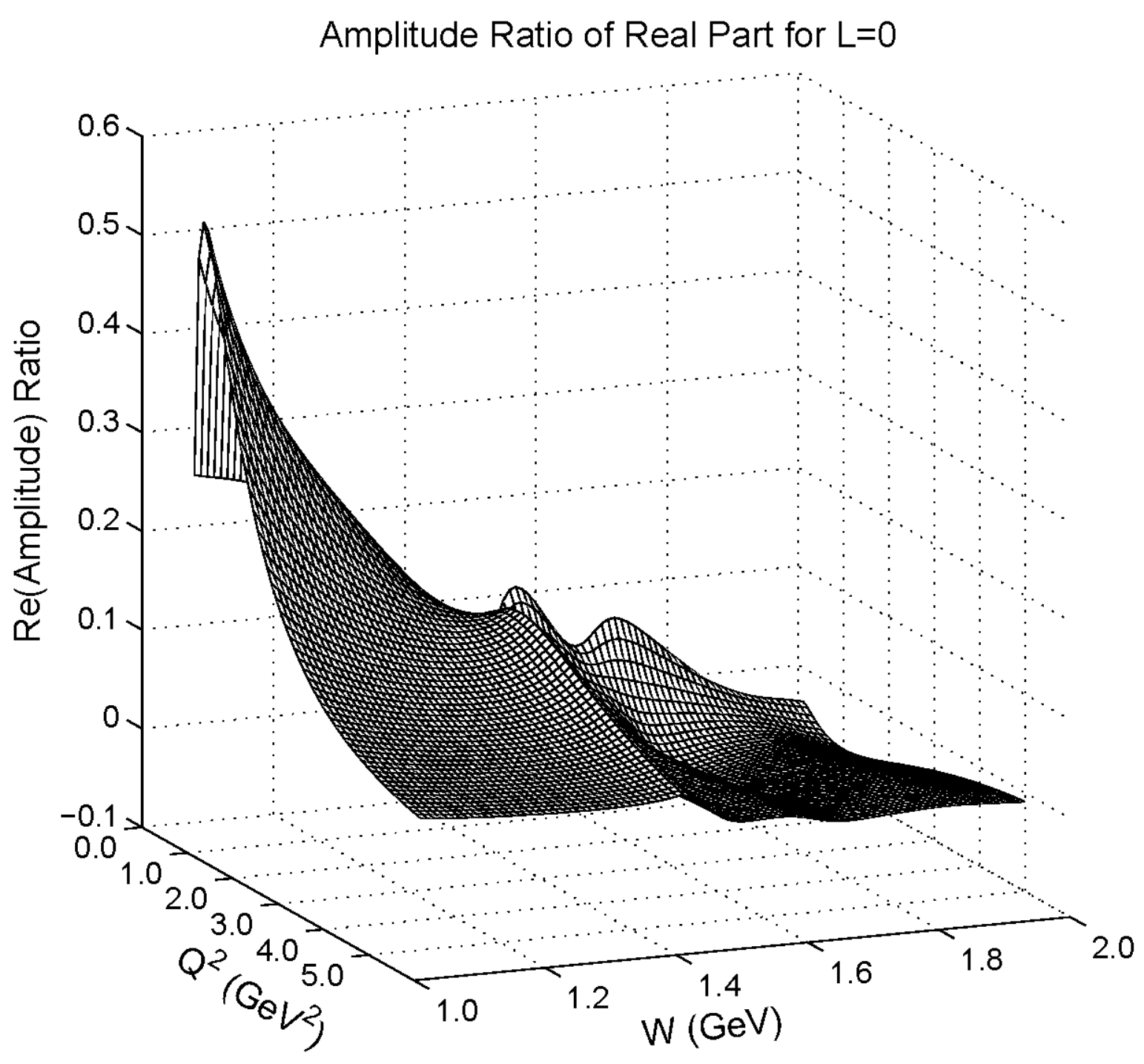}
\epsfxsize=0.44\textwidth\epsfbox{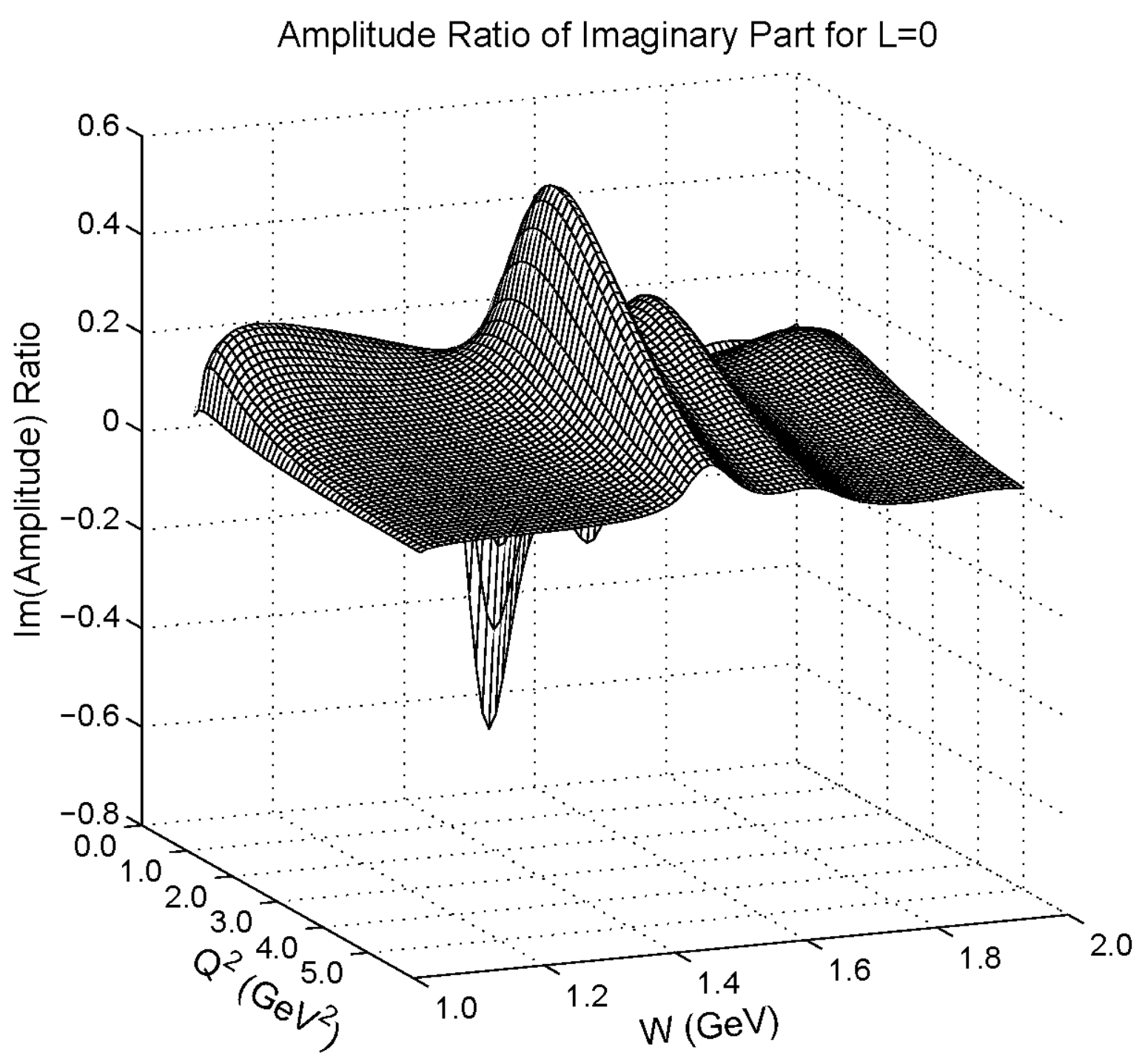}\\
\end{figure}

\begin{figure}[htp]
\caption{Magnetic multipole ($\pi^0$) data from MAID~2007.  The
l.h.s.\ ($M_{L-}$), r.h.s.\ ($M_{L+}$), and ratio of
relation~(\ref{M1}) for $L \geq 1$ are presented in separate rows,
with separate columns for the real and imaginary parts (except for the
$L = 4$ and $5$ imaginary parts, given as zero by MAID).}
\label{M1plot}
\epsfxsize=0.44\textwidth\epsfbox{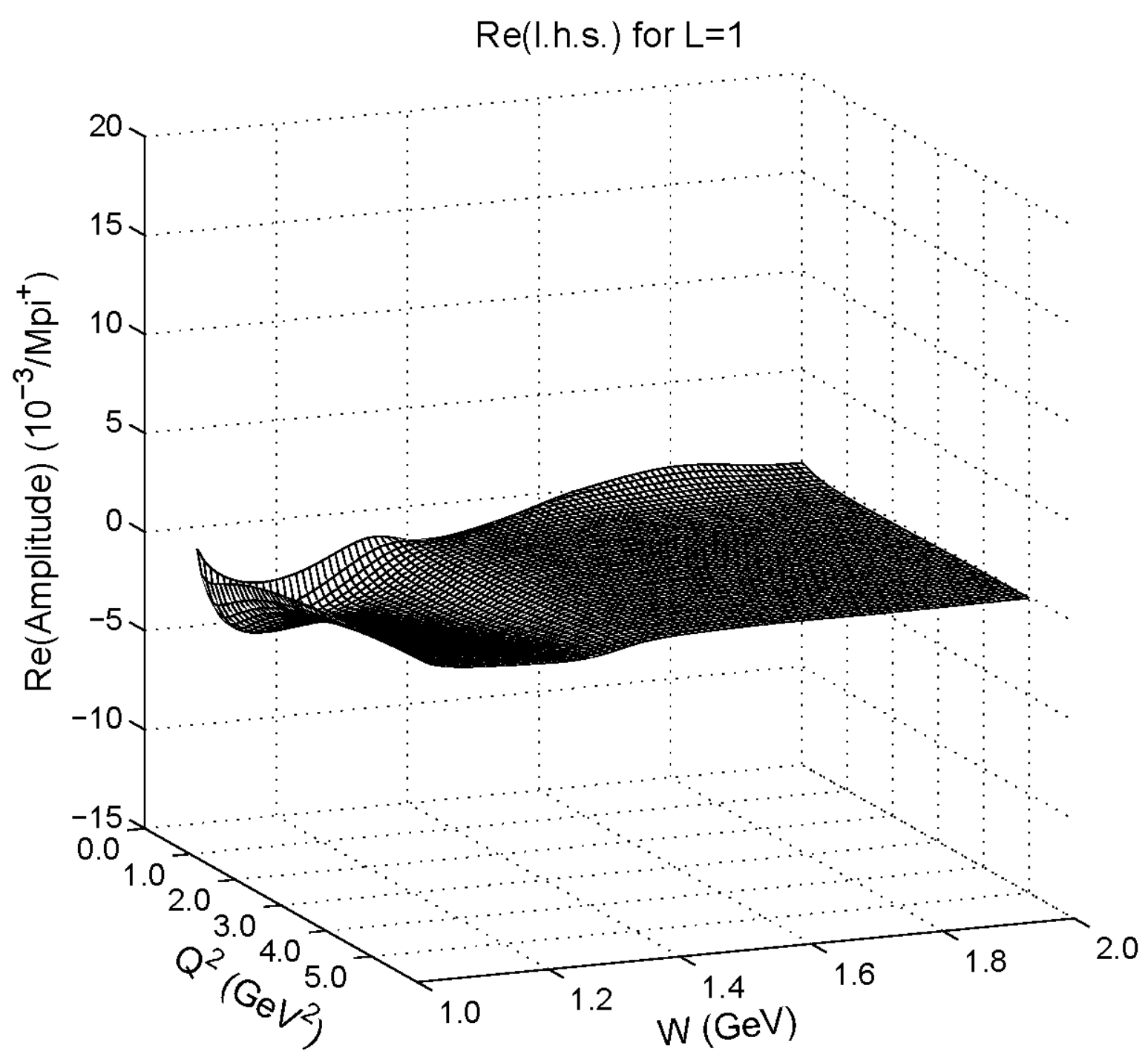}
\epsfxsize=0.44\textwidth\epsfbox{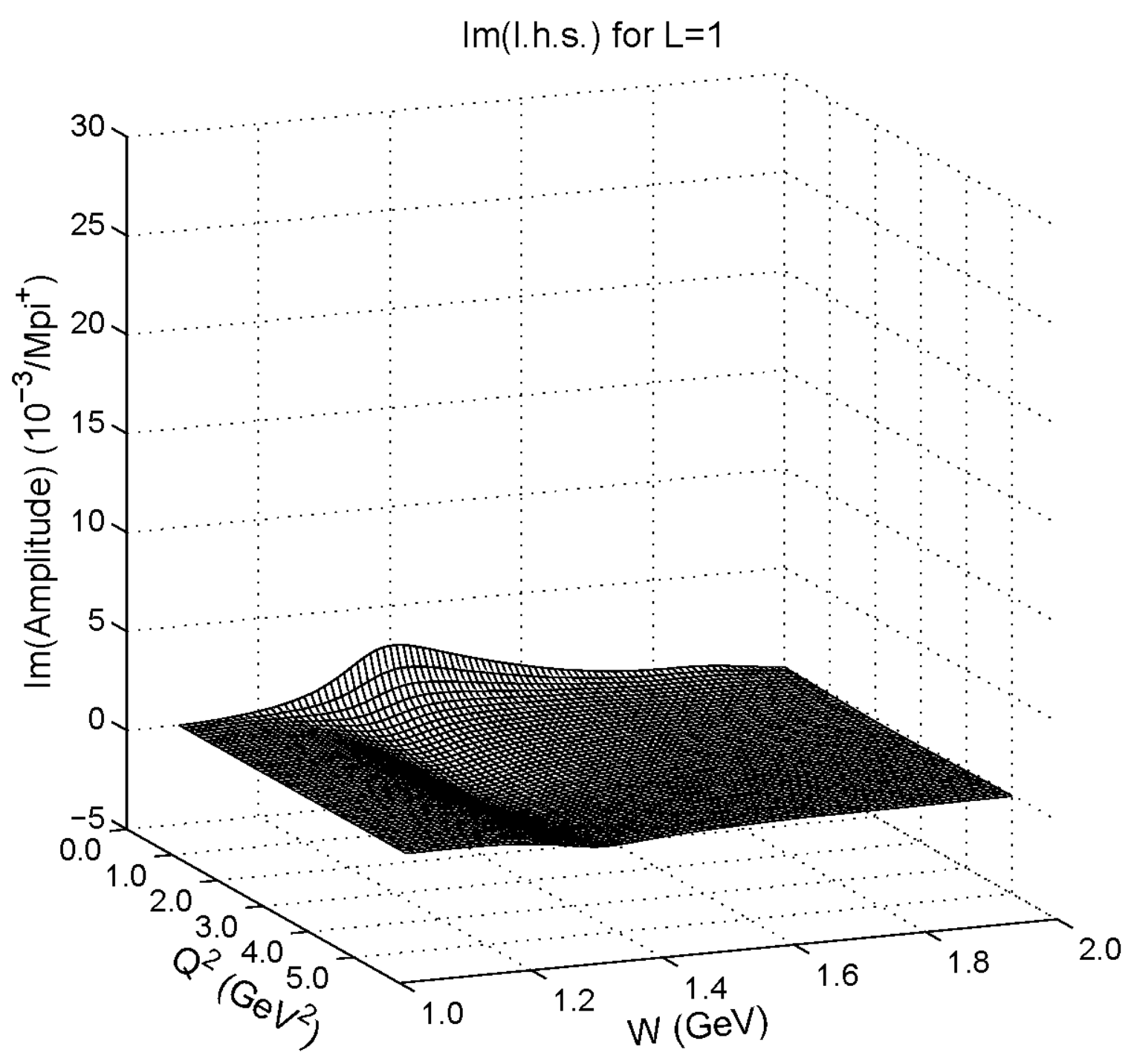}\\[1mm]
\epsfxsize=0.44\textwidth\epsfbox{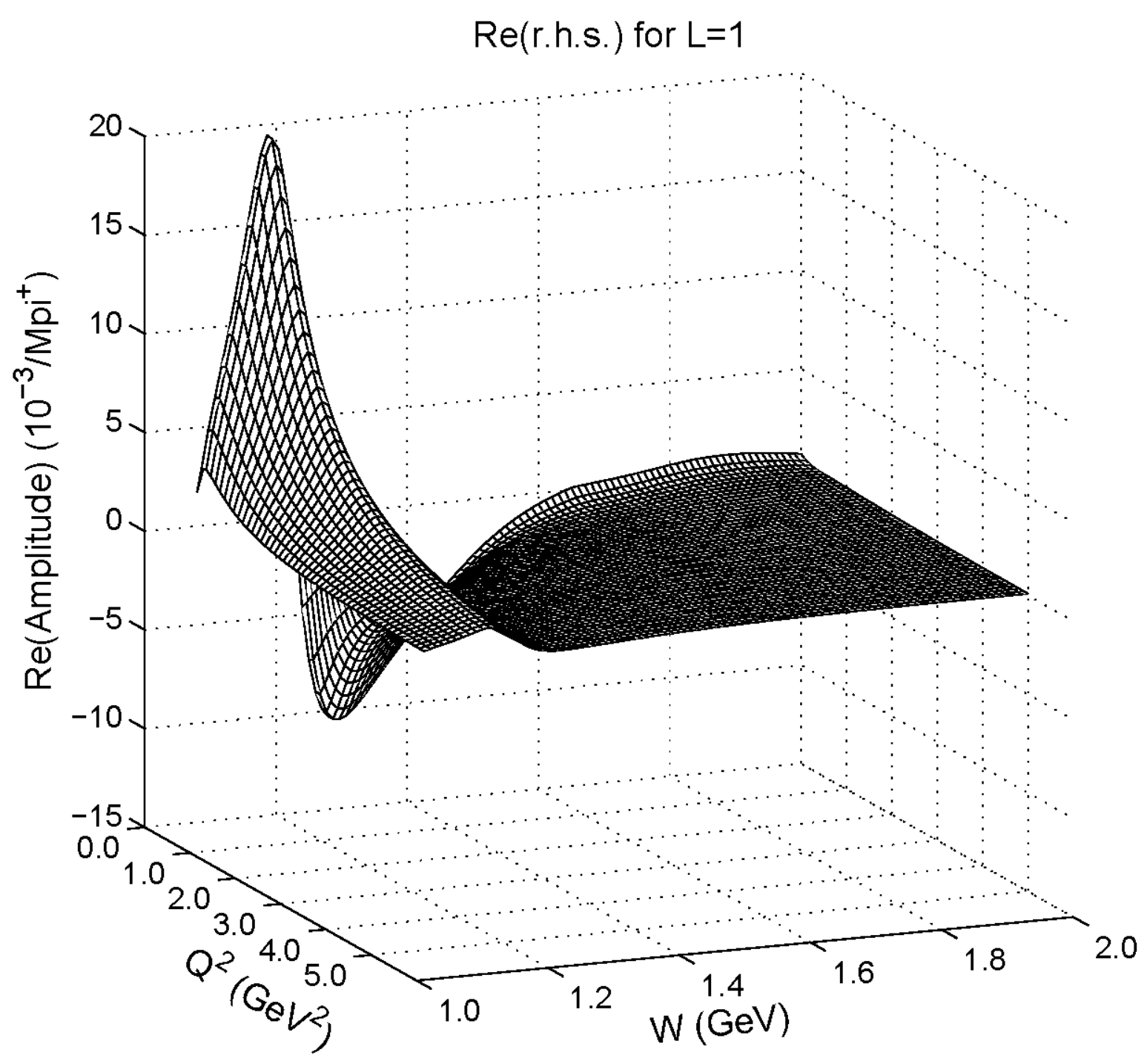}
\epsfxsize=0.44\textwidth\epsfbox{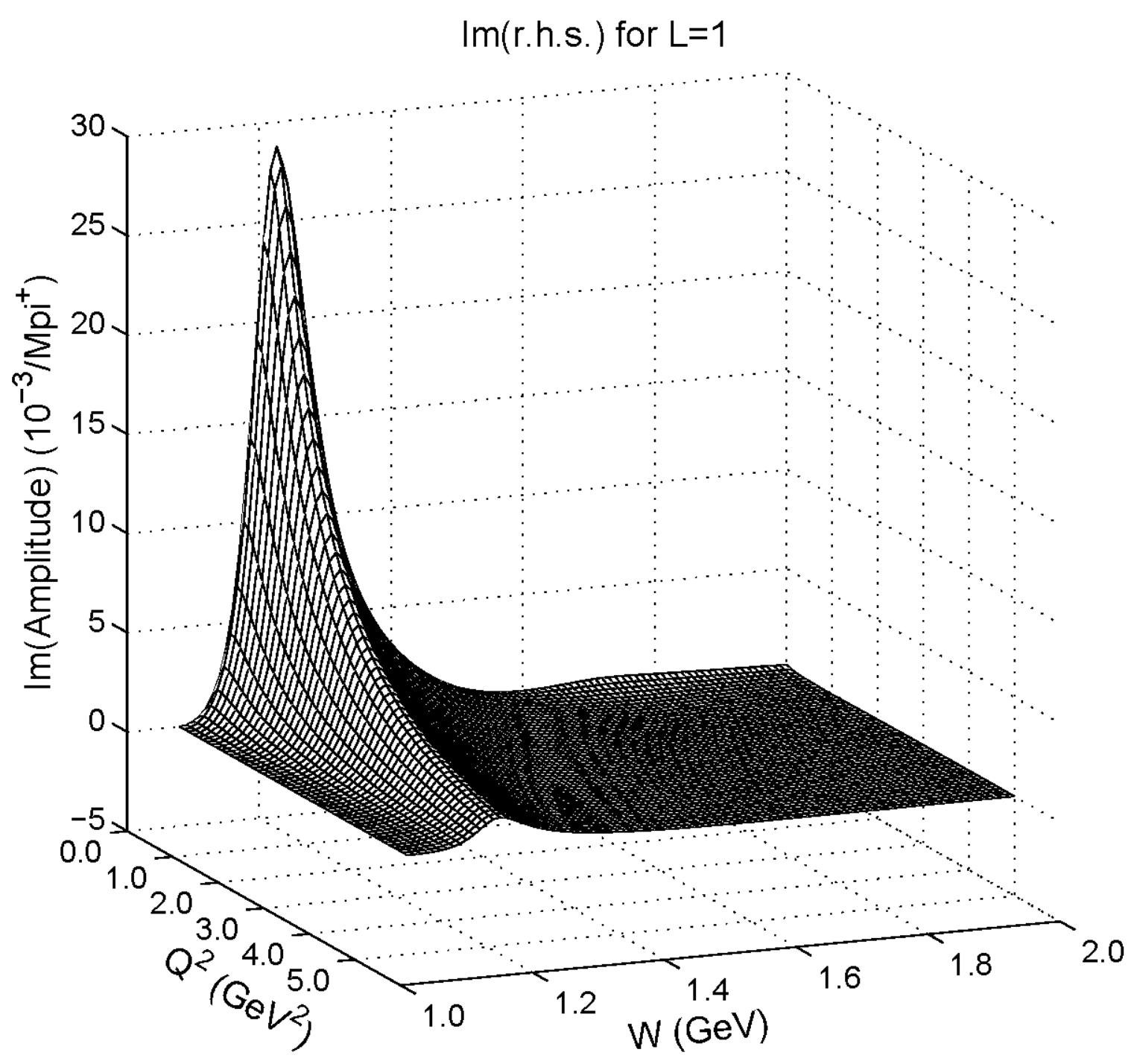}\\[1mm]
\epsfxsize=0.44\textwidth\epsfbox{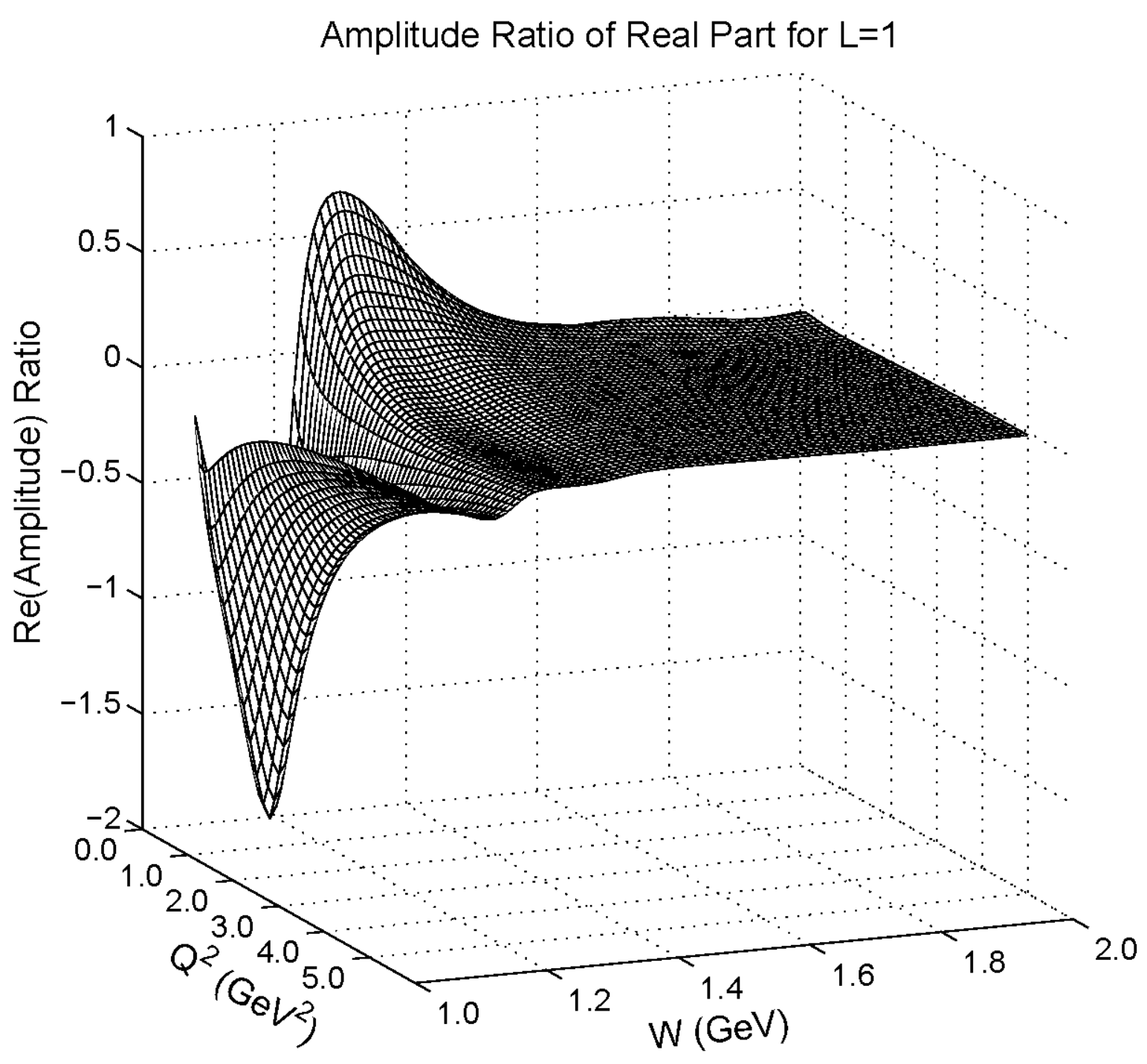}
\epsfxsize=0.44\textwidth\epsfbox{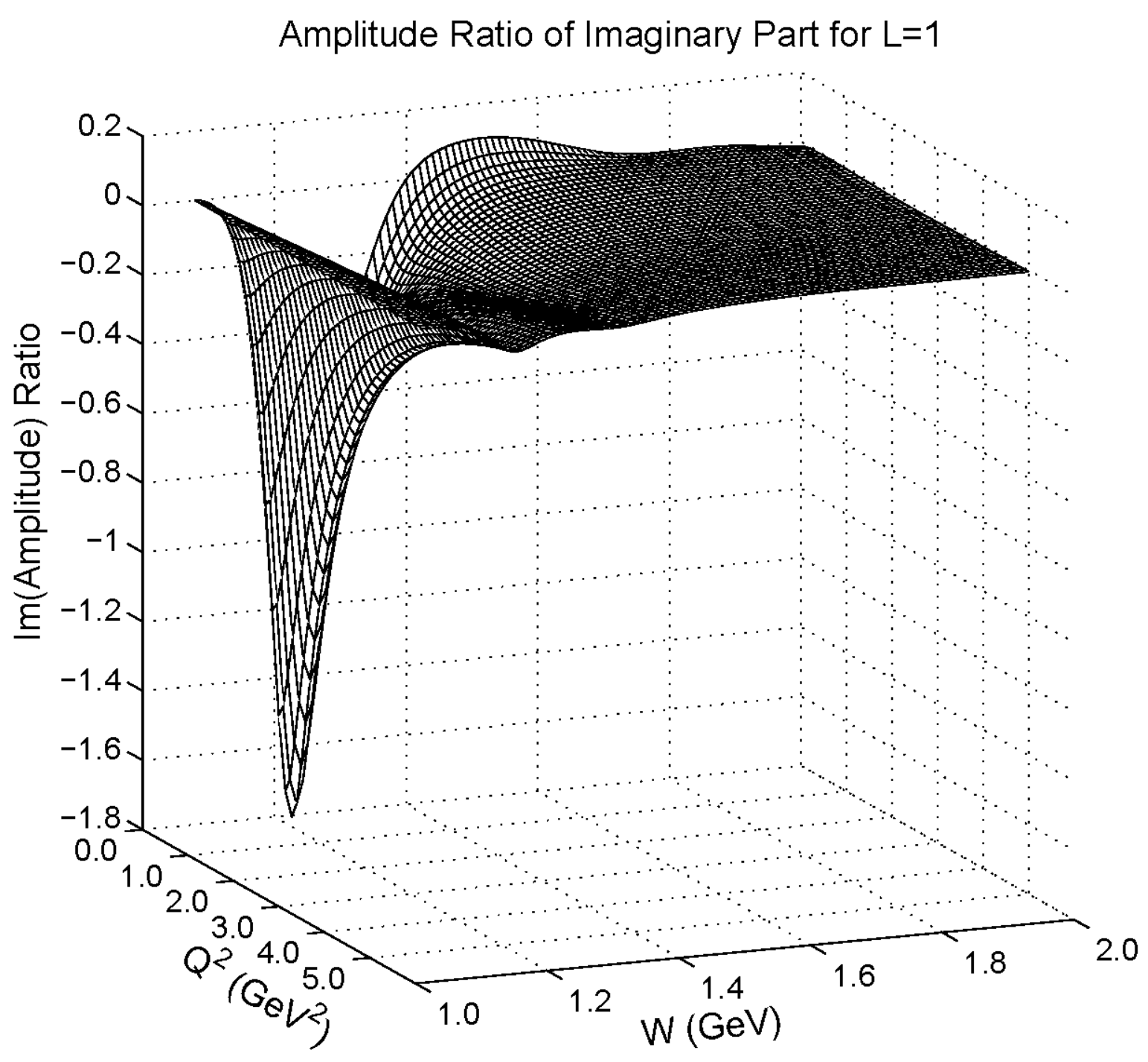}\\
\end{figure}

\begin{figure}[htp]
\caption{Magnetic multipole ($\pi^\pm$) data from MAID~2007.  Each
$L \geq 1$ fills a separate page for real and imaginary parts of
amplitudes (except for $L = 4$ and $5$ imaginary parts, given as zero
by MAID).  The l.h.s.\ of relation~(\ref{M2}) [or~(\ref{M3})] is
indicated in the first row, while the LO term [first r.h.s.\ of
relation~(\ref{M2})] and its ratio are presented in the left column,
and the (LO+NLO) combination [r.h.s.\ of relation~(\ref{M3})] and its
ratio are presented in the right column.}
\label{M2plot}
\epsfxsize=0.41\textwidth\epsfbox{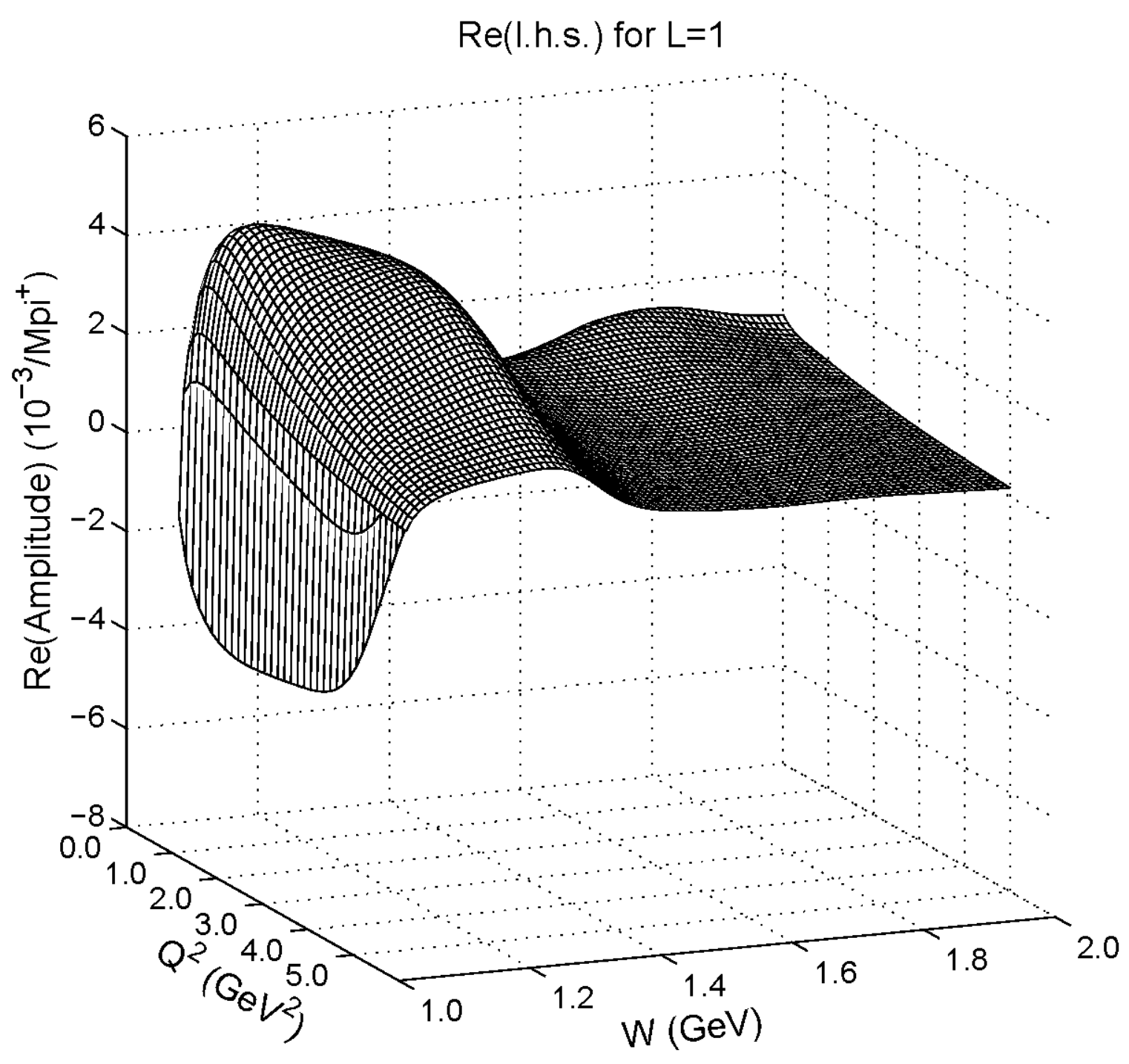}\\[1mm]
\epsfxsize=0.41\textwidth\epsfbox{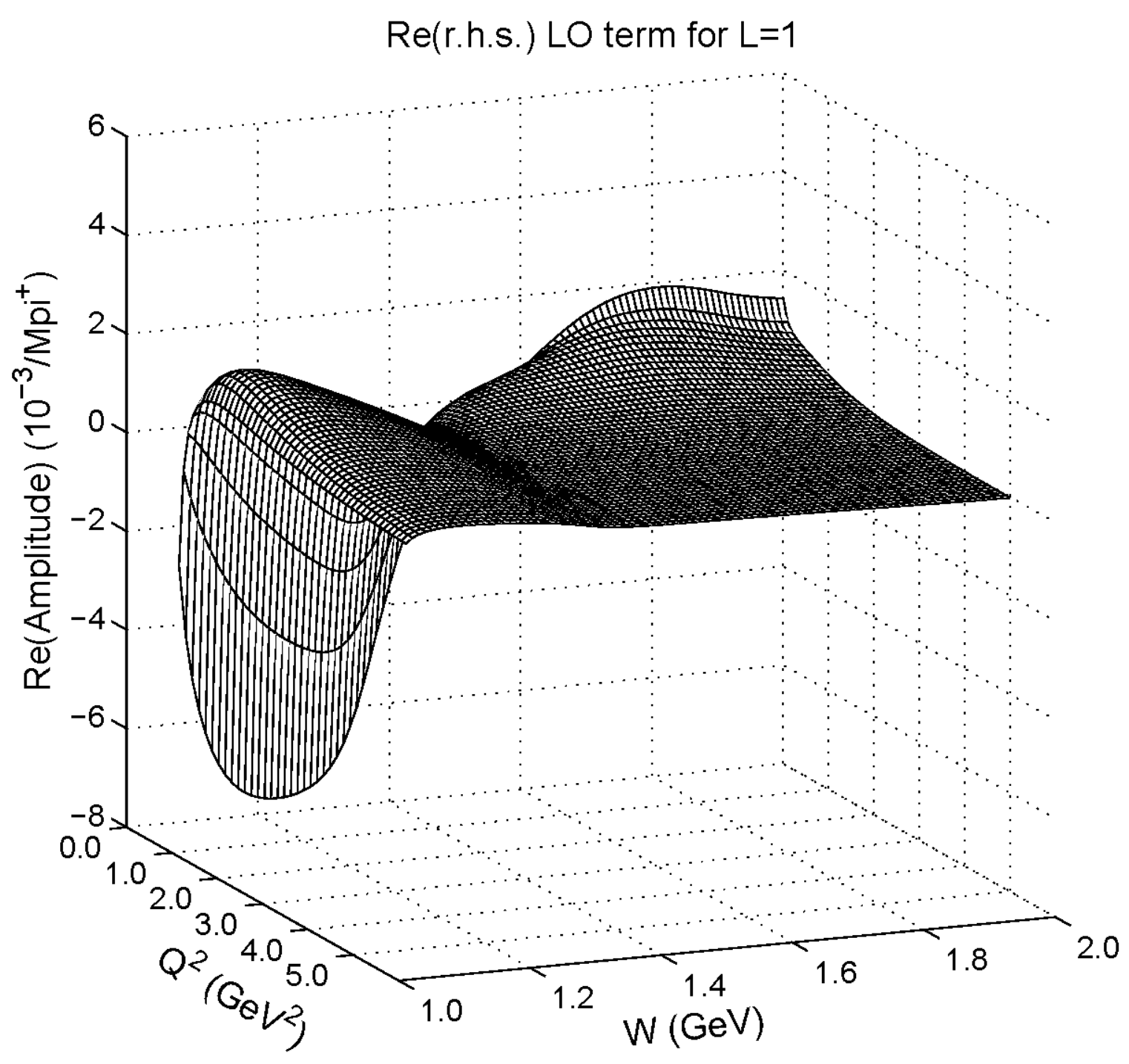}
\epsfxsize=0.41\textwidth\epsfbox{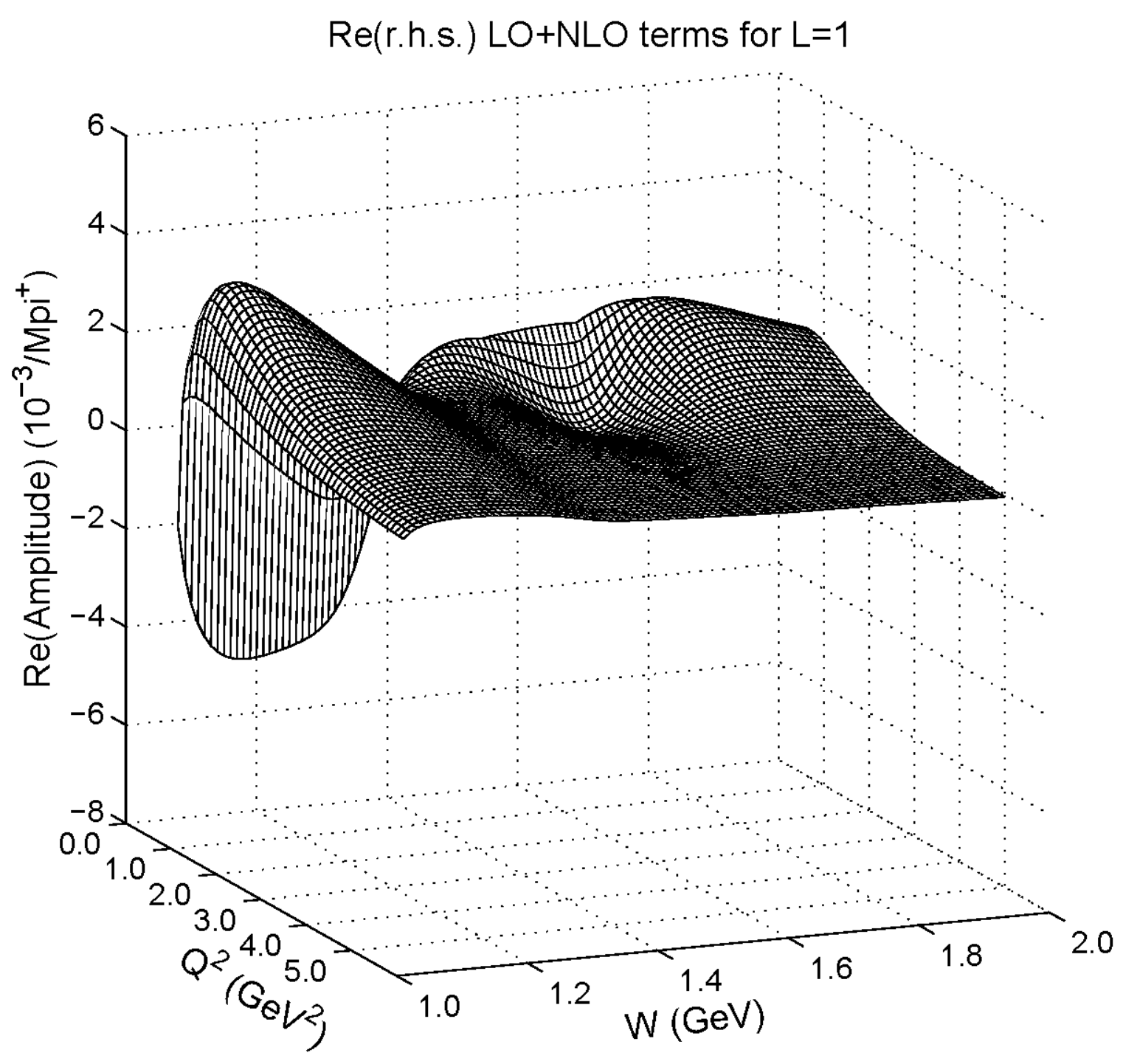}\\[1mm]
\epsfxsize=0.41\textwidth\epsfbox{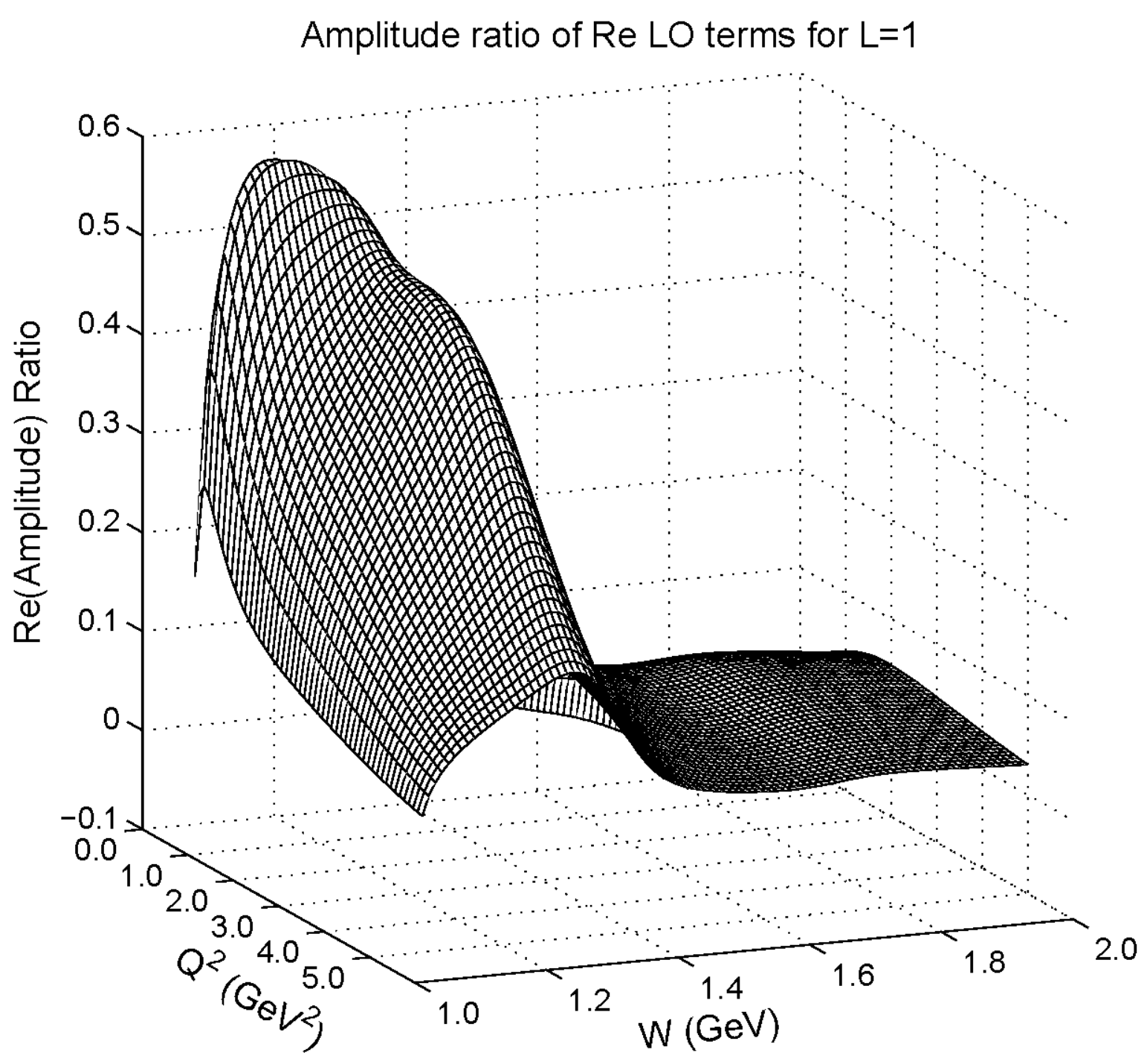}
\epsfxsize=0.41\textwidth\epsfbox{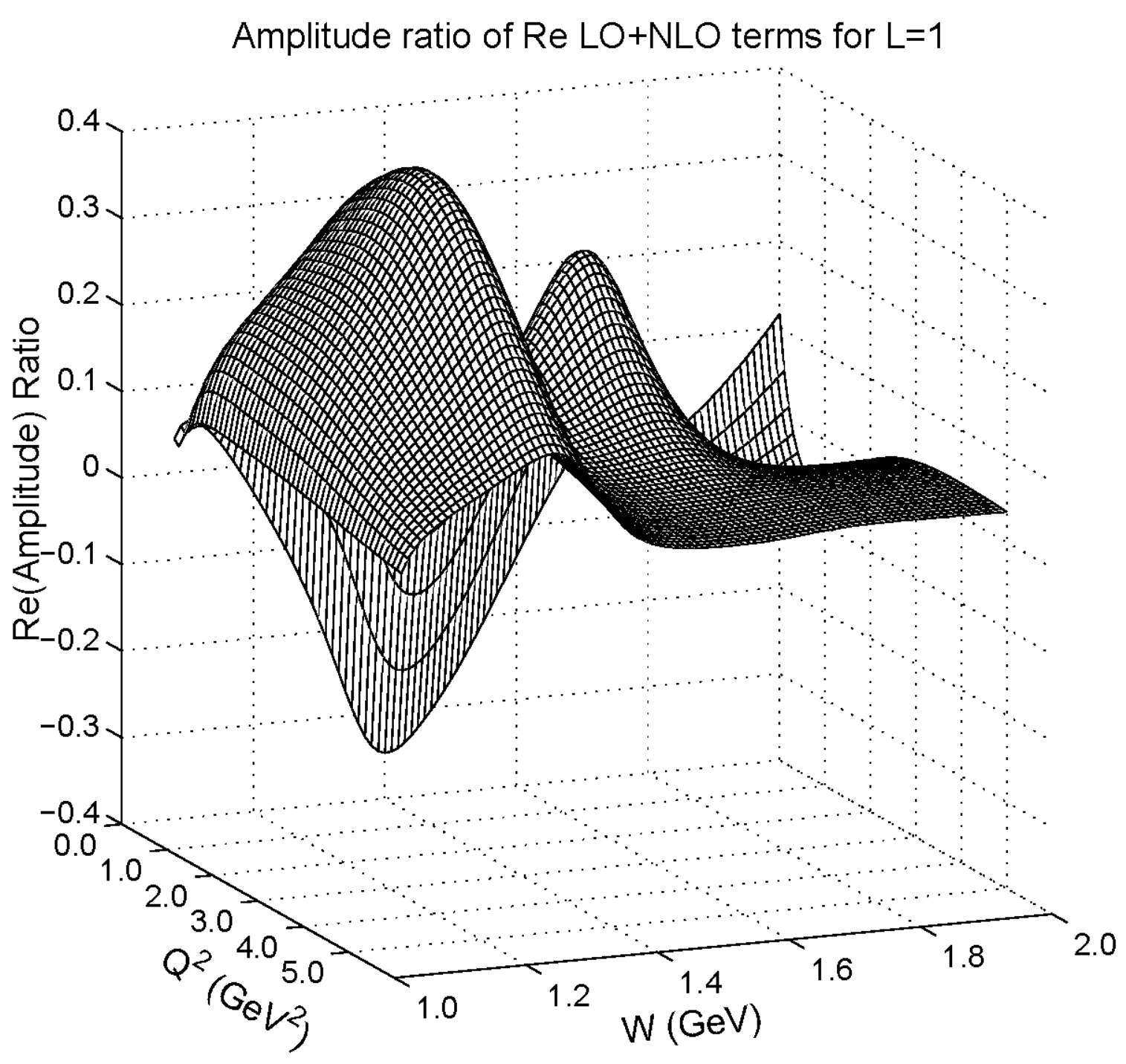}\\
\end{figure}

Figure~\ref{E-plot} tests the e multipole relations in Eq.~(\ref{E1}),
which compare amplitudes with $J \! = \! L \! - \!  1/2$, denoted as
$E_{L-}$ by MAID\@.  Figure~\ref{S-plot} does the same for $S_{L-}$
multipoles.  The $J \! = \! L \! + \! 1/2$ amplitudes for e and s
[relations in Eq.~(\ref{E2}) for multipoles $E_{L+}$ and $S_{L+}$]
appear in Figs.~\ref{E+plot} and \ref{S+plot}, respectively.  The m
multipole relations Eq.~(\ref{M1}) containing a $\pi^0$ and relating
$J \! = \!  L \! \pm 1/2$ amplitudes ($M_{L\pm}$) appear in
Fig.~\ref{M1plot}.  All of the relations presented in these first
figures are LO in the $1/N_c$ expansion.  Figure~\ref{M2plot} tests
the m multipole relations containing $\pi^\pm$, first for the $M_{L-}$
multipoles alone [the LO relation represented by the first expression
in Eq.~(\ref{M2})] and then for the NLO relation between $M_{L\pm}$
multipoles given in Eq.~(\ref{M3}).  We remind the reader that the
relations explored in Figs.~\ref{E-plot}--\ref{S+plot} and the LO
comparisons of Fig.~\ref{M2plot} are dominated by the isovector
component of the photon.

First observe that the limit $Q^2 \! \to \! 0$ corresponds to
scattering with an on-shell photon, {\it i.e.}, real pion
photoproduction.  The projection of each e and m multipole on the
$Q^2$ axis gives amplitude curves obtained in Ref.~\cite{Cohen:2004bk}
and analyzed there in terms of the multipole relations
Eqs.~(\ref{E1})--(\ref{E2}) and (\ref{M1})--(\ref{M3}) restricted to
$Q^2 \! = \! 0$.

The most prominent features in the amplitudes are of course the baryon
resonances, which are most apparent through large enhancements of the
imaginary parts at values of c.m.\ energy $W$ equal to a resonance
mass, but also noticeable from points where the real parts vanish.
The falloff of each amplitude with increasing $Q^2$ may be
interpreted as the canonical behavior of an $NN^*$ electromagnetic
transition form factor, with the on-shell photon ($Q^2 \! = \! 0$)
providing the least disruptive probe of the initial nucleon and hence
the most efficient probe for producing a resonance.  One finds,
however, interesting exceptions to this reasoning; in the imaginary
part of the $S_{2-}$ amplitudes (Fig.~\ref{S-plot}), one sees that the
$N(1520)$ produced in $\gamma^* p \! \to \! \pi^+ n$ (l.h.s.) peaks at
$Q^2 \!  = \! 0$, while that in $\gamma^* n \! \to \! \pi^- p$
(r.h.s.) peaks at $Q^2 \! = \! 0.6$~GeV$^2$.

The resonant behavior accounts for the largest source of discrepancies
of the $1/N_c$ relations.  Often one finds resonant behavior on both
sides of a given relation, but with greatly differing residues at the
peak; this is the case for the $N(1680)$ peak in each $E_{3-}$
amplitude of Fig.~\ref{E-plot} or each $S_{3-}$ amplitude of
Fig.~\ref{S-plot}.  Even the sign of the $N(1720)$ residue is
different between the two $E_{1+}$ amplitudes in Fig.~\ref{E+plot},
while the $N(1720)$ appears to be absent altogether from the r.h.s.\
plot for $S_{1+}$ in Fig.~\ref{S+plot} although the amplitudes
otherwise appear very similar.  These are interesting anomalies
arising even in amplitudes for which the value of $J$ on both sides of
the amplitude relation are the same, so that the same resonance (or
more accurately, different members of an isomultiplet) appears on both
sides.  For Fig.~\ref{M1plot} and the NLO relation of
Fig.~\ref{M2plot}, $J$ differs on the two sides of the equation,
meaning that distinct resonances appear.  In the $L \! = \!  1$
amplitudes of Fig.~\ref{M1plot}, for example, the $N(1440)$ resonance
appearing in the $M_{1-}$ (l.h.s.) amplitude is broad and shallow,
while the $\Delta(1232)$ forms a huge peak in the $M_{1+}$ (r.h.s.)
amplitude, with just a hint of the $N(1440)$'s true large-$N_c$
partner~\cite{Cohen03}, the $\Delta(1600)$, perhaps just visible.

In fact, the $\Delta(1232)$ should be eliminated from a large $N_c$
analysis of baryon resonances, because it is actually a degenerate
partner to the nucleon whose width vanishes as $N_c \! \to \! \infty$,
unlike the true resonances whose widths remain finite in this
limit~\cite{Cohen:2003fv}.  It is only due to the numerical accident
that the chiral limit ($m_\pi \! \to \! 0$) is more closely achieved
in our $N_c \! = \! 3$ universe than is the $1/N_c \! \to \! 0$
limit~\cite{Cohen:2006up} that the $\Delta(1232)$ in our universe
decays to $\pi N$.  However, true resonant ``shifted degenerate''
pairs such as $N(1440)$ and $\Delta(1600)$ are particularly
interesting because their masses differ by an amount that is
$O(1/N_c^2)$ relative to their average~\cite{Cohen03}, about
100--150~MeV.  Another example of this effect appears in the $L \! =
\! 2$ amplitudes in Fig.~\ref{M1plot}, with the $N(1520)$ [and perhaps
also the $N(1700)$ or $\Delta(1700)$] appearing in the $M_{2-}$
(l.h.s.) and $N(1675)$ with a much smaller residue appearing in the
$M_{2+}$ (r.h.s.).  In fact, precisely the pair of $N(1520)$ and
$N(1675)$ are considered as degenerate resonances in
Ref.~\cite{Cohen:2004bk}, where it is shown that the on-resonance
couplings for the different channels behave exactly as expected from
the $1/N_c$ expansion for $N_c \! = \! 3$.  Another ``shifted
degenerate'' pair in Fig.~\ref{M1plot} appears to be $N(1680)$ in the
$M_{3-}$ (l.h.s.)  amplitude and $\Delta(1950)$ in the $M_{3+}$
(r.h.s.) amplitude; the $\Delta(1950)$ is not apparent in the plots of
Ref.~\cite{Cohen:2004bk}, which employs the older MAID~2003 analysis
rather than the MAID~2007 variant used here.

In the case of the NLO relations Eq.~(\ref{M3}) considered in
Fig.~\ref{M2plot}, the LO terms are $M_{L-}$ and the NLO terms are
$M_{L+}$, so entirely new resonances can appear at NLO\@.  A curious
example occurs for $L \! = \! 1$; the broad peak appearing for
$\gamma^* p \! \to \! \pi^+ n$ (l.h.s.), probably due to the
$N(1440)$, is unmatched by a contribution in $\gamma^* n \! \to \!
\pi^- p$ in the $M_{1-}$ amplitudes, and moreover, the $N(1720)$
appears prominently in the $M_{1+}$ NLO corrections.  On the other
hand, the NLO contributions sometimes indicate a shifted degenerate
pair; this effect occurs for the already-mentioned $N(1520)$ and
$N(1675)$ pair occurring at $L \! = \! 2$~\cite{Cohen:2004bk}.

In regions away from resonances and for amplitudes not exhibiting any
obvious resonances, the agreement tends to be rather better.  This is
certainly true for the $L \! = \! 4$ and 5 amplitudes, but also for
larger values of $Q^2$ and for values of $W$ close to threshold as
well as values at least 50--100~MeV from resonant peaks.  Of the
highest quality are the isovector-dominated relations in
Figs.~\ref{E-plot}--\ref{S+plot}, for which the same value of $J$
appearing on either side of the equation means that resonances
appearing on the two sides of the relation belong to an isomultiplet.
On the other hand, the benefits of the NLO correction in
Fig.~\ref{M2plot} remain ambiguous; for example, some improvement
appears in the real part of the $L \! = \! 1$ relation, but the
quality degrades in some kinematic regions for $L \! = \! 2$, and
scarcely any change due to the NLO terms is noticeable for $L \! \ge
\! 3$.

\section{Conclusions} \label{concl}

The $1/N_c$ expansion, which has provided so much qualitative and
semiquantitative guidance to understanding baryons in general, and
the baryon resonance spectrum in particular, appears in the case of
electroproduction to produce more ambiguous results.  On one hand, it
gives a natural explanation for the dominance of isovector over
isoscalar amplitudes, and it provides a definite set of linear
relations between multipole amplitudes that are expected to hold at
all values of c.m.\ energy $W$ and photon virtuality $Q^2$.  For
values of $W$ not near resonance masses as well as for larger values
of $Q^2$, the agreement tends to be in accordance with the
expectations of the $1/N_c$ expansion.  Even in the resonant region,
one often sees evidence of the ``shifted degenerate'' resonances
carrying different quantum numbers [such as $N(1520)$ and $N(1675)$]
that have related couplings.  However, just as many cases exist in
which the amplitudes in the resonant region do not entirely conform to
naive $1/N_c$ expectations, both at leading and subleading order.  The
specific reason that a given large $N_c$ relation for pion
electroproduction works surprisingly well or surprisingly poorly in
the resonant region remains a challenge for future research.

%
\begin{figure}[htp]
\epsfxsize=0.48\textwidth\epsfbox{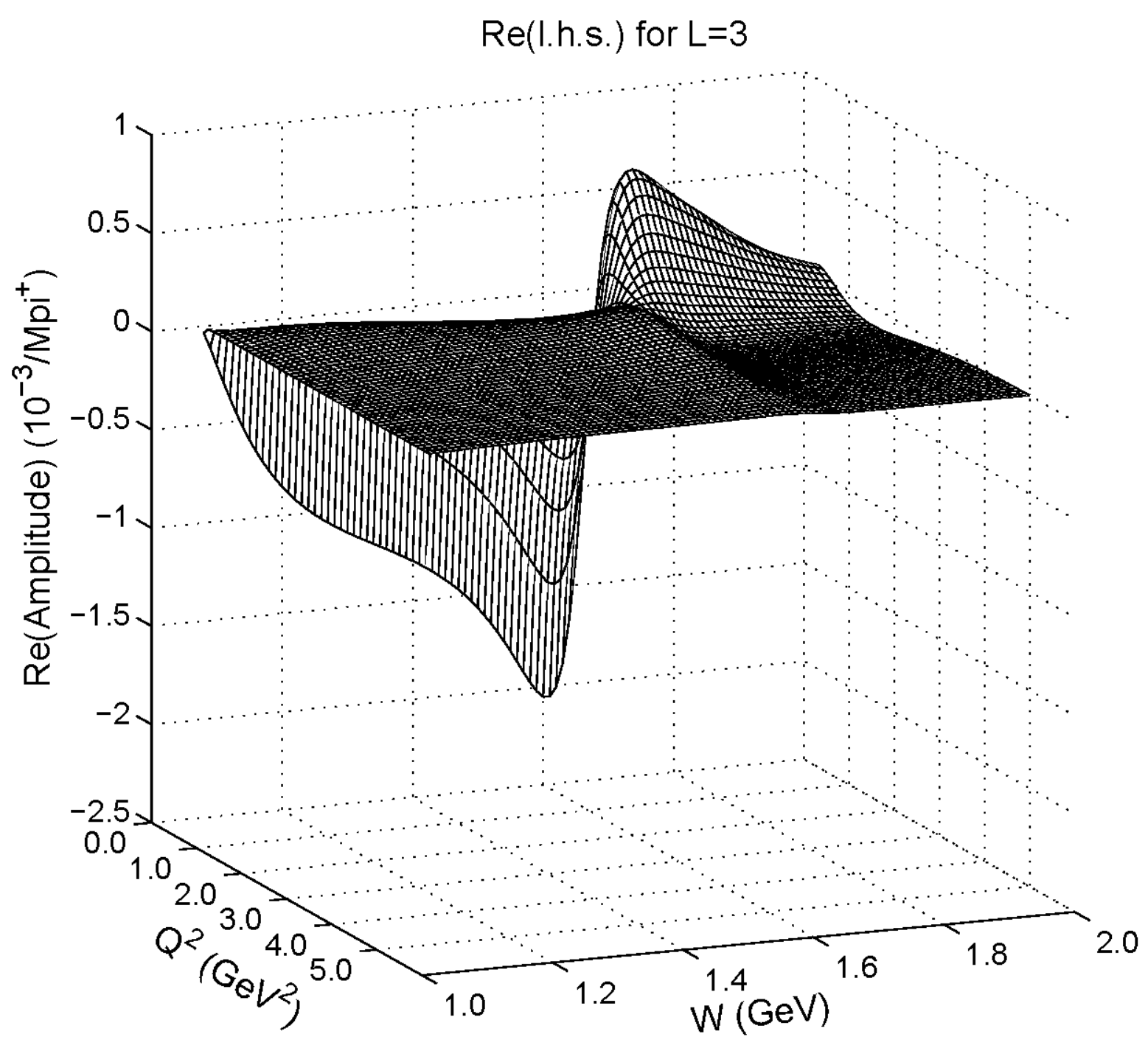}
\epsfxsize=0.48\textwidth\epsfbox{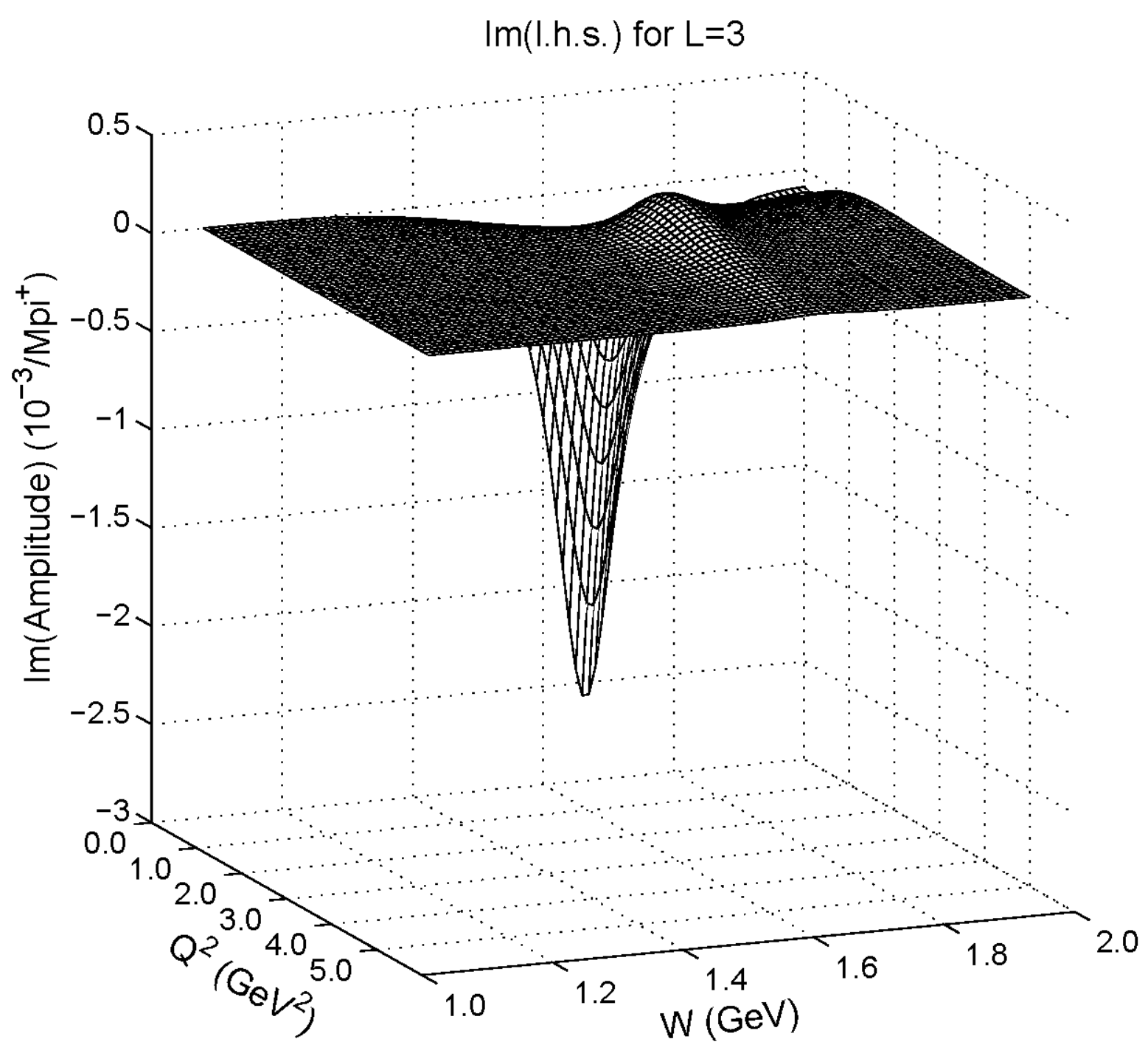}\\[1mm]
\epsfxsize=0.48\textwidth\epsfbox{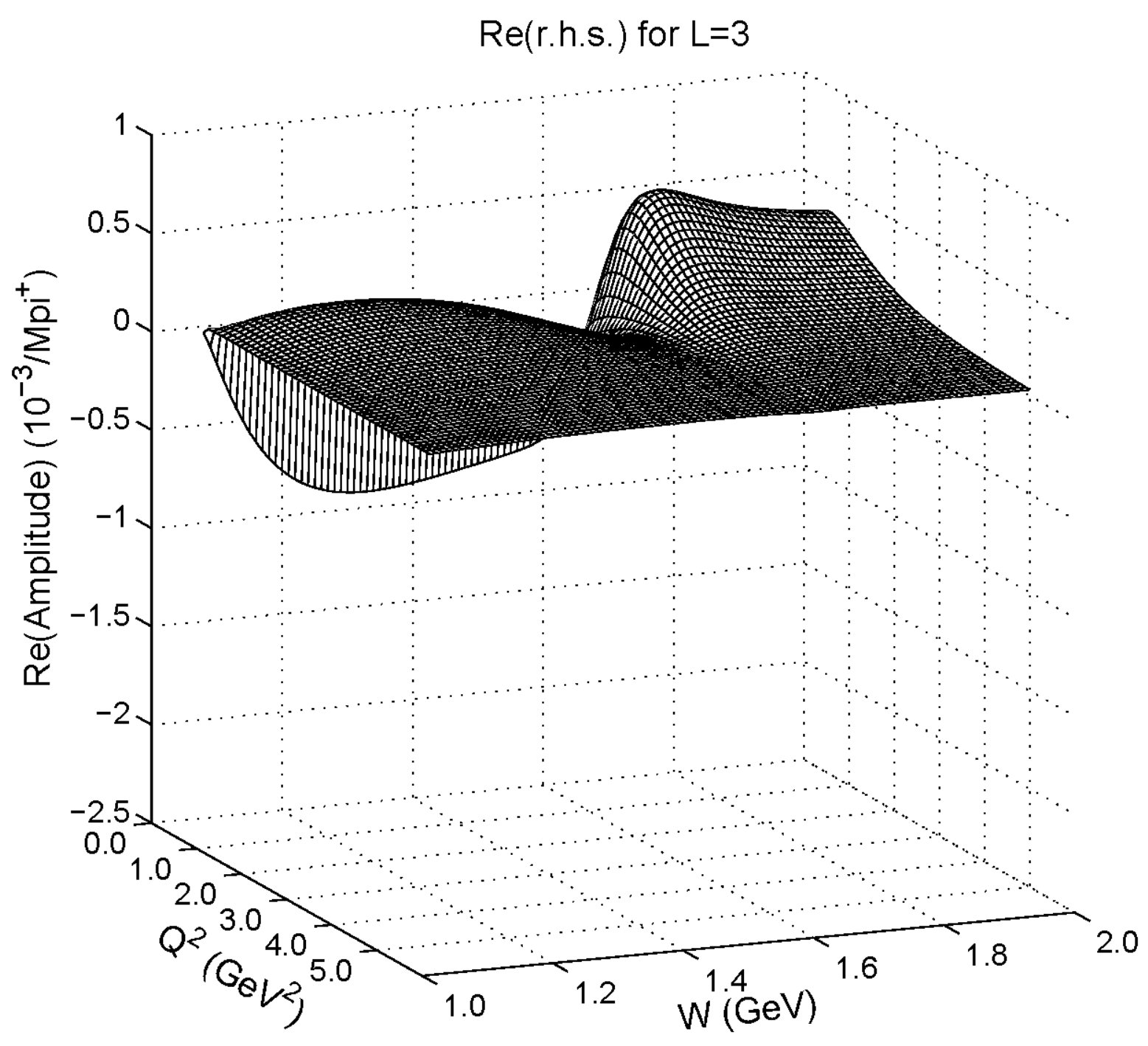}
\epsfxsize=0.48\textwidth\epsfbox{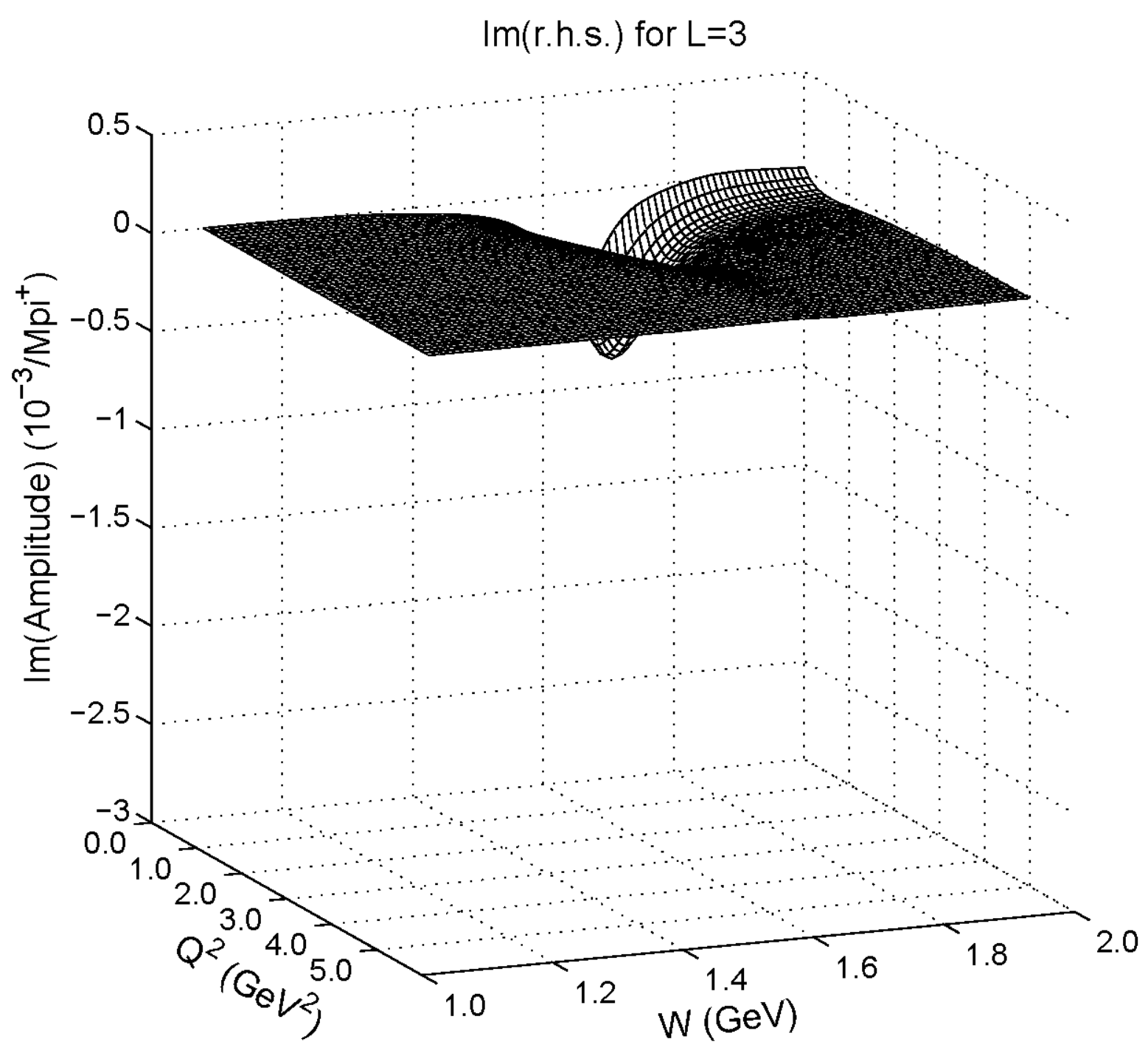}\\[1mm]
\epsfxsize=0.48\textwidth\epsfbox{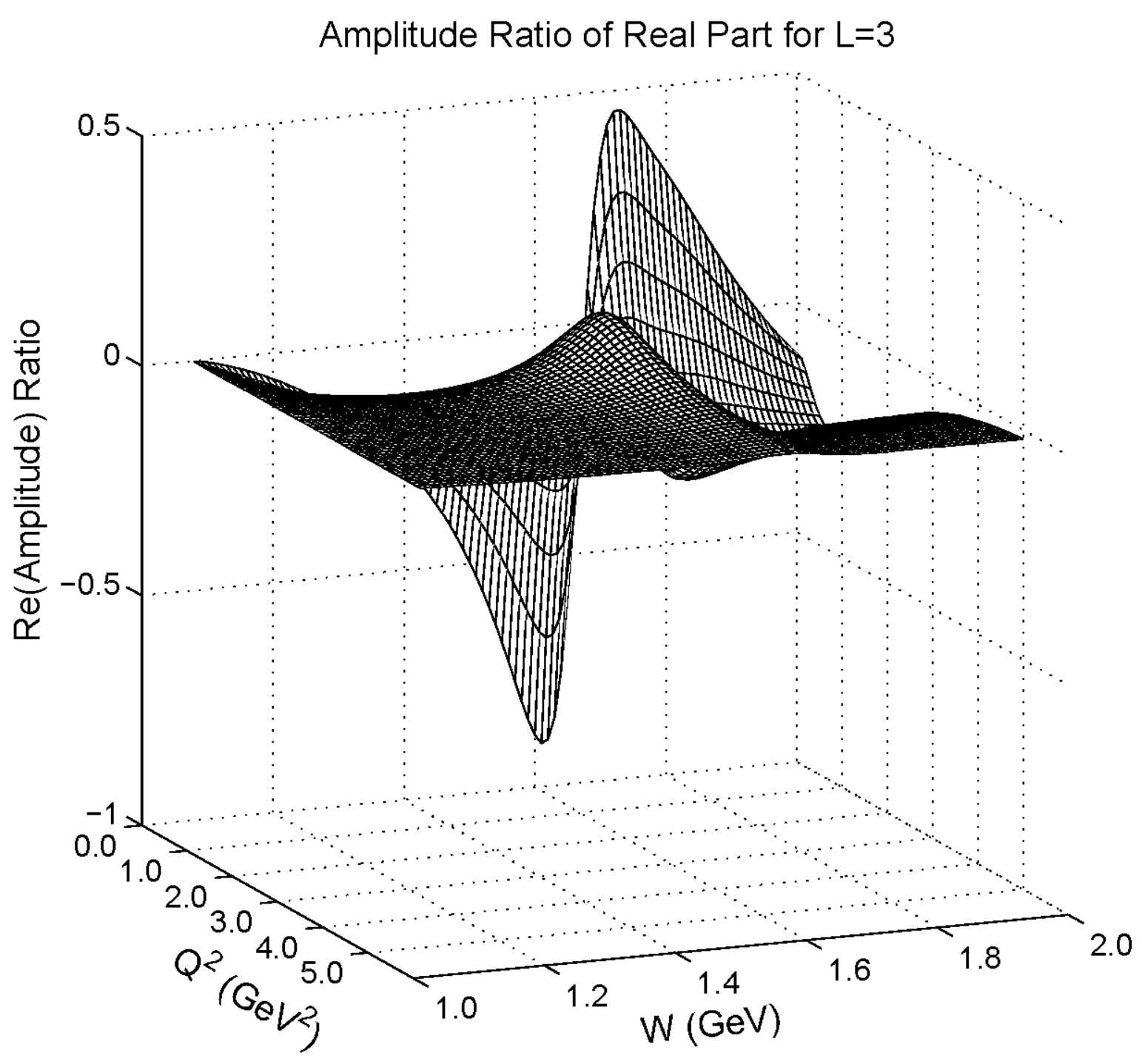}
\epsfxsize=0.48\textwidth\epsfbox{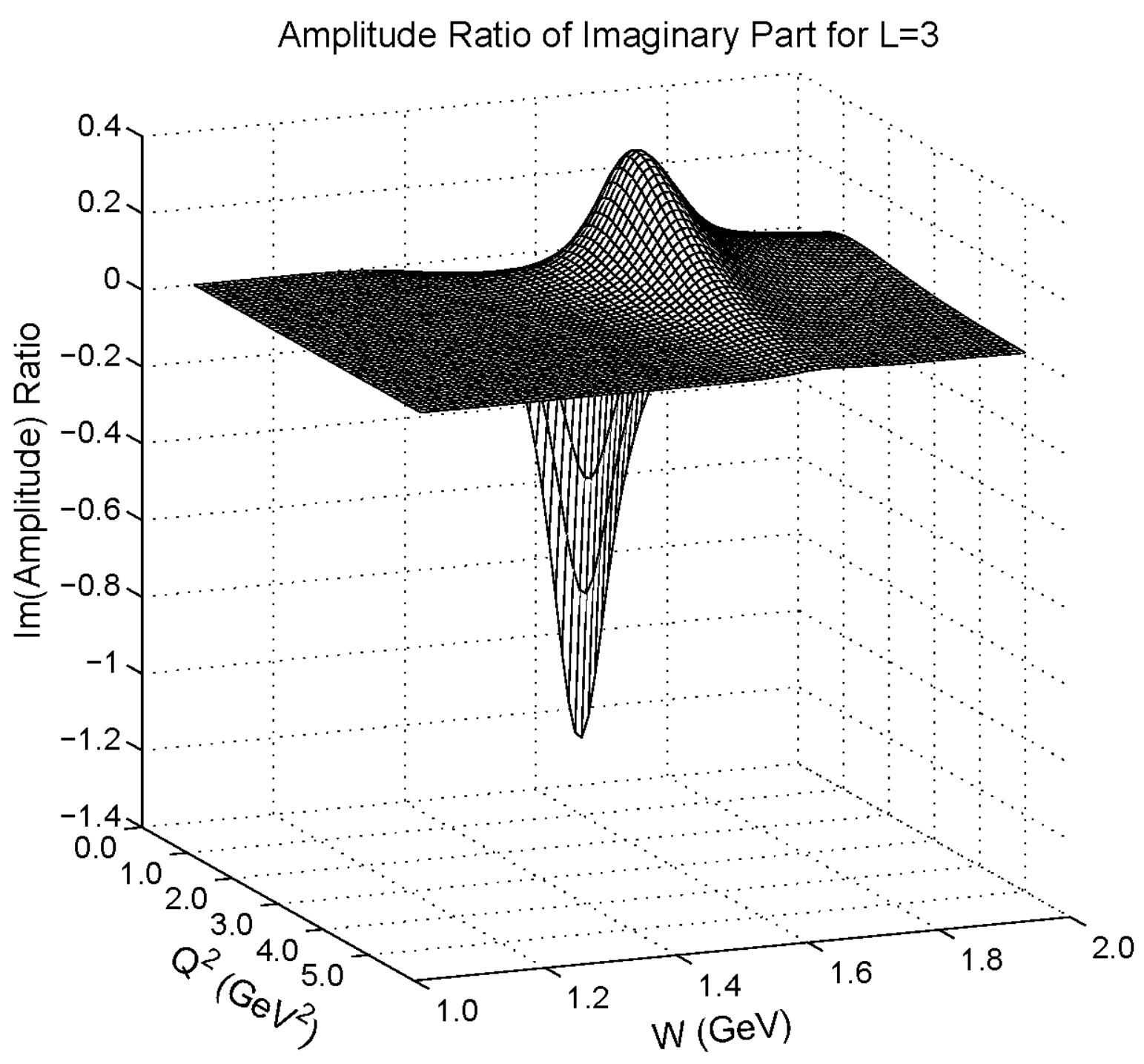}\\
\end{figure}
\begin{figure}[htp]
\epsfxsize=0.48\textwidth\epsfbox{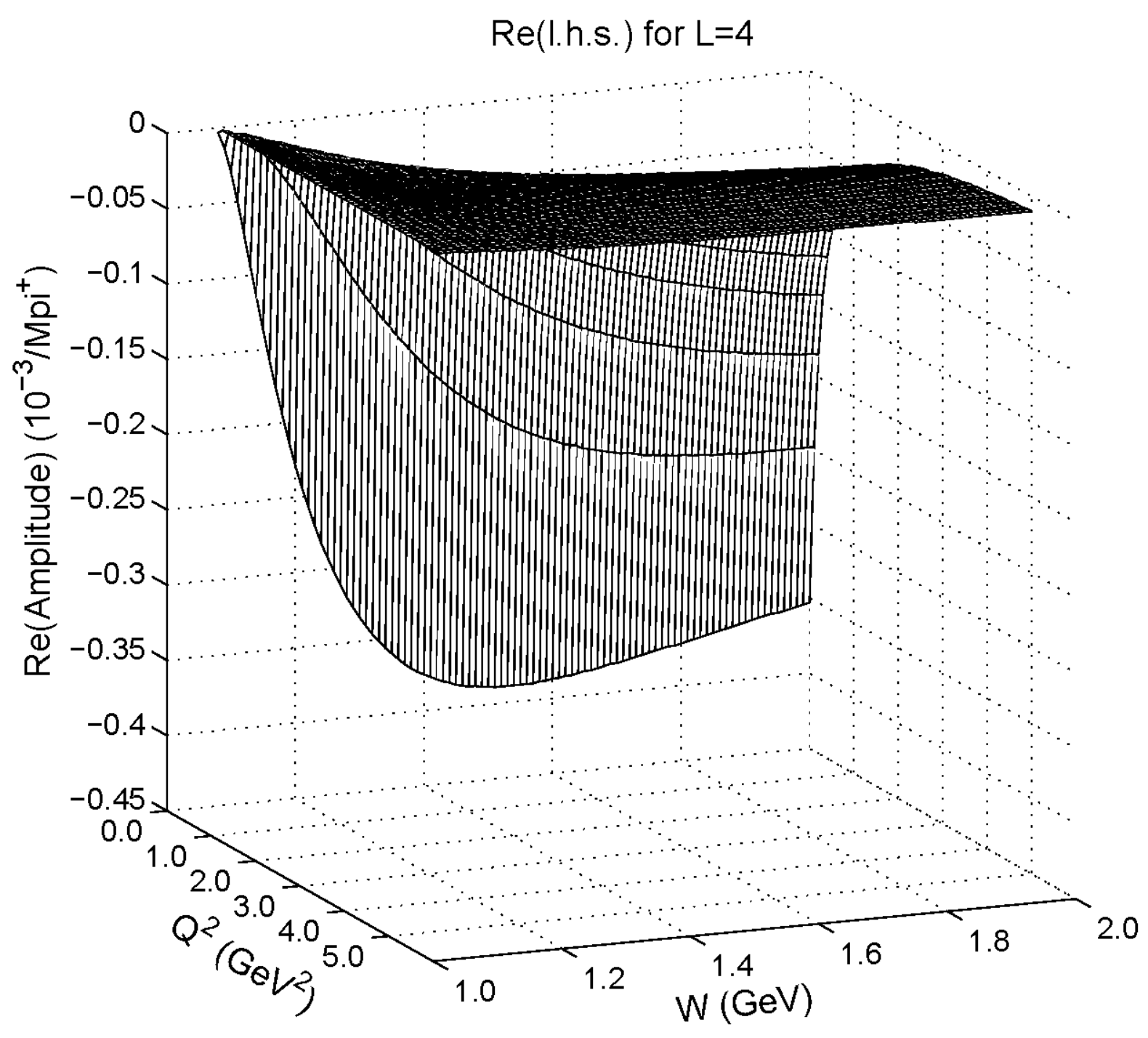}
\epsfxsize=0.48\textwidth\epsfbox{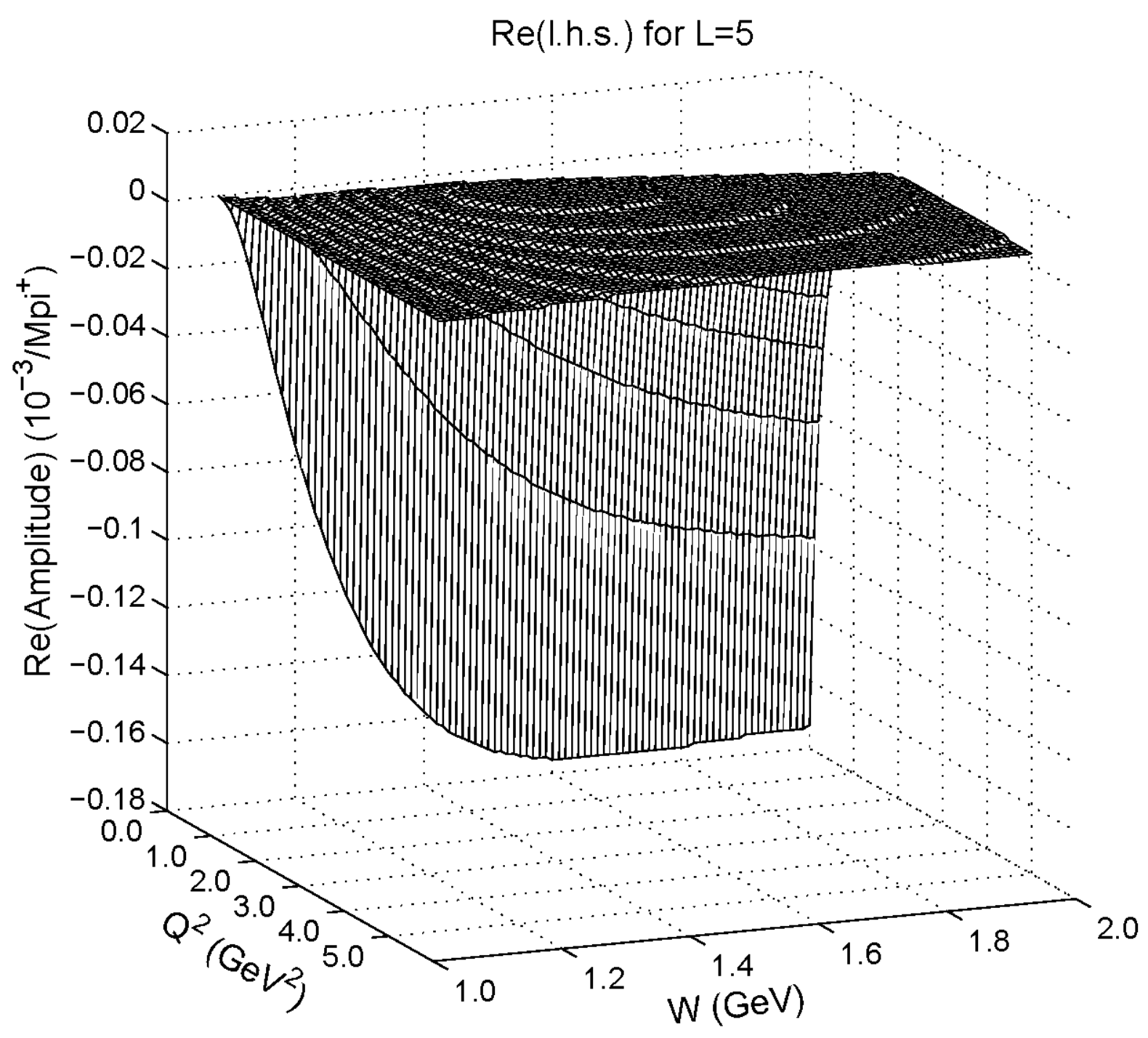}\\[1mm]
\epsfxsize=0.48\textwidth\epsfbox{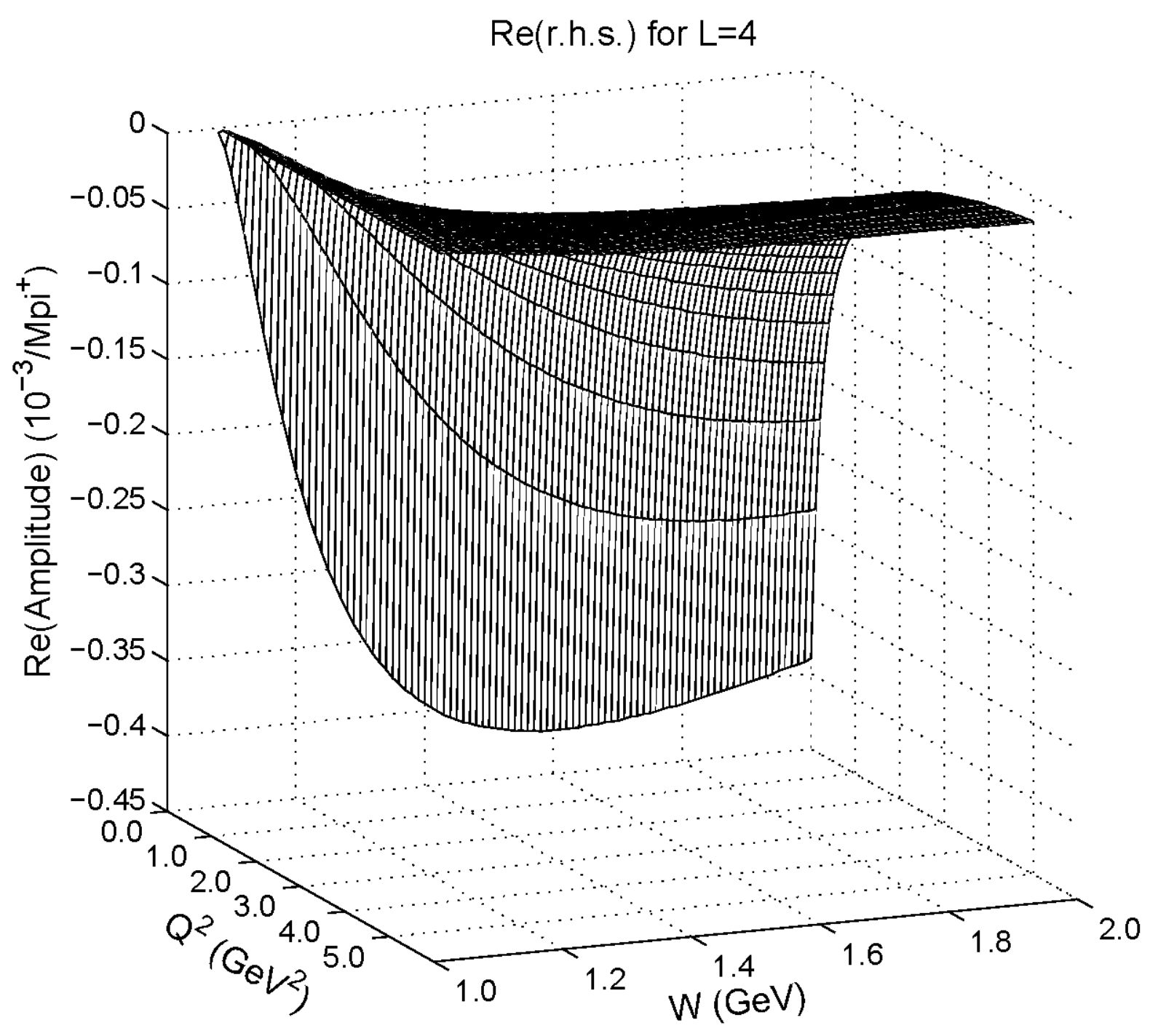}
\epsfxsize=0.48\textwidth\epsfbox{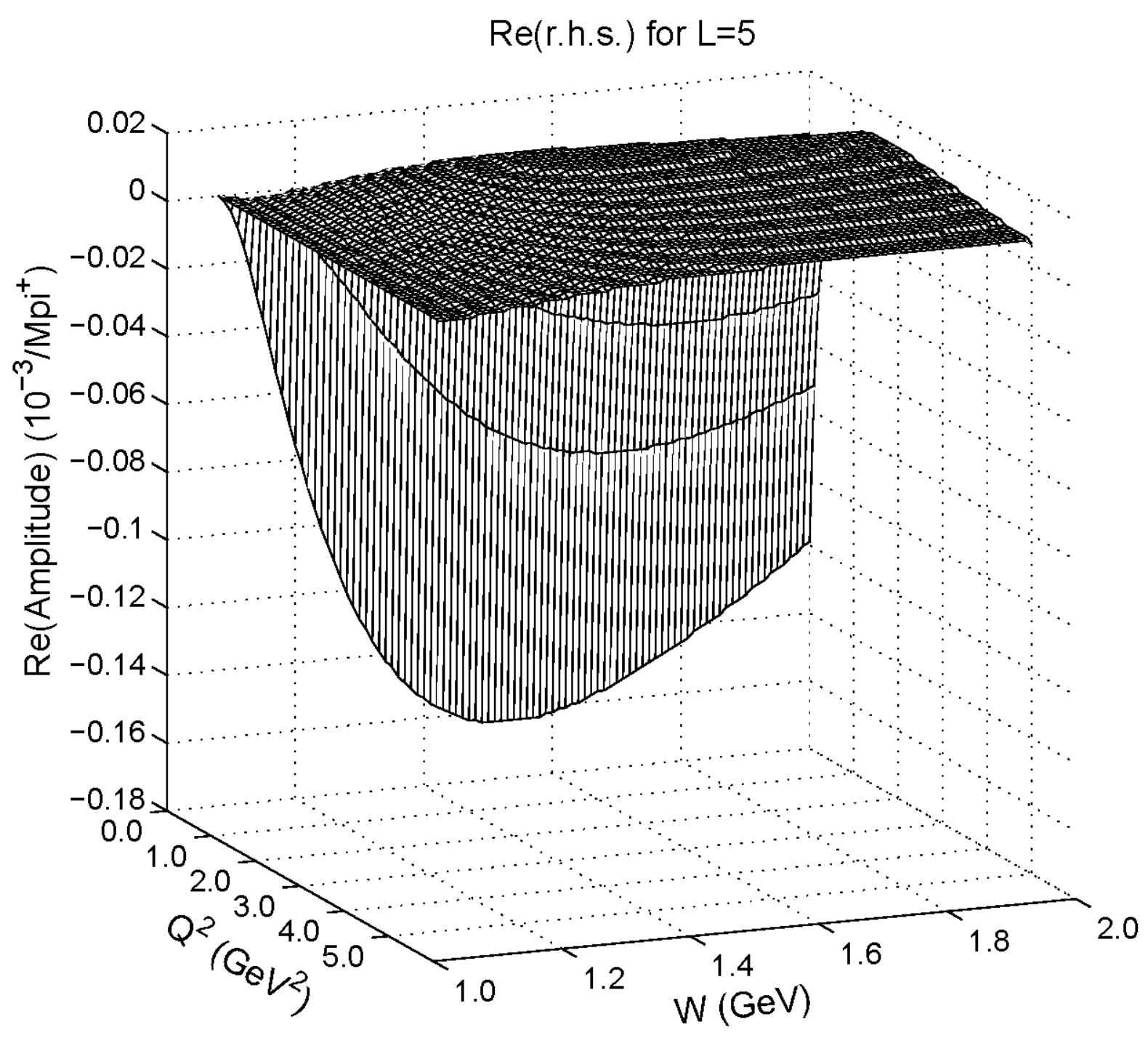}\\[1mm]
\epsfxsize=0.48\textwidth\epsfbox{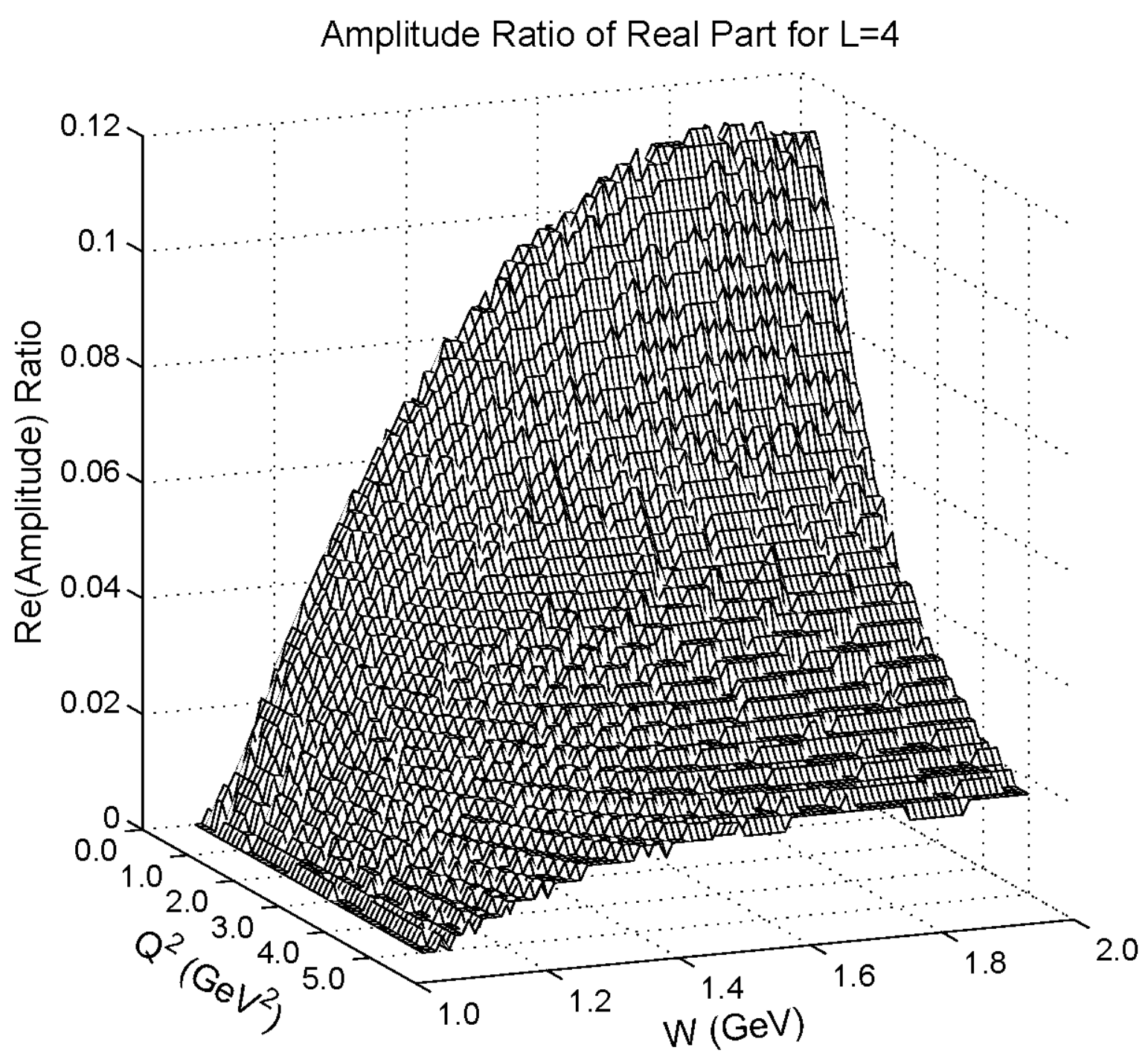}
\epsfxsize=0.48\textwidth\epsfbox{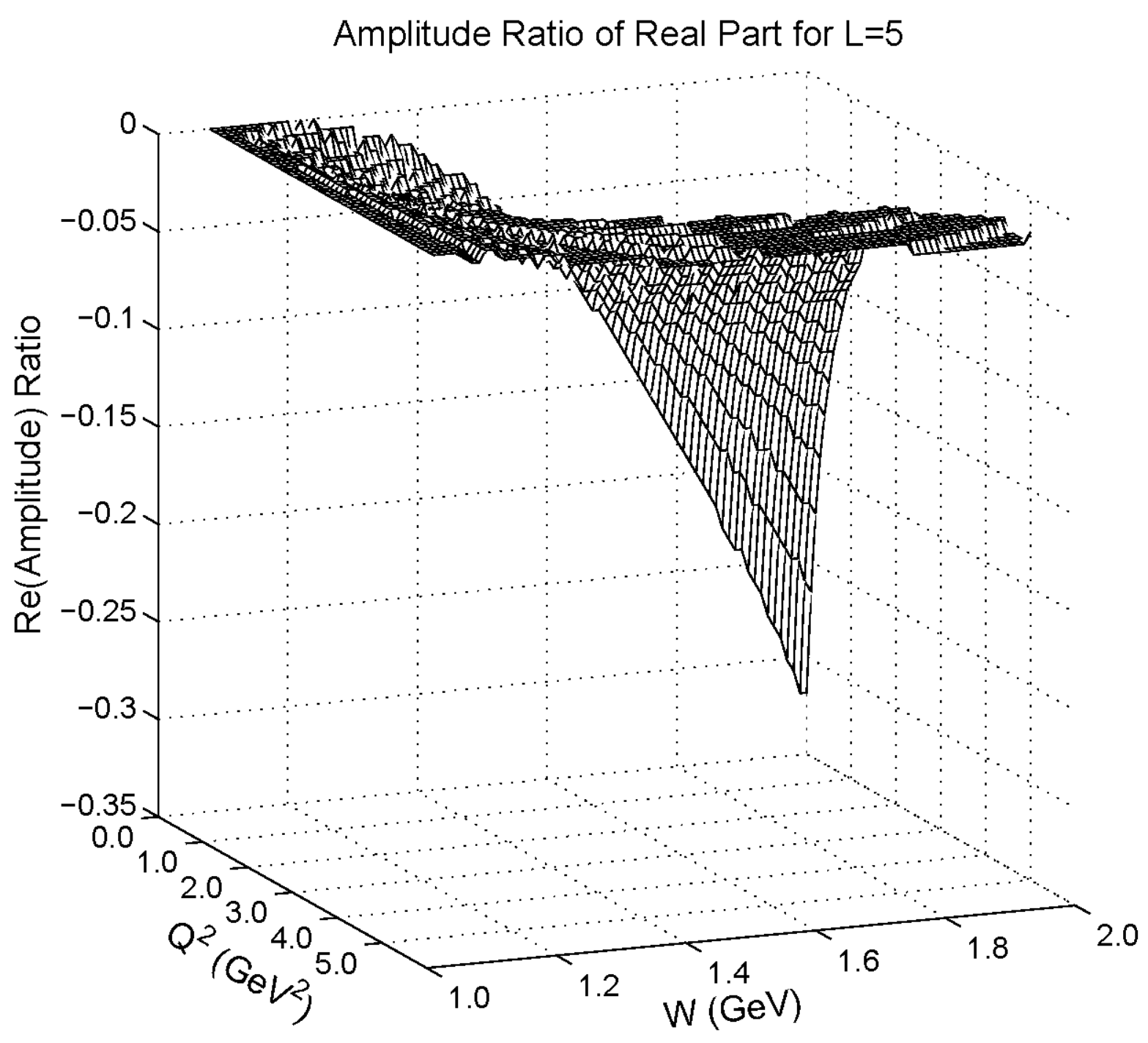}
\end{figure}
%
%
%
\begin{figure}[htp]
\epsfxsize=0.48\textwidth\epsfbox{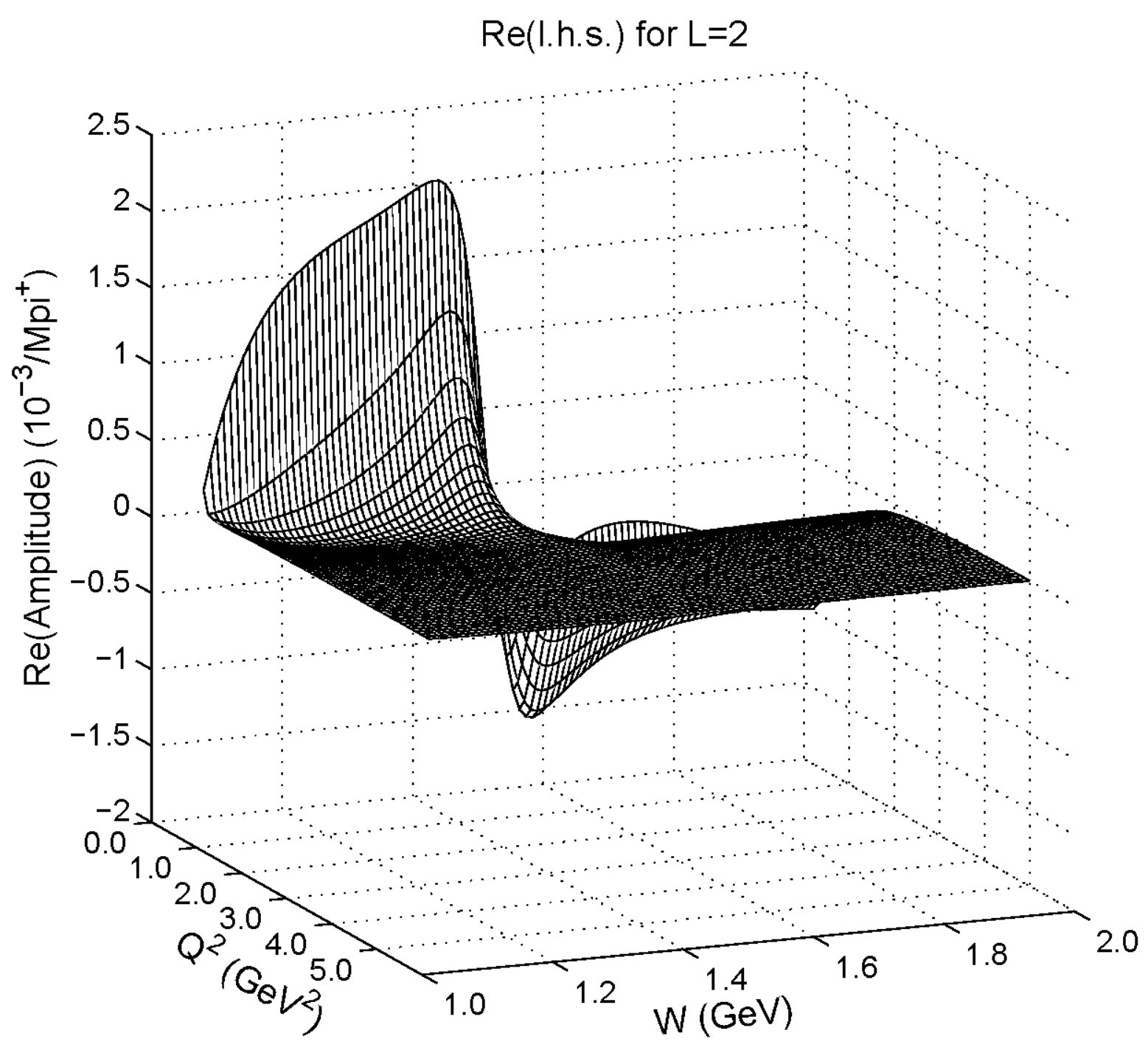}
\epsfxsize=0.48\textwidth\epsfbox{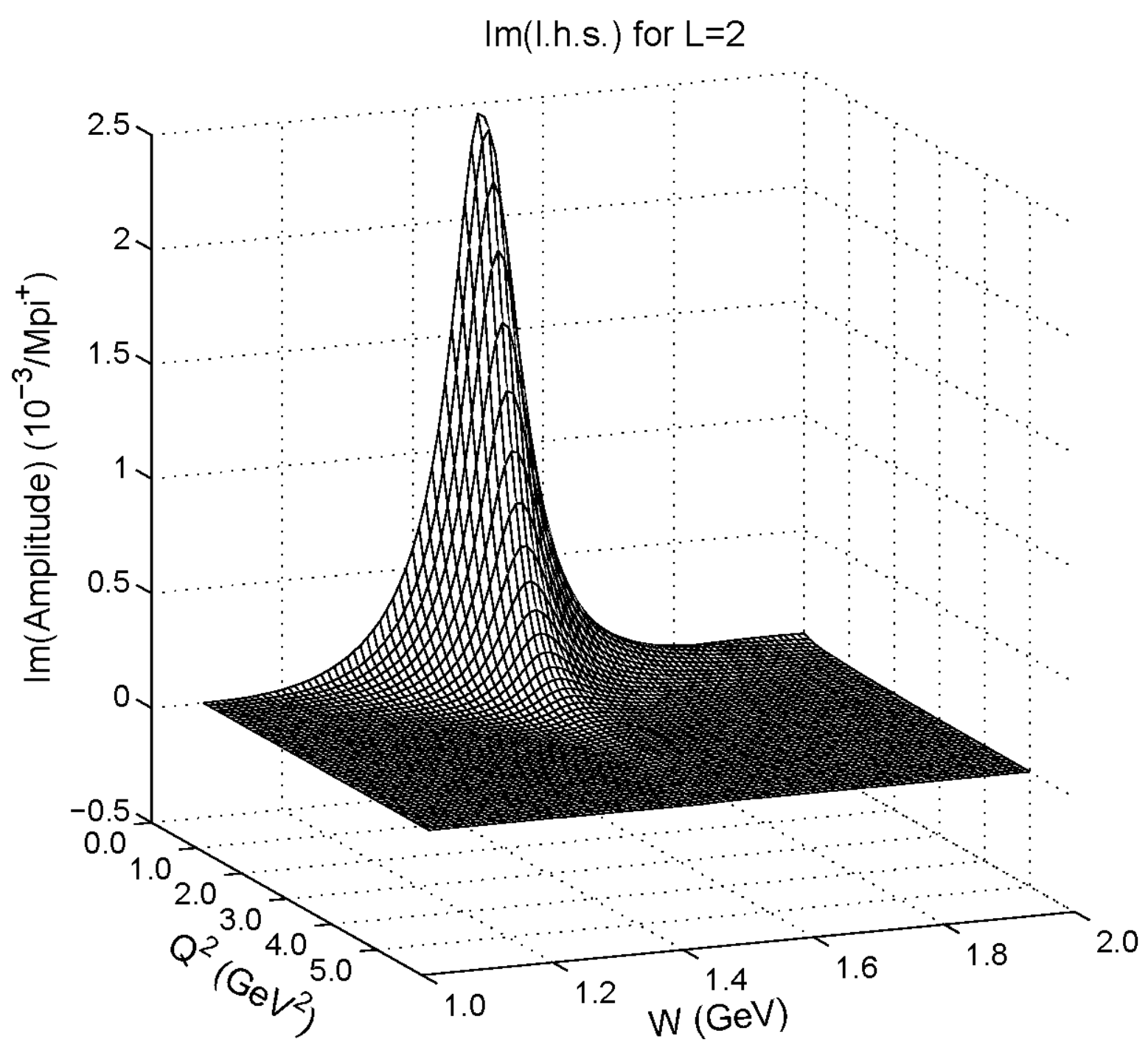}\\[1mm]
\epsfxsize=0.48\textwidth\epsfbox{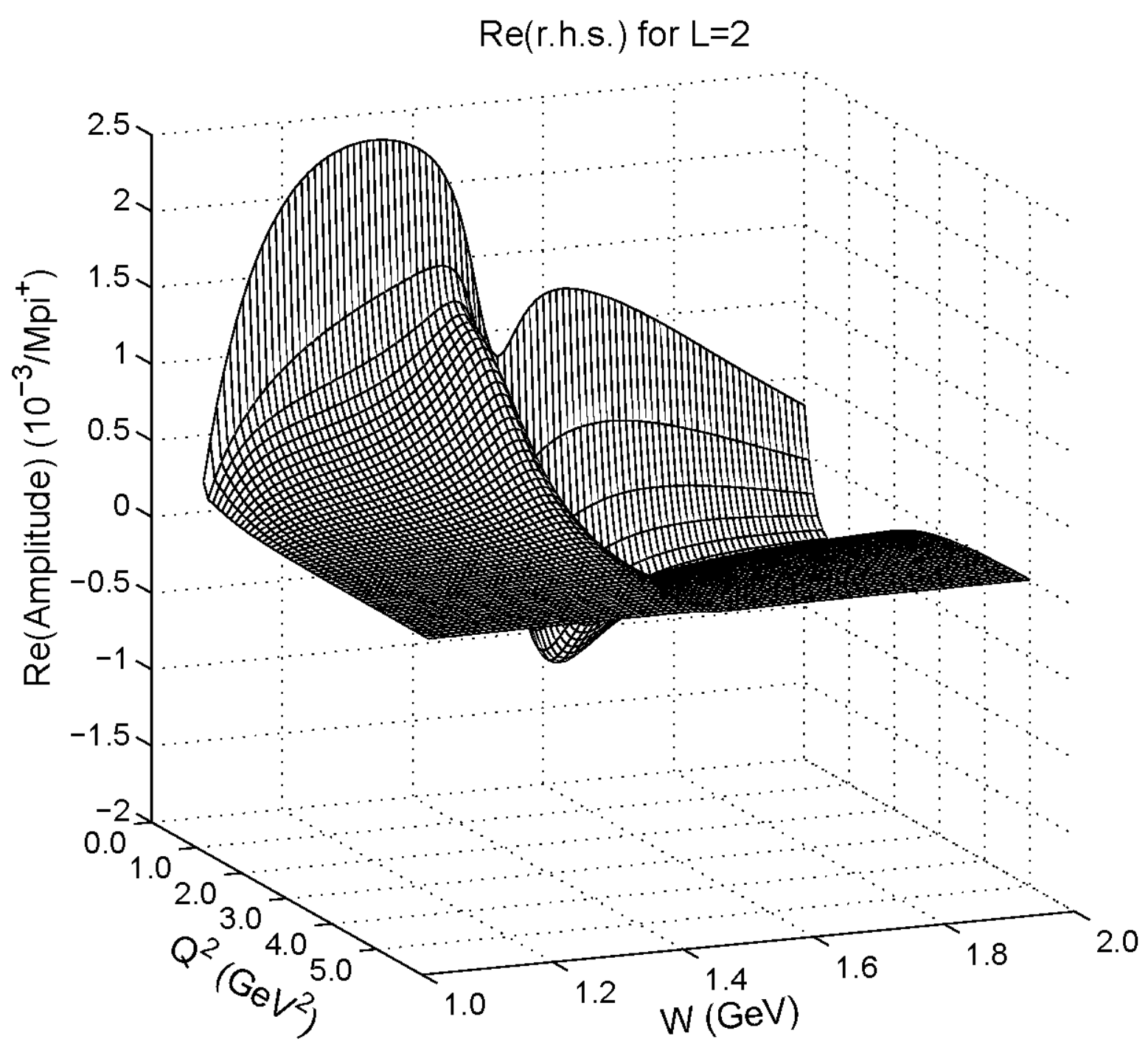}
\epsfxsize=0.48\textwidth\epsfbox{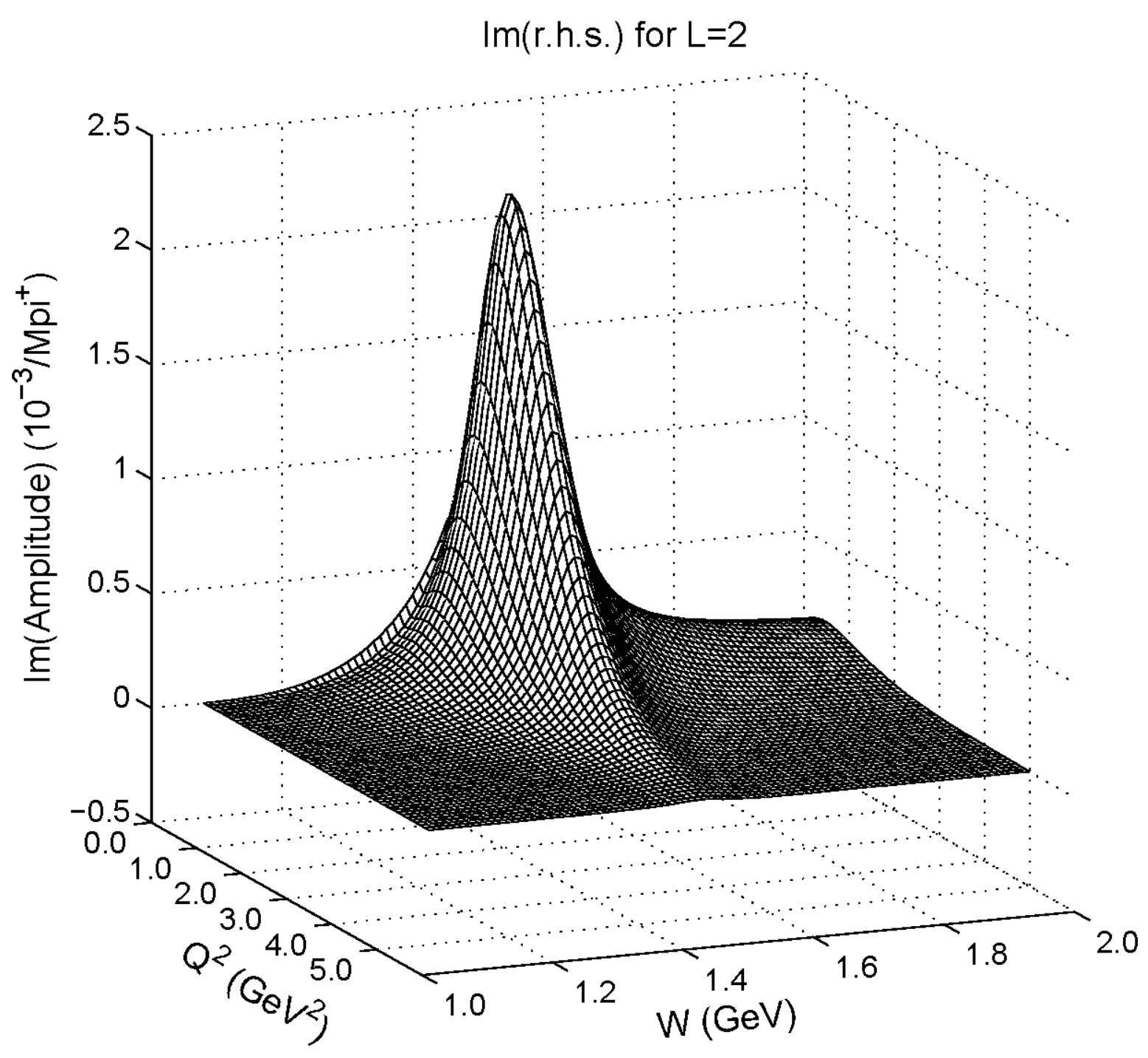}\\[1mm]
\epsfxsize=0.48\textwidth\epsfbox{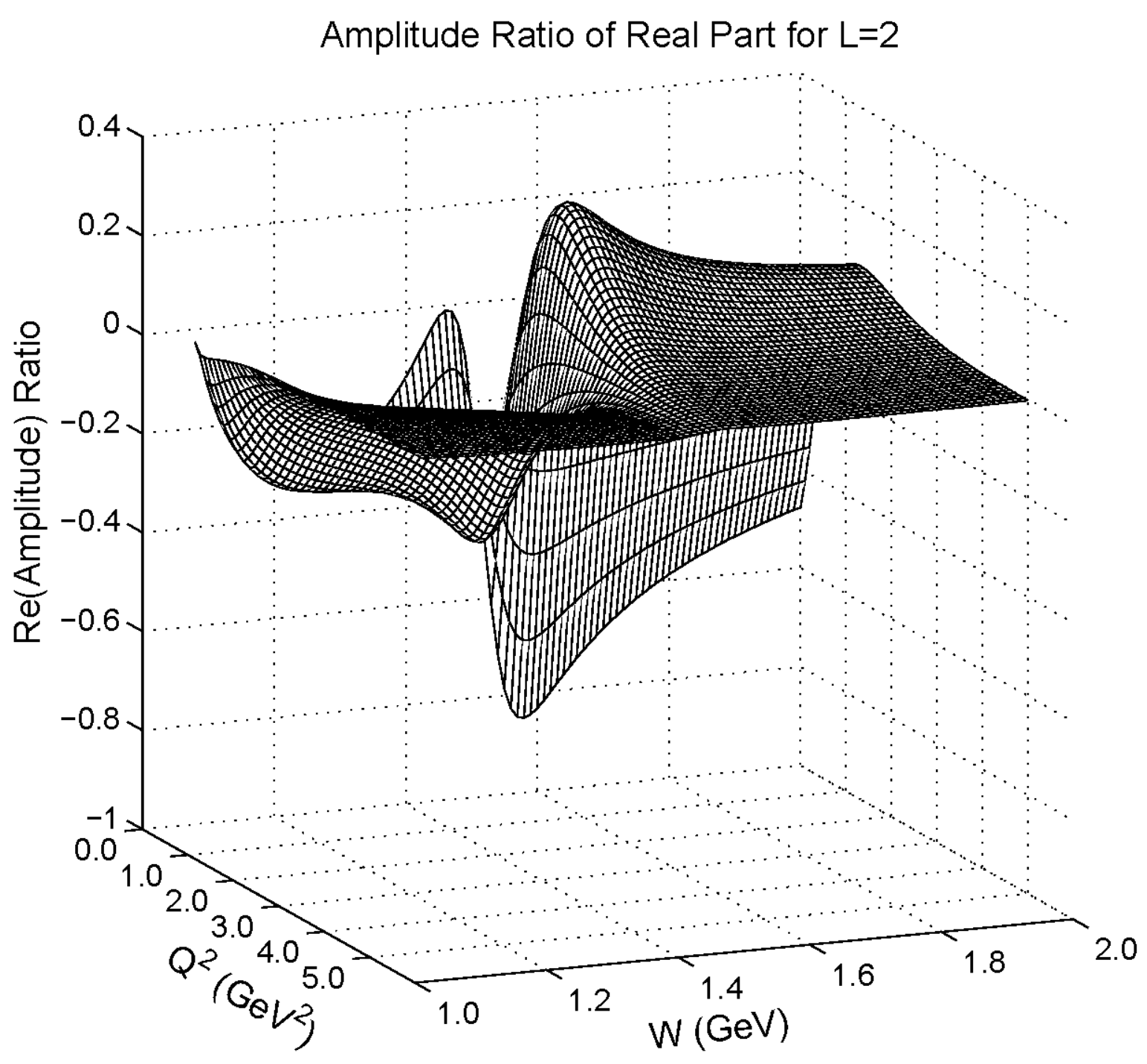}
\epsfxsize=0.48\textwidth\epsfbox{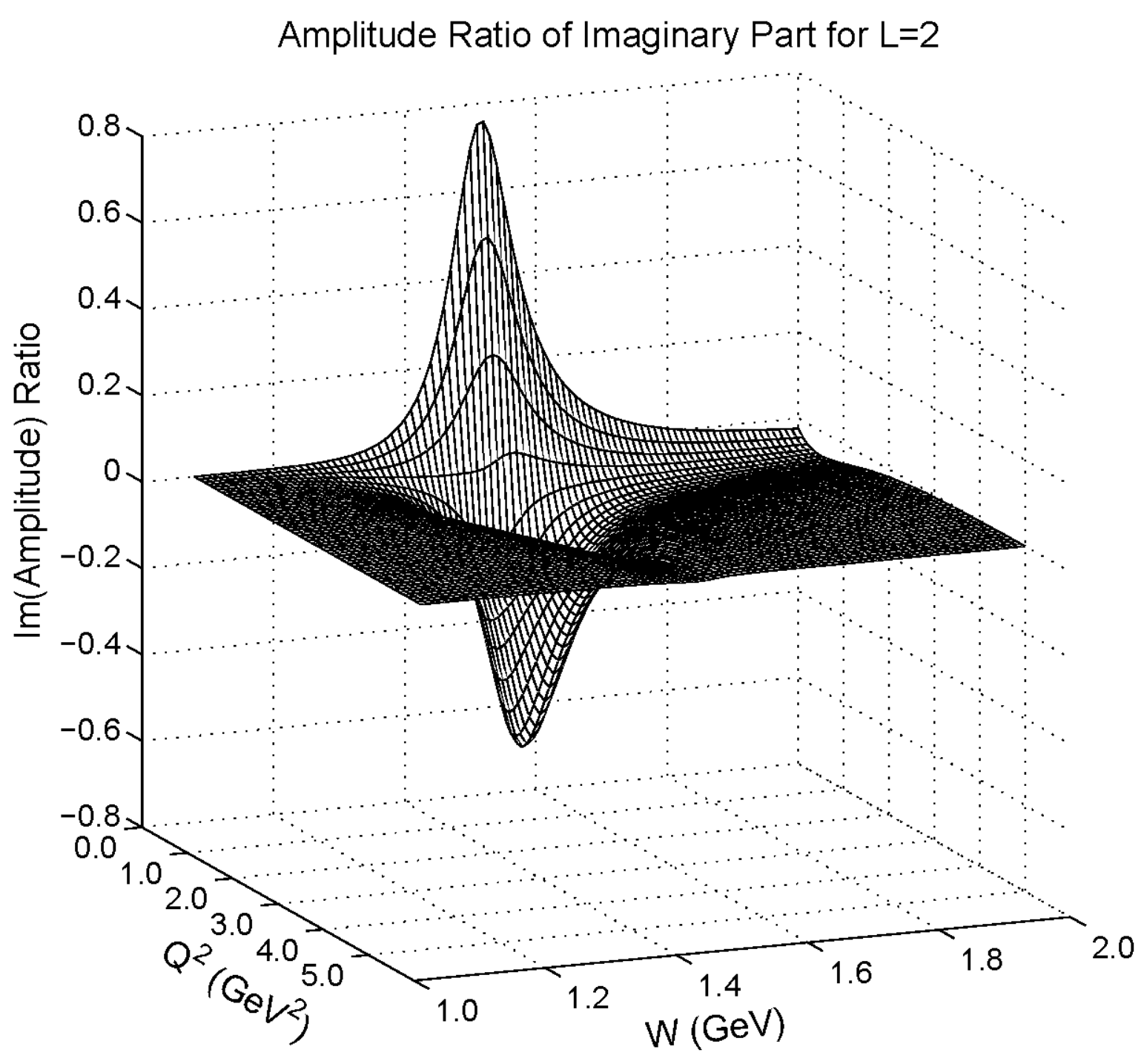}\\
\end{figure}
\begin{figure}[htp]
\epsfxsize=0.48\textwidth\epsfbox{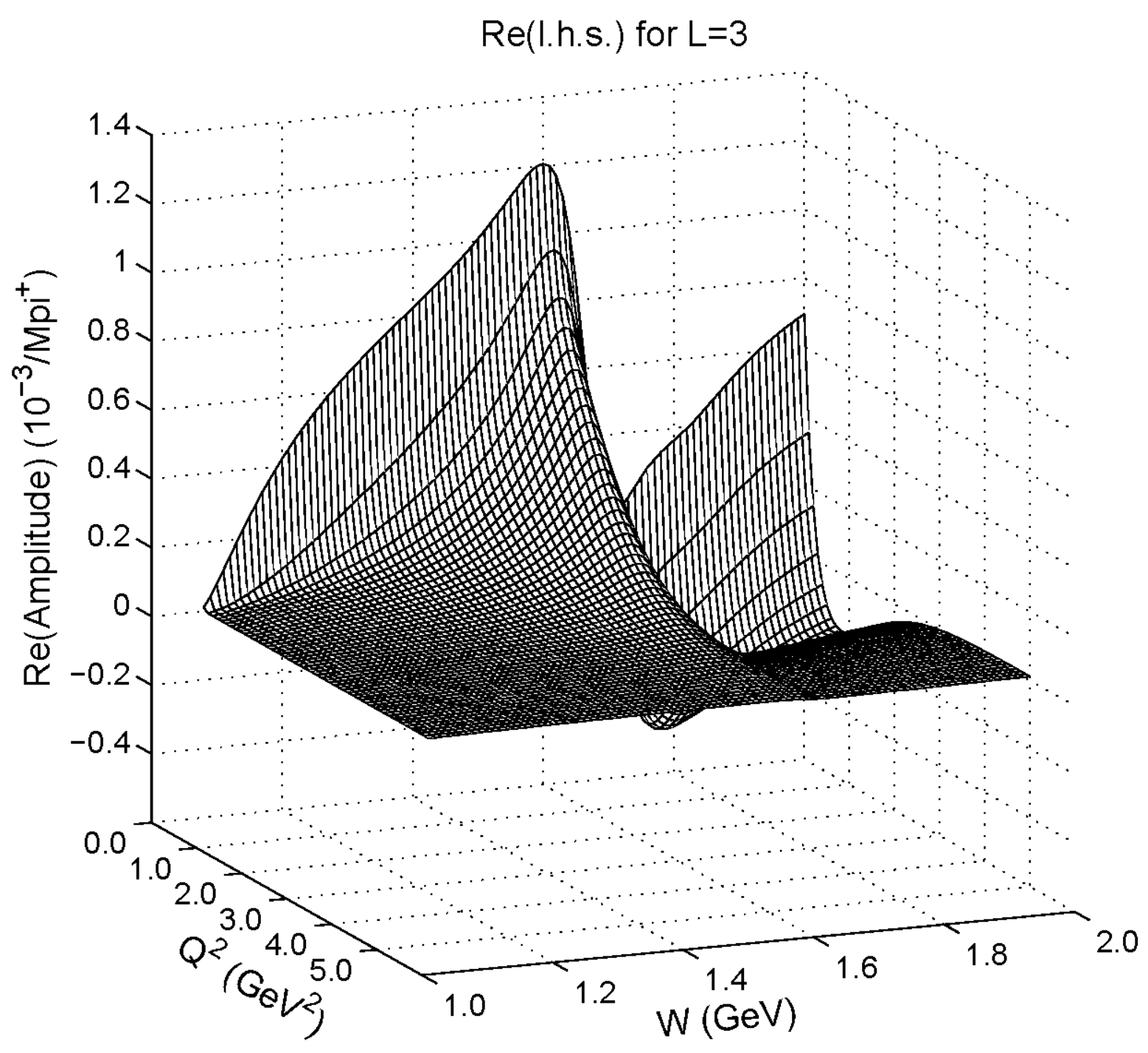}
\epsfxsize=0.48\textwidth\epsfbox{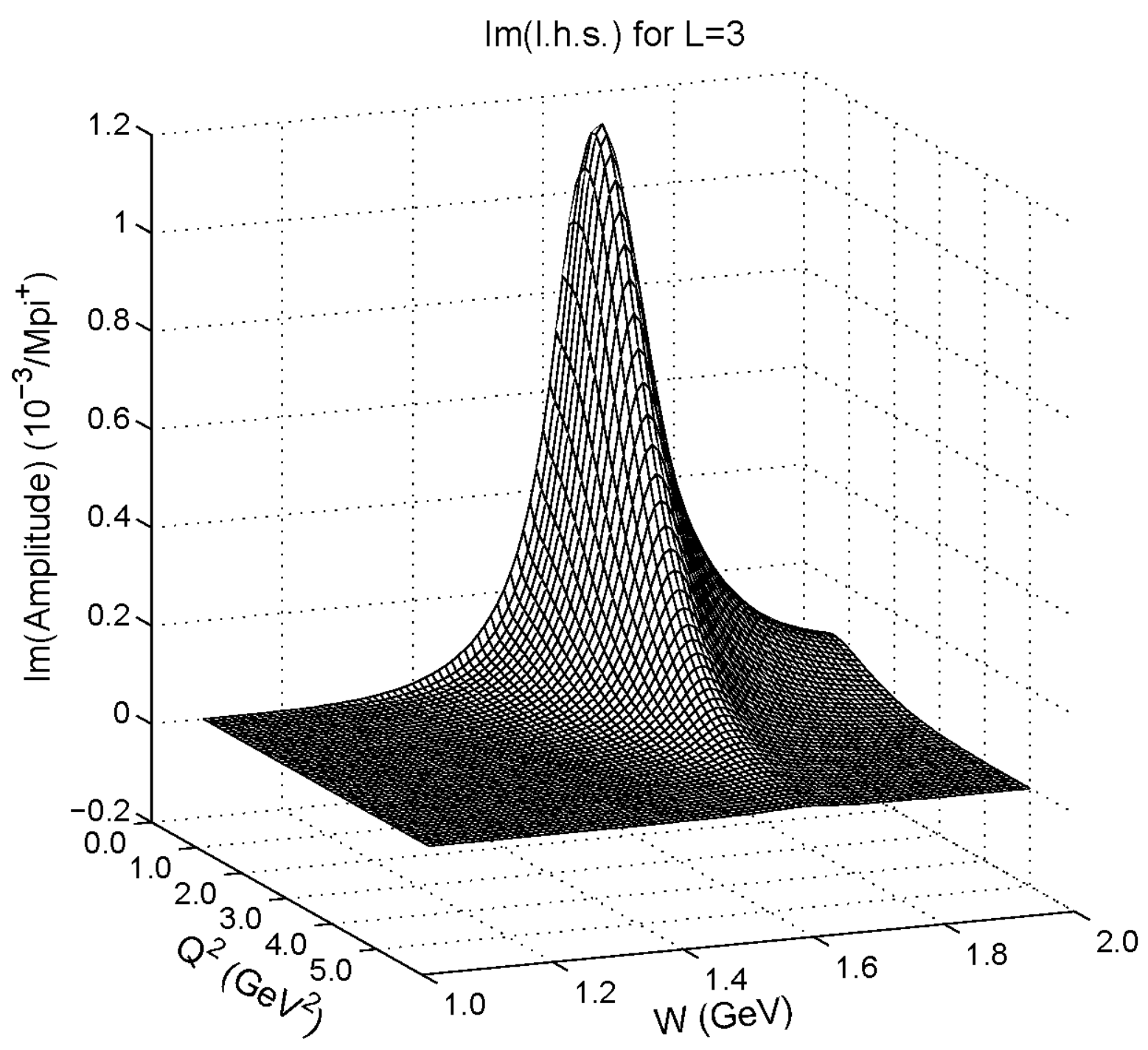}\\[1mm]
\epsfxsize=0.48\textwidth\epsfbox{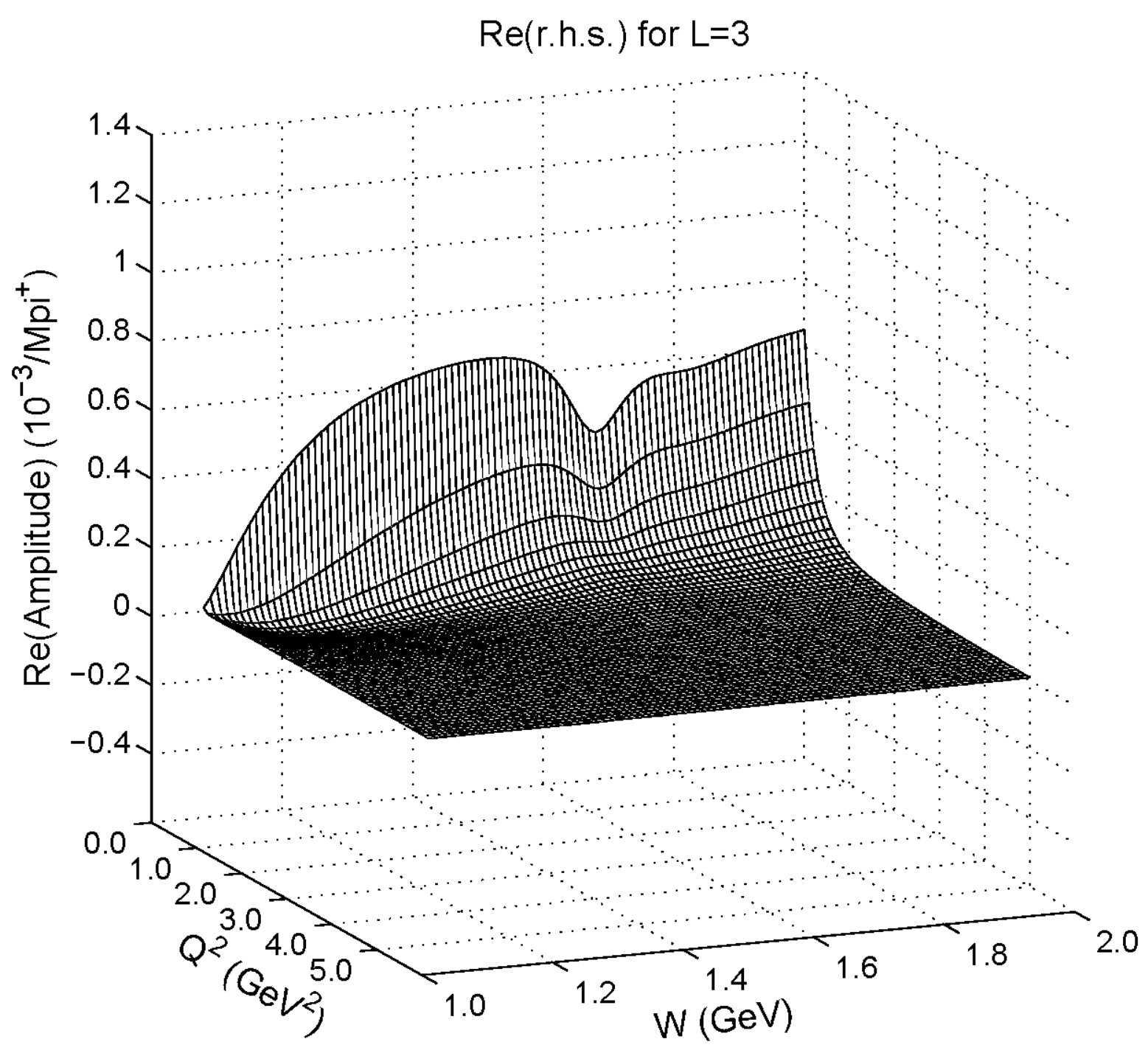}
\epsfxsize=0.48\textwidth\epsfbox{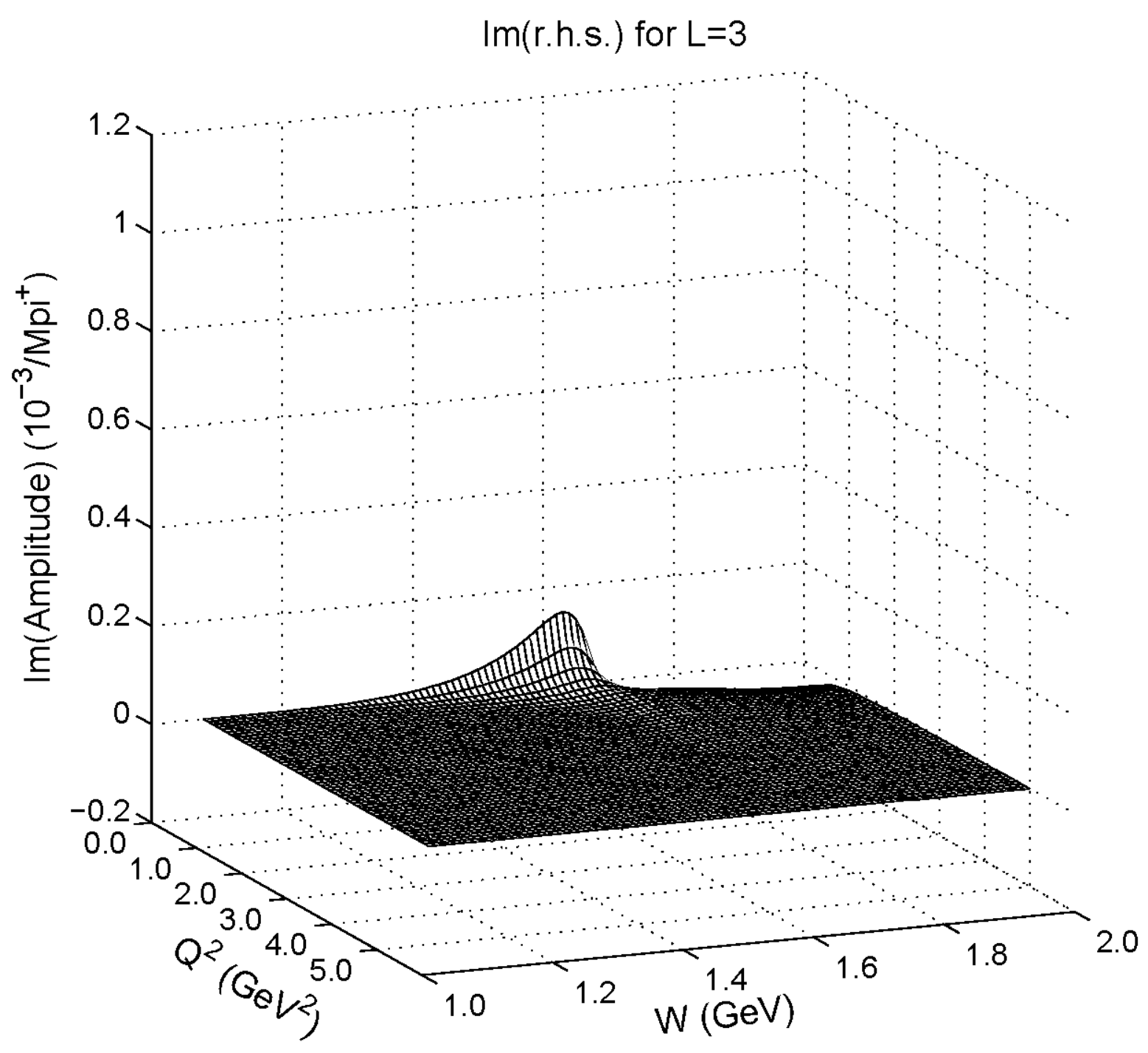}\\[1mm]
\epsfxsize=0.48\textwidth\epsfbox{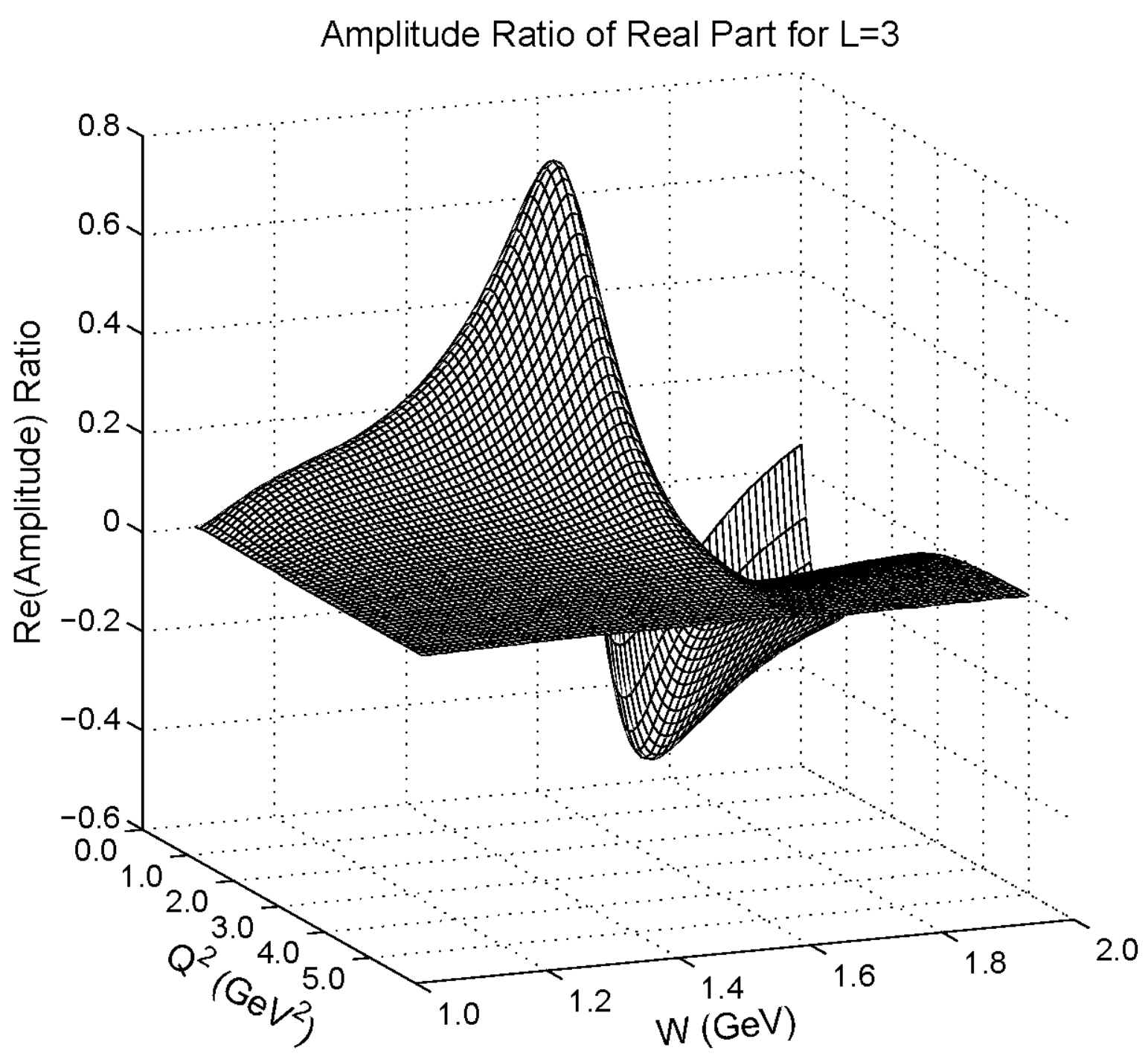}
\epsfxsize=0.48\textwidth\epsfbox{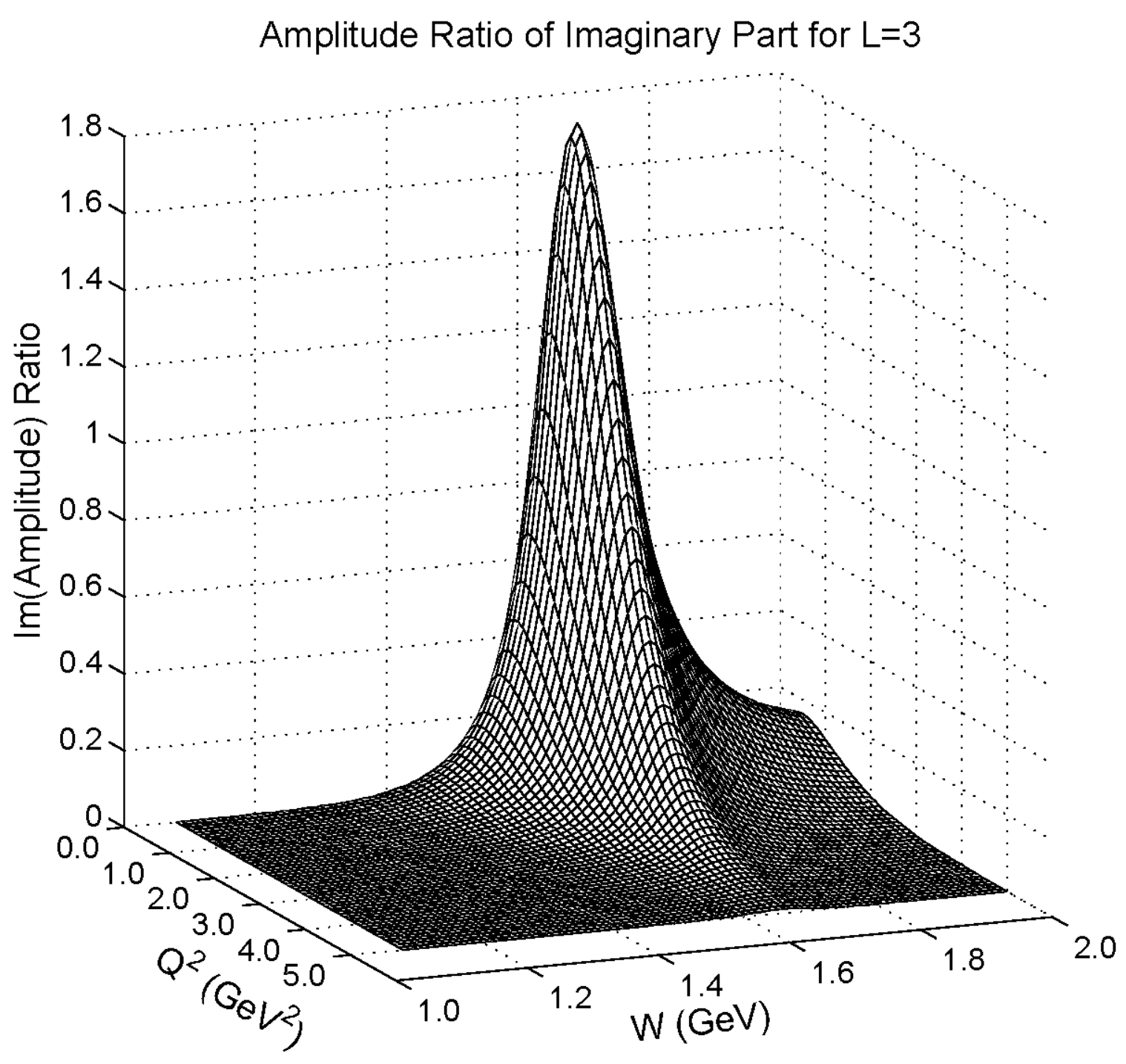}\\
\end{figure}
\begin{figure}[htp]
\epsfxsize=0.48\textwidth\epsfbox{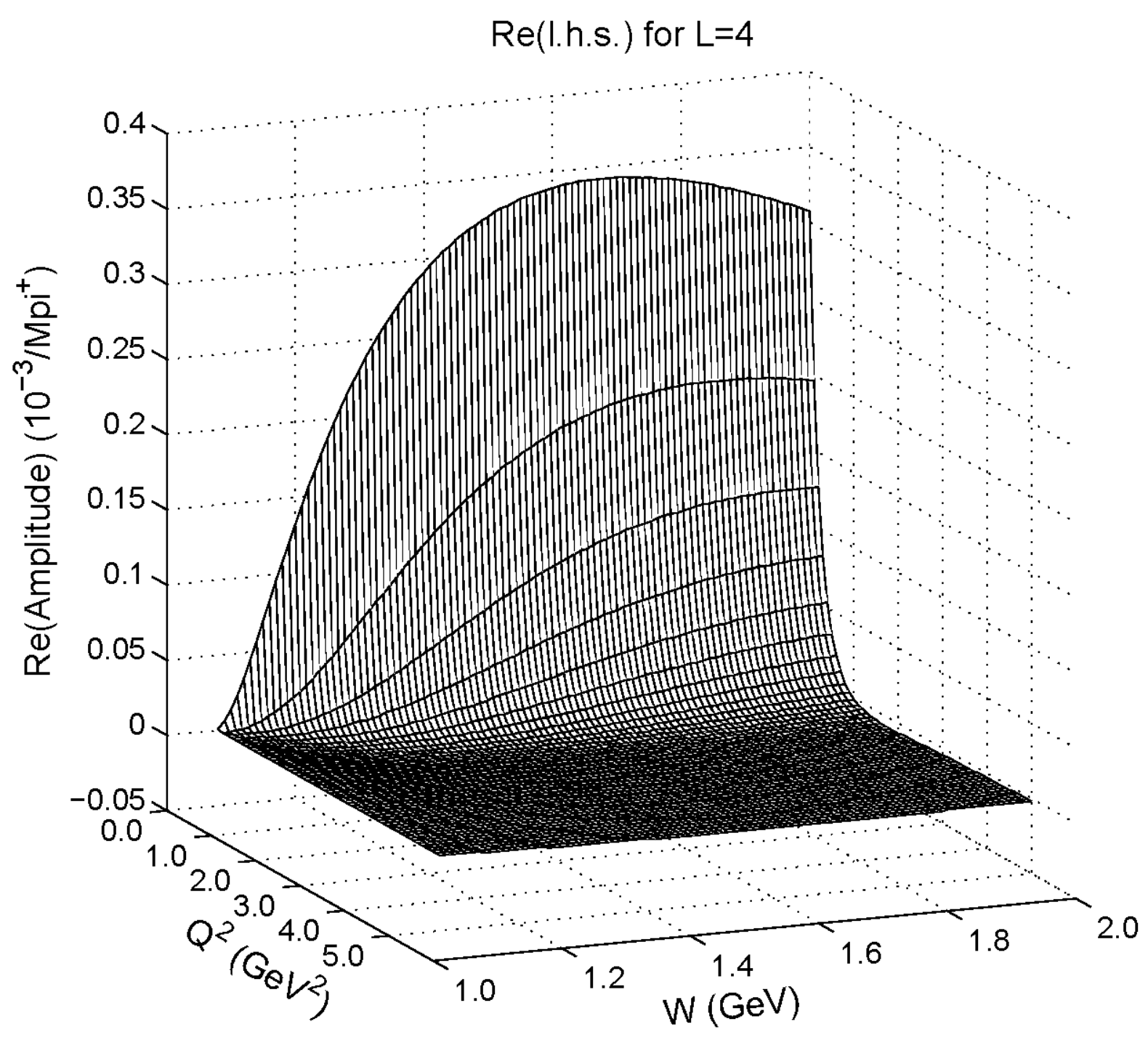}
\epsfxsize=0.48\textwidth\epsfbox{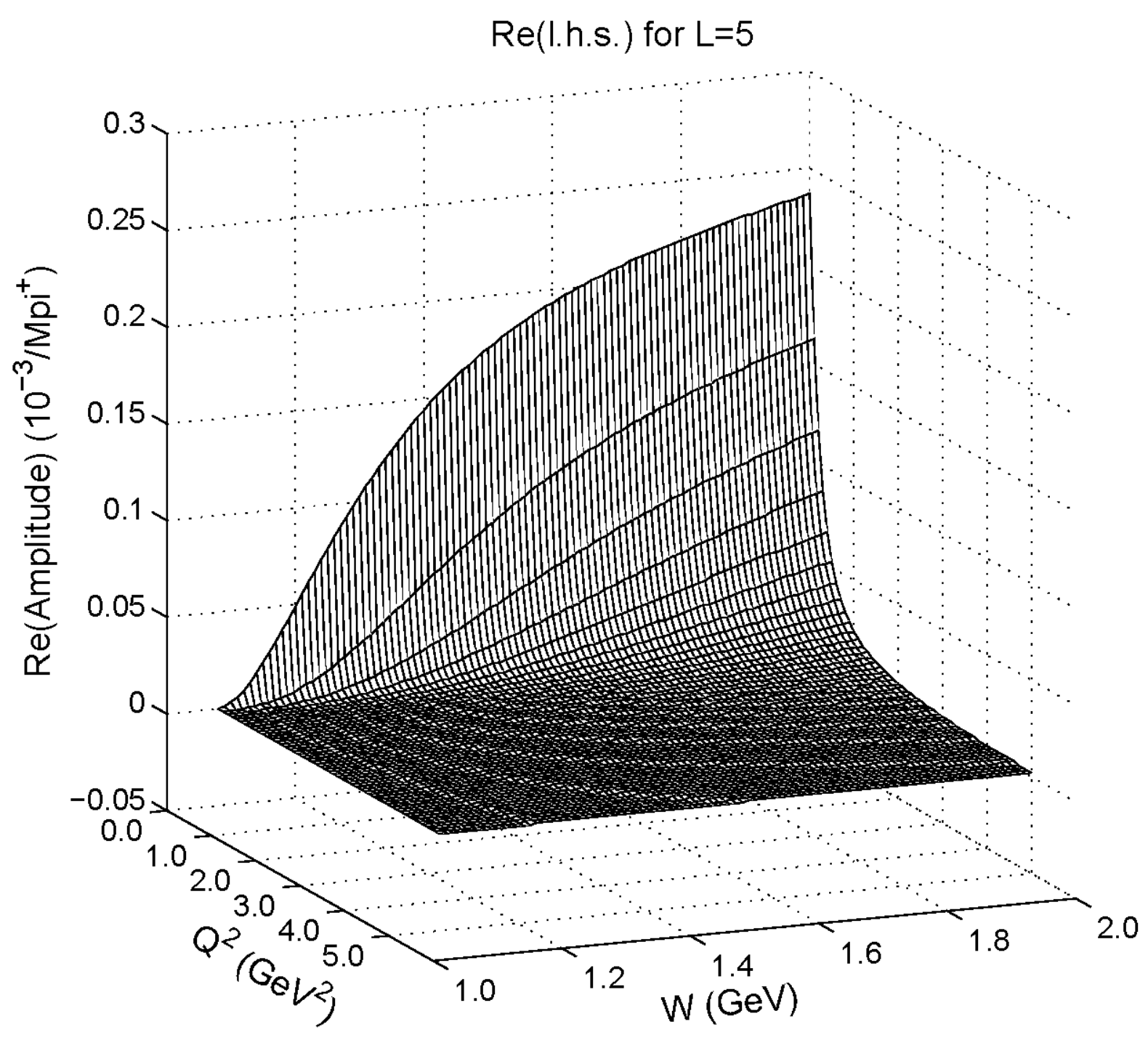}\\[1mm]
\epsfxsize=0.48\textwidth\epsfbox{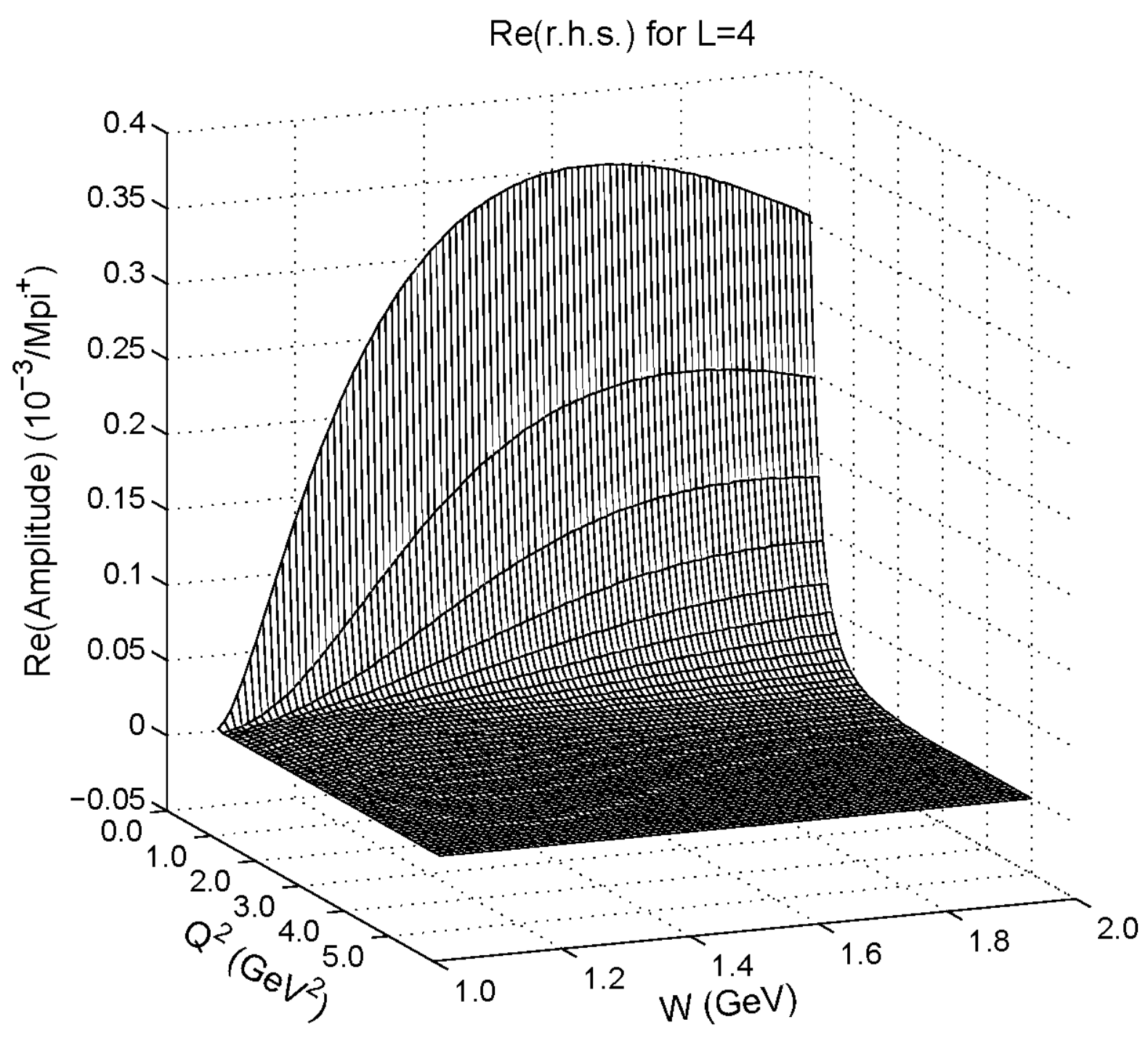}
\epsfxsize=0.48\textwidth\epsfbox{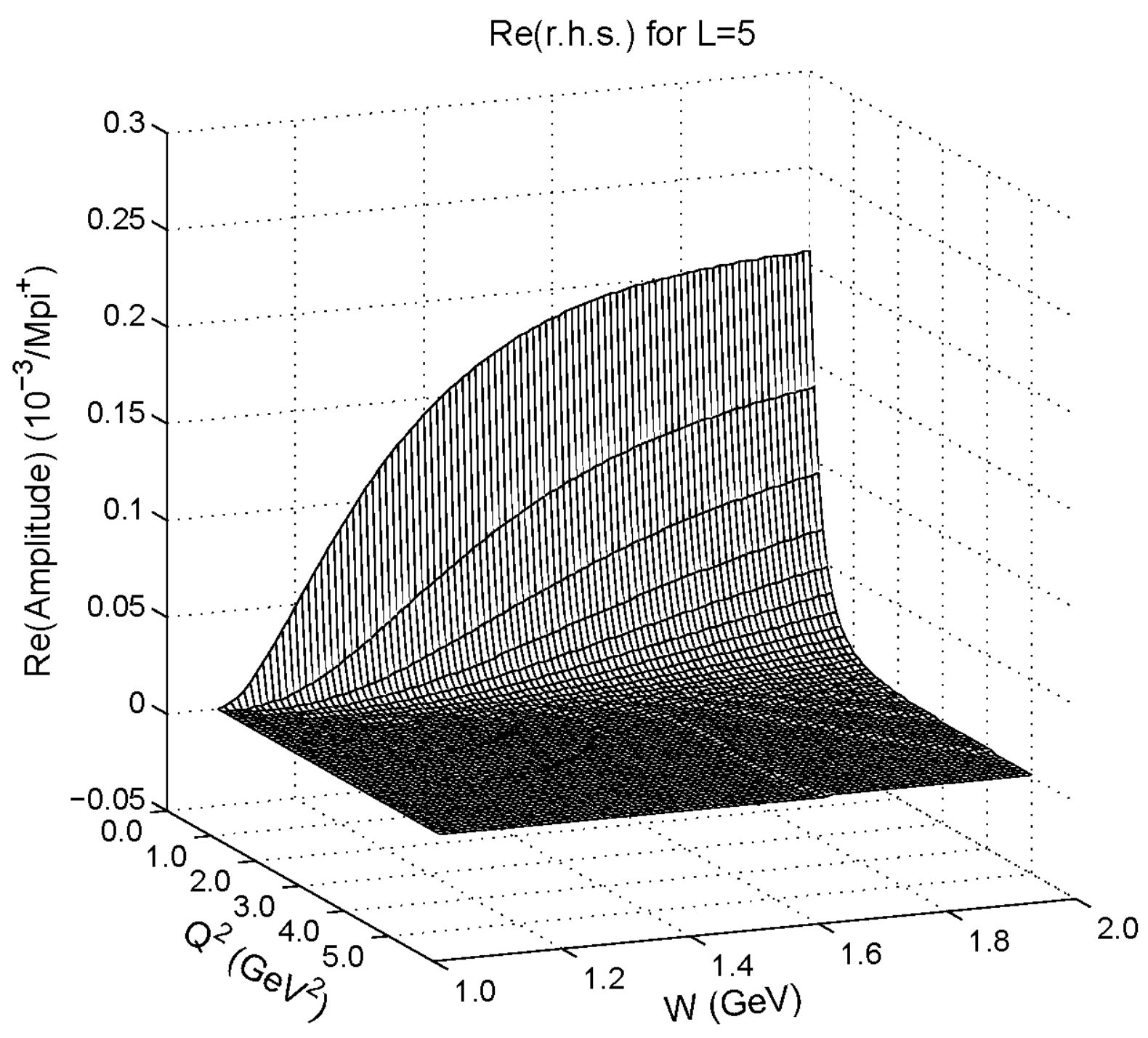}\\[1mm]
\epsfxsize=0.48\textwidth\epsfbox{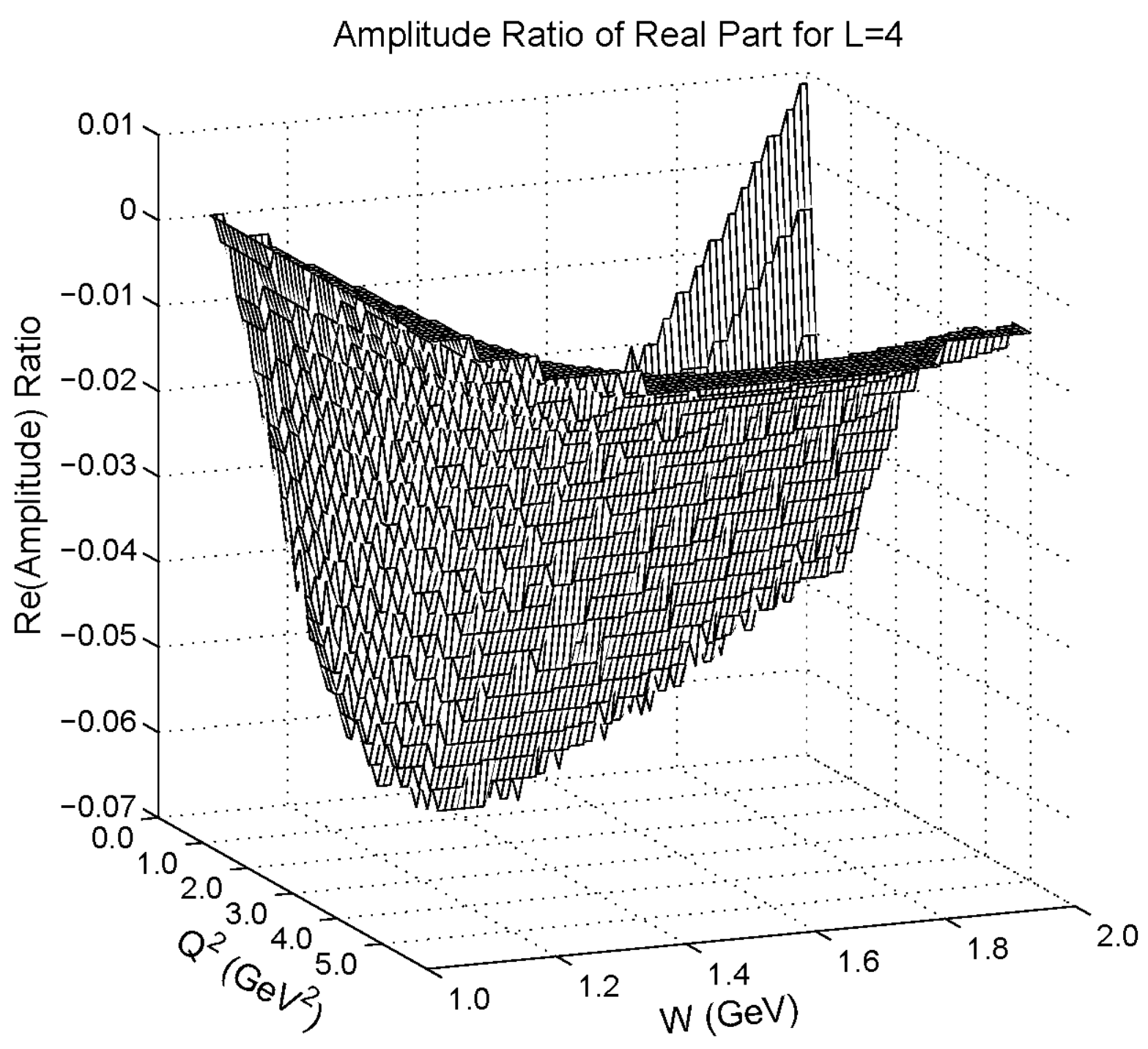}
\epsfxsize=0.48\textwidth\epsfbox{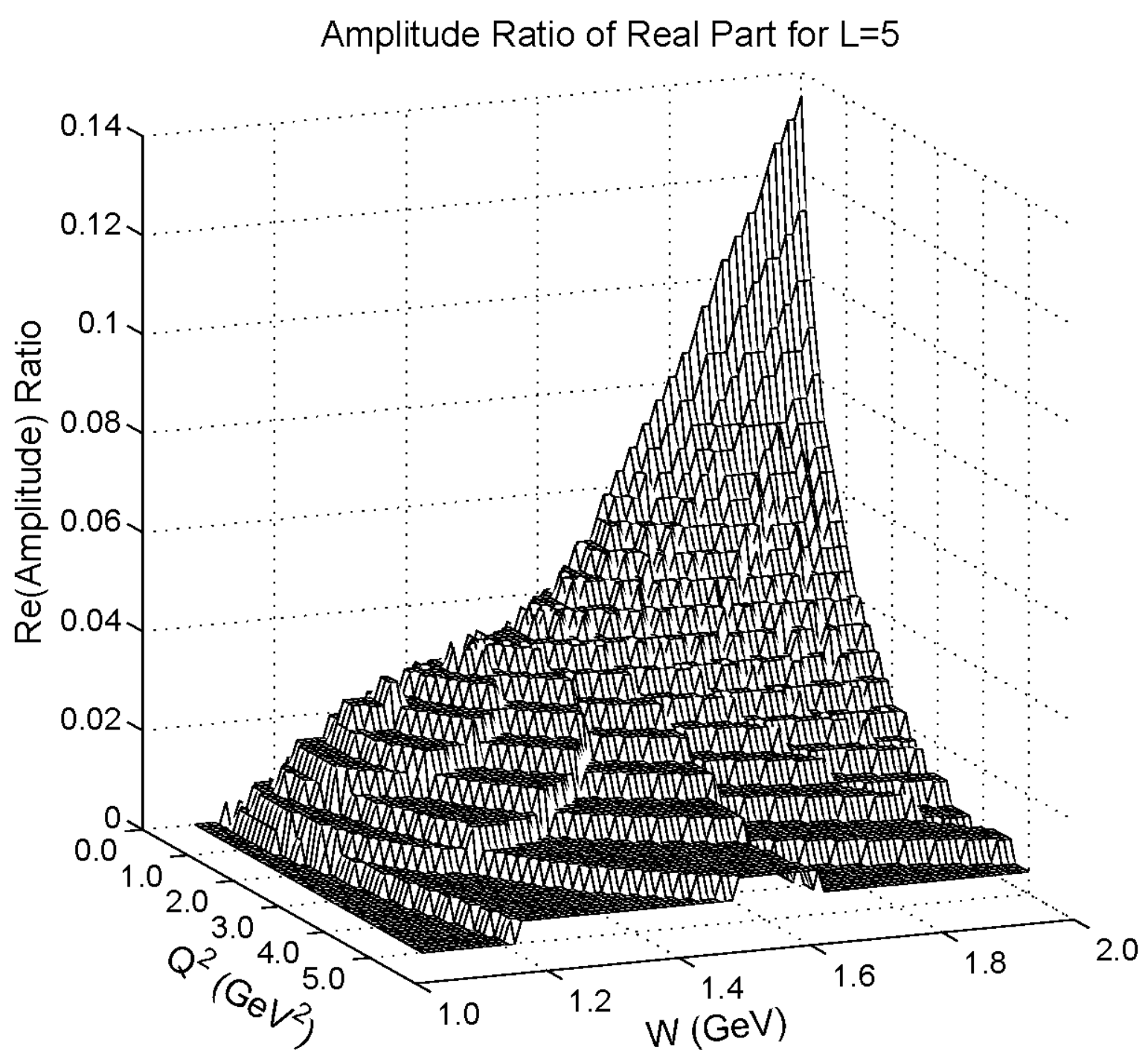}
\end{figure}
%

%
\begin{figure}[htp]
\epsfxsize=0.48\textwidth\epsfbox{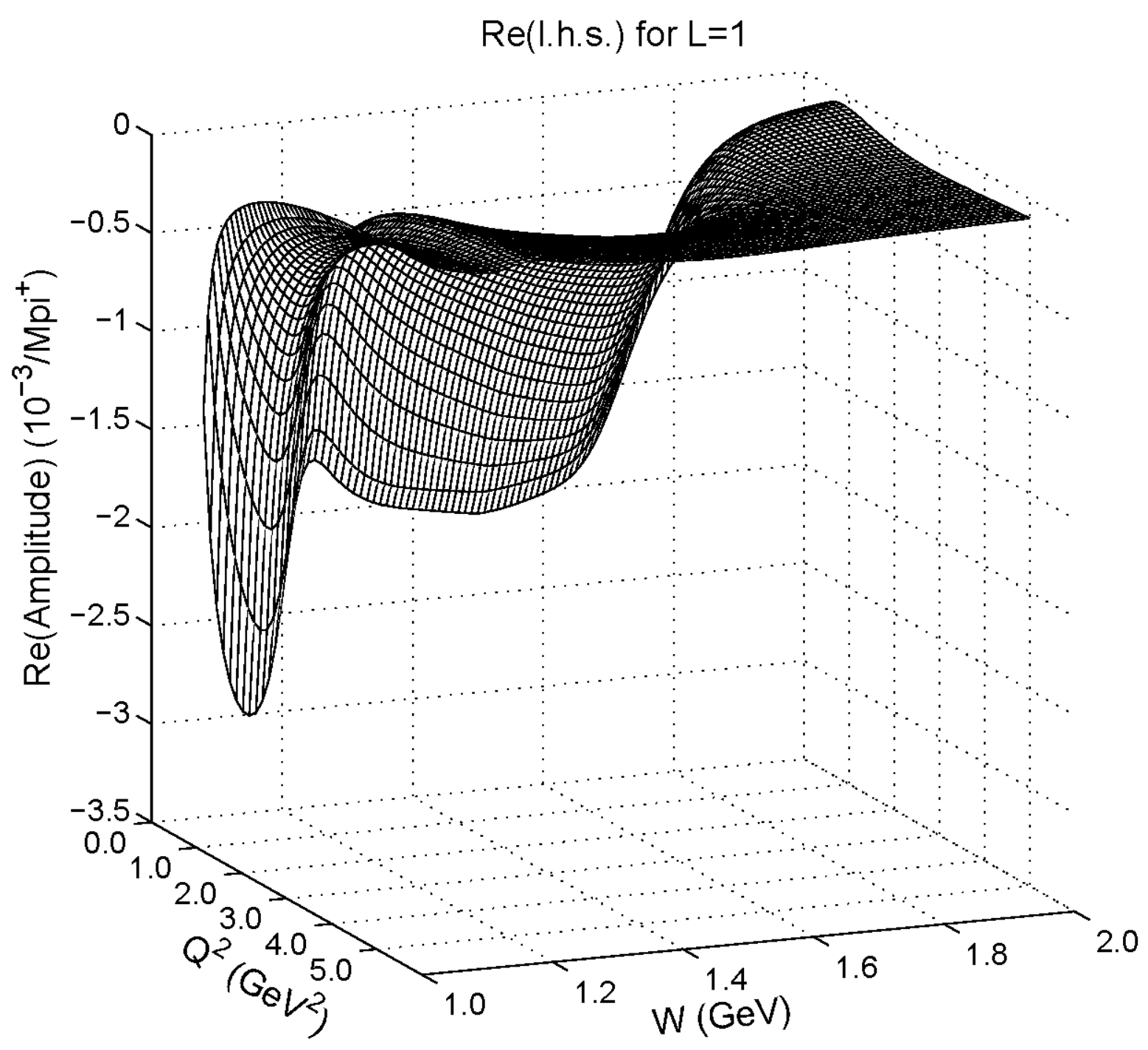}
\epsfxsize=0.48\textwidth\epsfbox{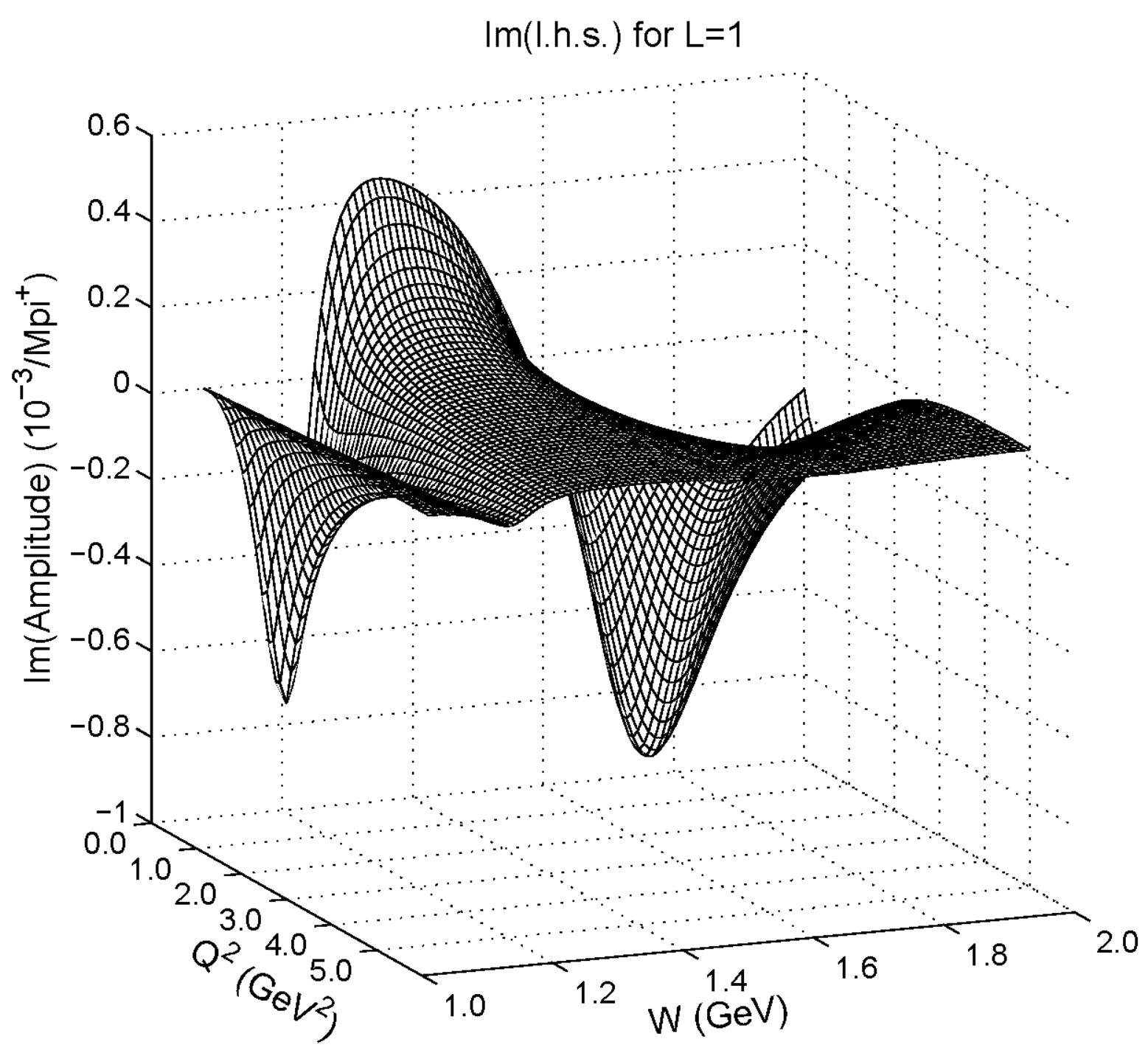}\\[1mm]
\epsfxsize=0.48\textwidth\epsfbox{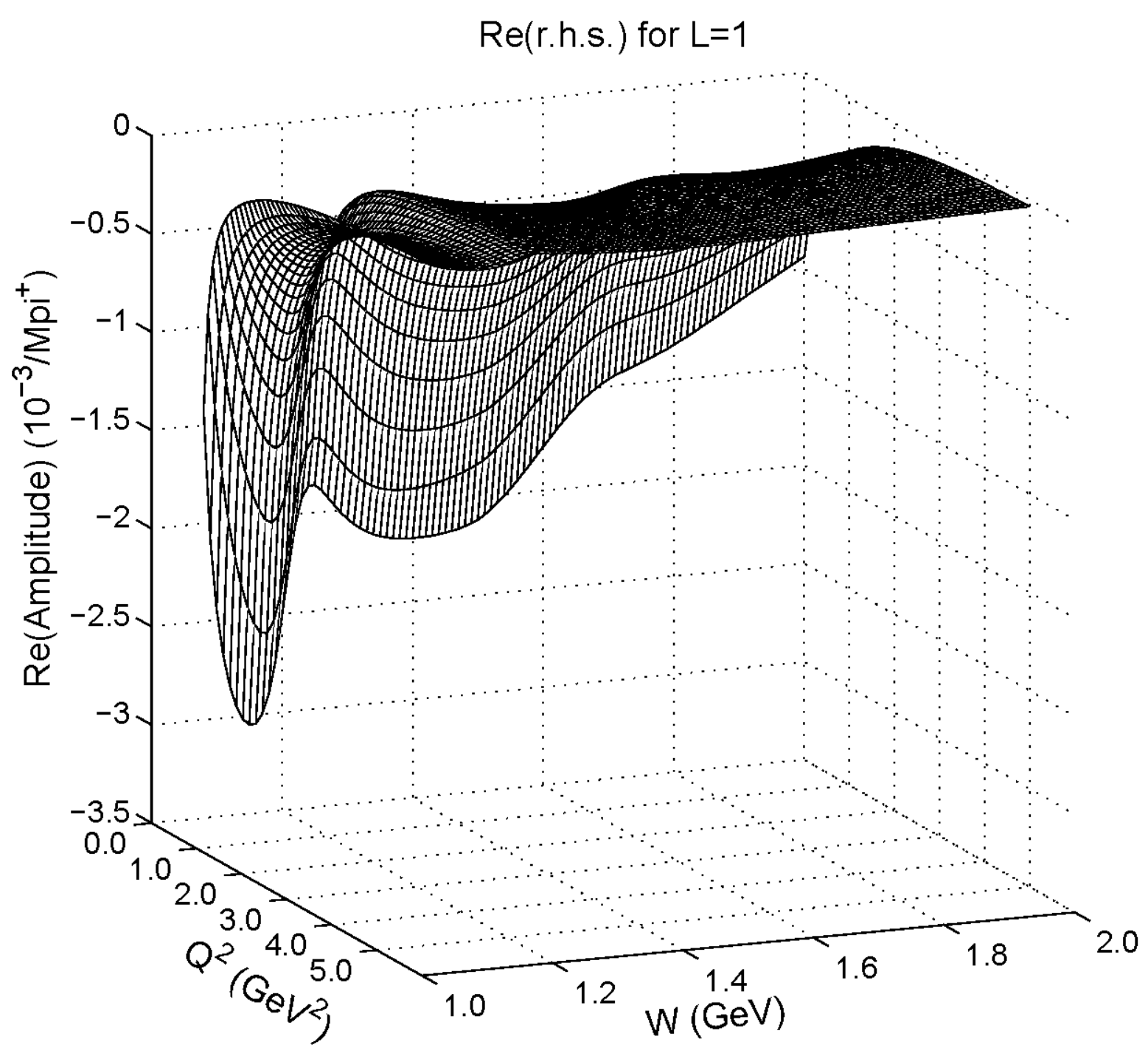}
\epsfxsize=0.48\textwidth\epsfbox{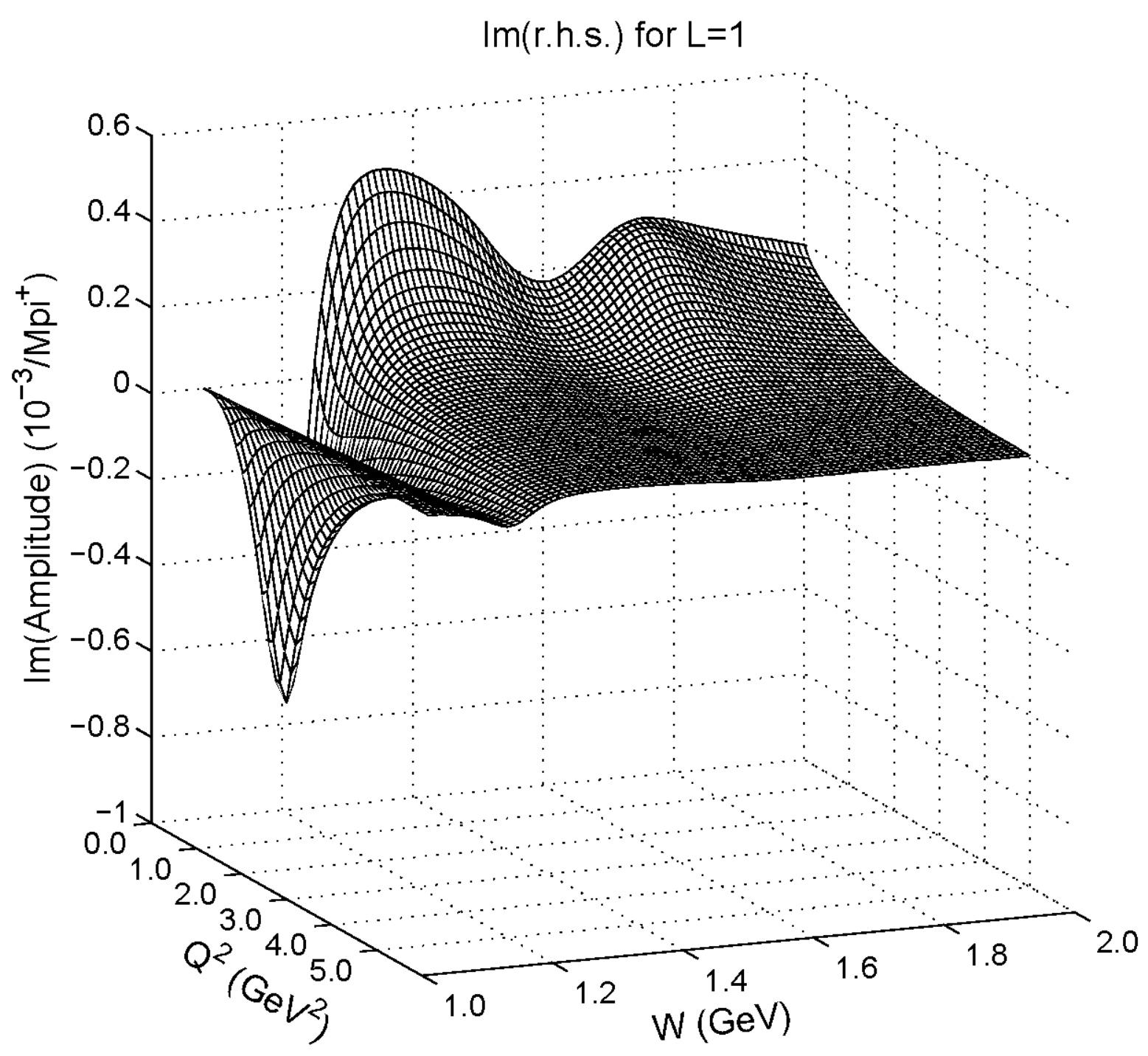}\\[1mm]
\epsfxsize=0.48\textwidth\epsfbox{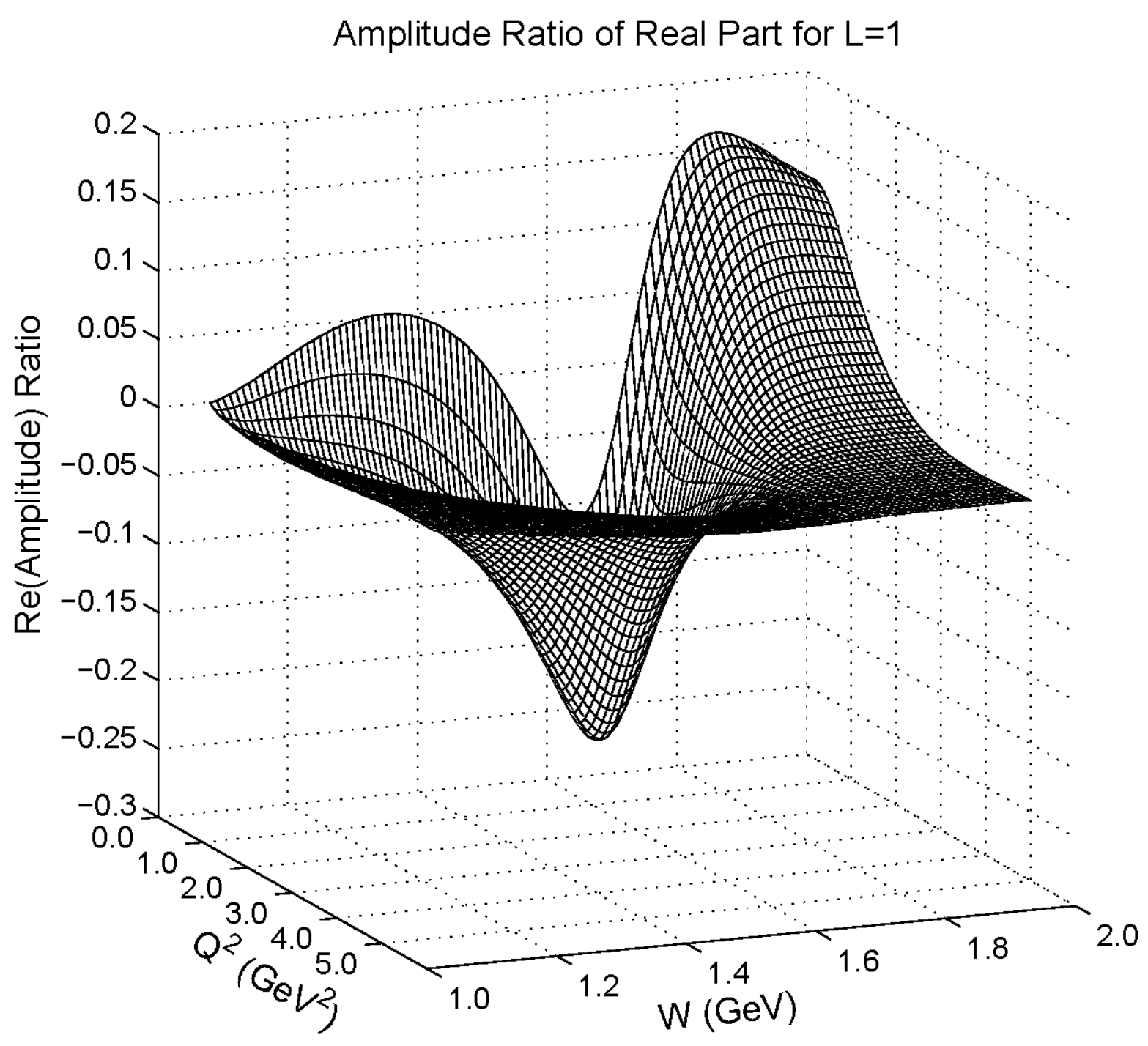}
\epsfxsize=0.48\textwidth\epsfbox{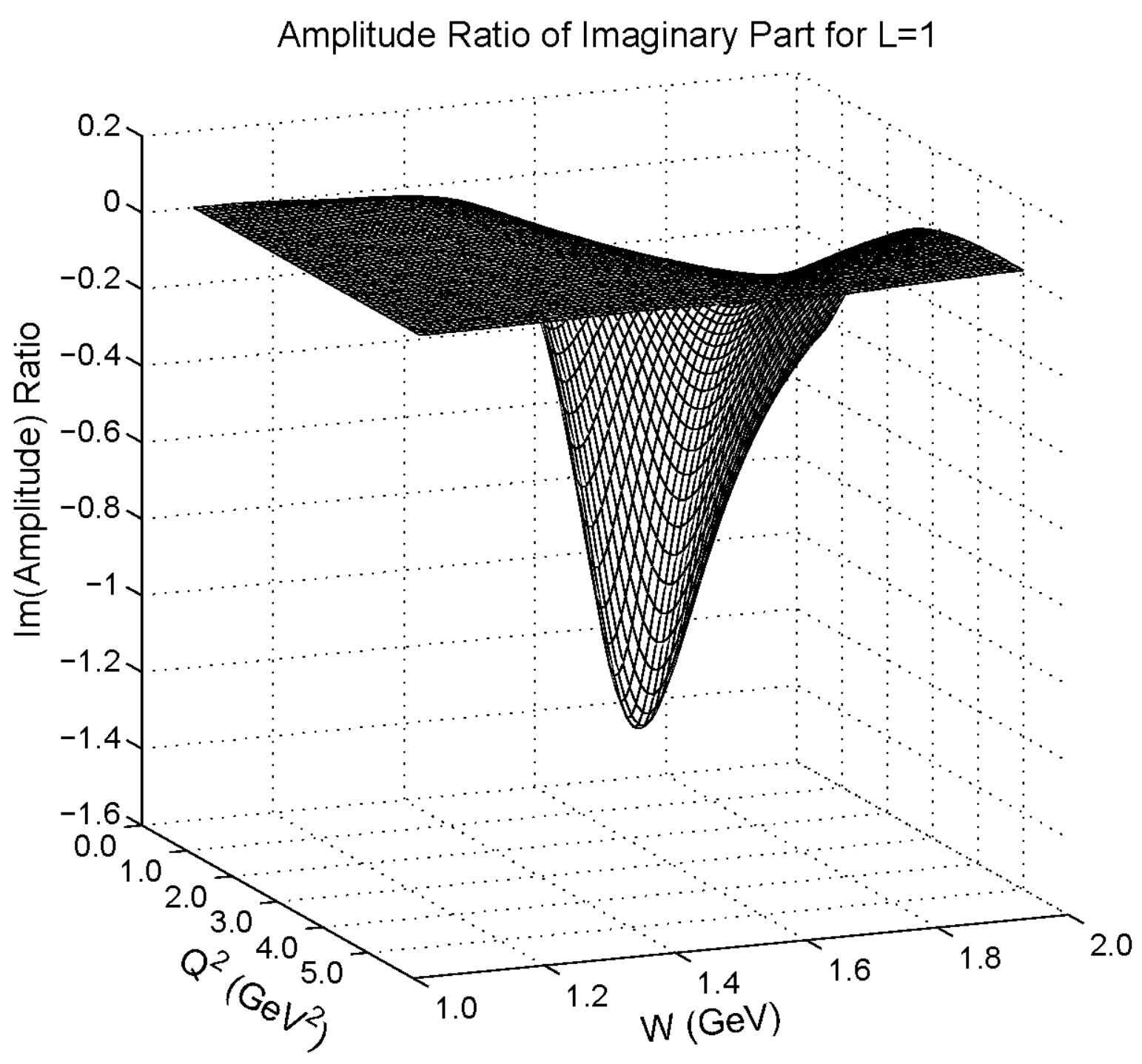}\\
\end{figure}
\begin{figure}[htp]
\epsfxsize=0.48\textwidth\epsfbox{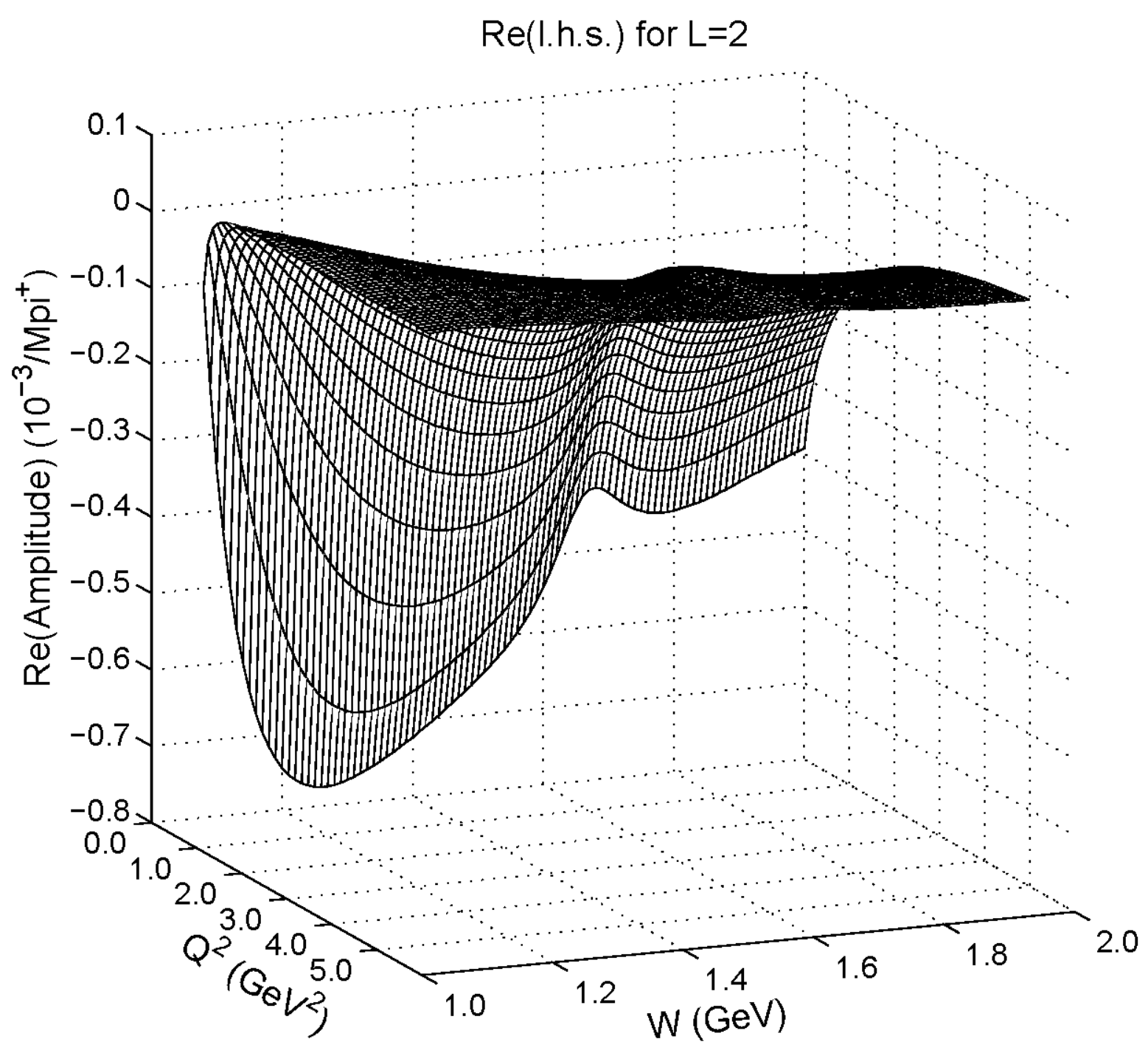}
\epsfxsize=0.48\textwidth\epsfbox{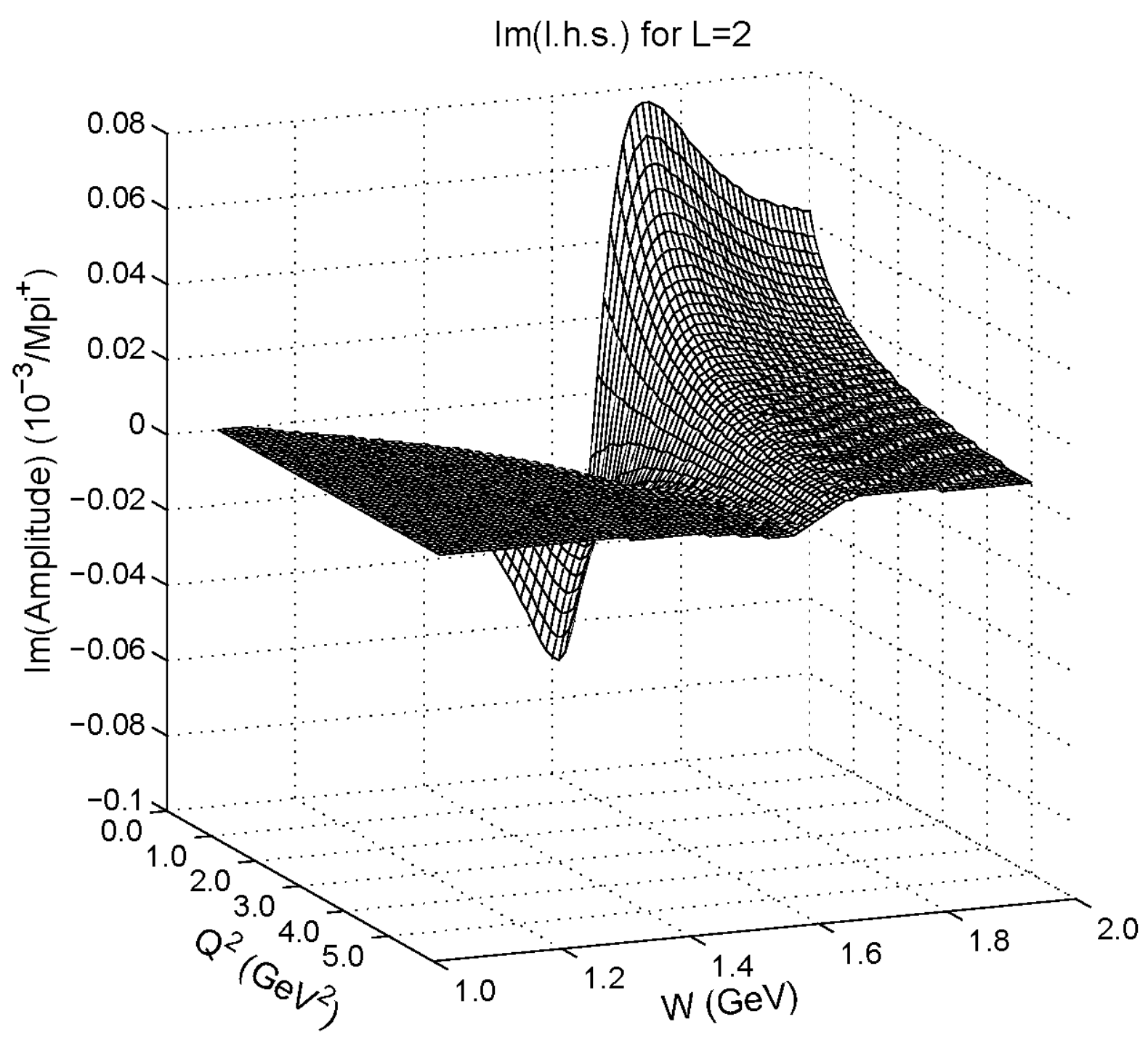}\\[1mm]
\epsfxsize=0.48\textwidth\epsfbox{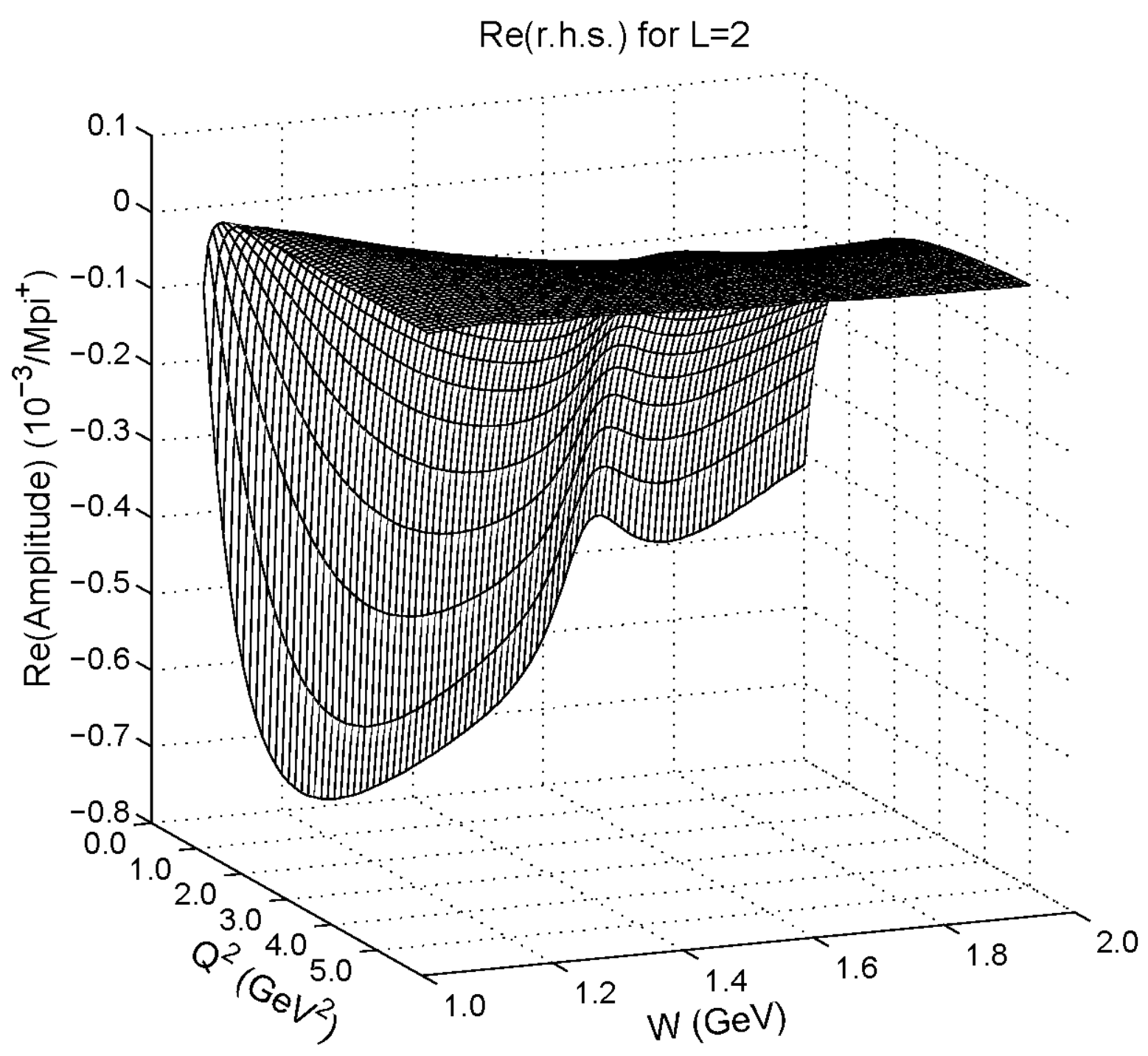}
\epsfxsize=0.48\textwidth\epsfbox{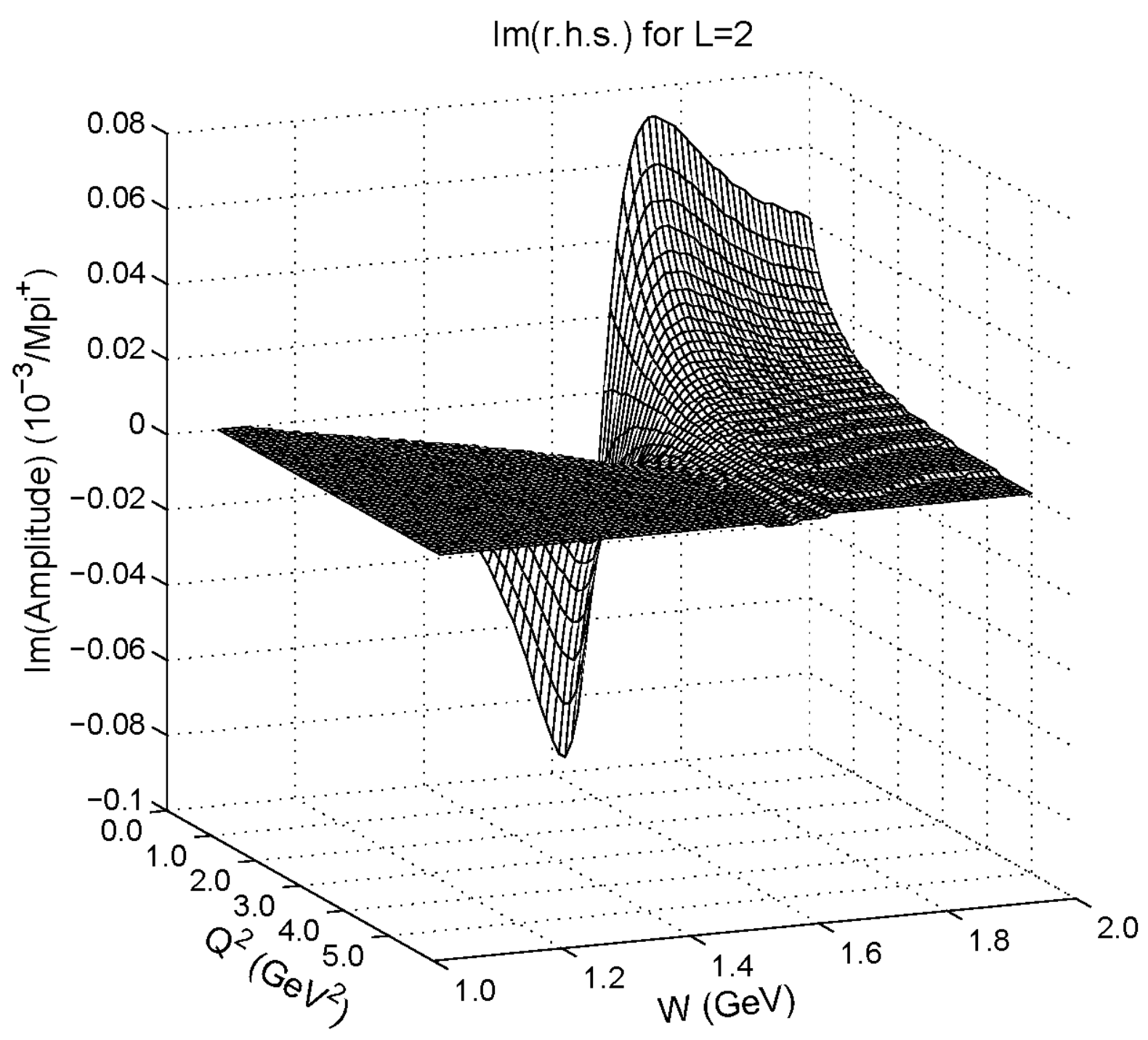}\\[1mm]
\epsfxsize=0.48\textwidth\epsfbox{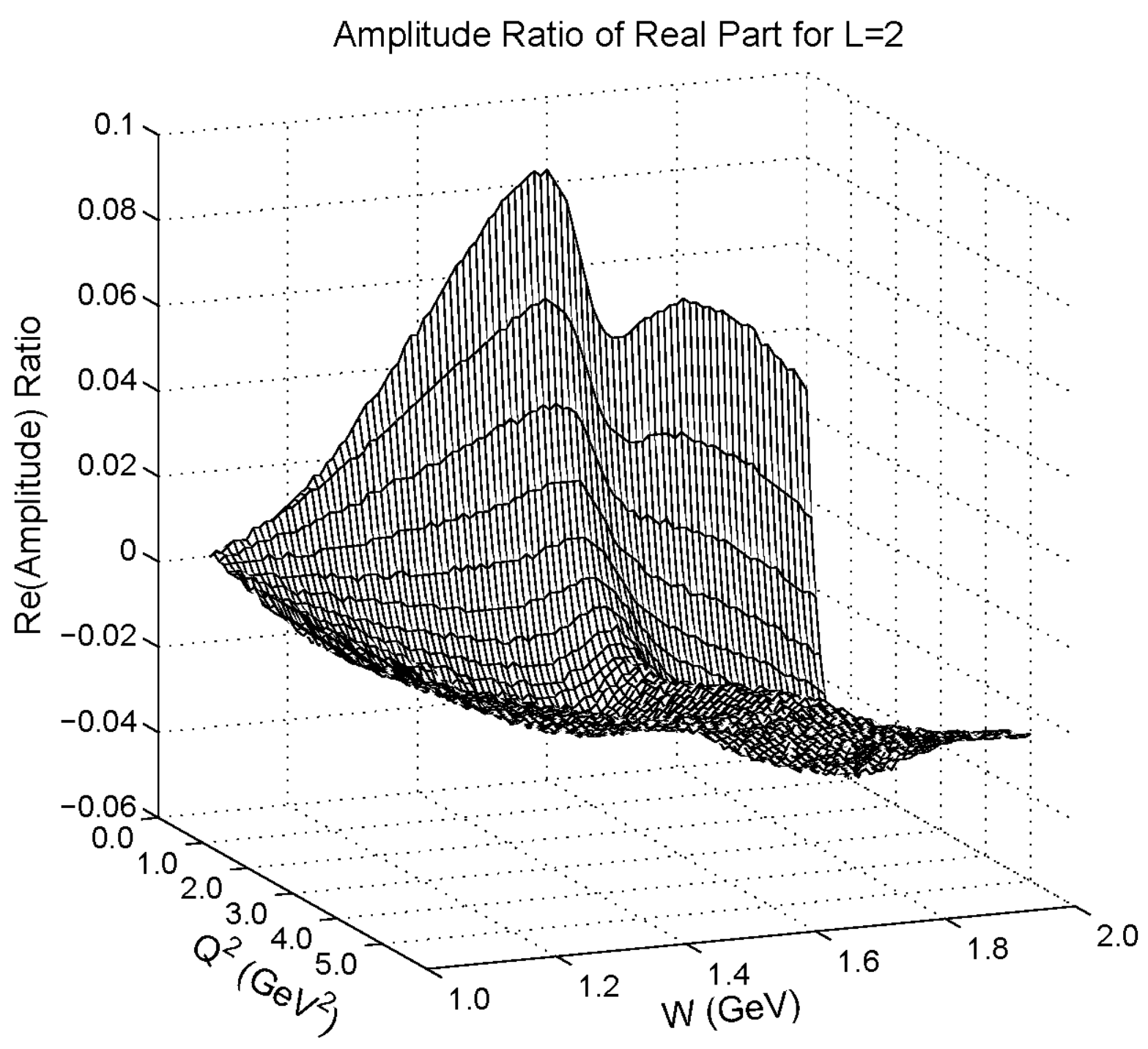}
\epsfxsize=0.48\textwidth\epsfbox{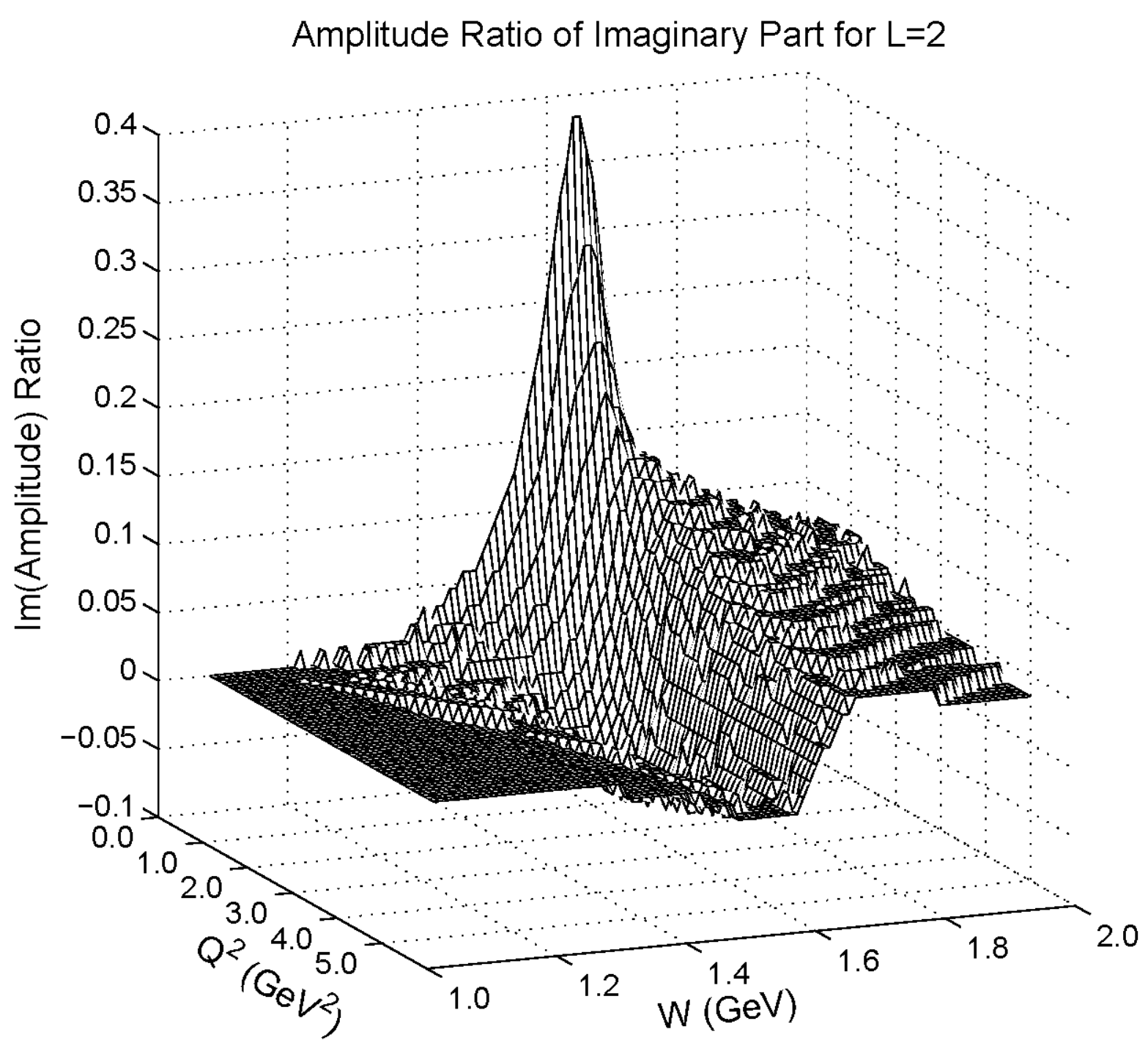}\\
\end{figure}
\begin{figure}[htp]
\epsfxsize=0.48\textwidth\epsfbox{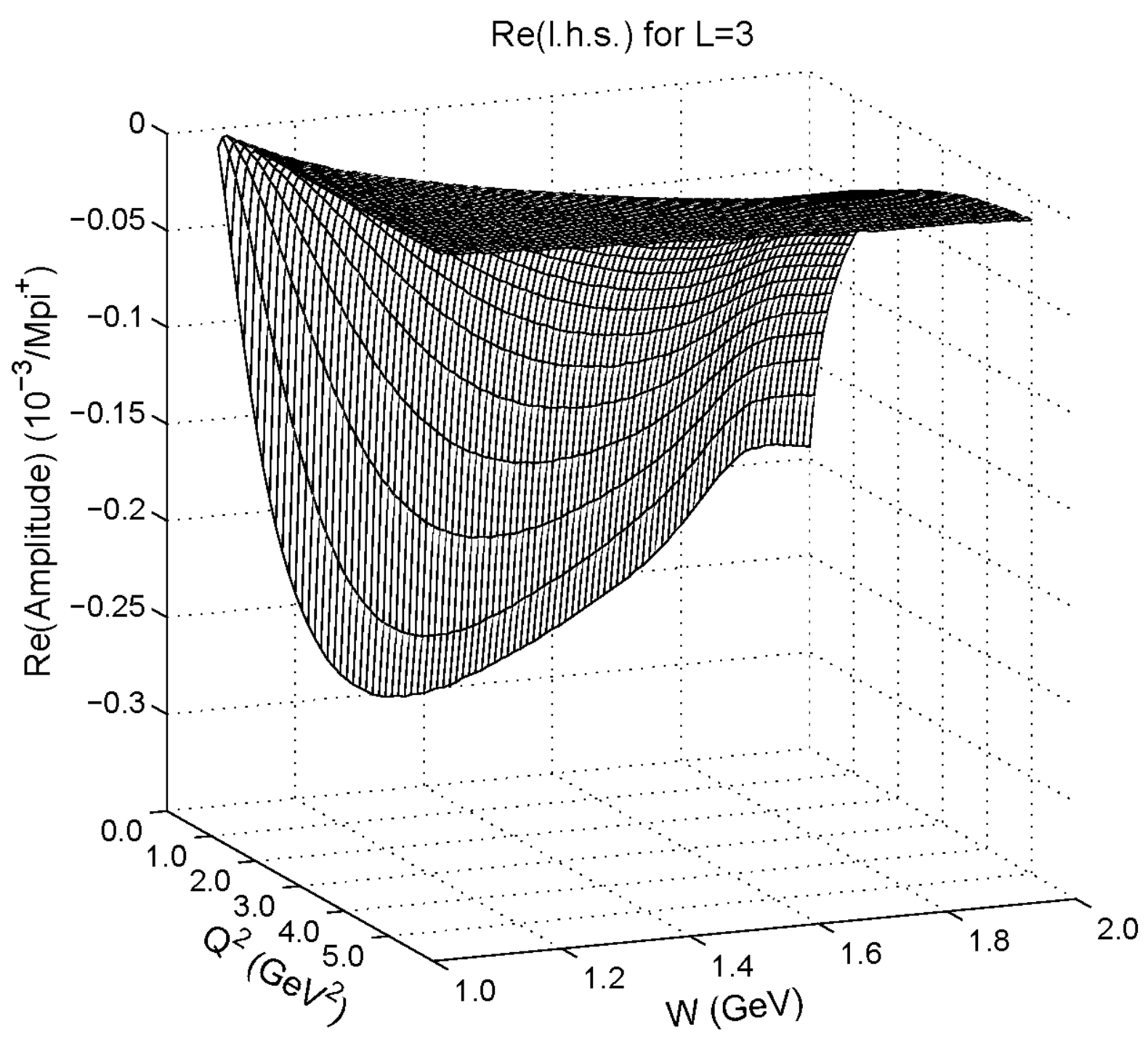}
\epsfxsize=0.48\textwidth\epsfbox{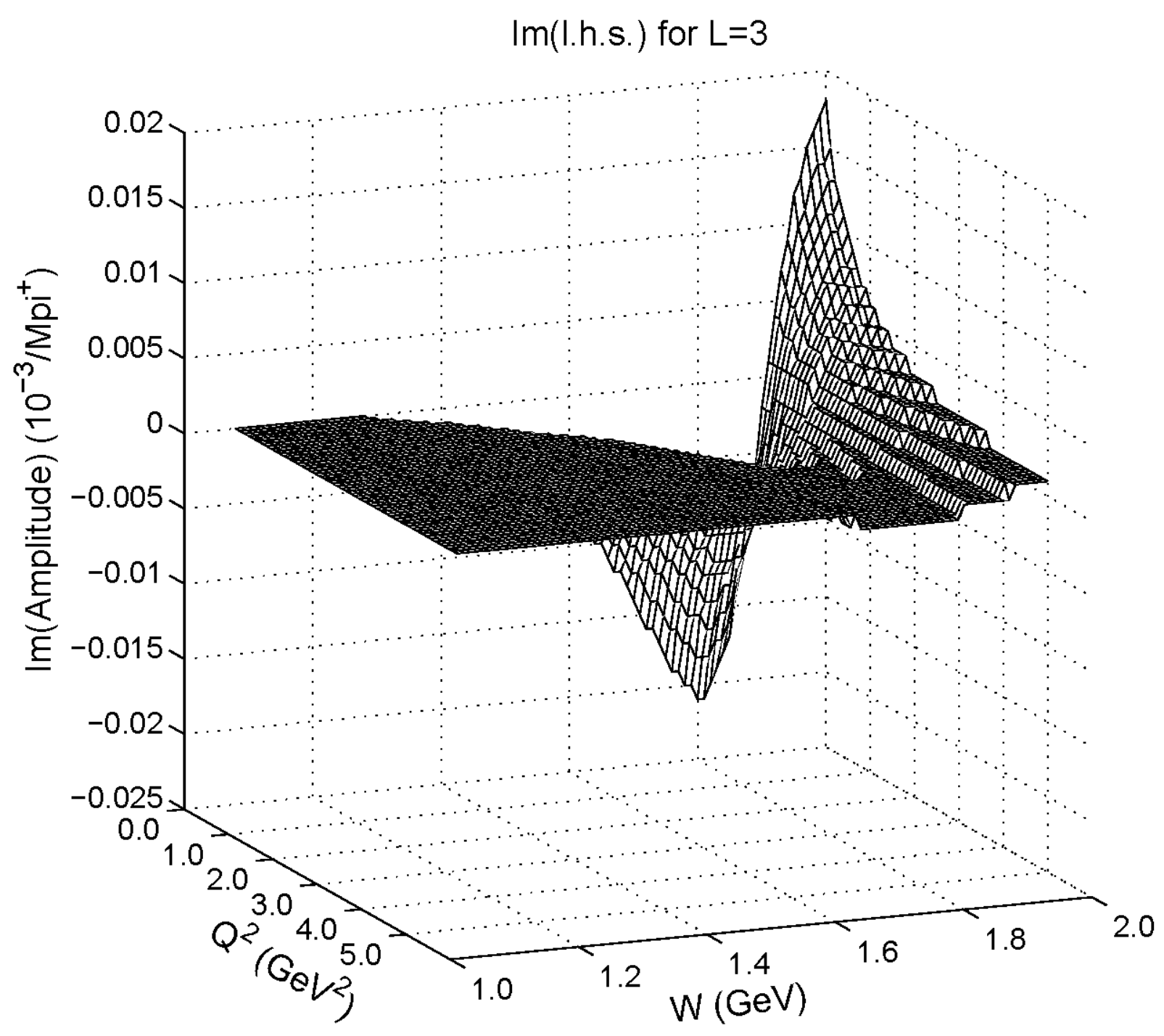}\\[1mm]
\epsfxsize=0.48\textwidth\epsfbox{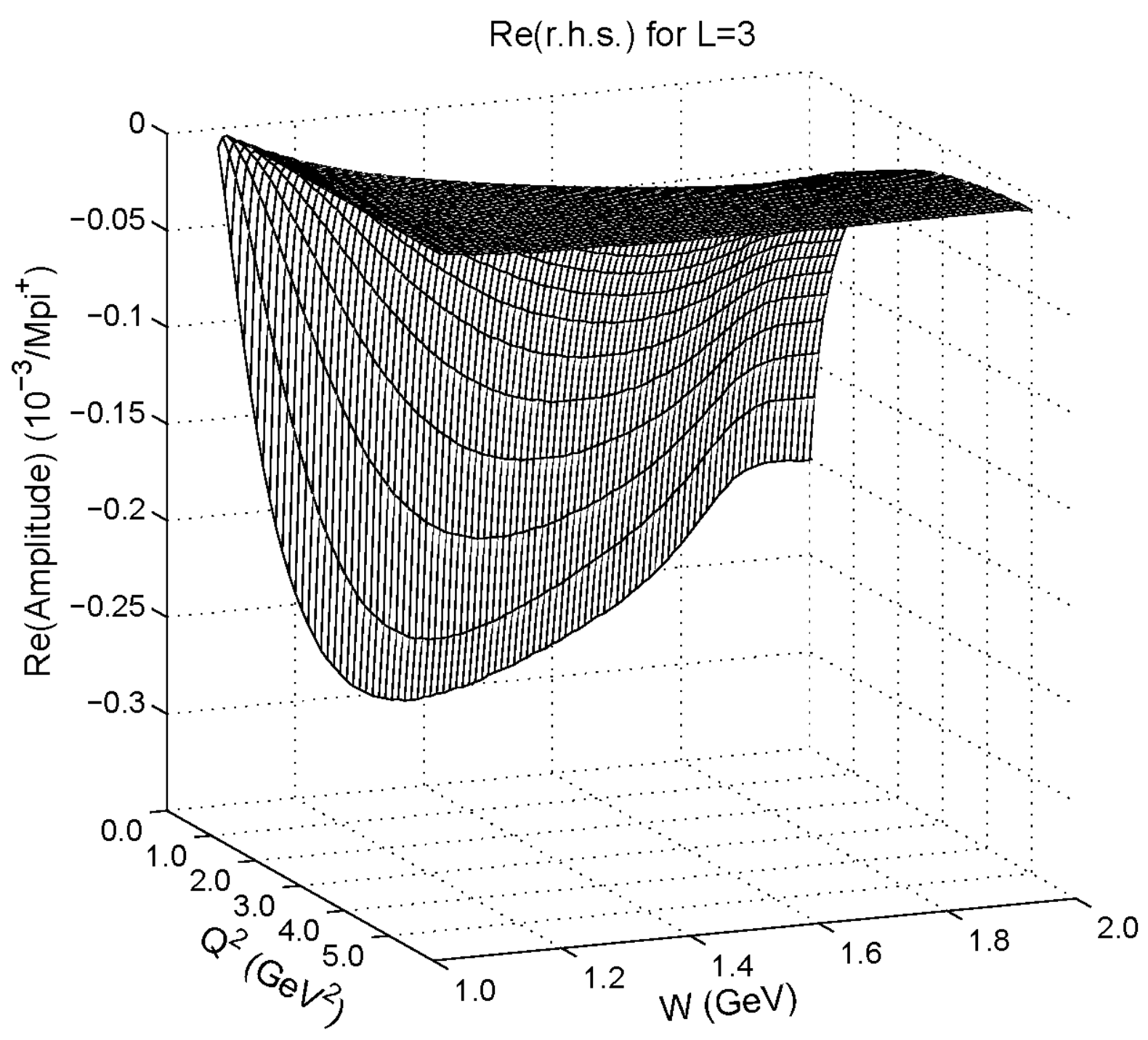}
\epsfxsize=0.48\textwidth\epsfbox{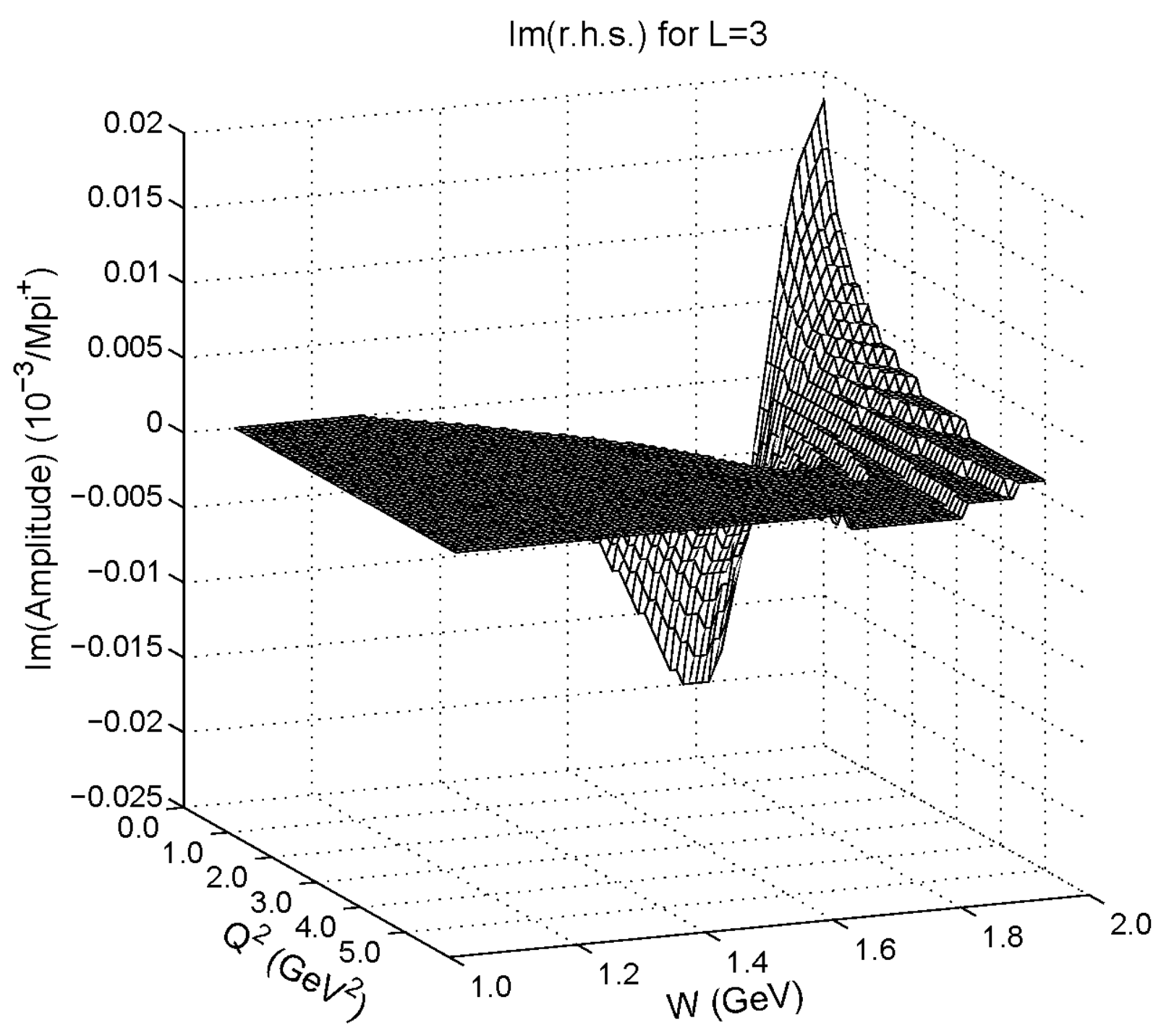}\\[1mm]
\epsfxsize=0.48\textwidth\epsfbox{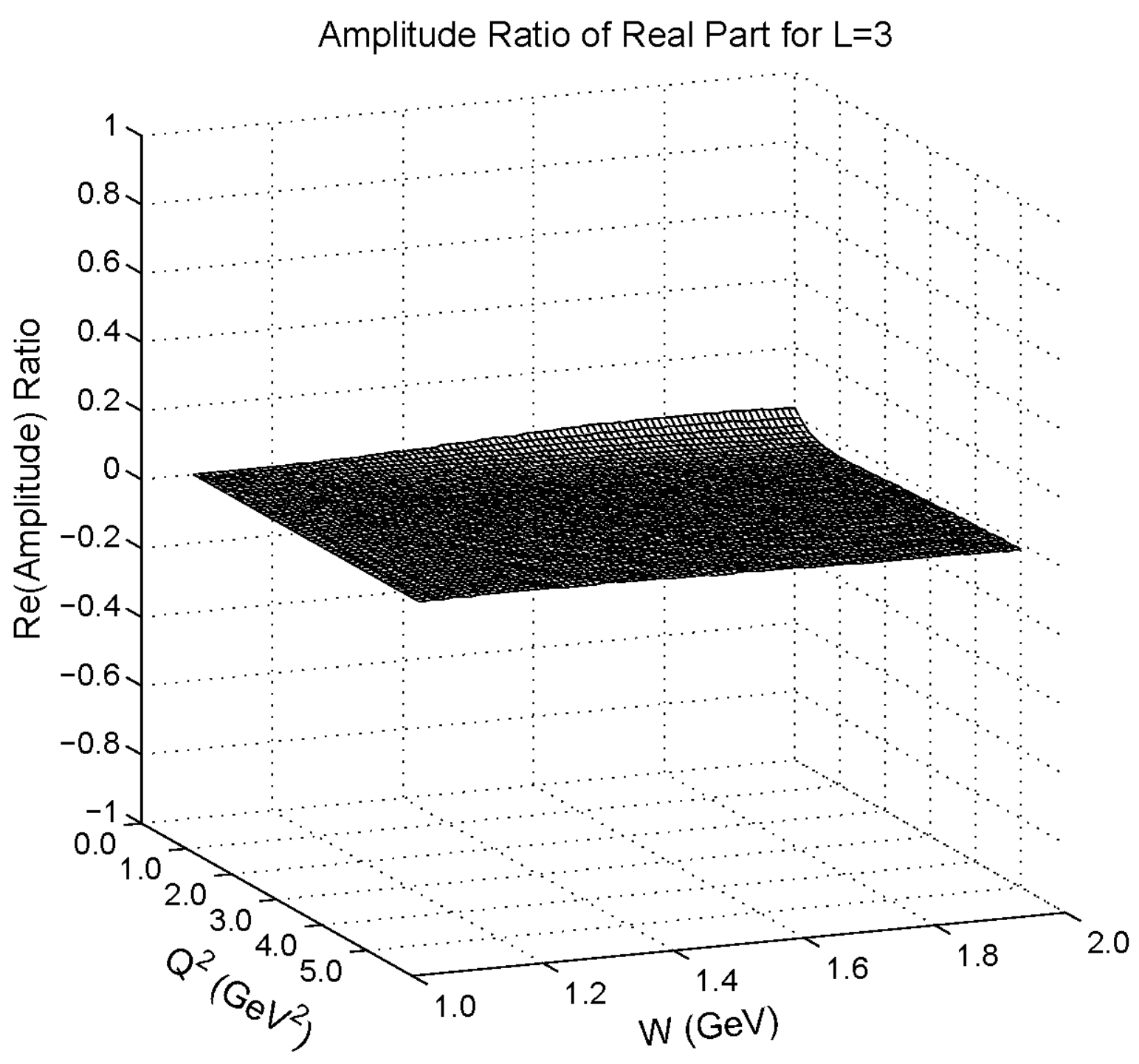}
\epsfxsize=0.48\textwidth\epsfbox{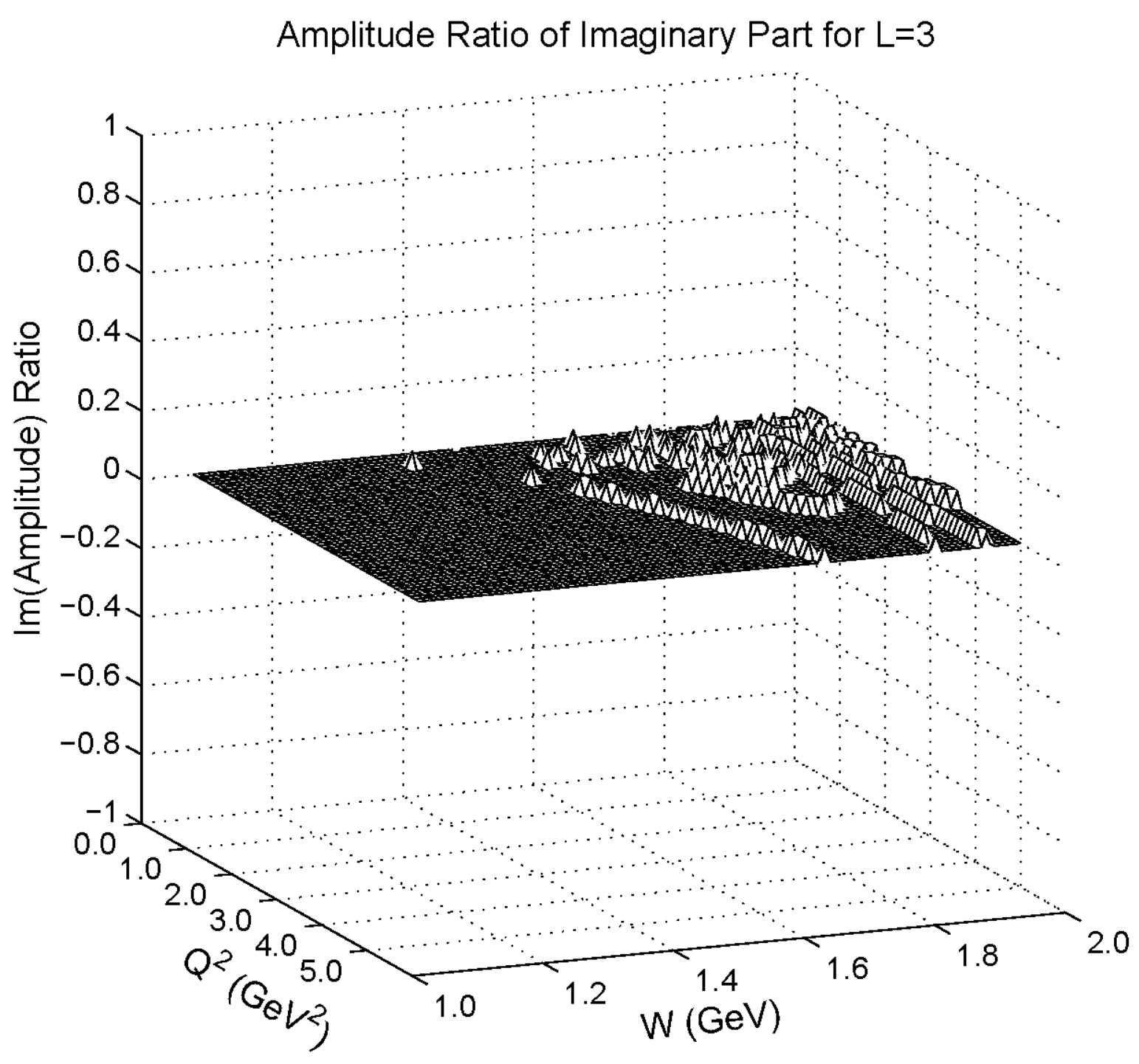}\\
\end{figure}
\begin{figure}[htp]
\epsfxsize=0.48\textwidth\epsfbox{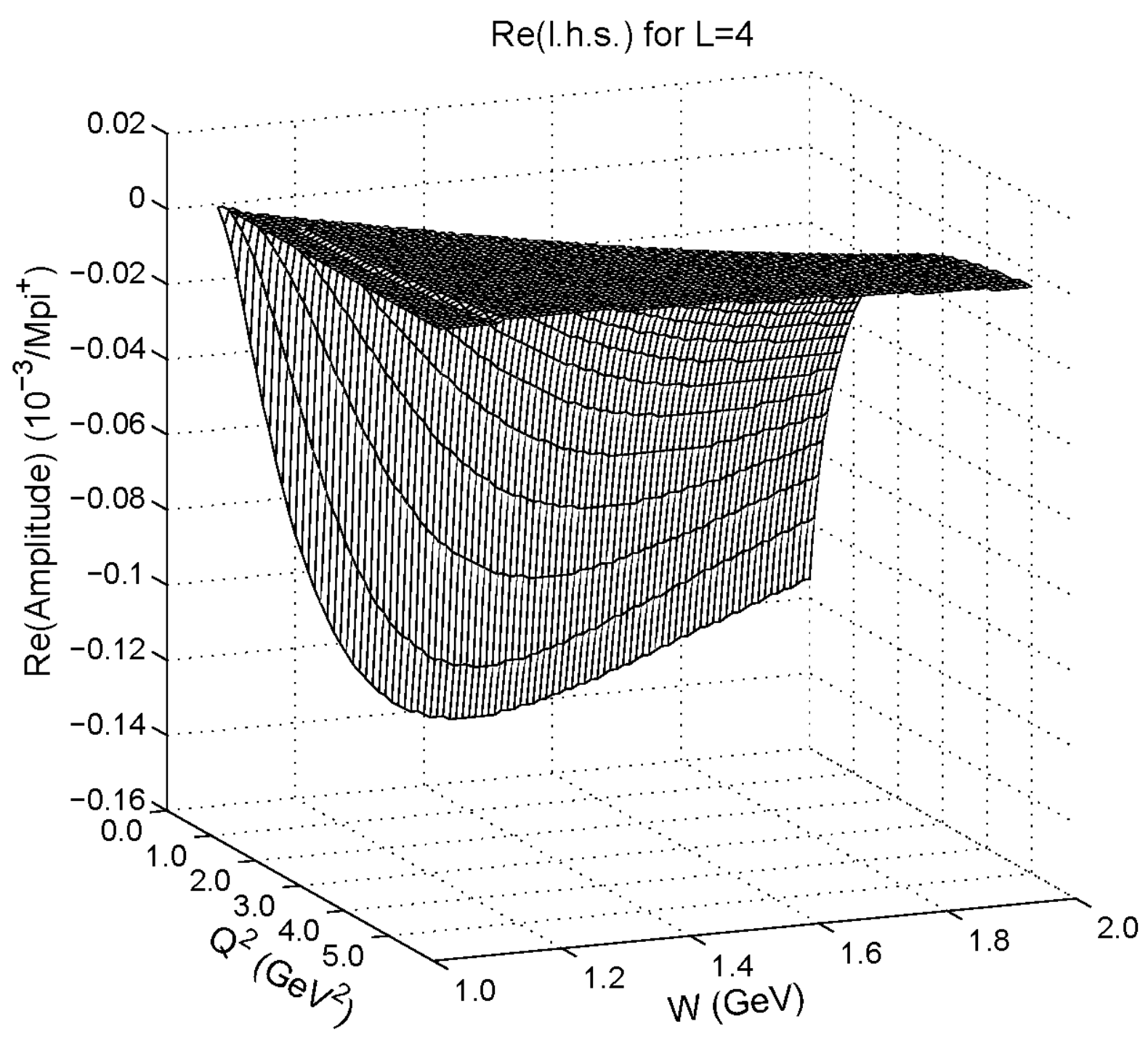}
\epsfxsize=0.48\textwidth\epsfbox{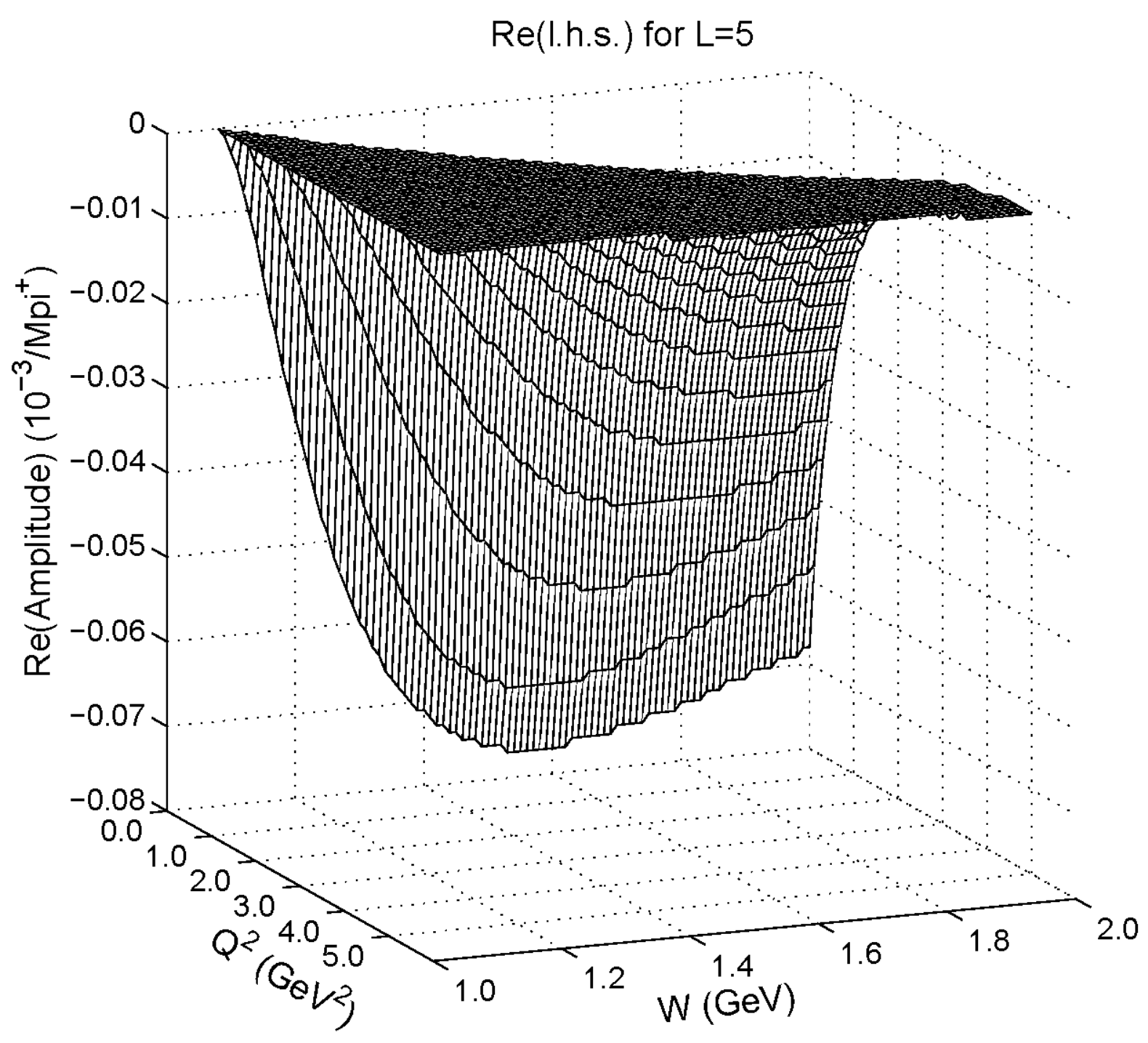}\\[1mm]
\epsfxsize=0.48\textwidth\epsfbox{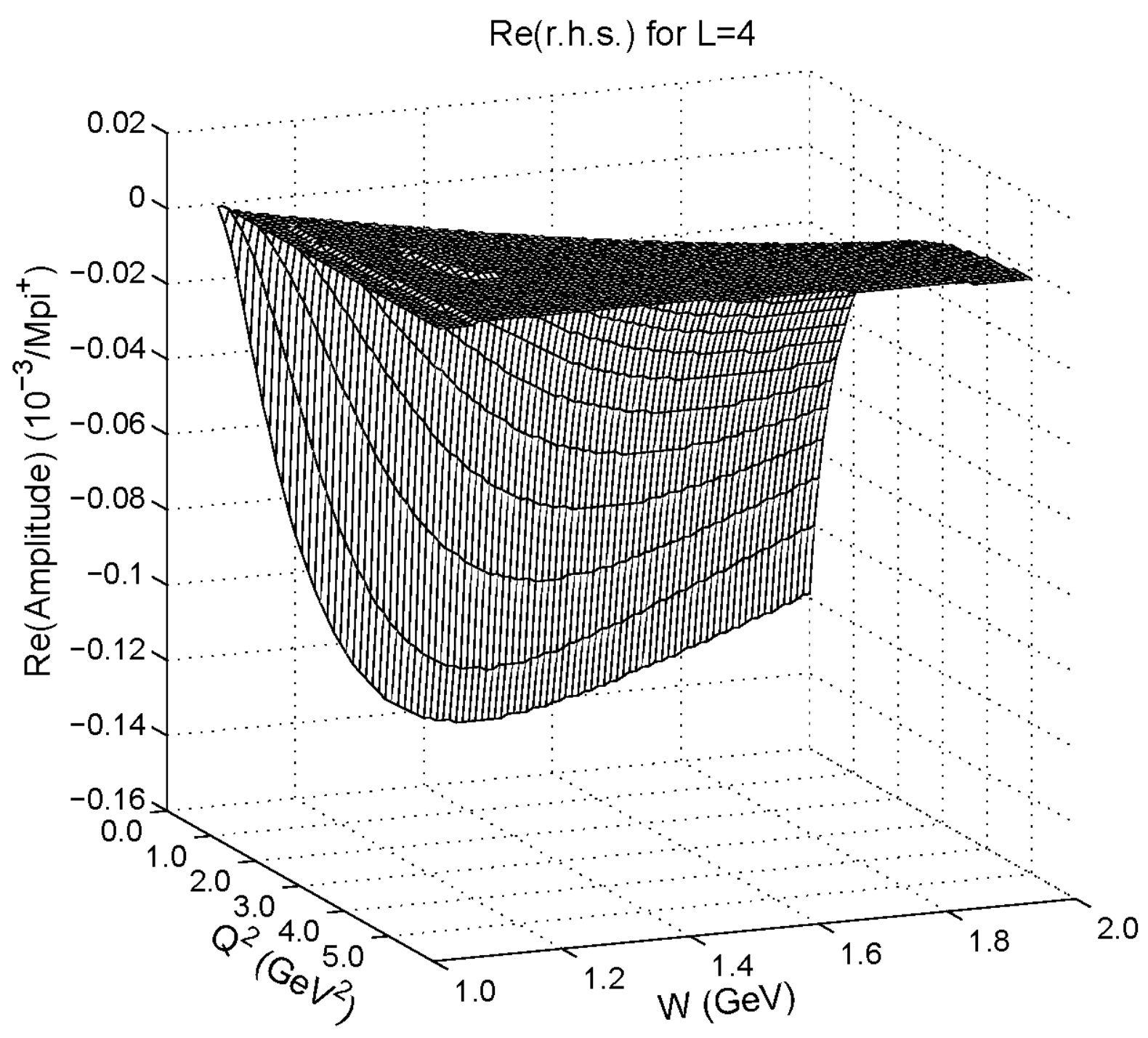}
\epsfxsize=0.48\textwidth\epsfbox{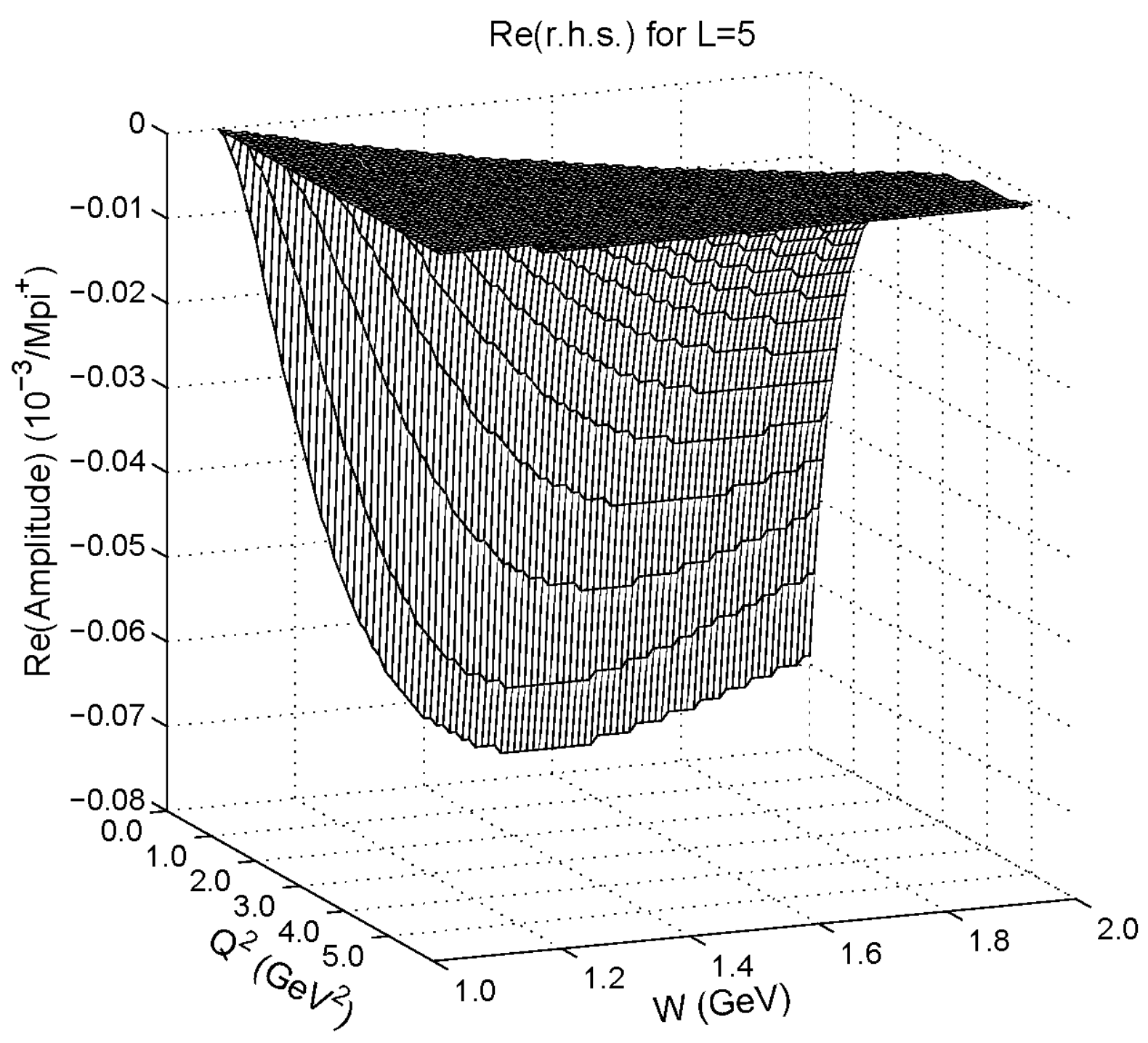}\\[1mm]
\epsfxsize=0.48\textwidth\epsfbox{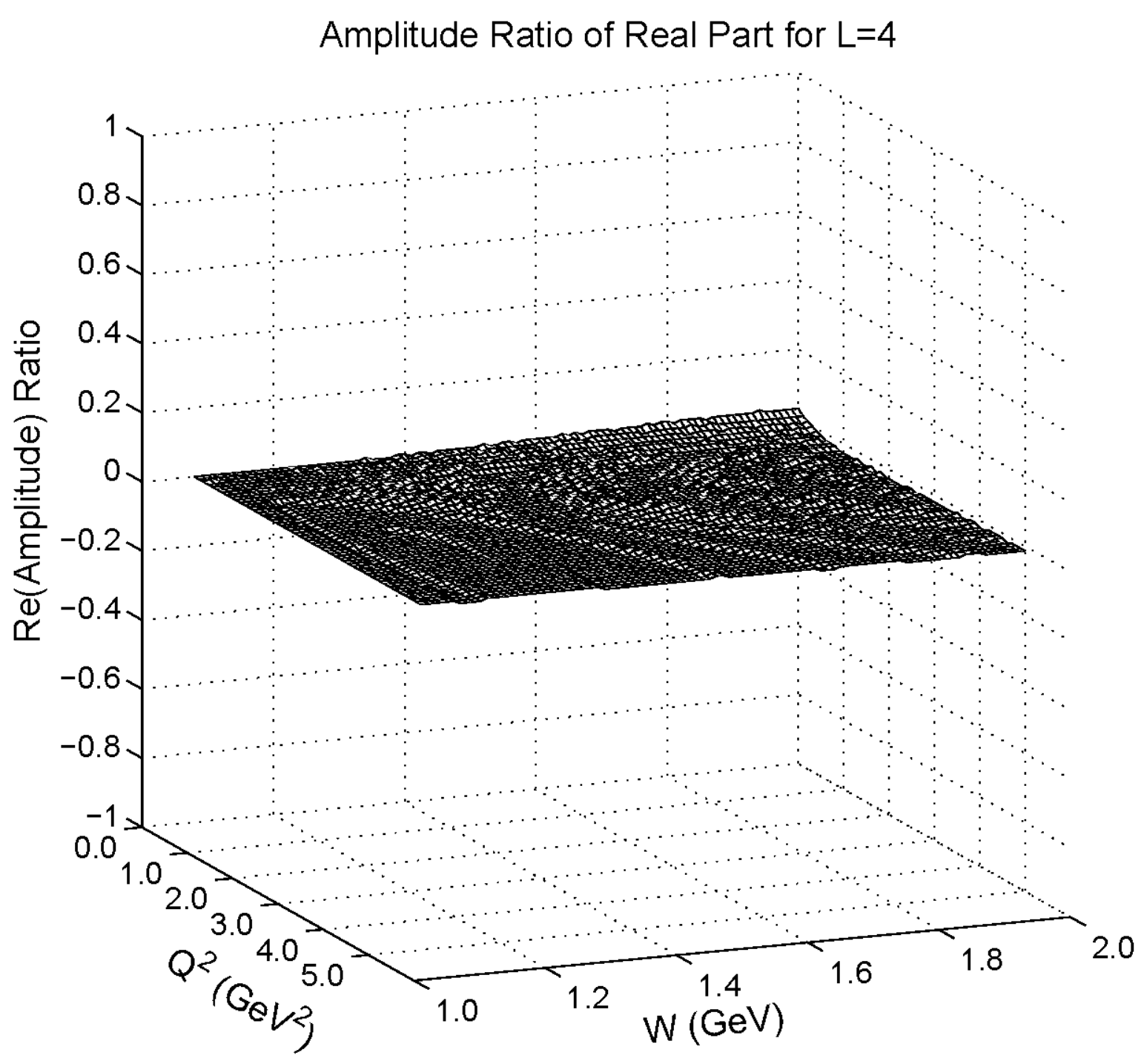}
\epsfxsize=0.48\textwidth\epsfbox{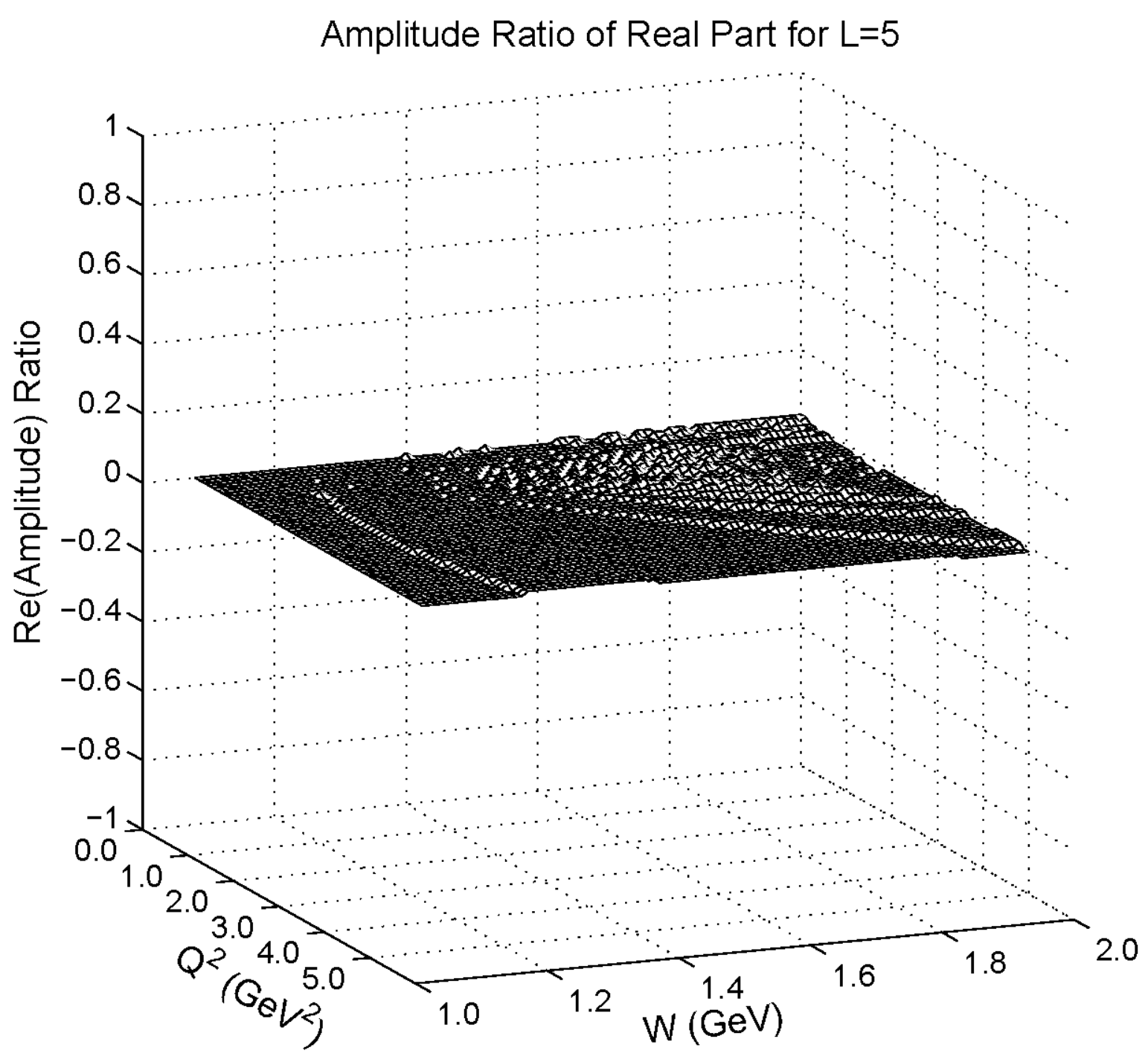}
\end{figure}
%
%
%
\begin{figure}[htp]
\epsfxsize=0.48\textwidth\epsfbox{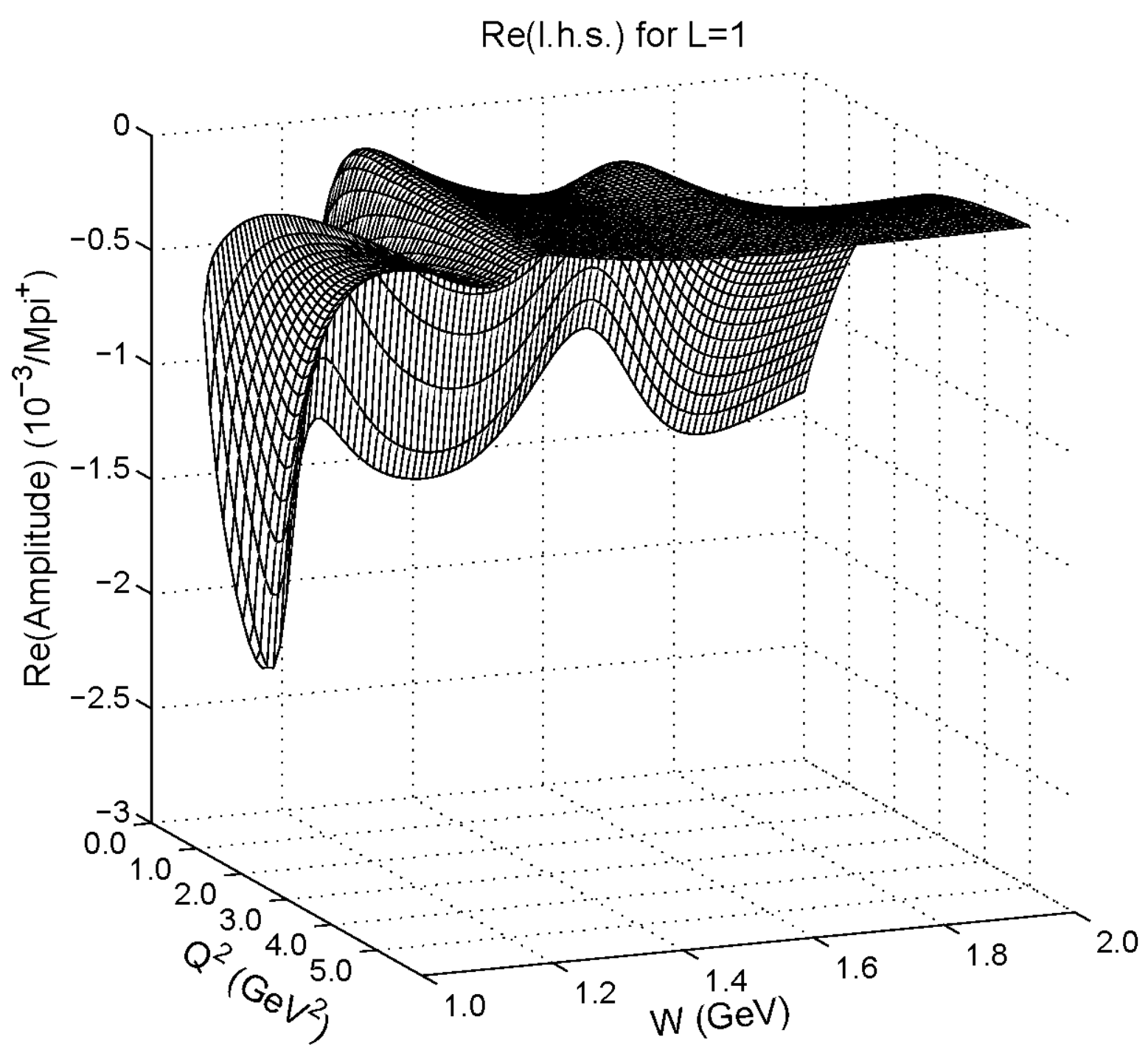}
\epsfxsize=0.48\textwidth\epsfbox{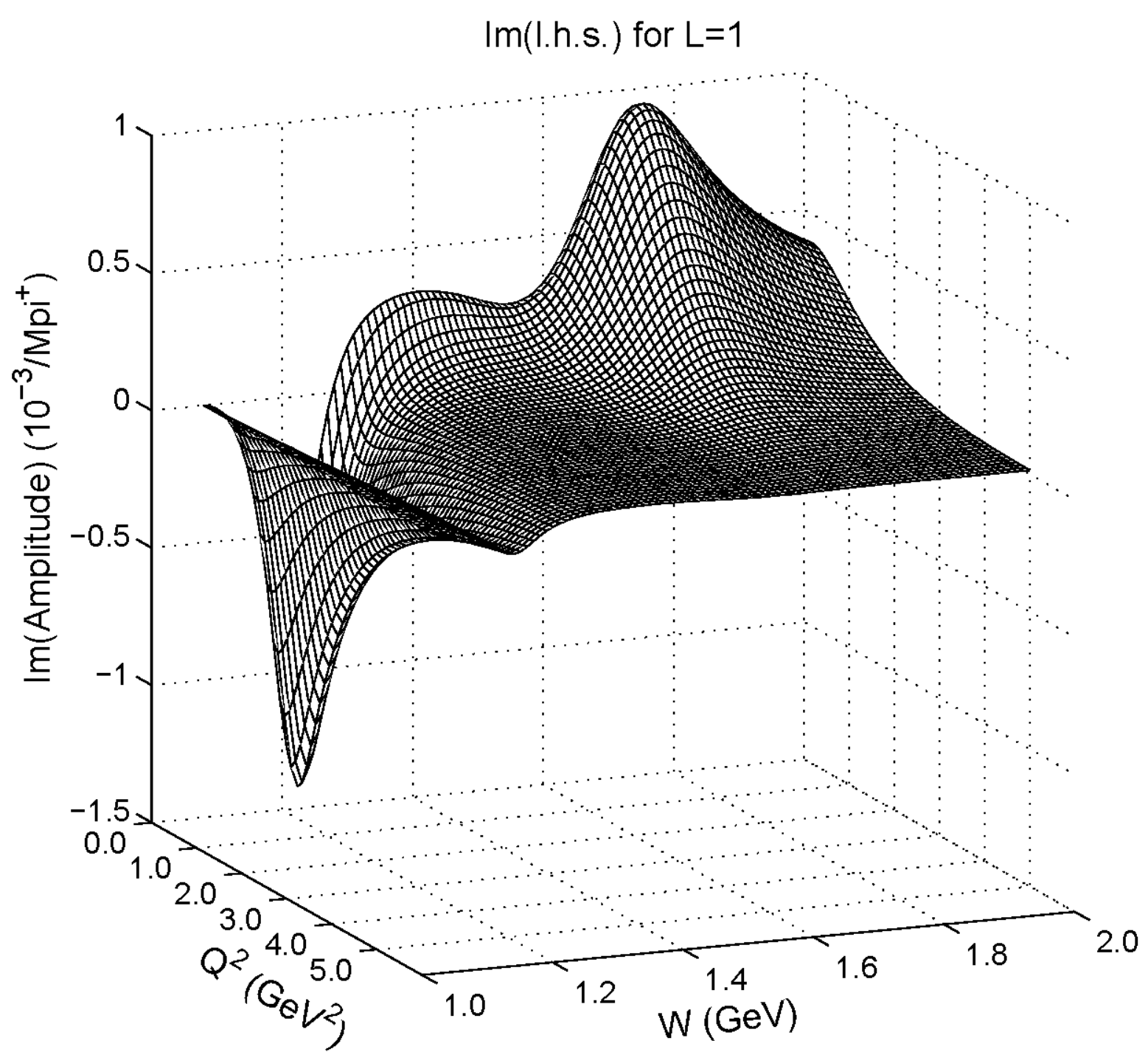}\\[1mm]
\epsfxsize=0.48\textwidth\epsfbox{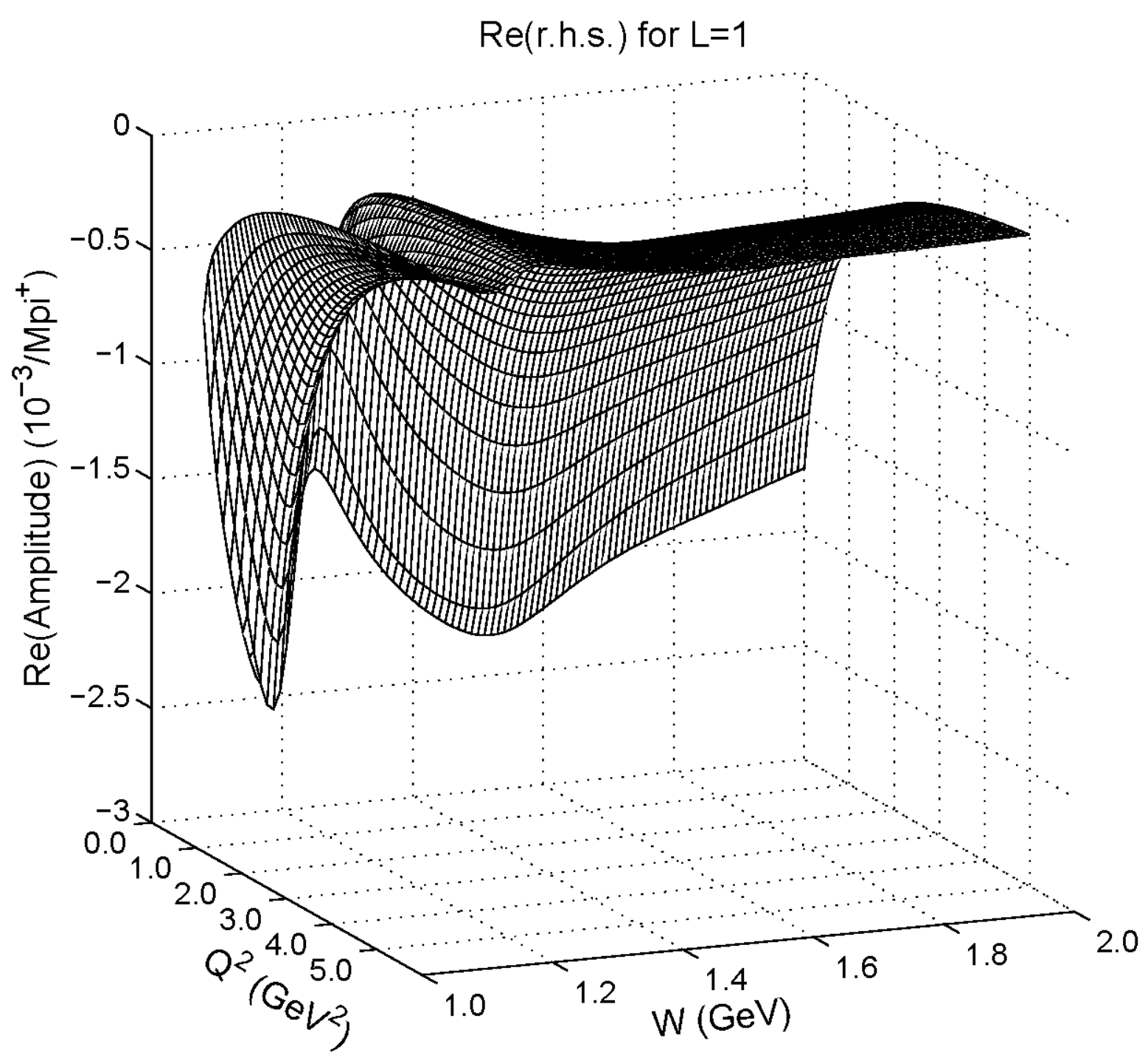}
\epsfxsize=0.48\textwidth\epsfbox{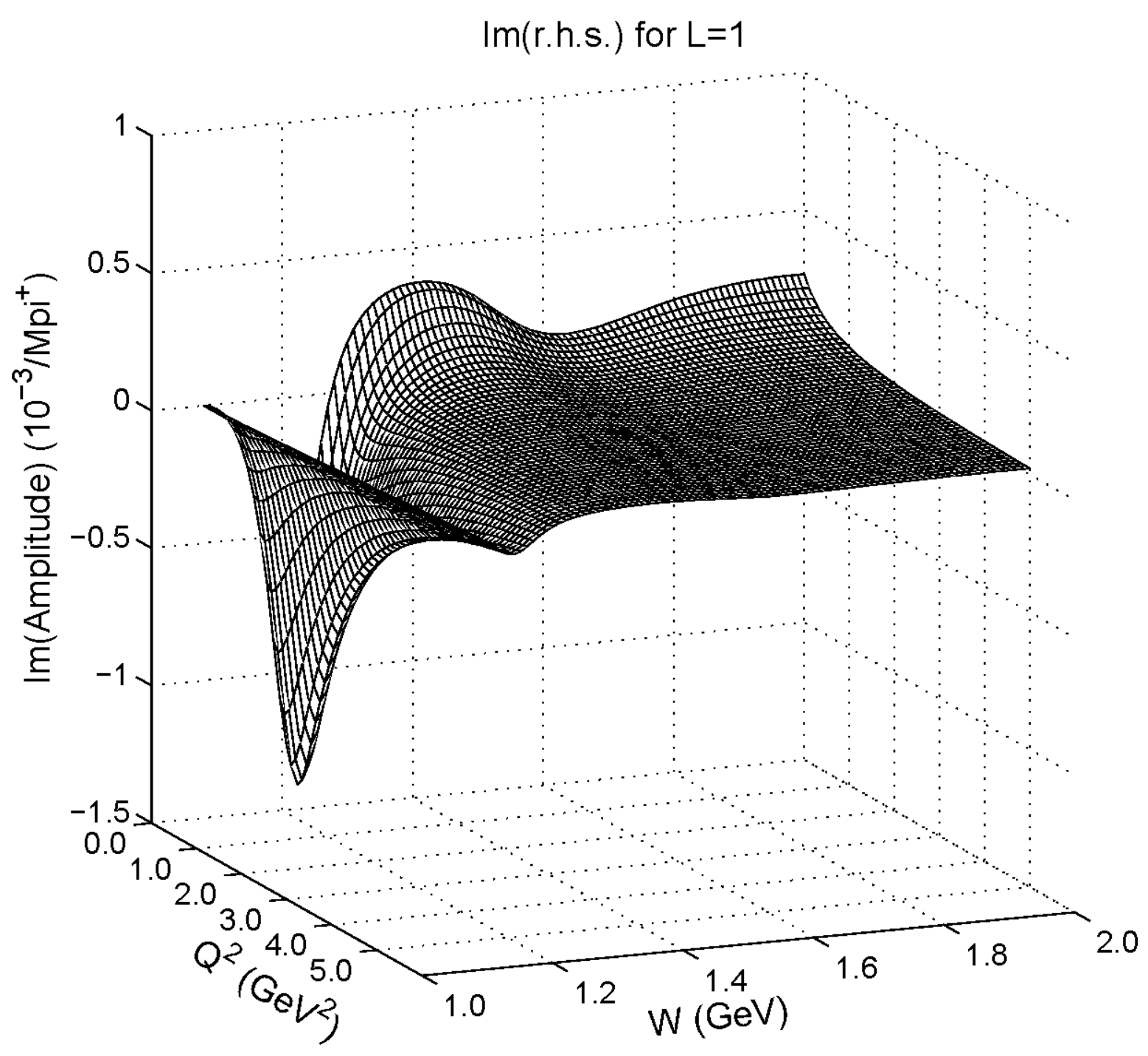}\\[1mm]
\epsfxsize=0.48\textwidth\epsfbox{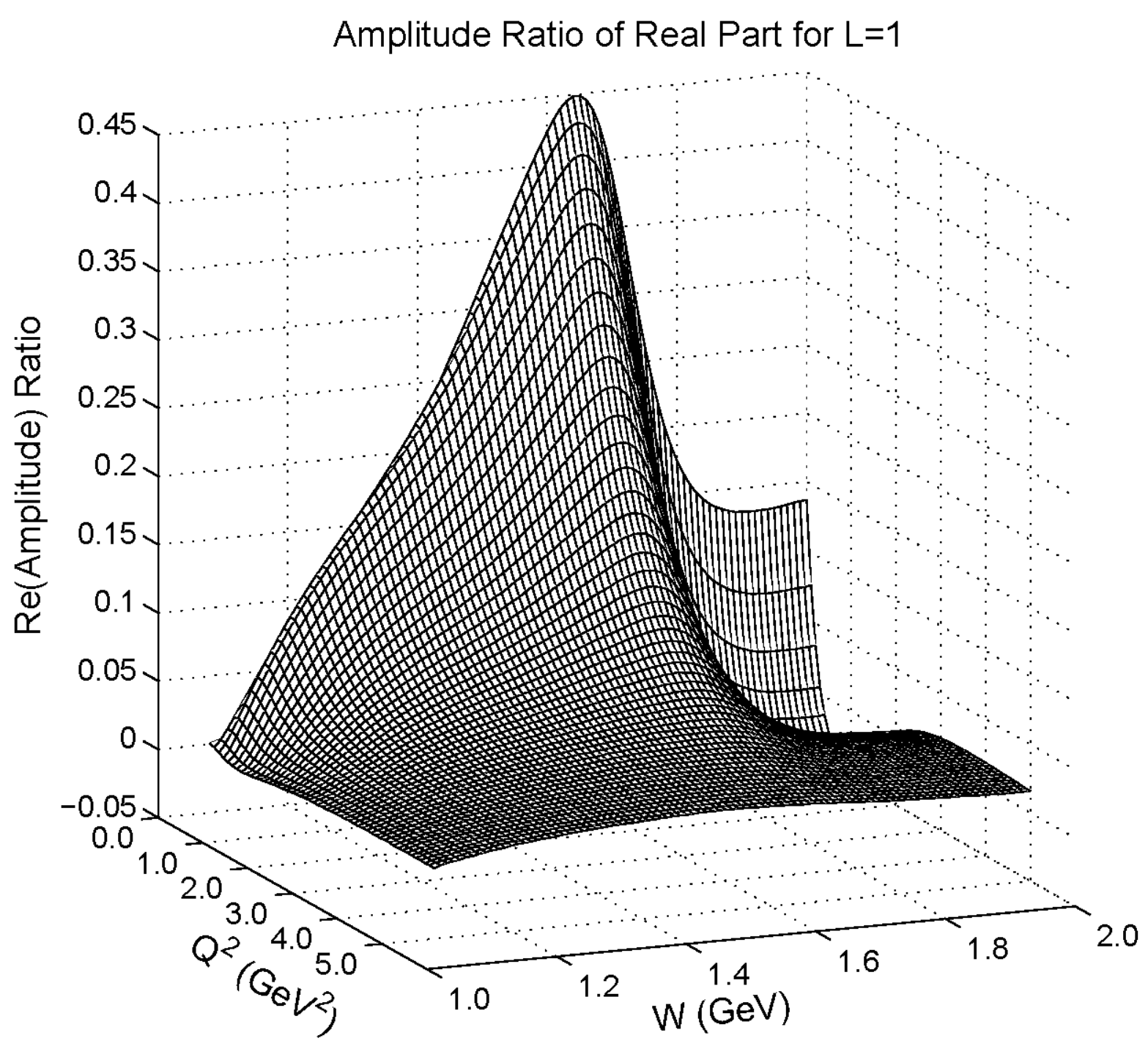}
\epsfxsize=0.48\textwidth\epsfbox{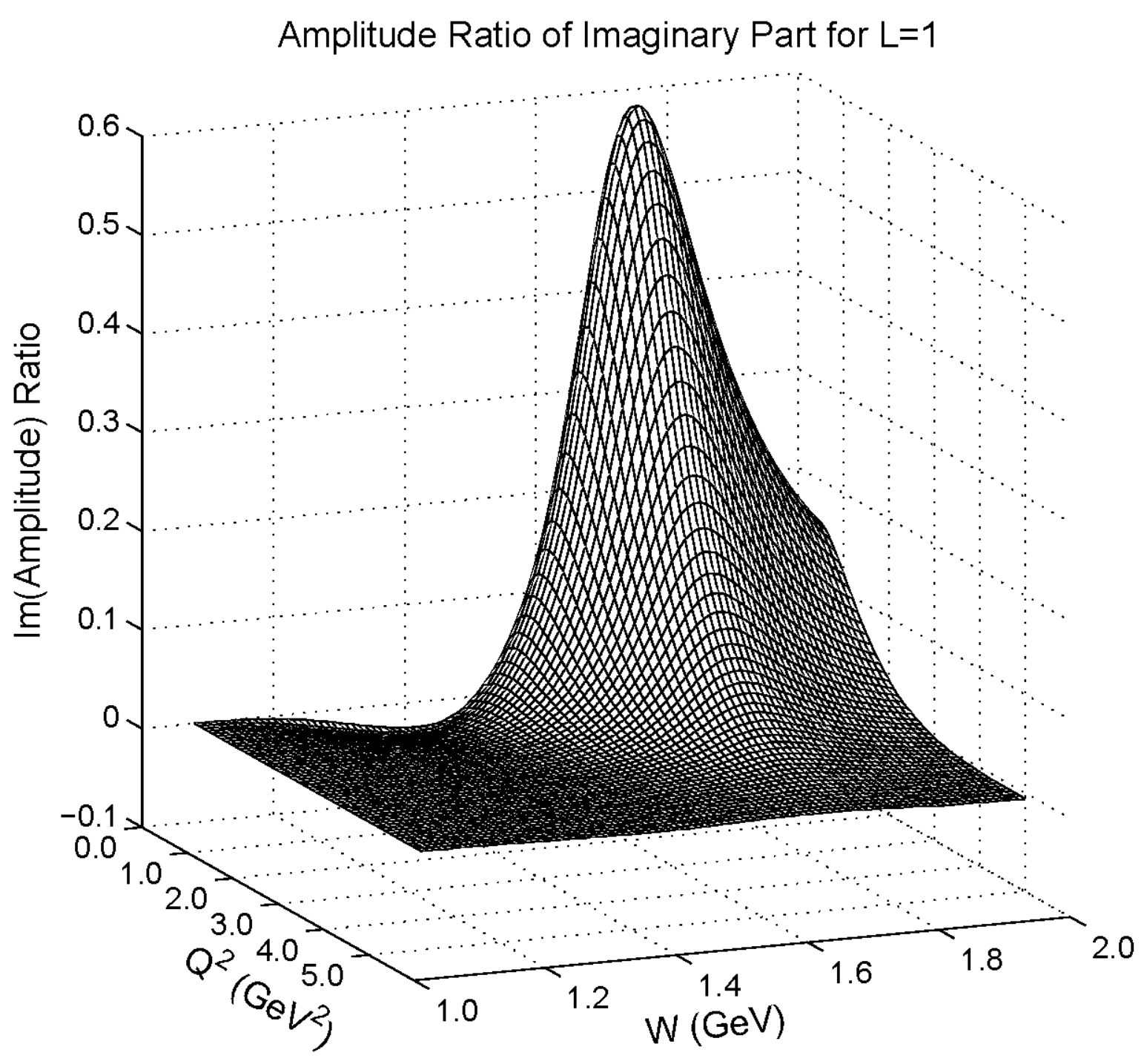}\\
\end{figure}
\begin{figure}[htp]
\epsfxsize=0.48\textwidth\epsfbox{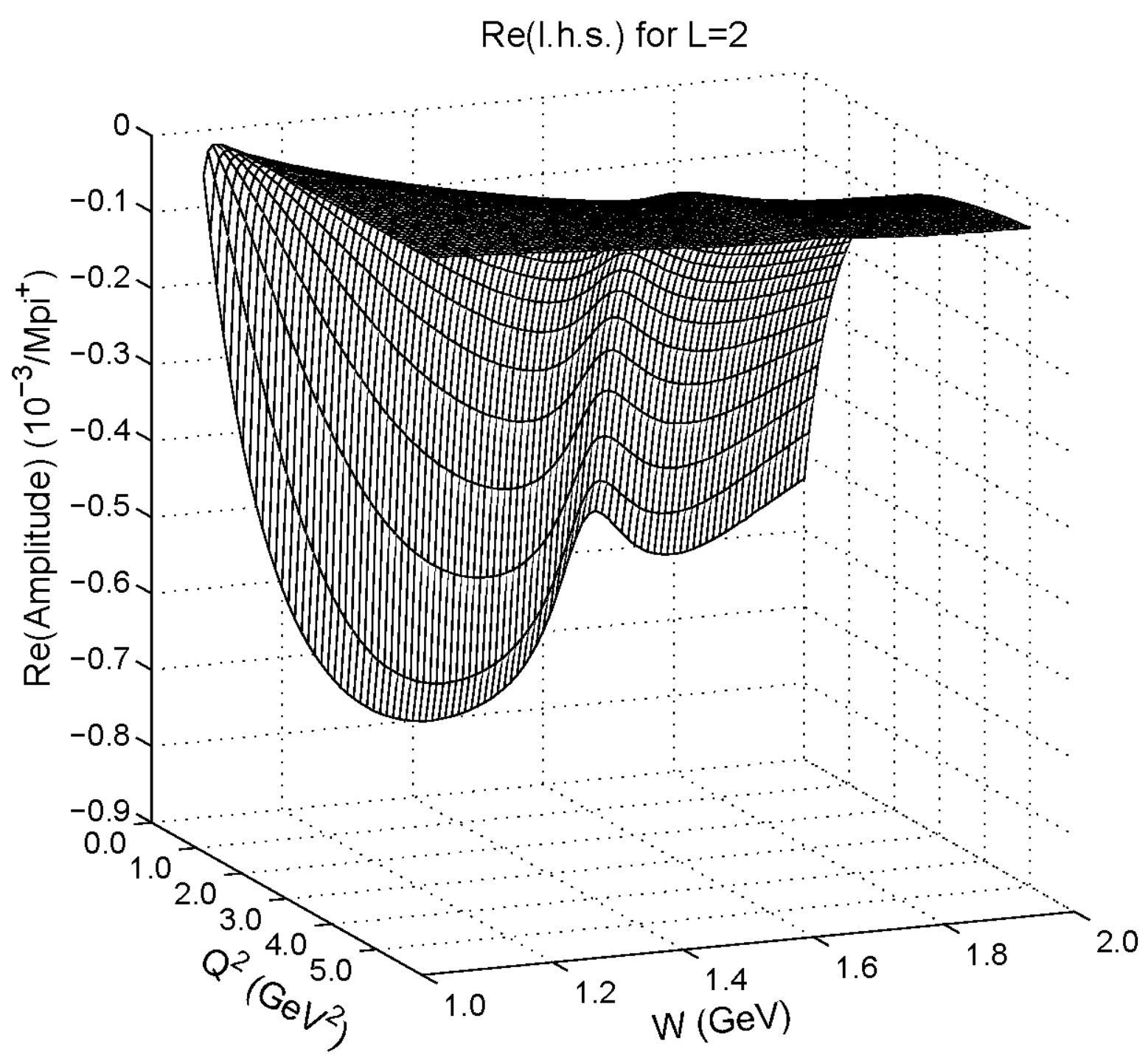}
\epsfxsize=0.48\textwidth\epsfbox{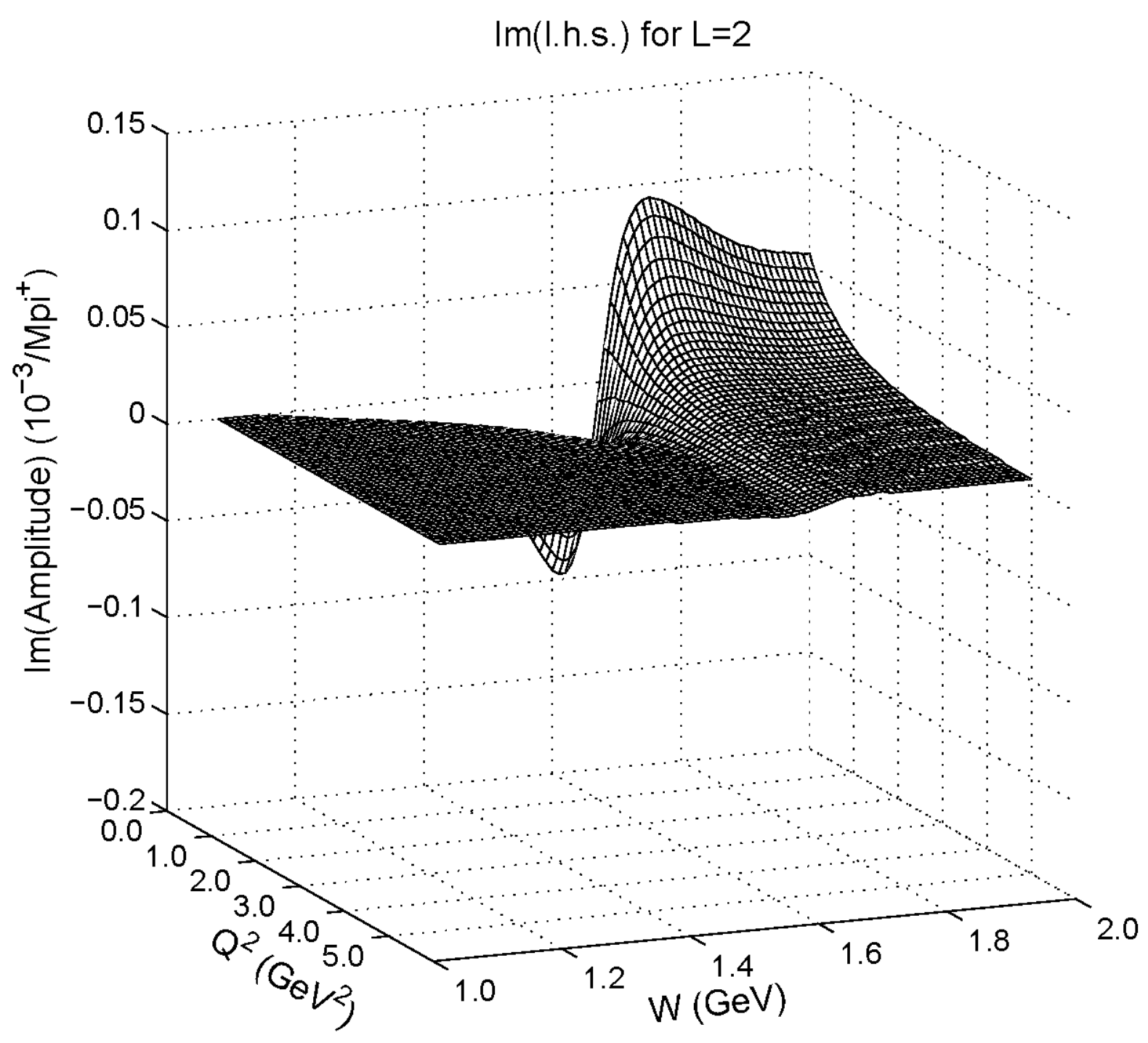}\\[1mm]
\epsfxsize=0.48\textwidth\epsfbox{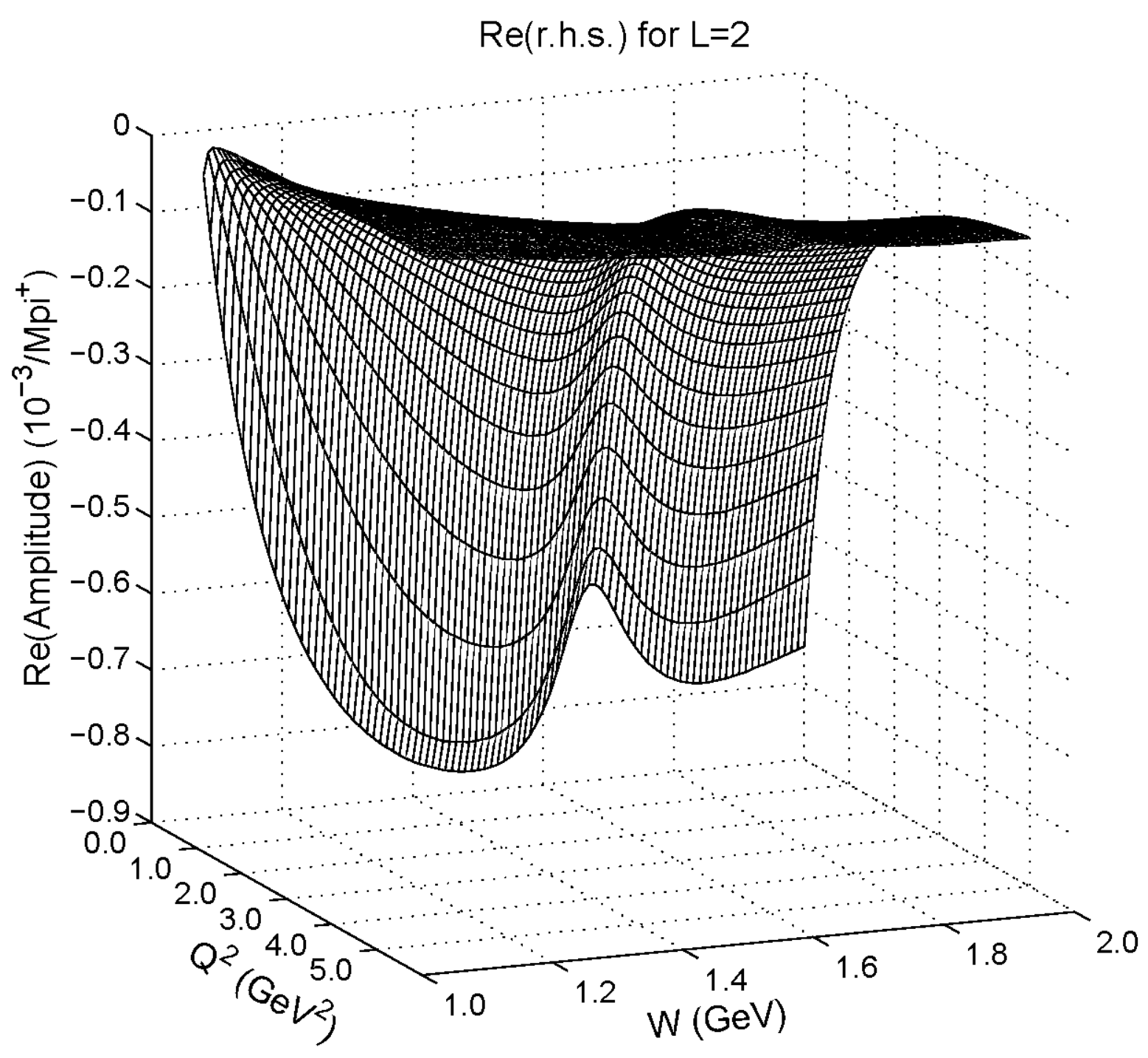}
\epsfxsize=0.48\textwidth\epsfbox{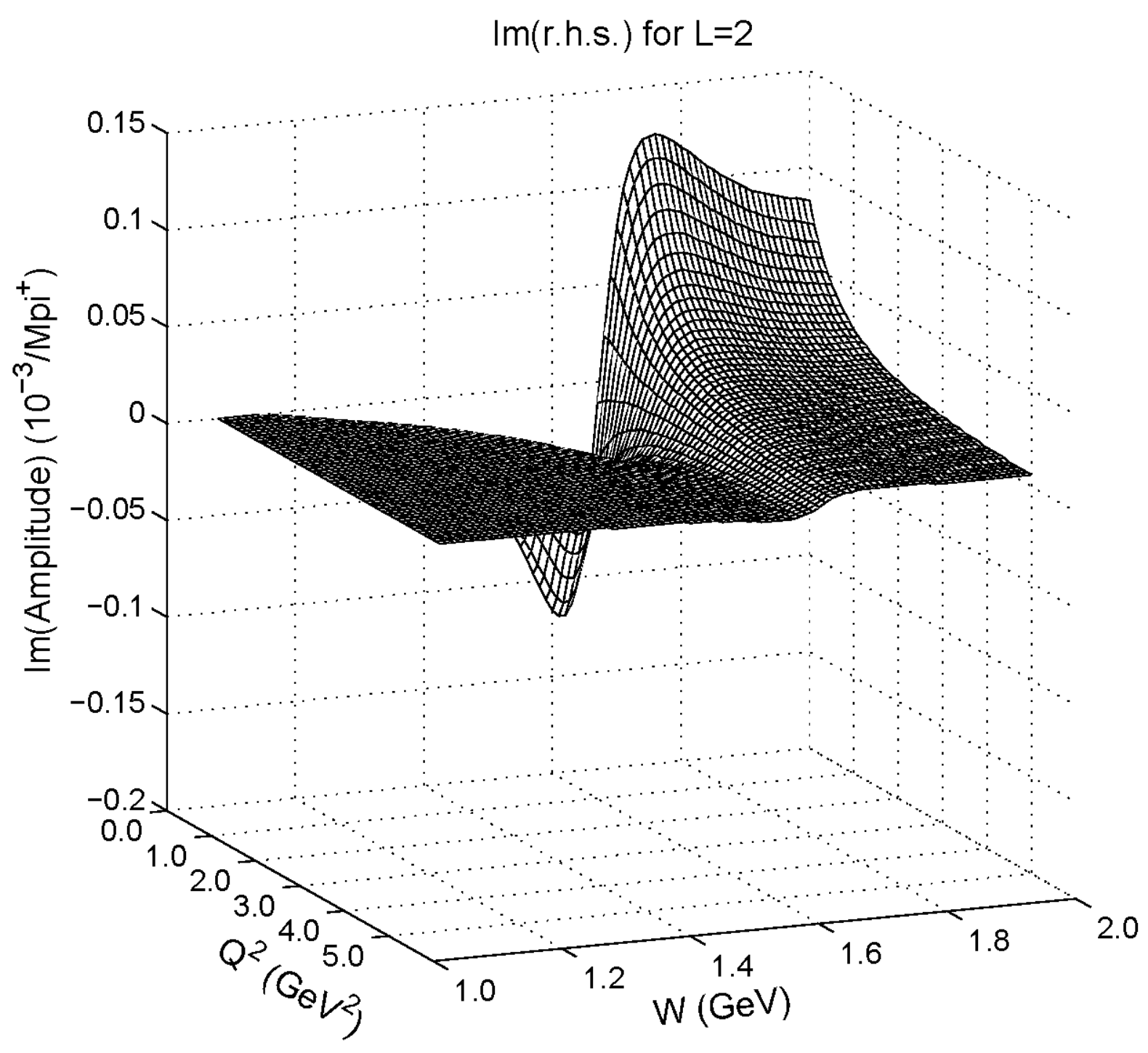}\\[1mm]
\epsfxsize=0.48\textwidth\epsfbox{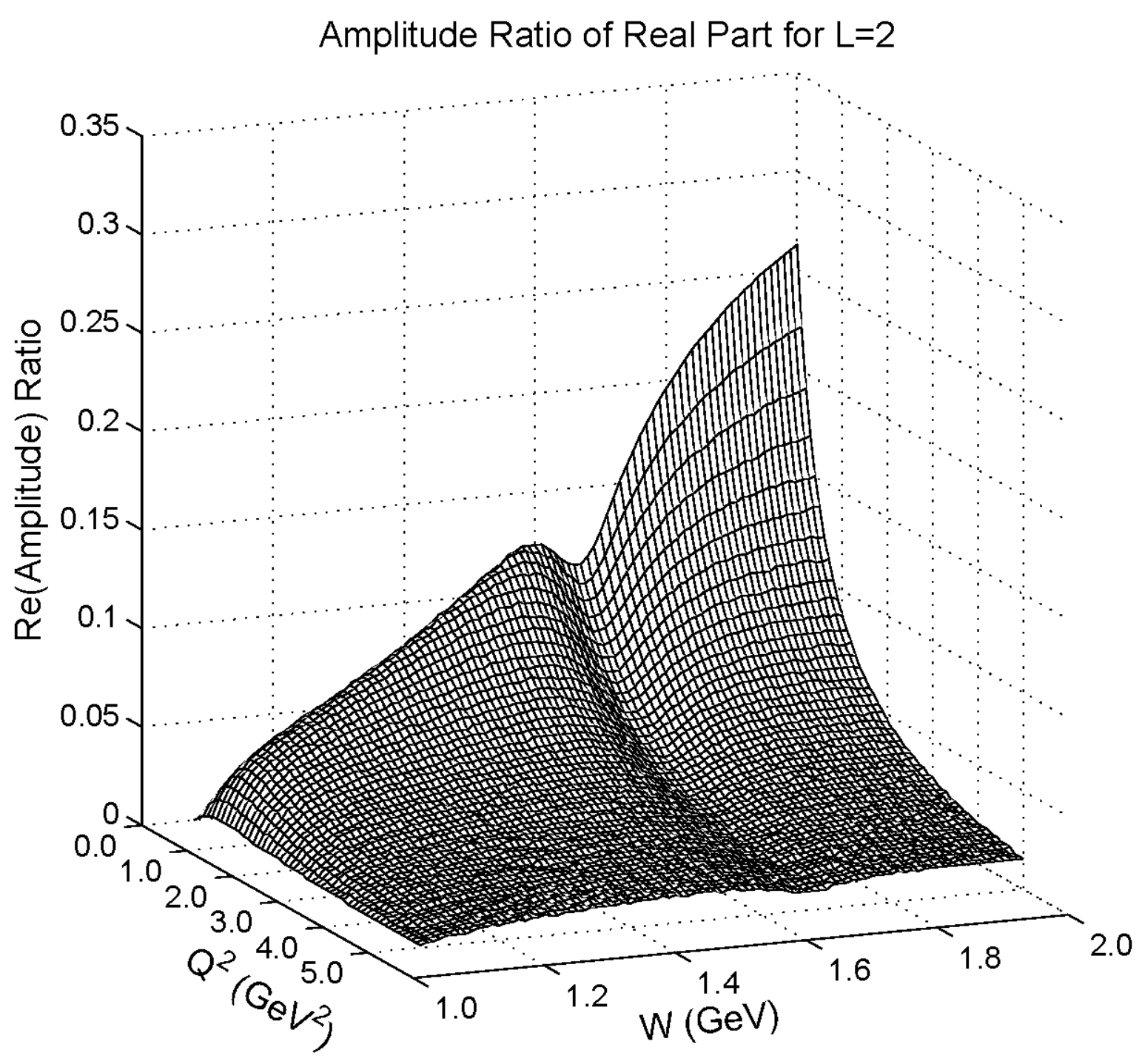}
\epsfxsize=0.48\textwidth\epsfbox{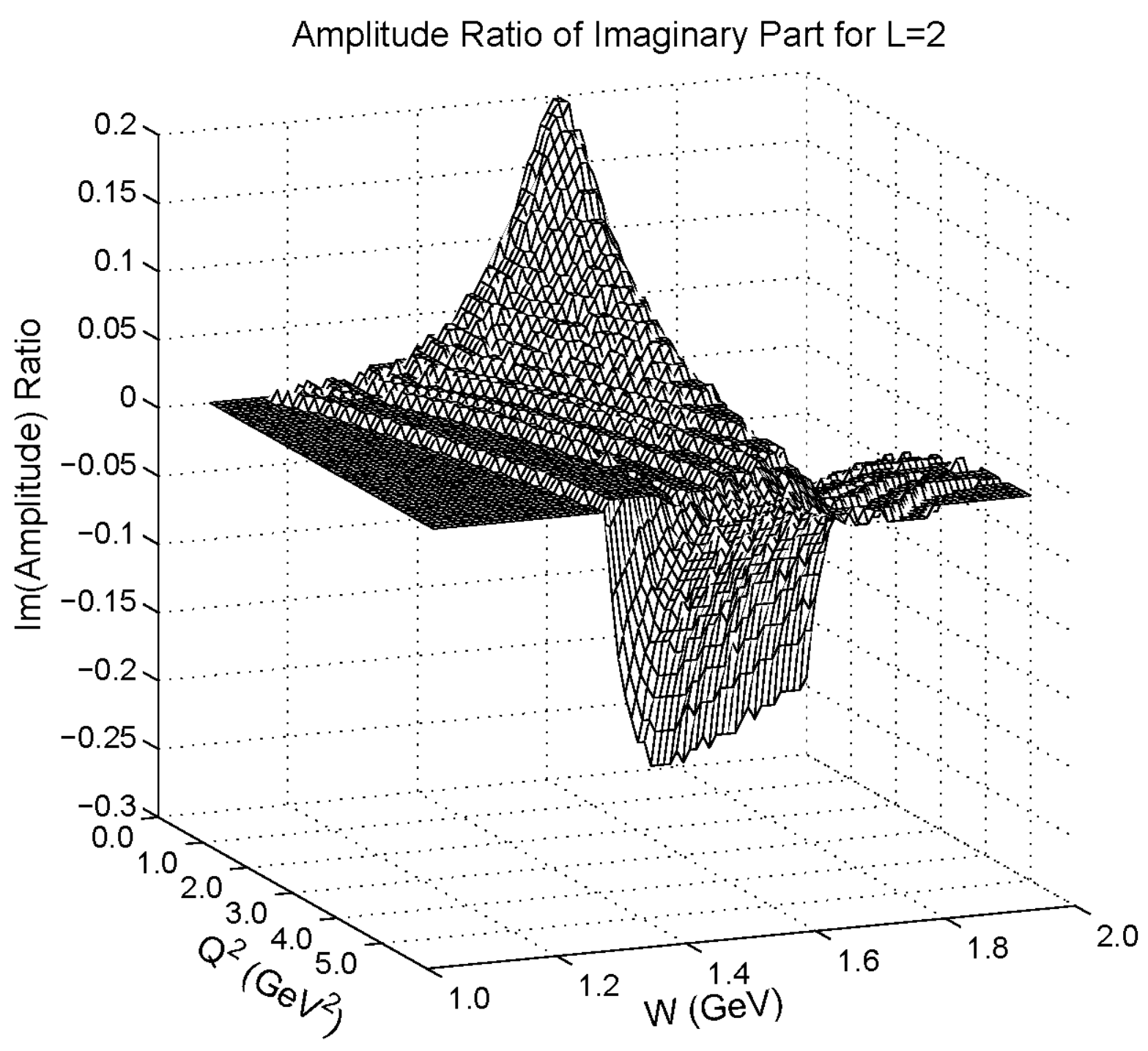}\\
\end{figure}
\begin{figure}[htp]
\epsfxsize=0.48\textwidth\epsfbox{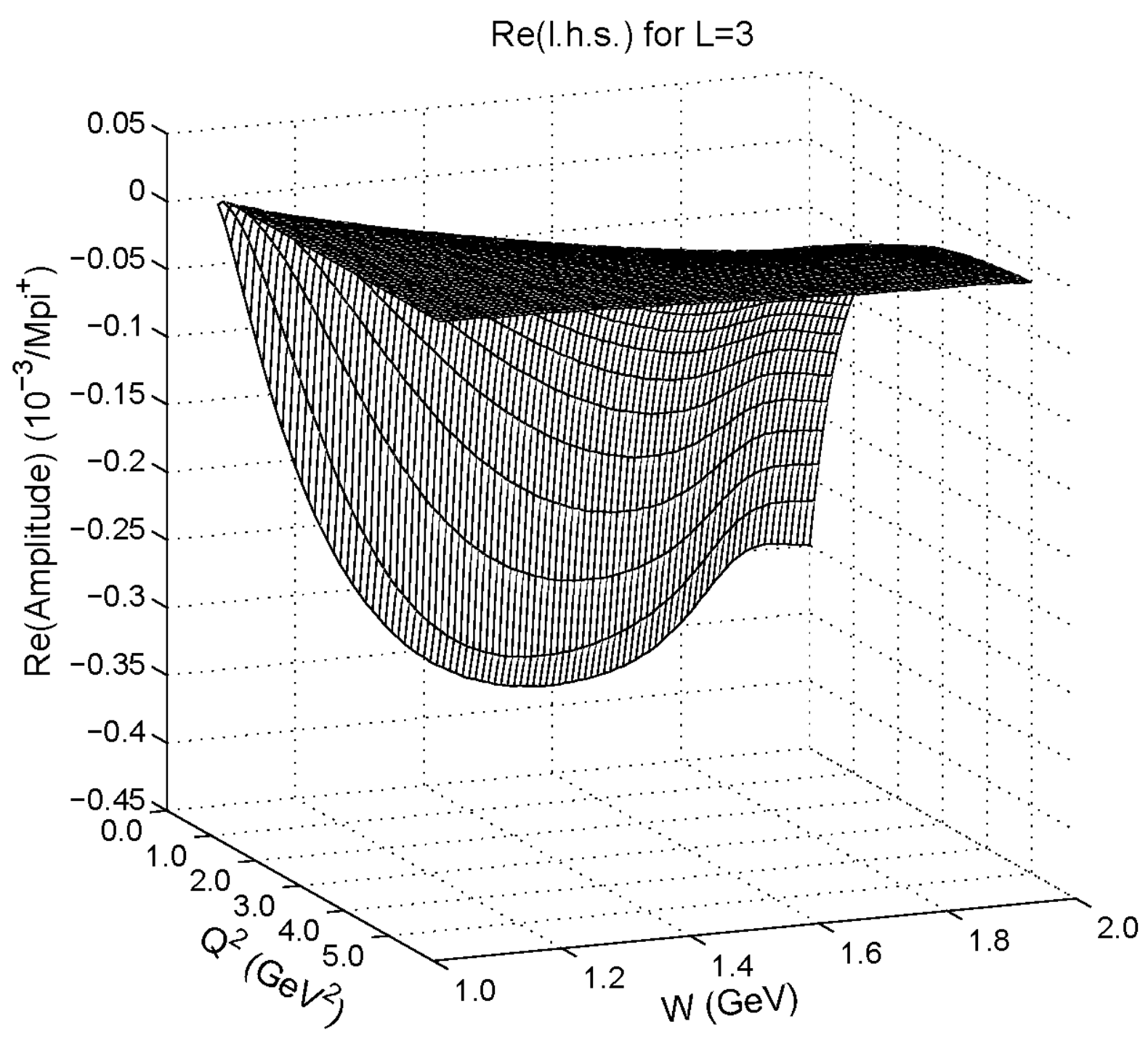}
\epsfxsize=0.48\textwidth\epsfbox{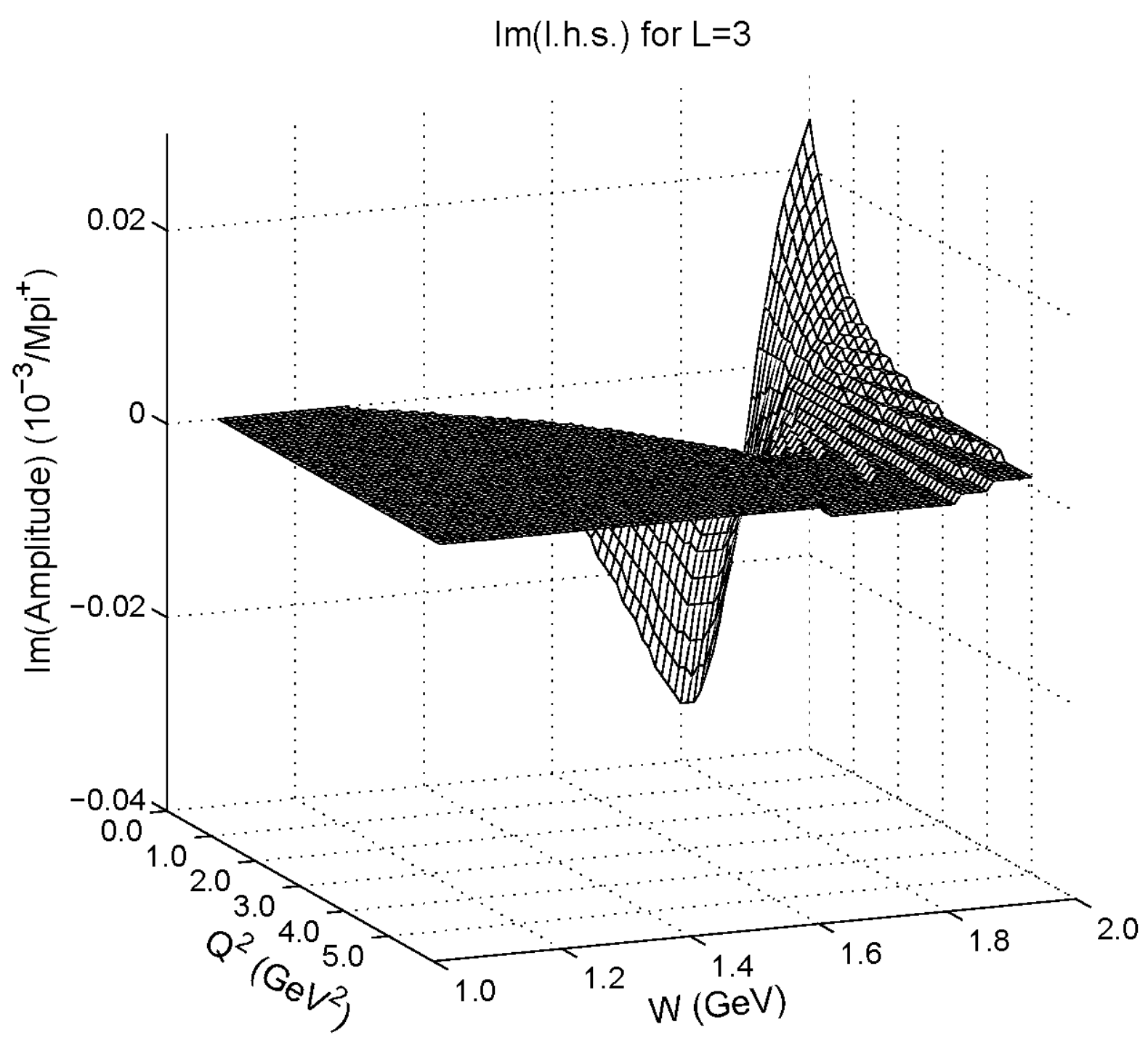}\\[1mm]
\epsfxsize=0.48\textwidth\epsfbox{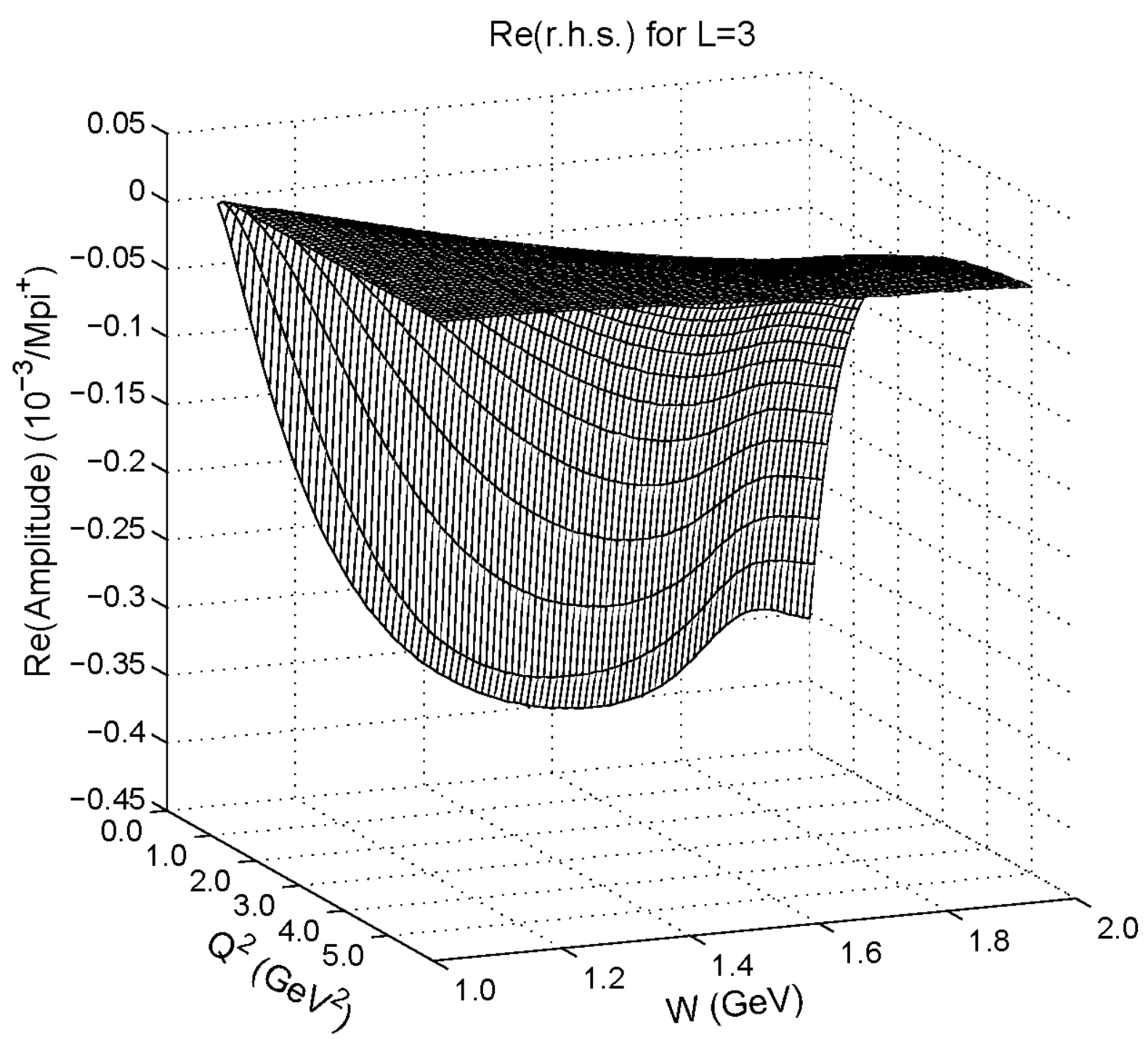}
\epsfxsize=0.48\textwidth\epsfbox{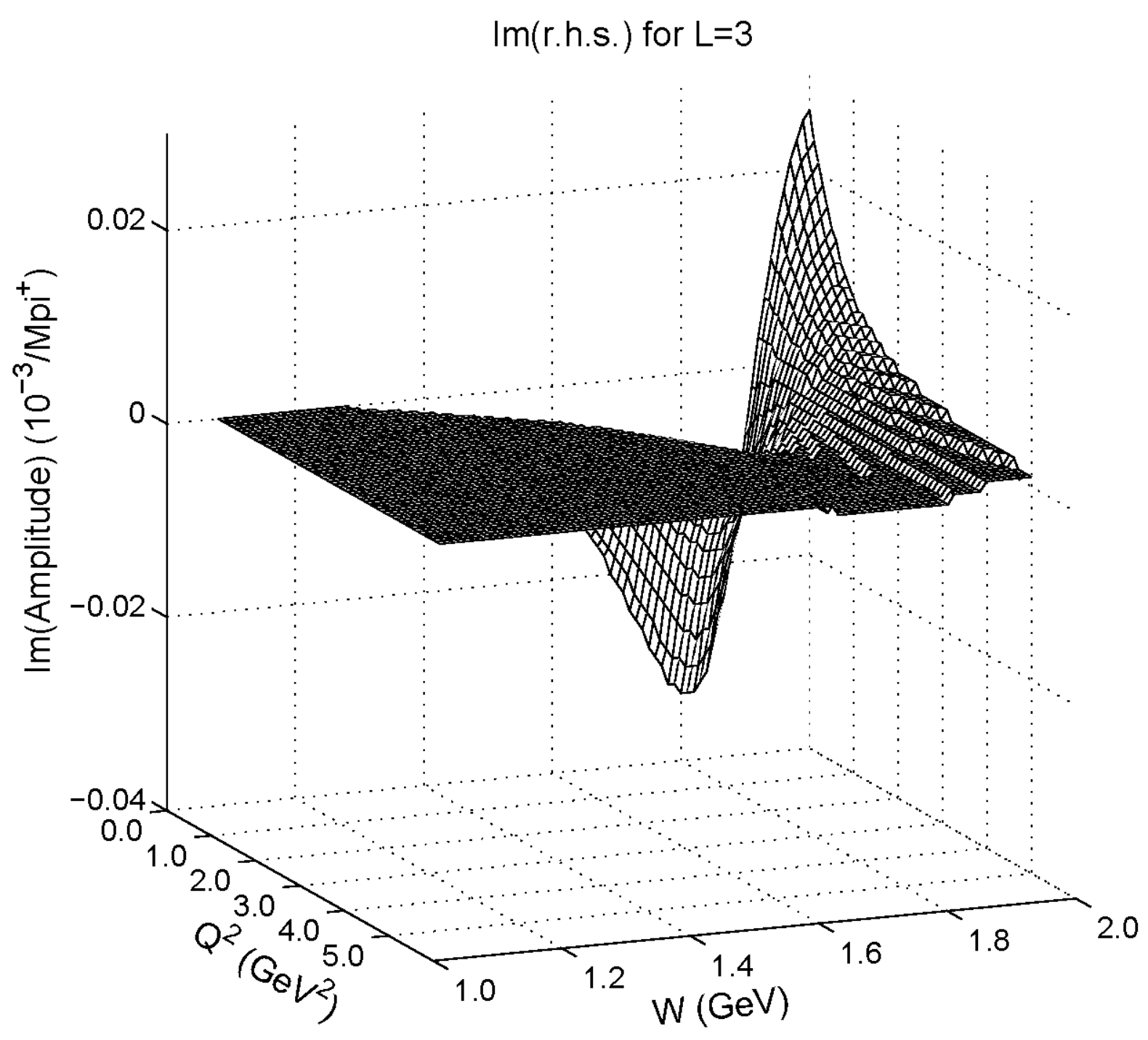}\\[1mm]
\epsfxsize=0.48\textwidth\epsfbox{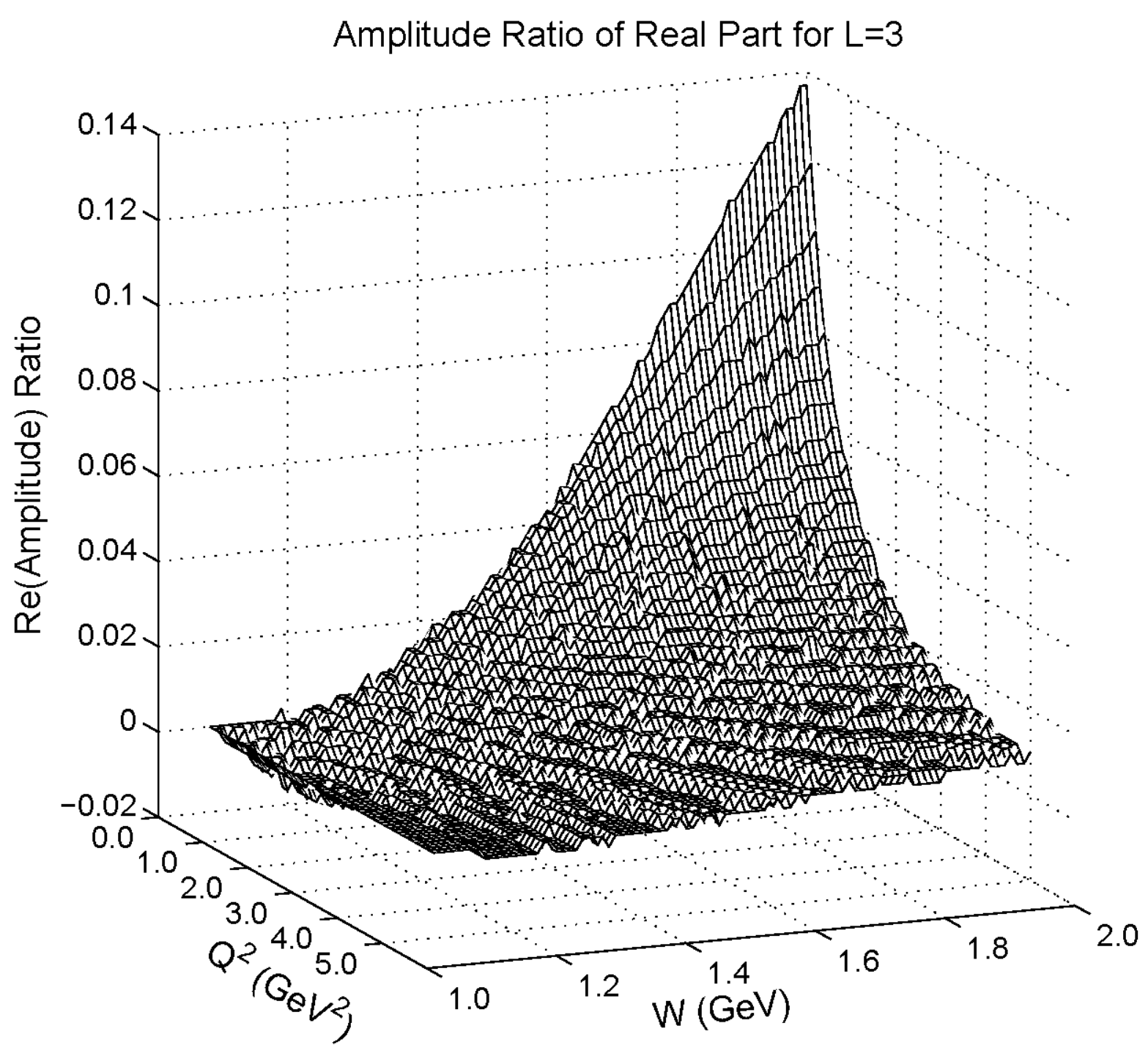}
\epsfxsize=0.48\textwidth\epsfbox{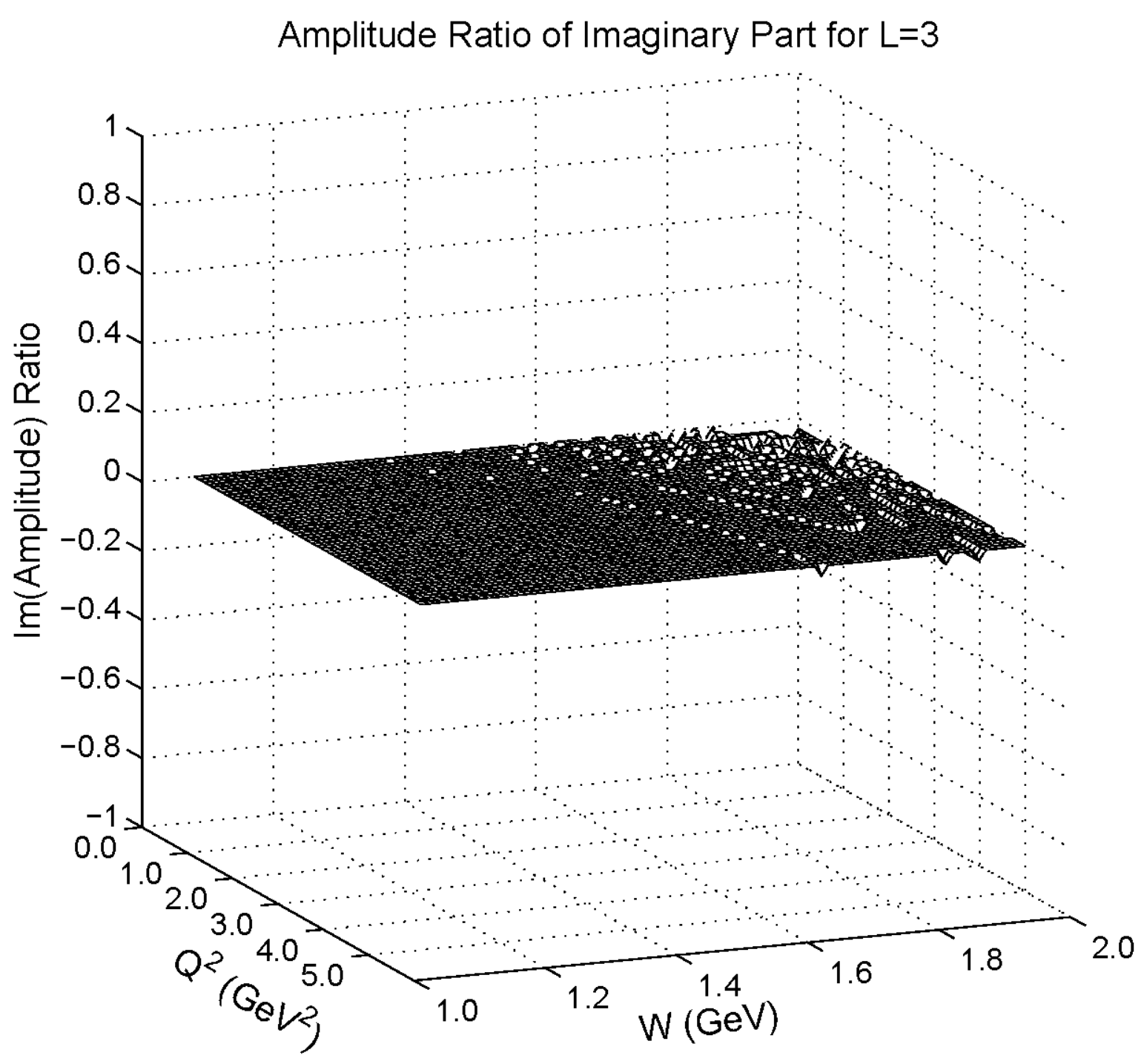}\\
\end{figure}
\begin{figure}[htp]
\epsfxsize=0.48\textwidth\epsfbox{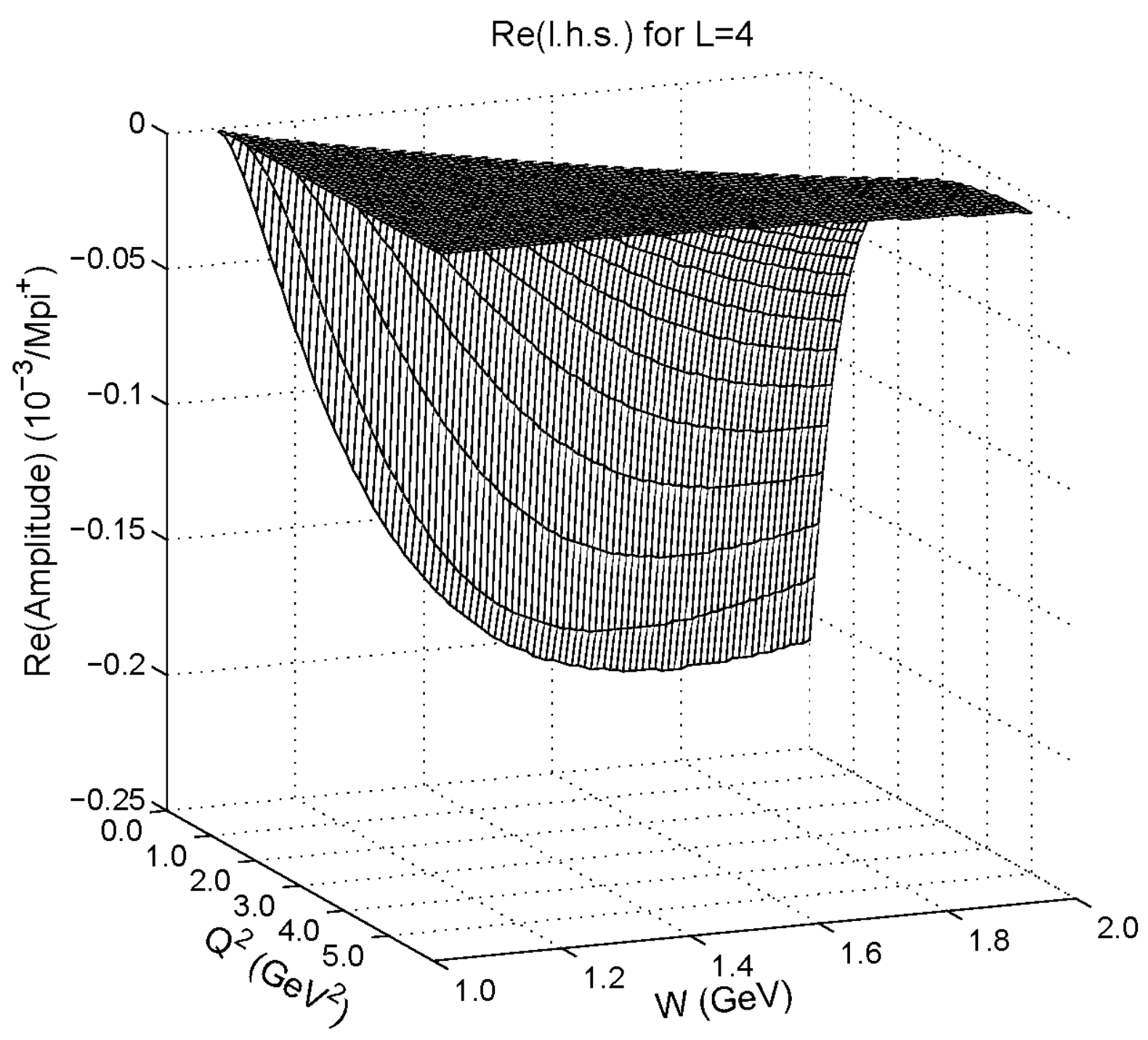}
\epsfxsize=0.48\textwidth\epsfbox{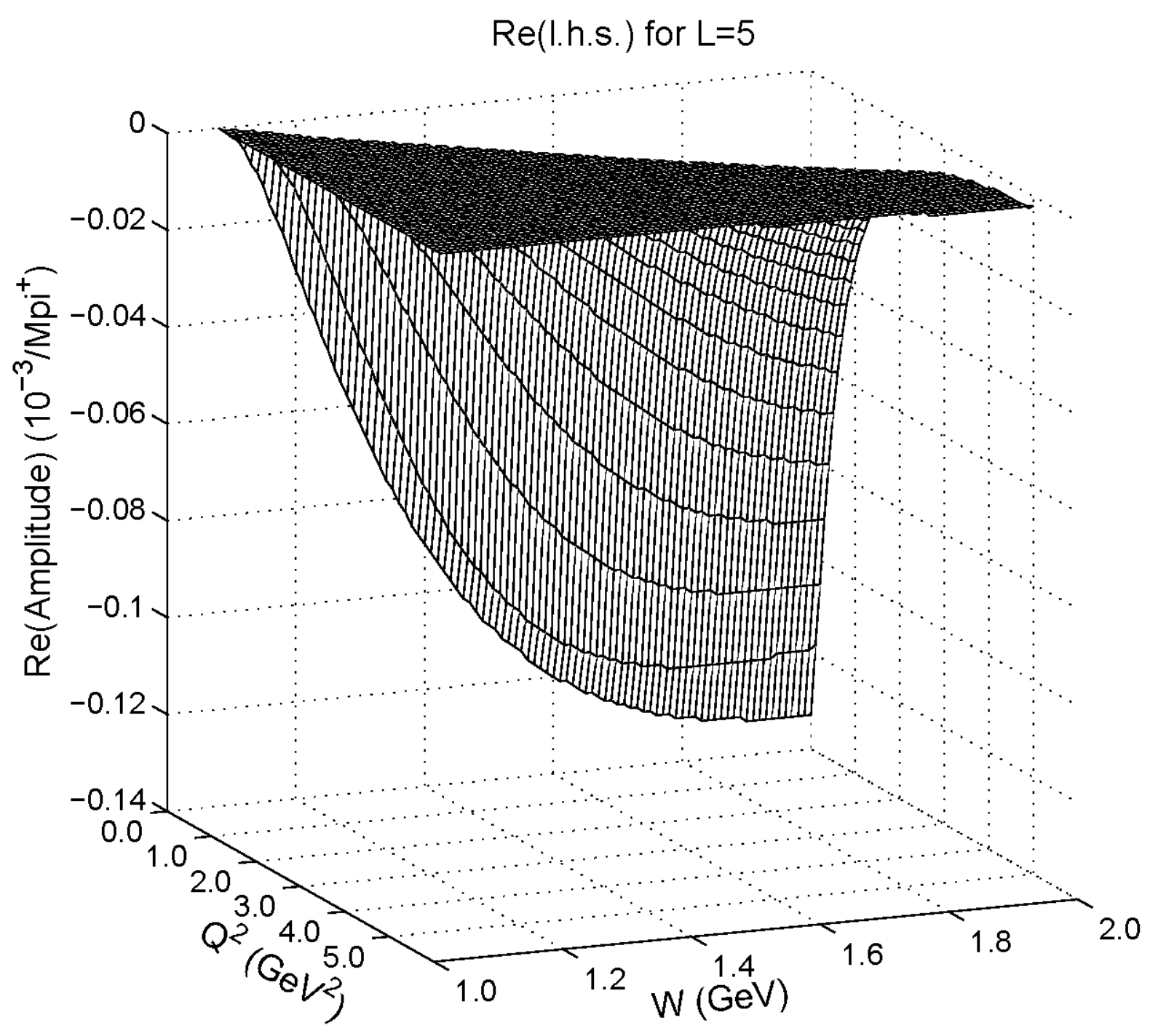}\\[1mm]
\epsfxsize=0.48\textwidth\epsfbox{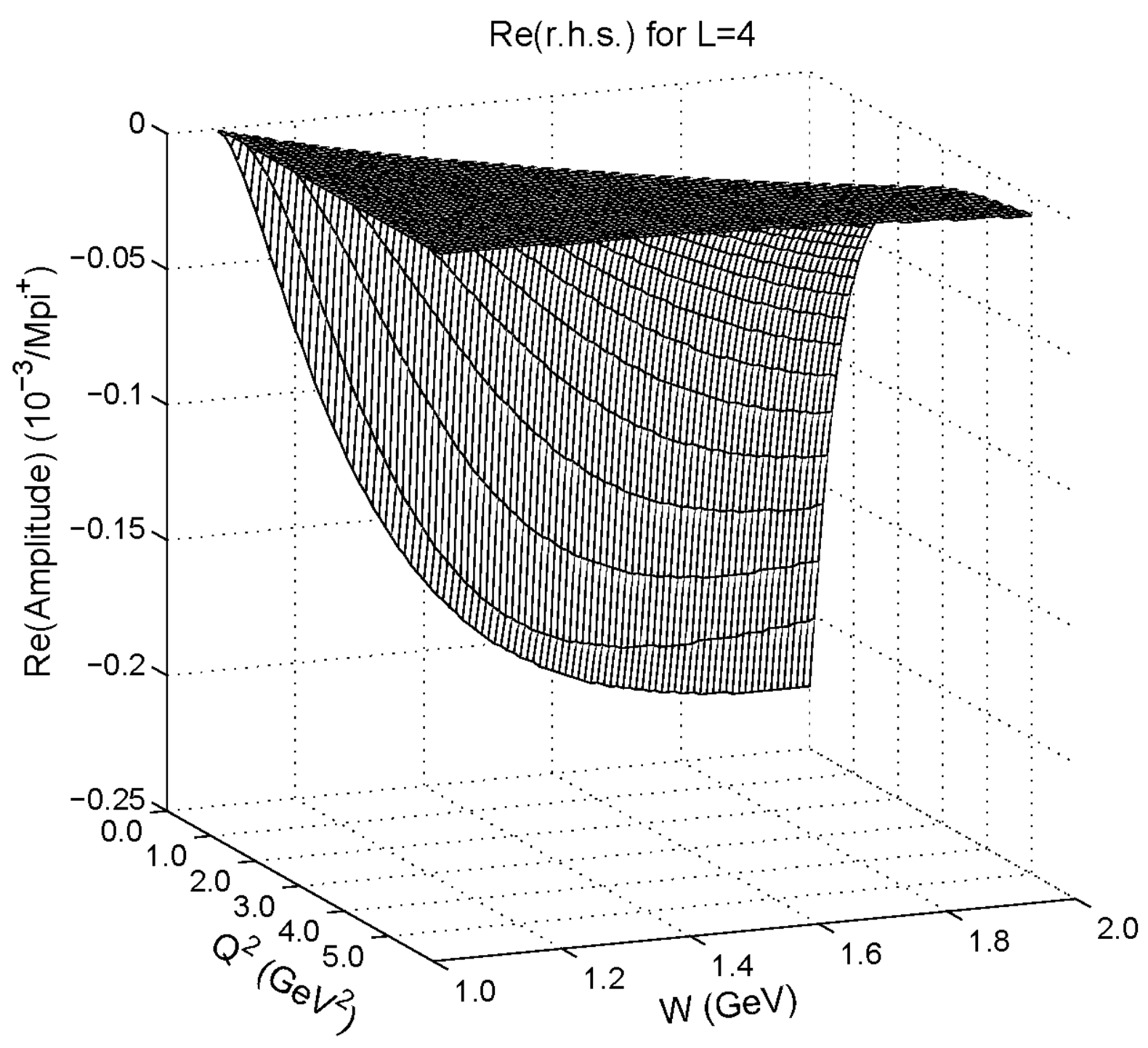}
\epsfxsize=0.48\textwidth\epsfbox{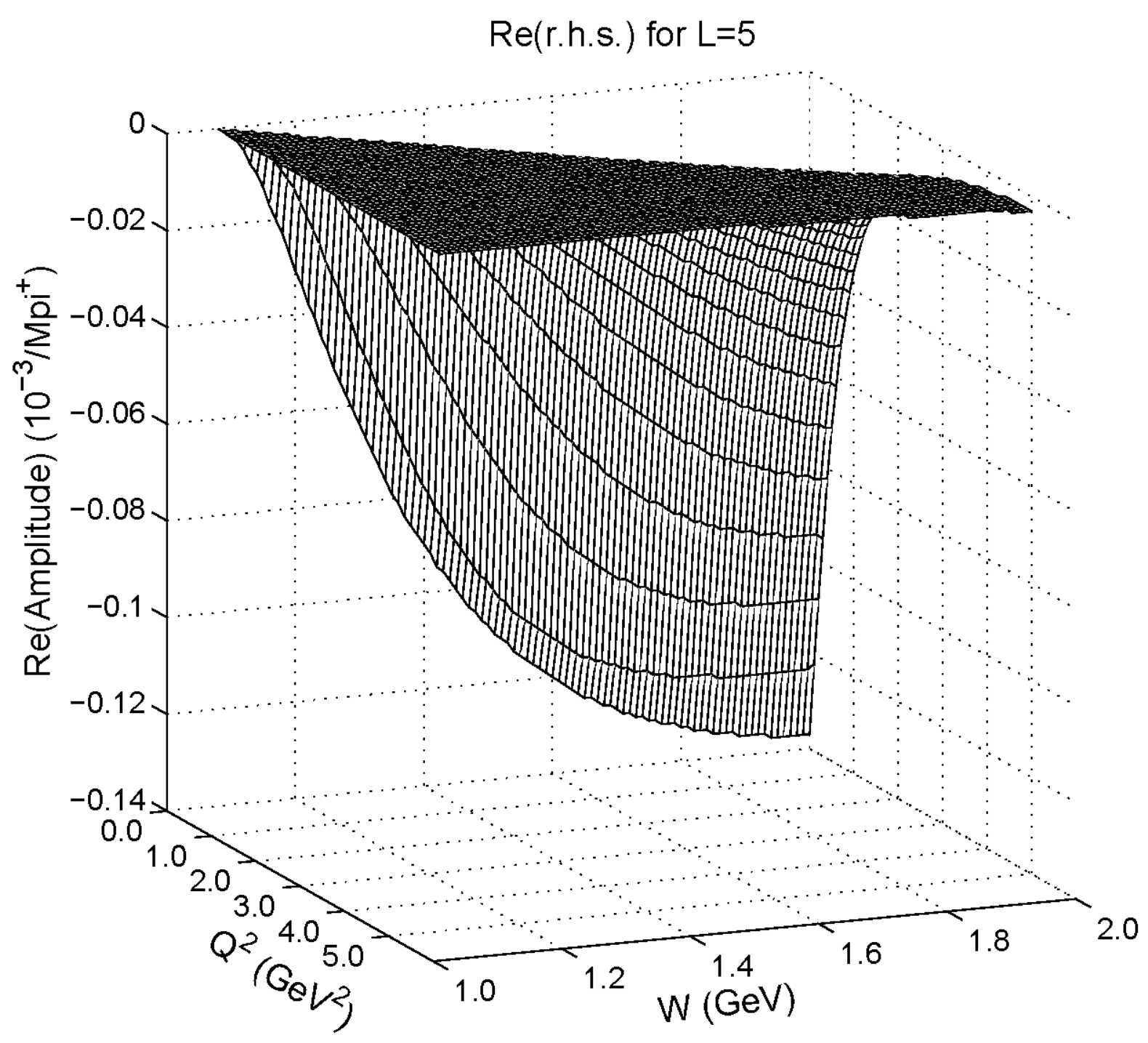}\\[1mm]
\epsfxsize=0.48\textwidth\epsfbox{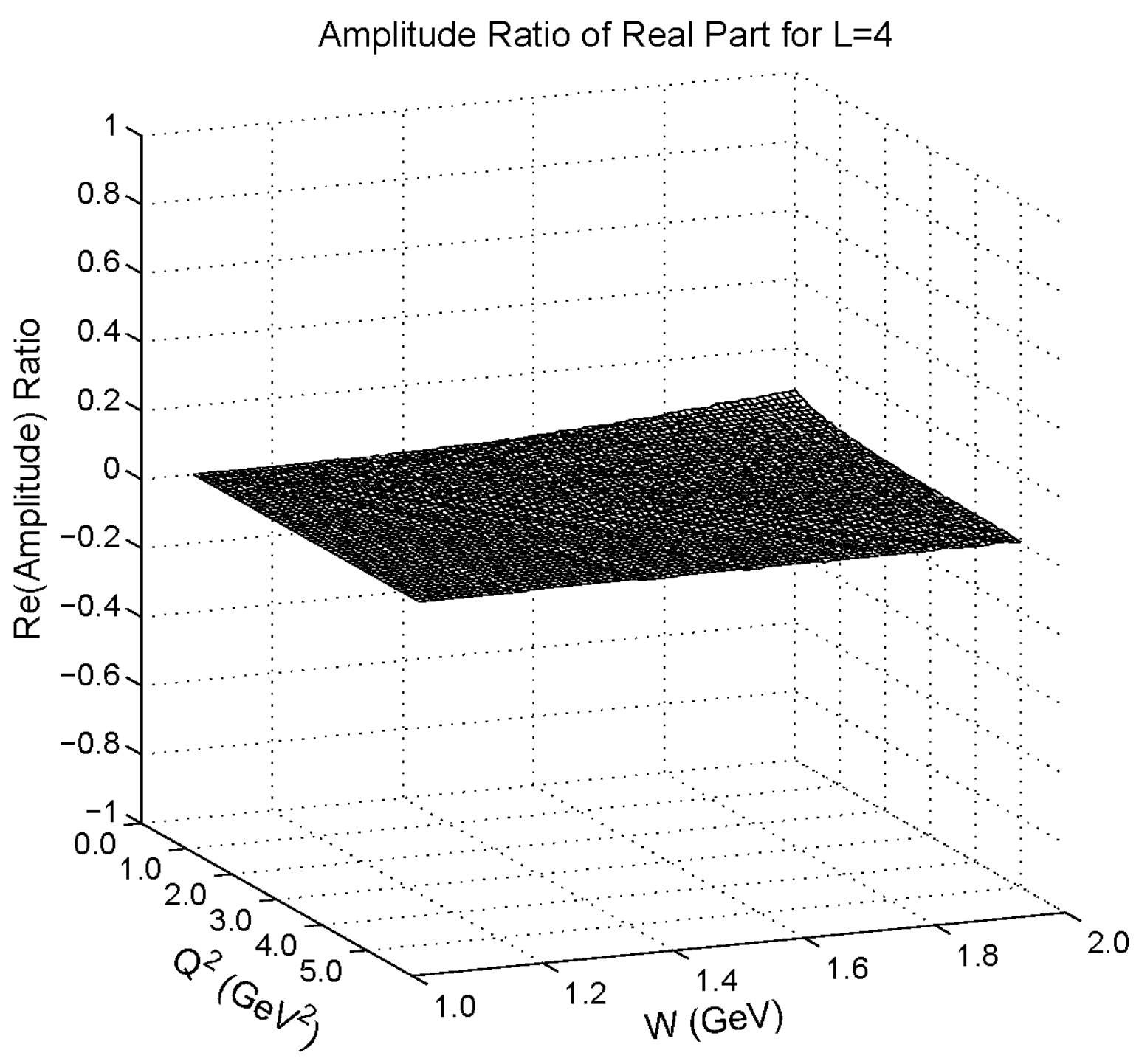}
\epsfxsize=0.48\textwidth\epsfbox{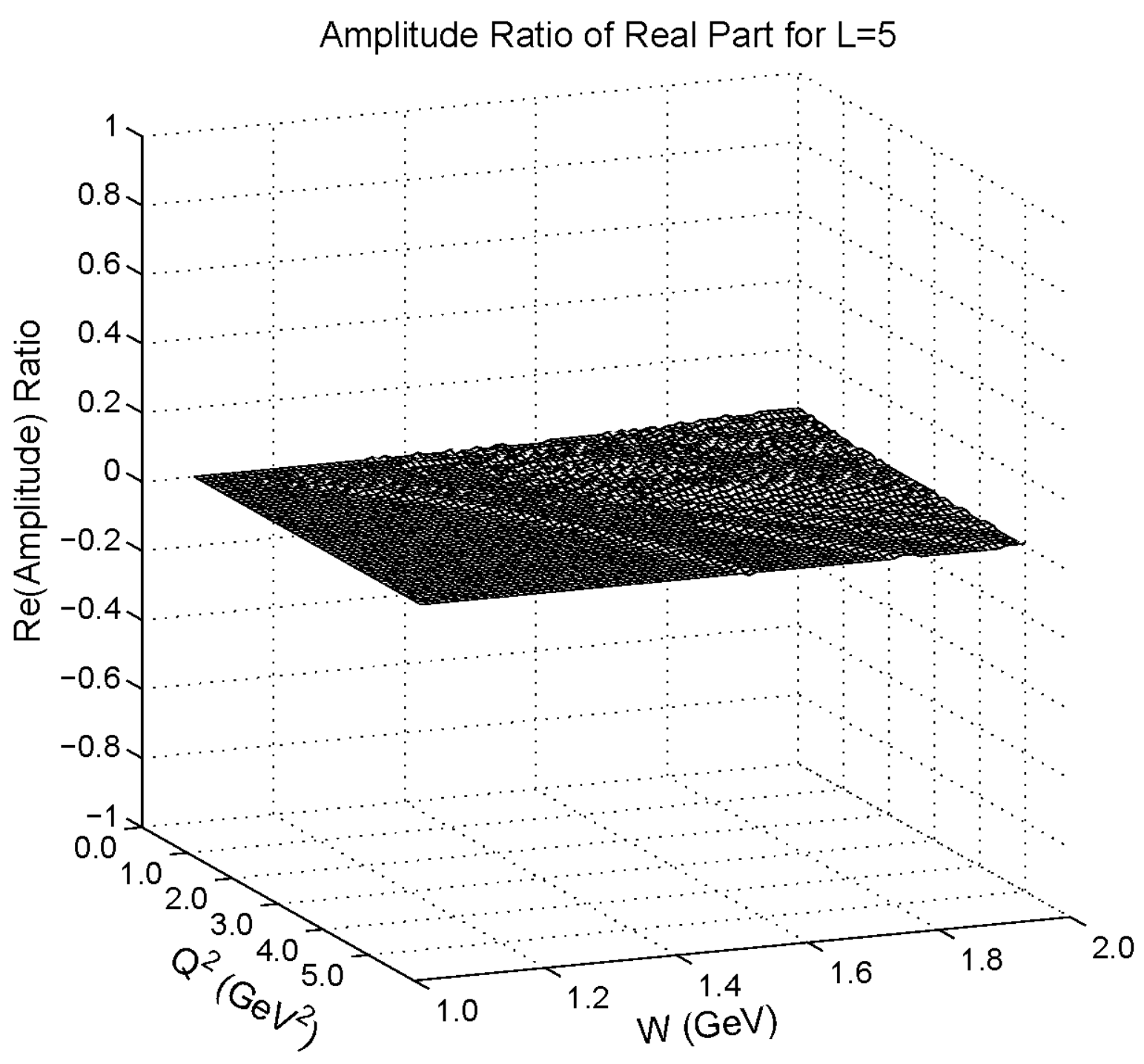}
\end{figure}
%
%
%
\begin{figure}[htp]
\epsfxsize=0.48\textwidth\epsfbox{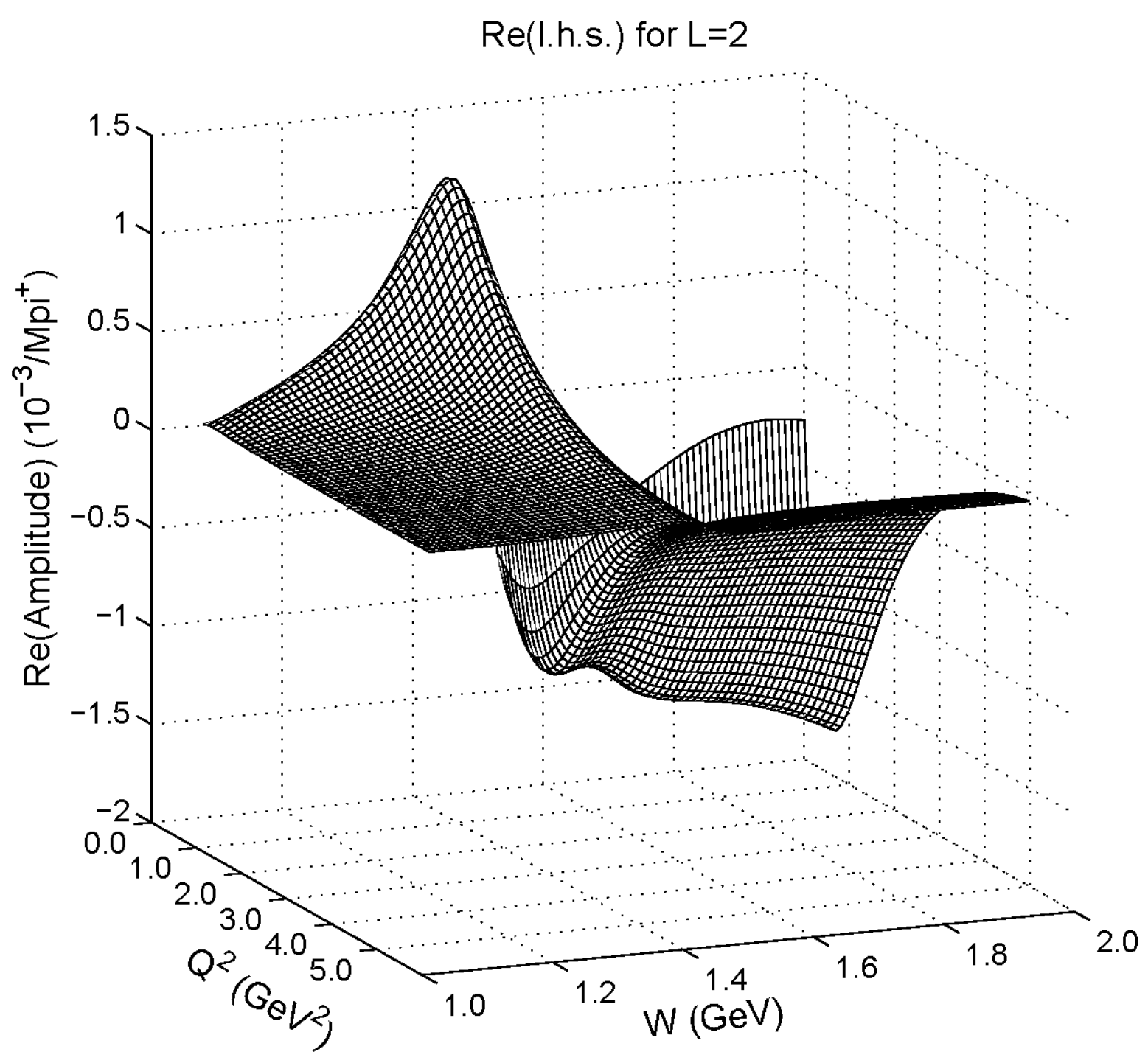}
\epsfxsize=0.48\textwidth\epsfbox{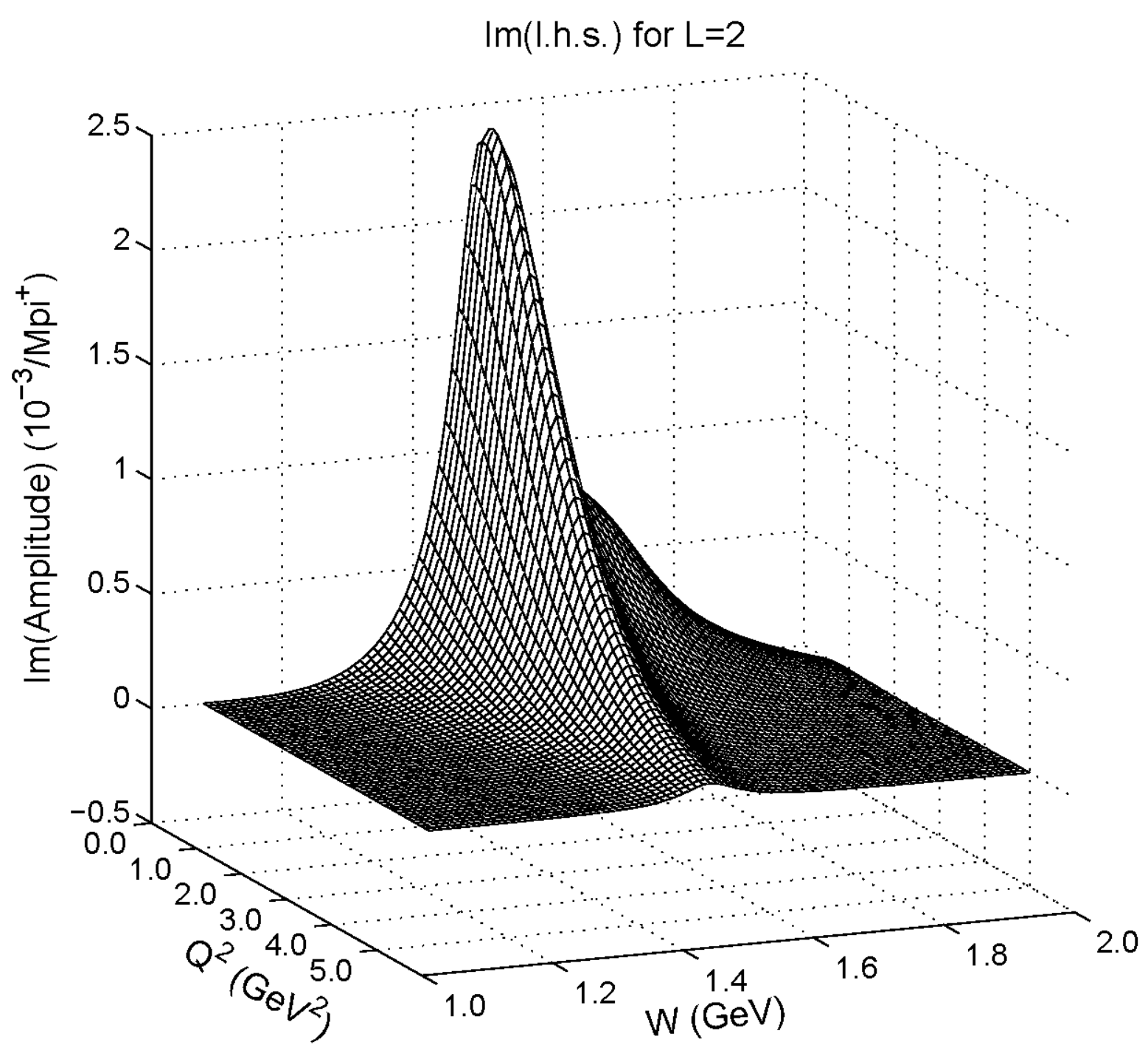}\\[1mm]
\epsfxsize=0.48\textwidth\epsfbox{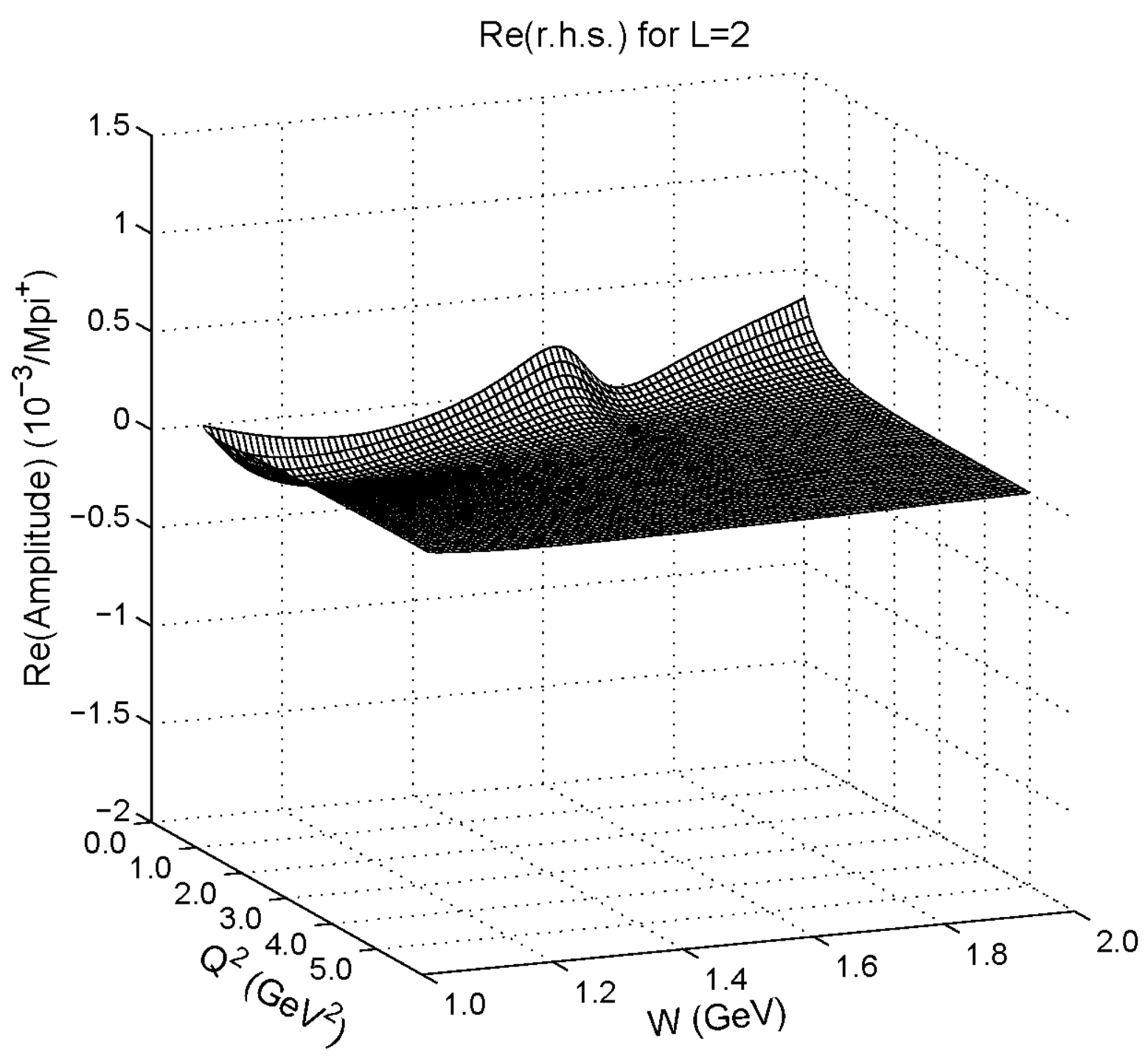}
\epsfxsize=0.48\textwidth\epsfbox{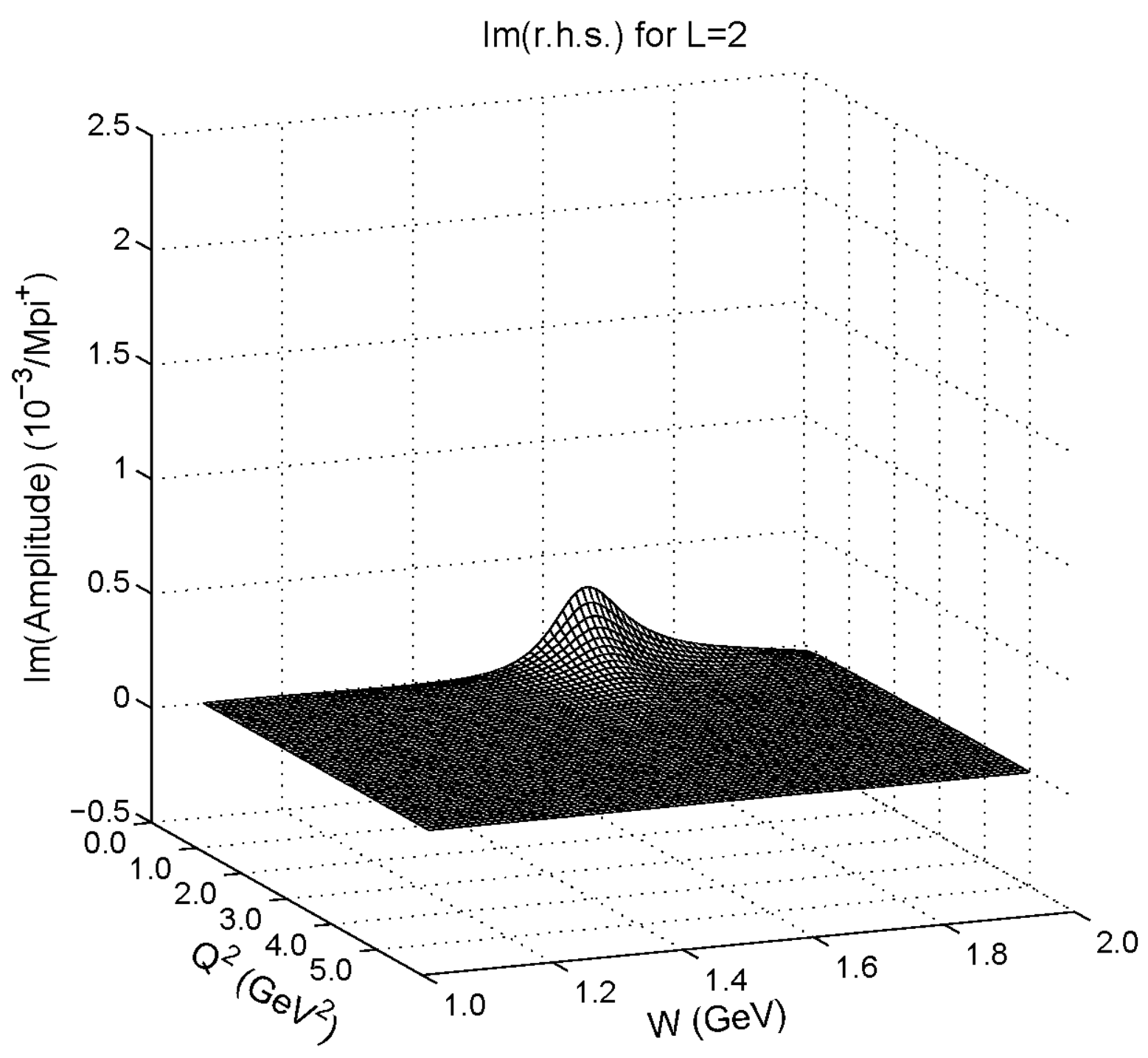}\\[1mm]
\epsfxsize=0.48\textwidth\epsfbox{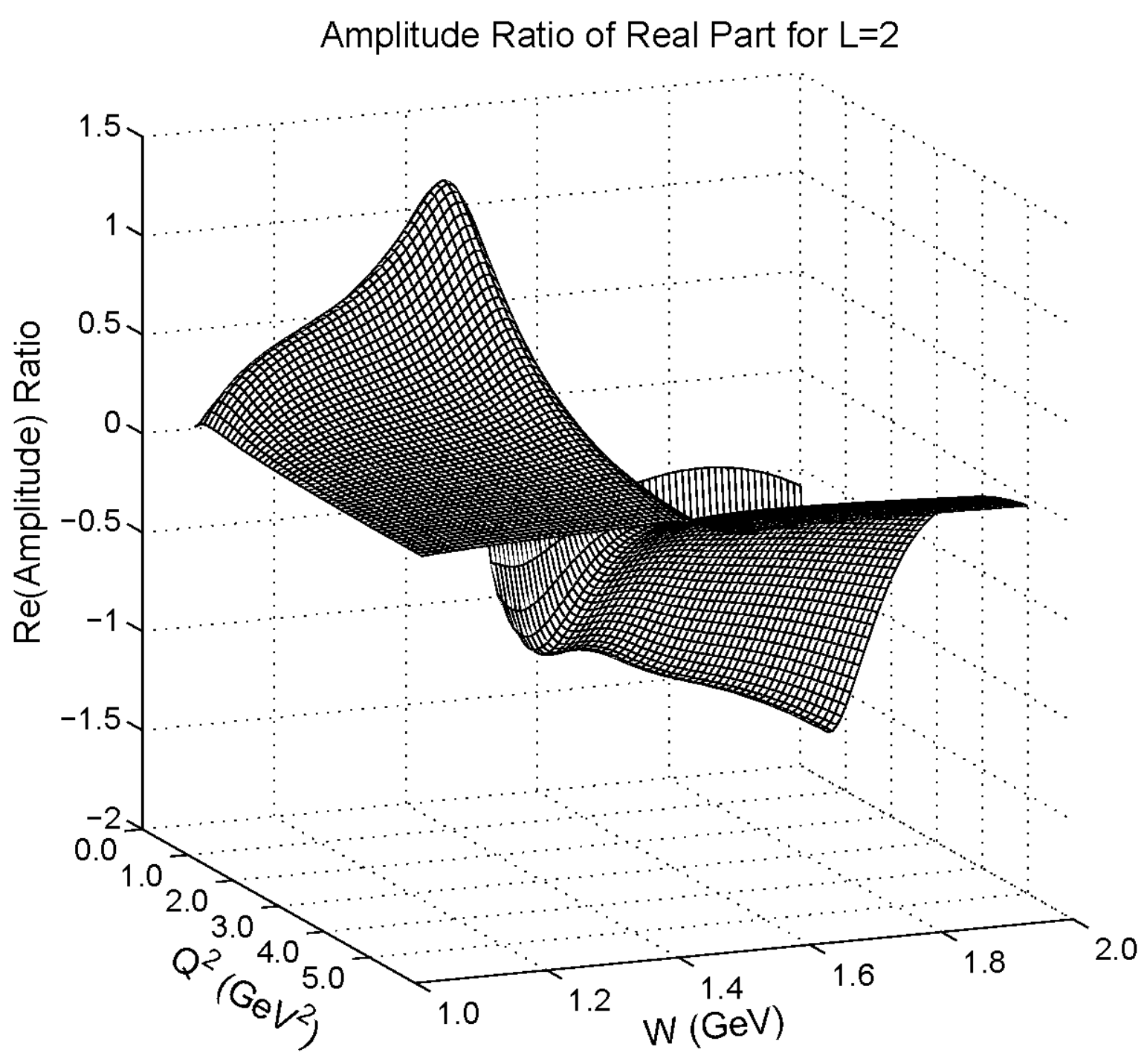}
\epsfxsize=0.48\textwidth\epsfbox{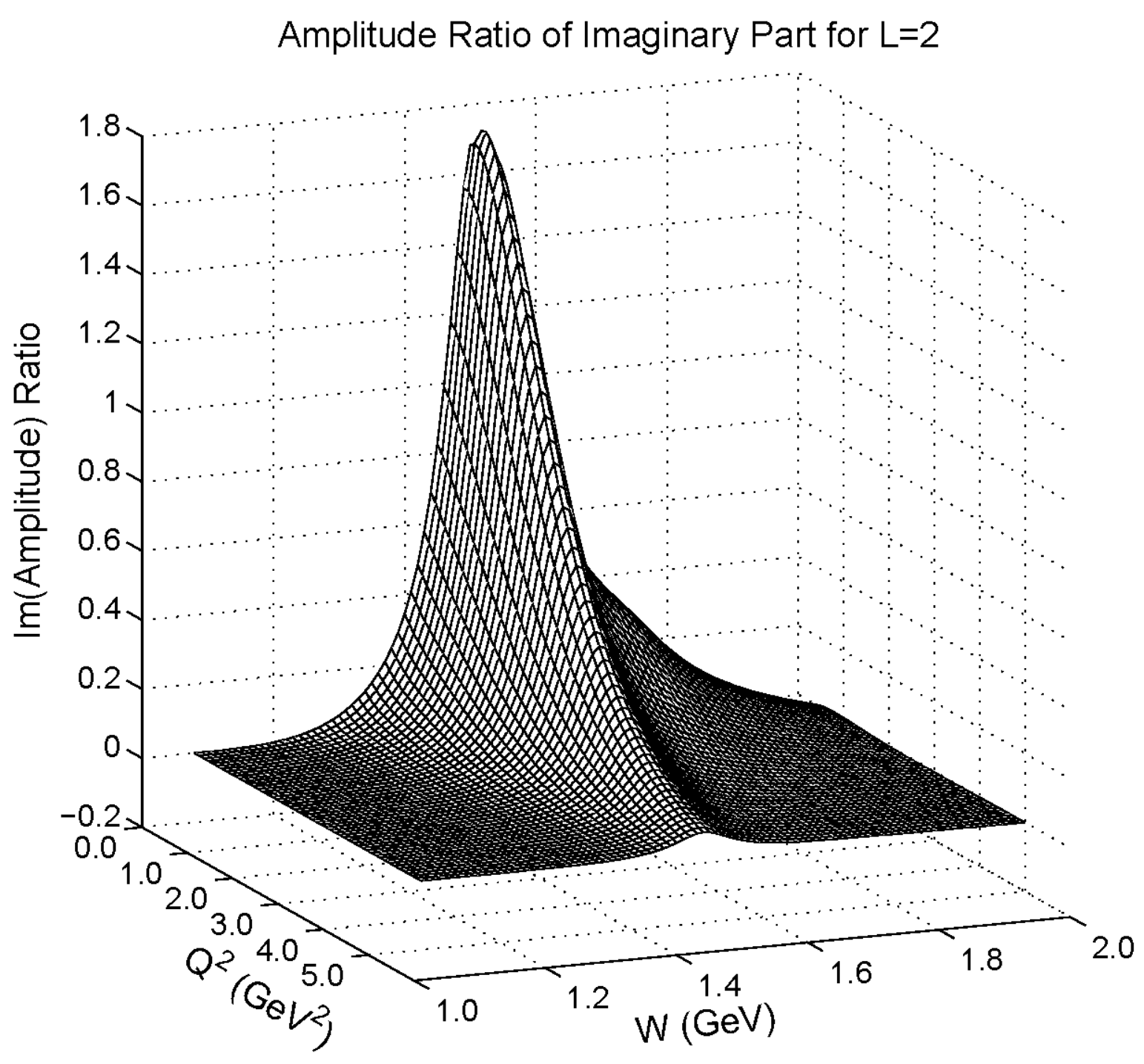}\\
\end{figure}
\begin{figure}[htp]
\epsfxsize=0.48\textwidth\epsfbox{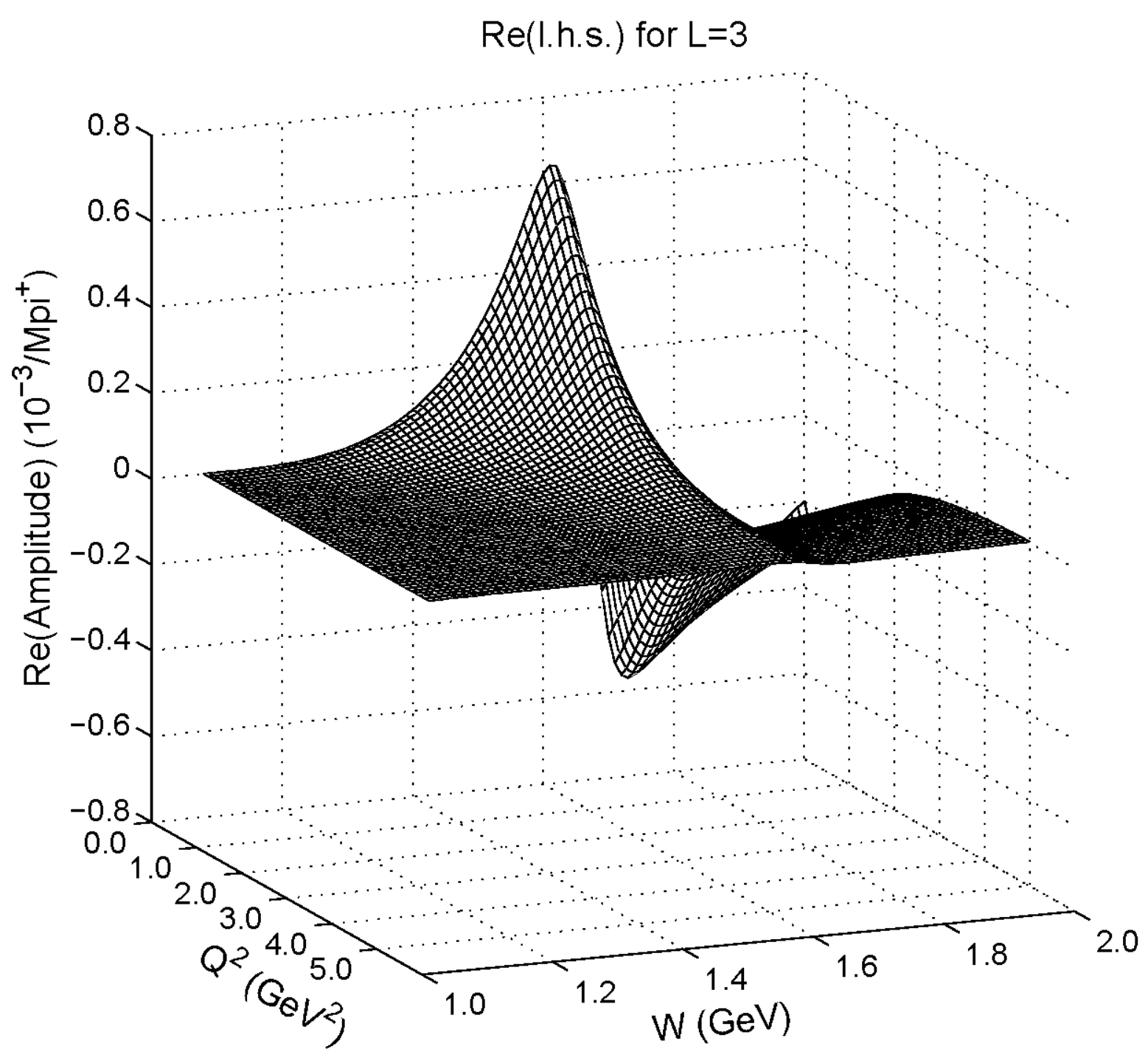}
\epsfxsize=0.48\textwidth\epsfbox{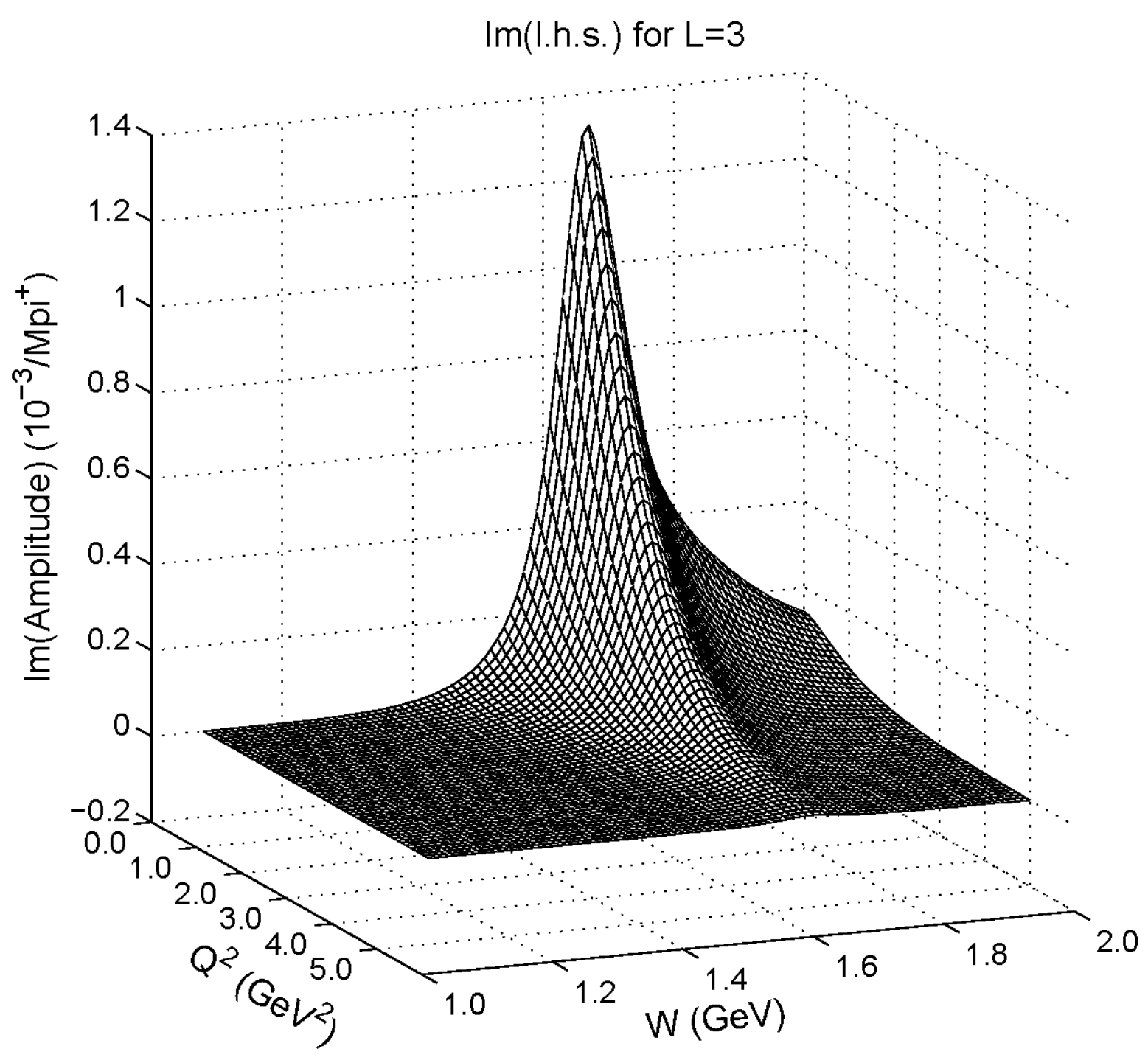}\\[1mm]
\epsfxsize=0.48\textwidth\epsfbox{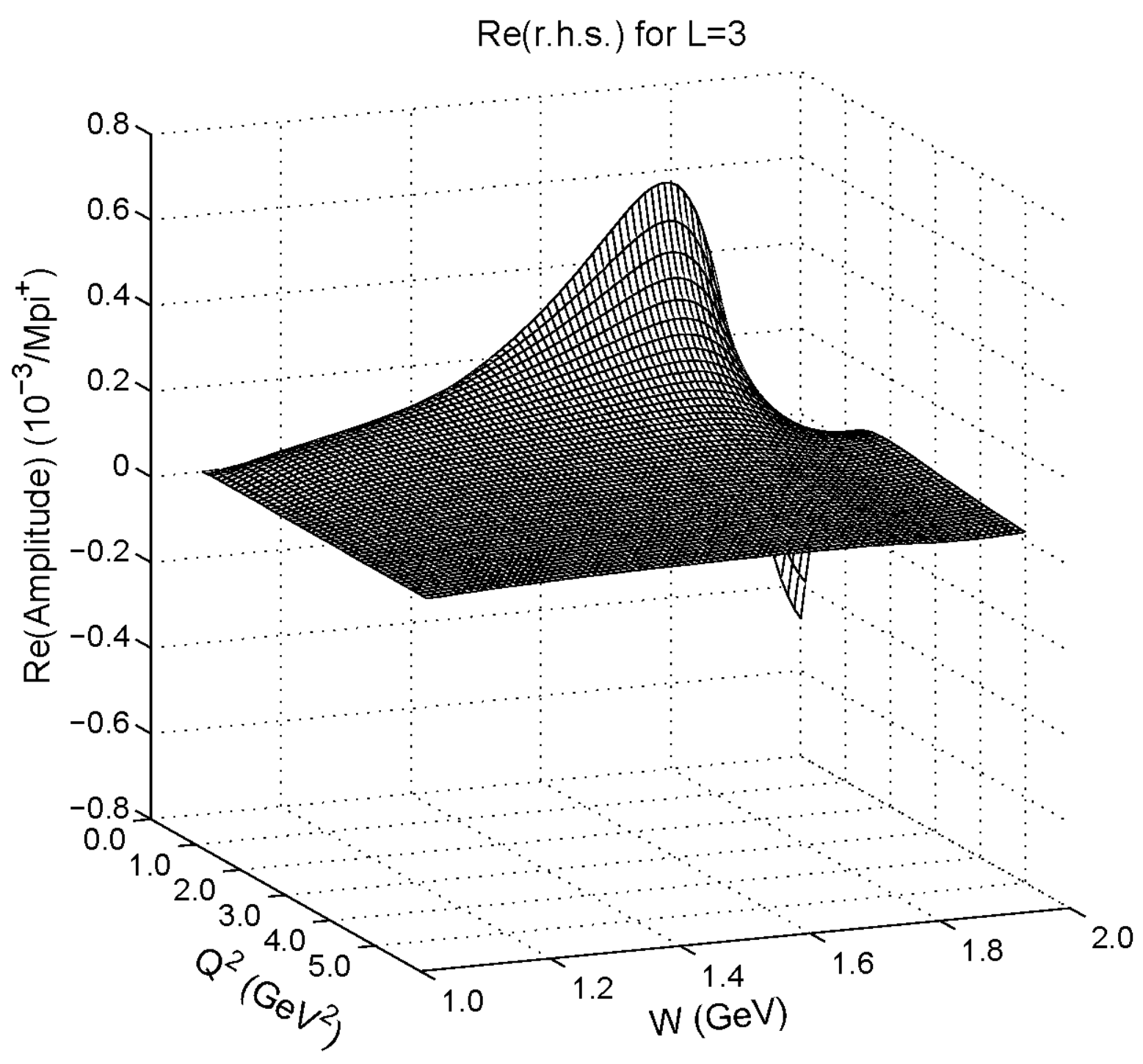}
\epsfxsize=0.48\textwidth\epsfbox{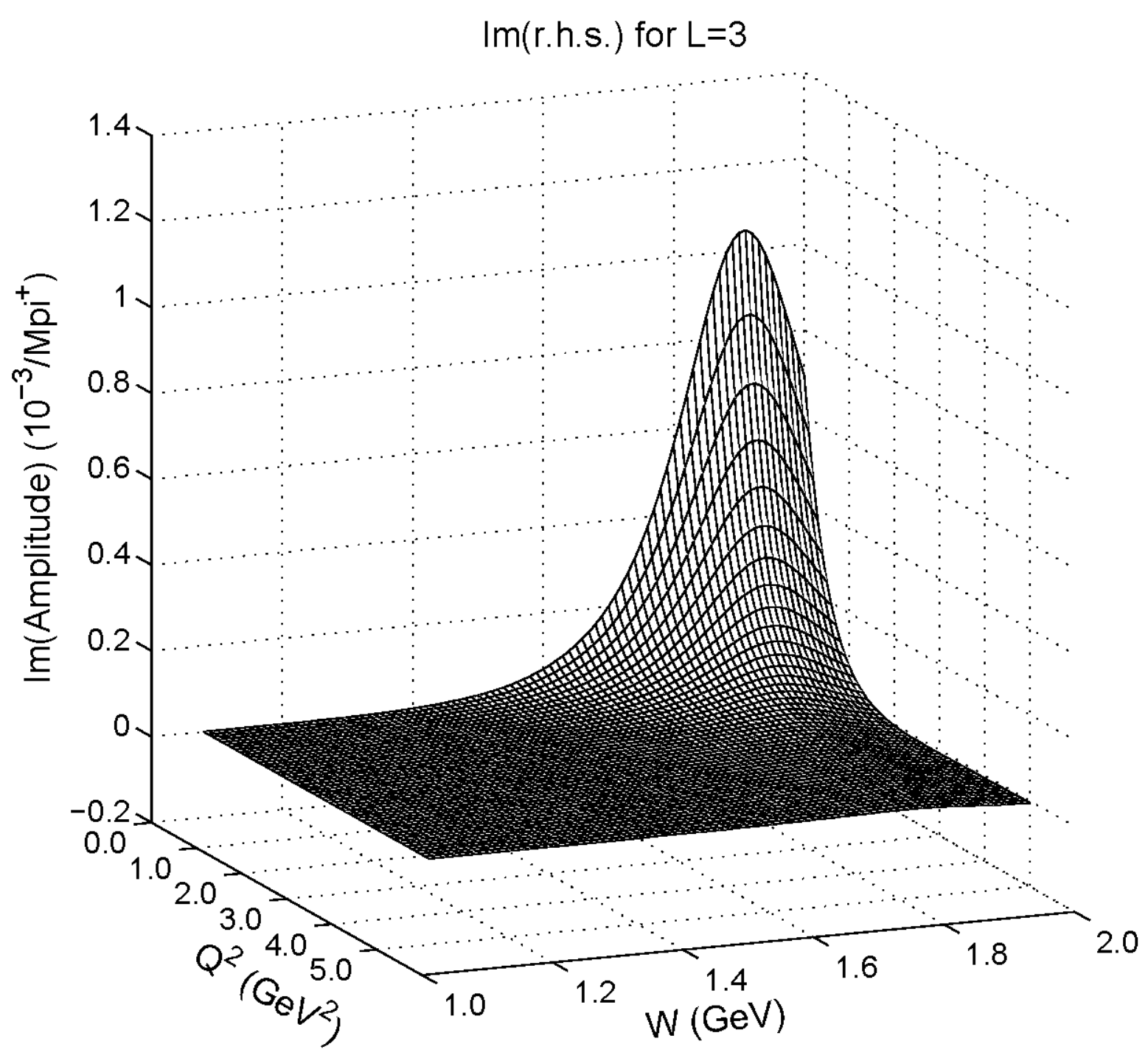}\\[1mm]
\epsfxsize=0.48\textwidth\epsfbox{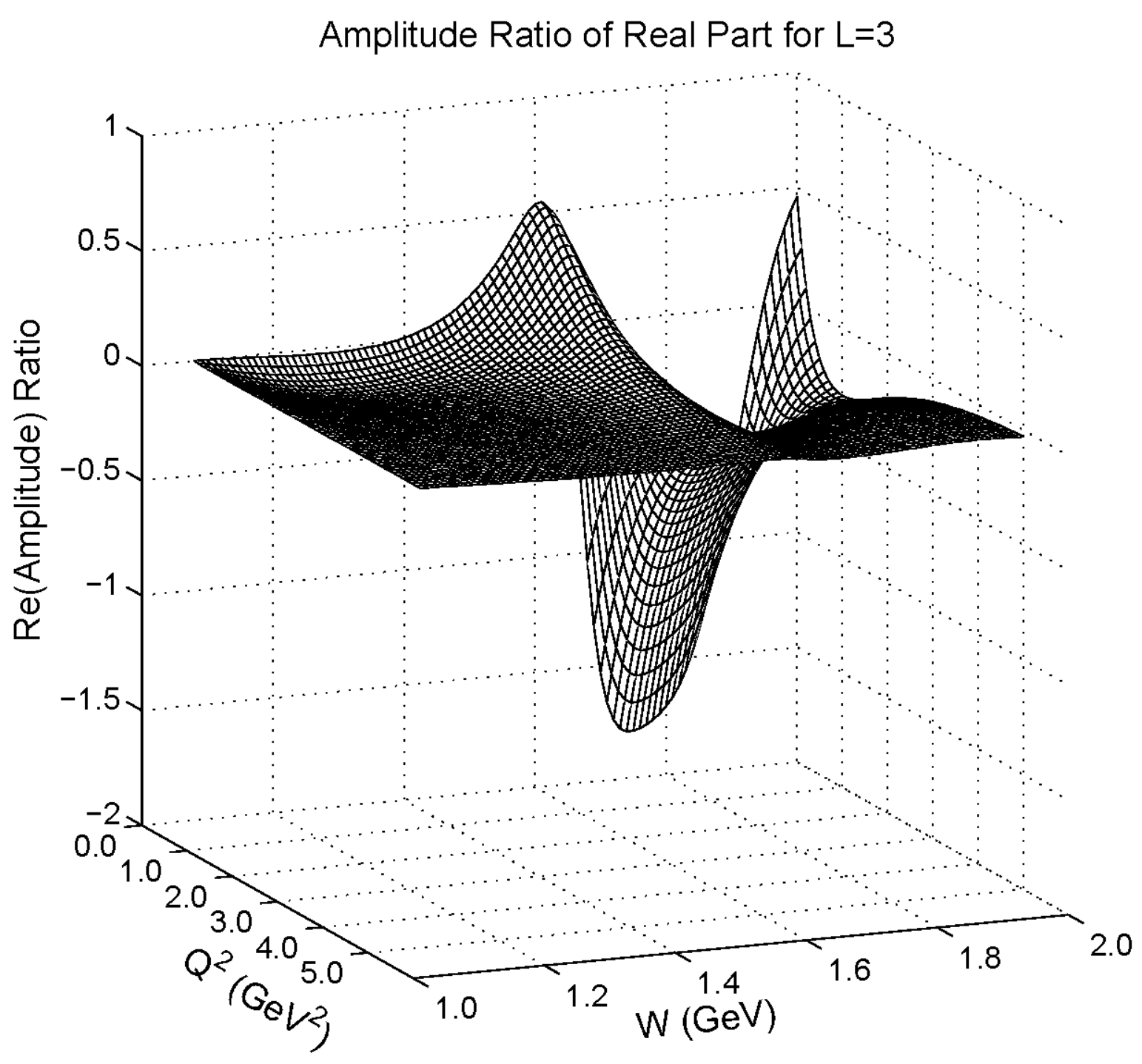}
\epsfxsize=0.48\textwidth\epsfbox{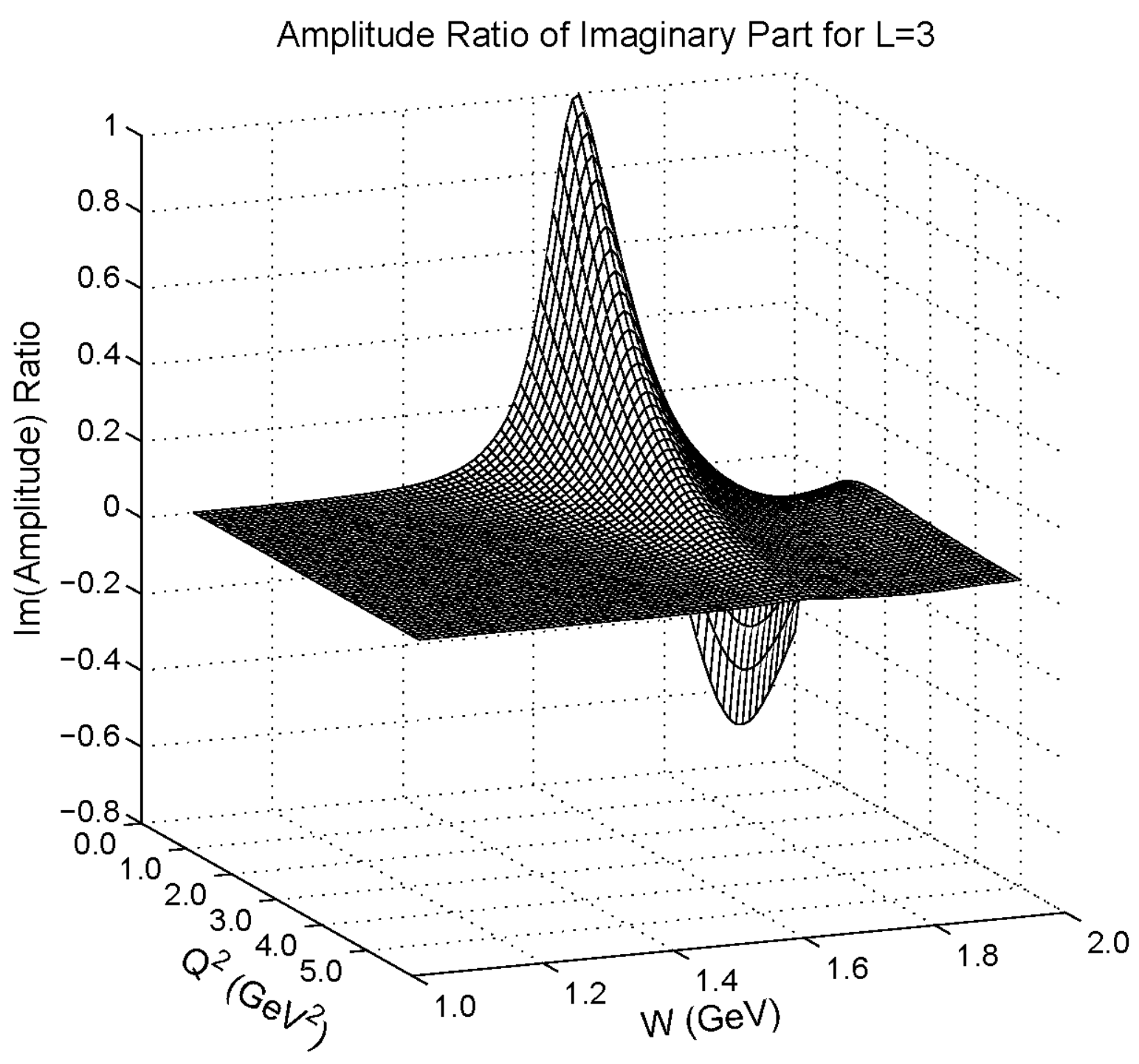}\\
\end{figure}
\begin{figure}[htp]
\epsfxsize=0.48\textwidth\epsfbox{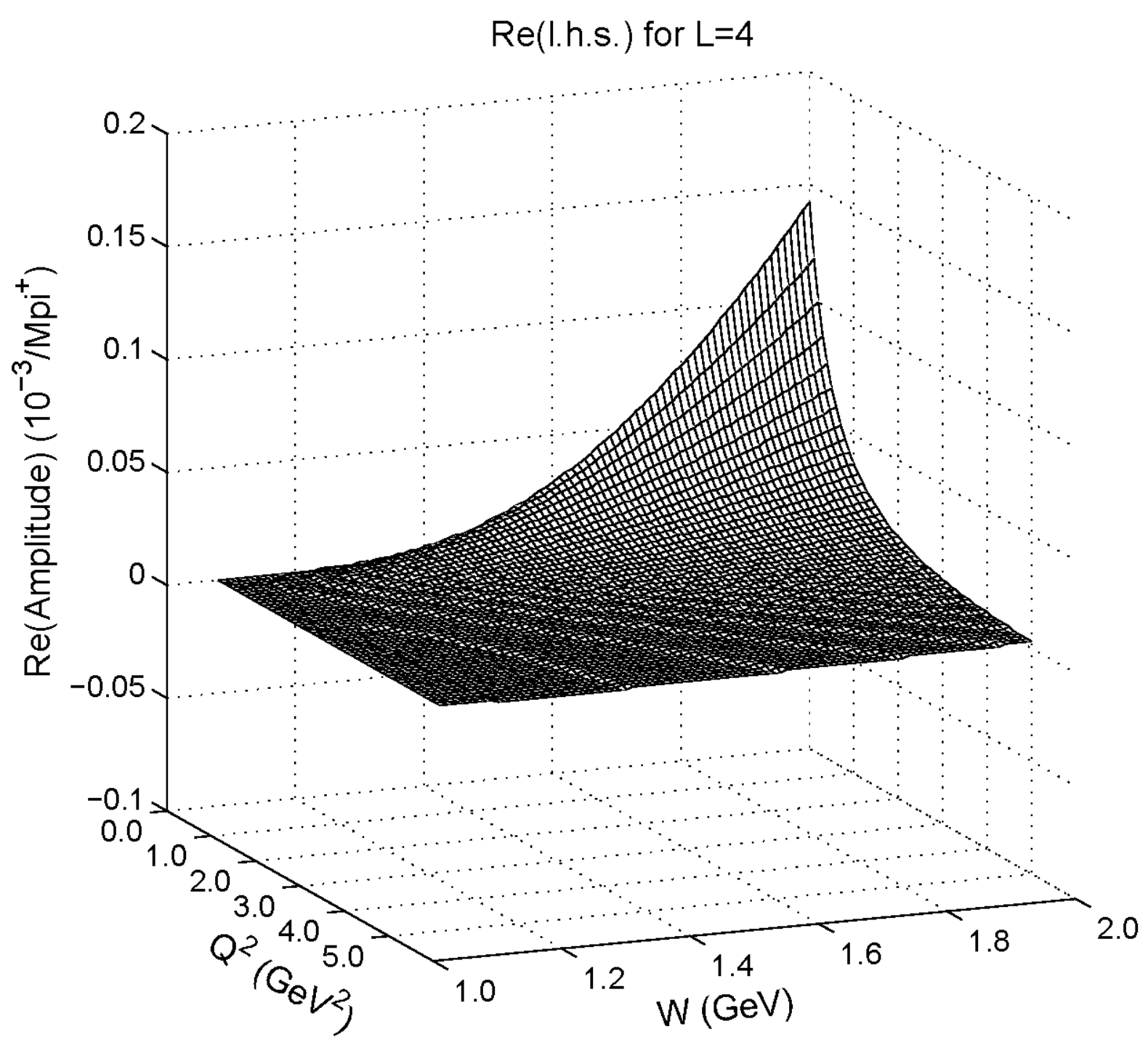}
\epsfxsize=0.48\textwidth\epsfbox{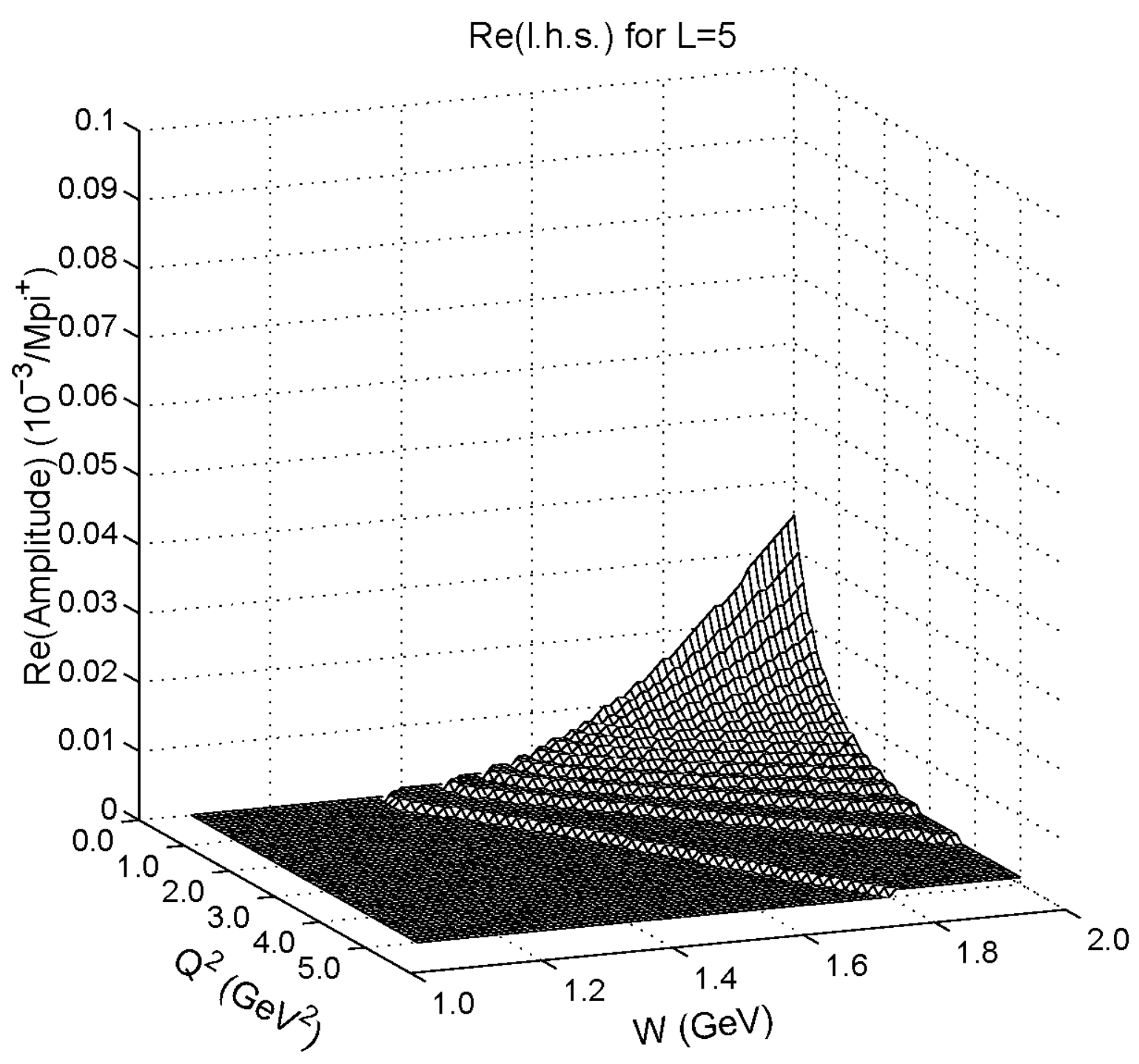}\\[1mm]
\epsfxsize=0.48\textwidth\epsfbox{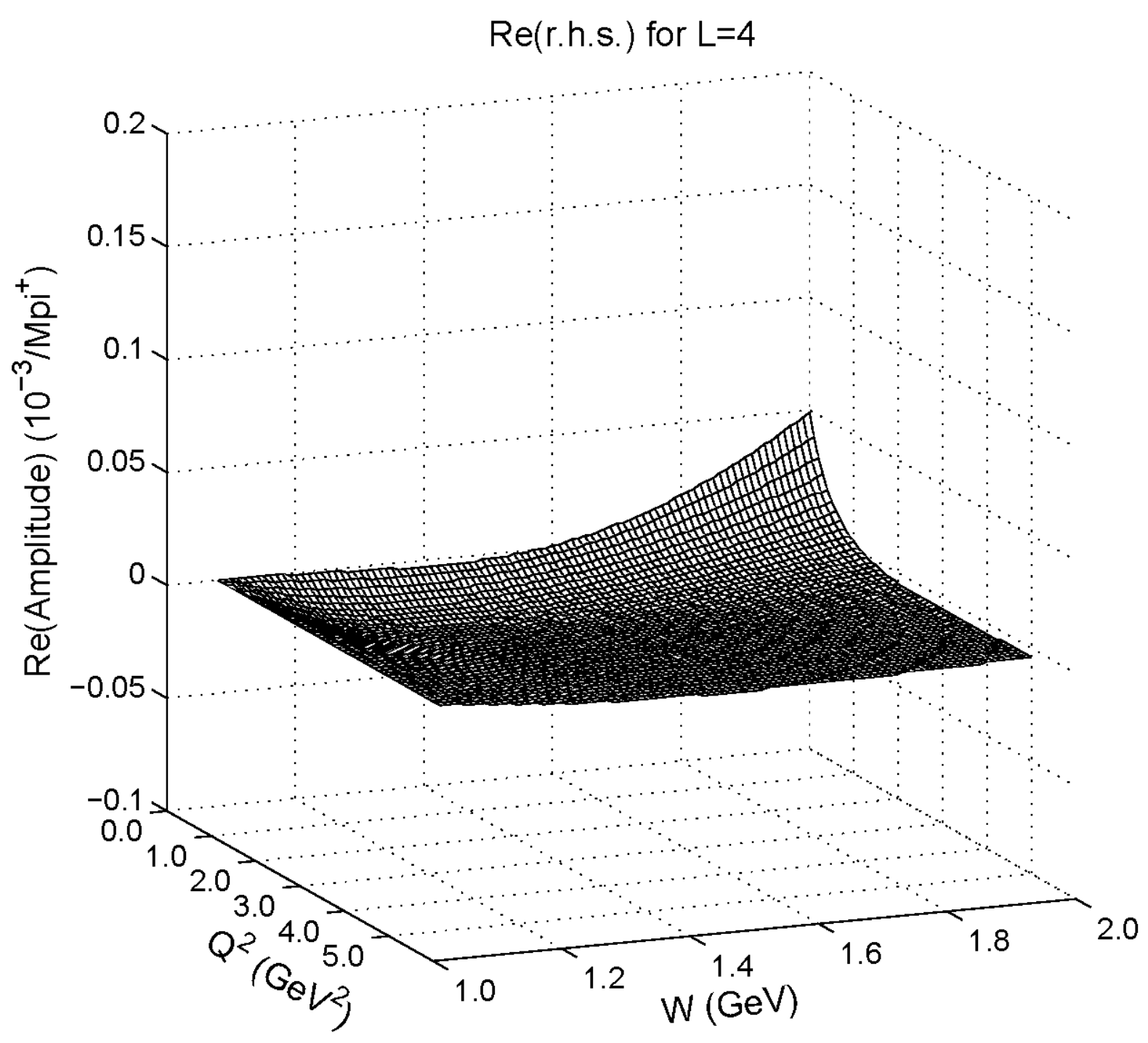}
\epsfxsize=0.48\textwidth\epsfbox{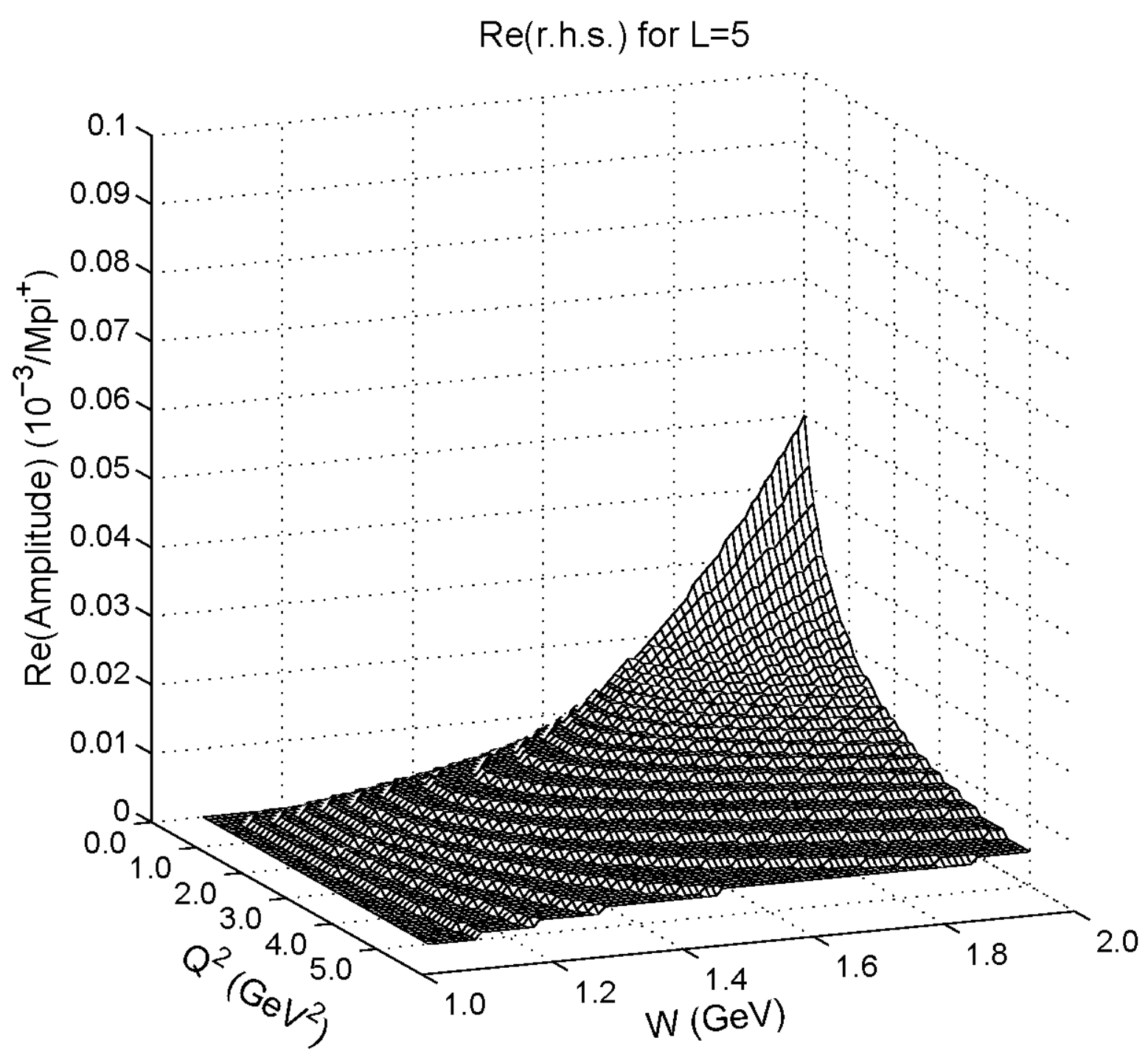}\\[1mm]
\epsfxsize=0.48\textwidth\epsfbox{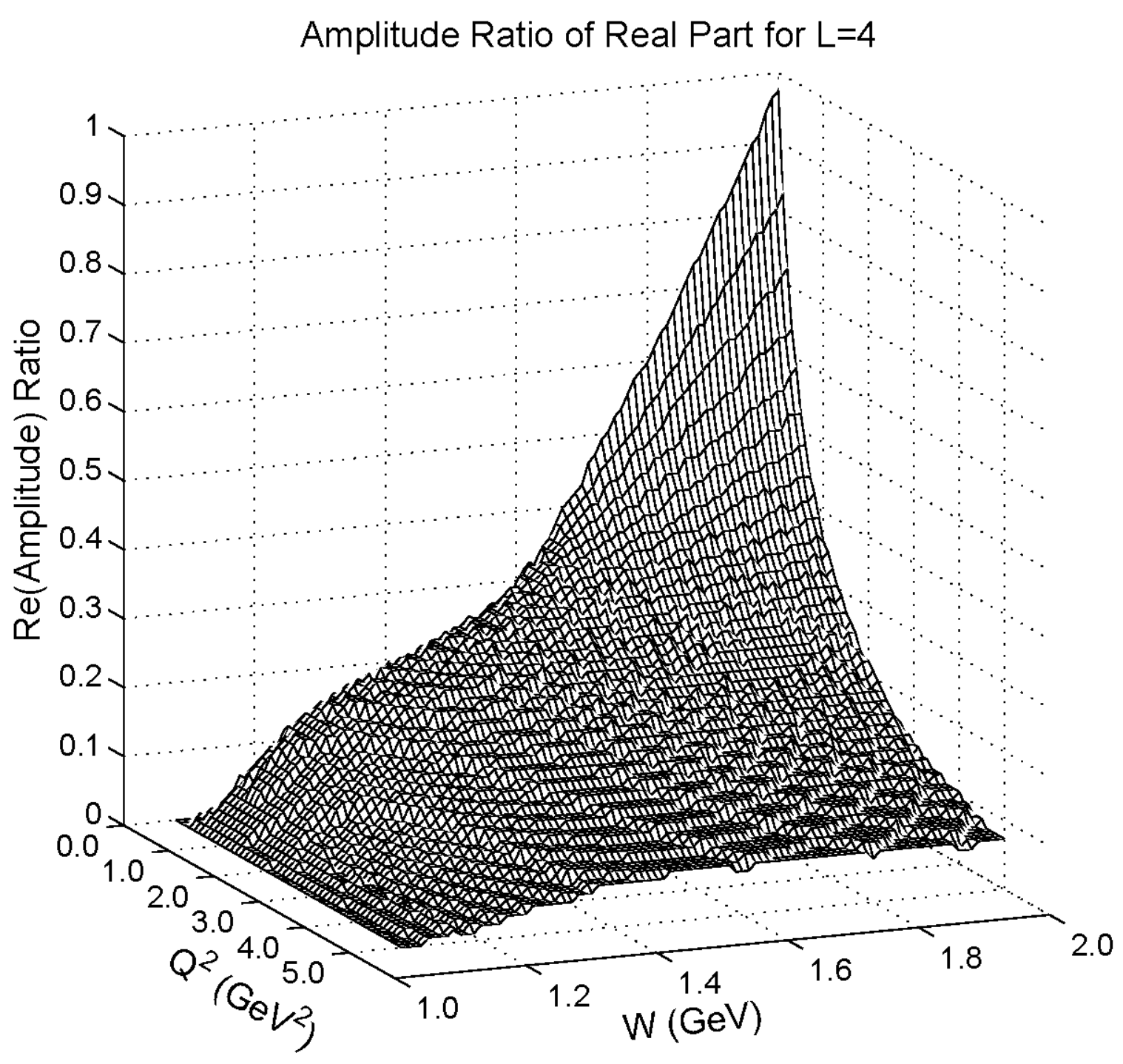}
\epsfxsize=0.48\textwidth\epsfbox{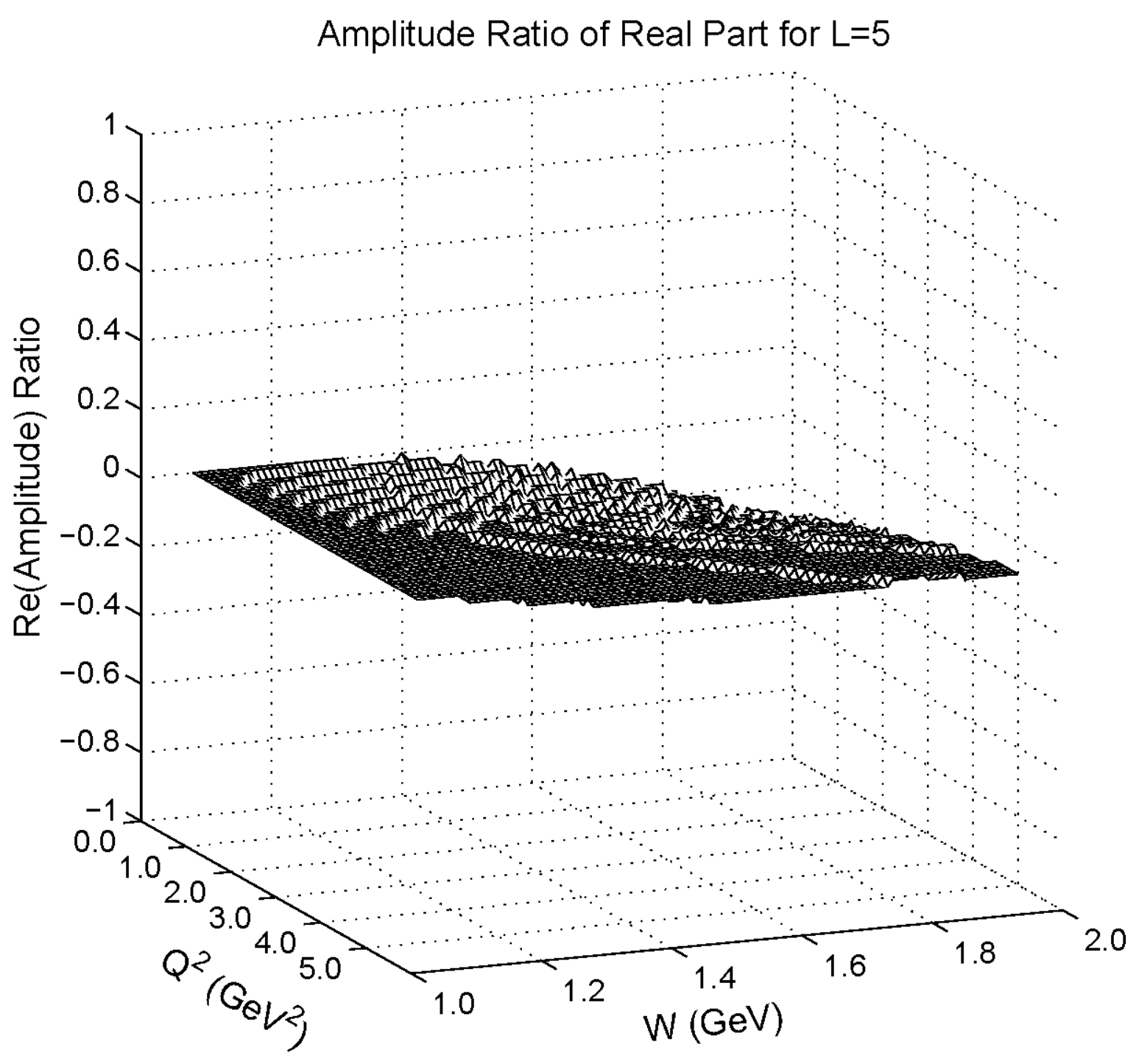}
\end{figure}
%
%
%
\begin{figure}[htp]
\epsfxsize=0.48\textwidth\epsfbox{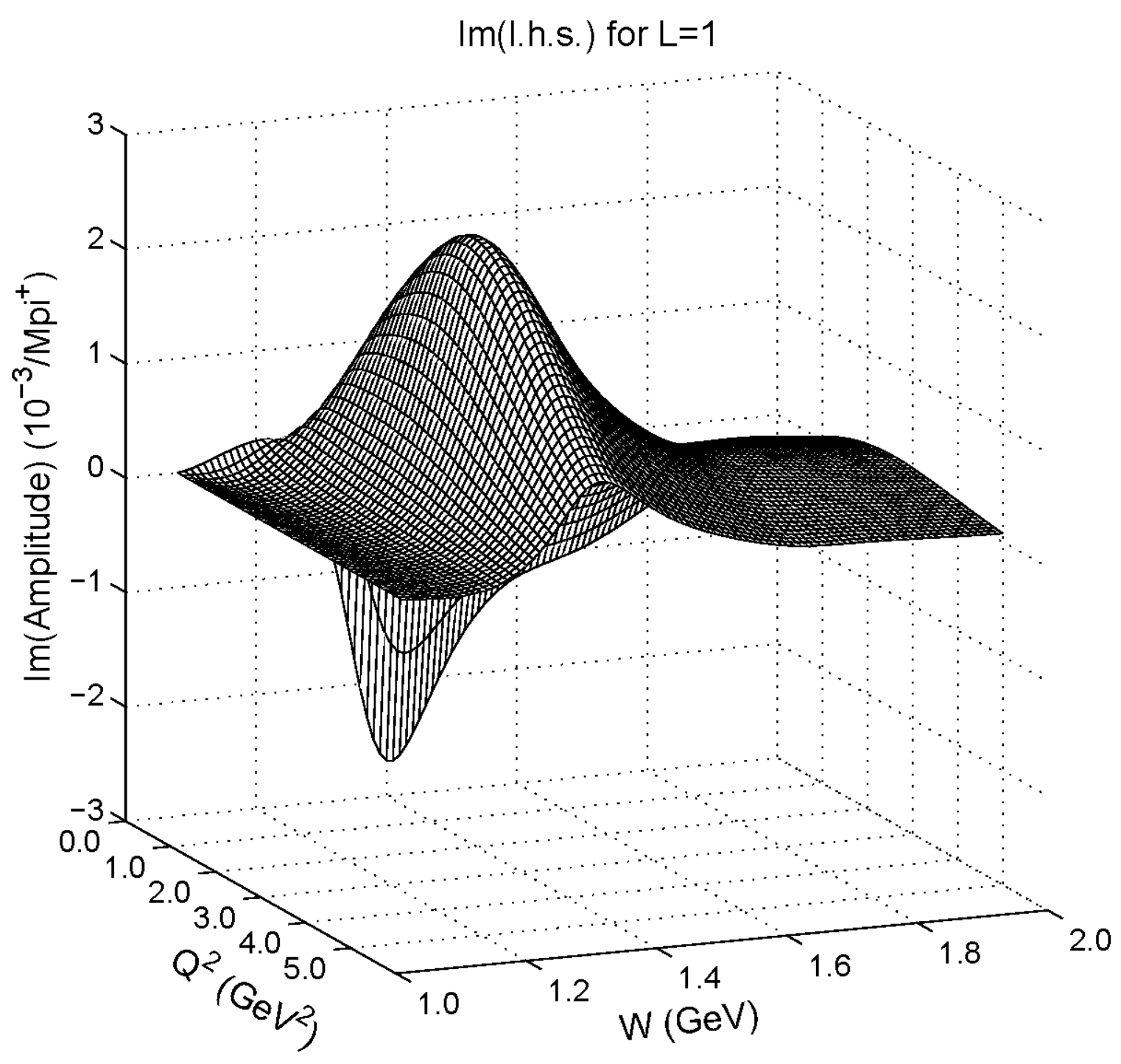}\\[1mm]
\epsfxsize=0.48\textwidth\epsfbox{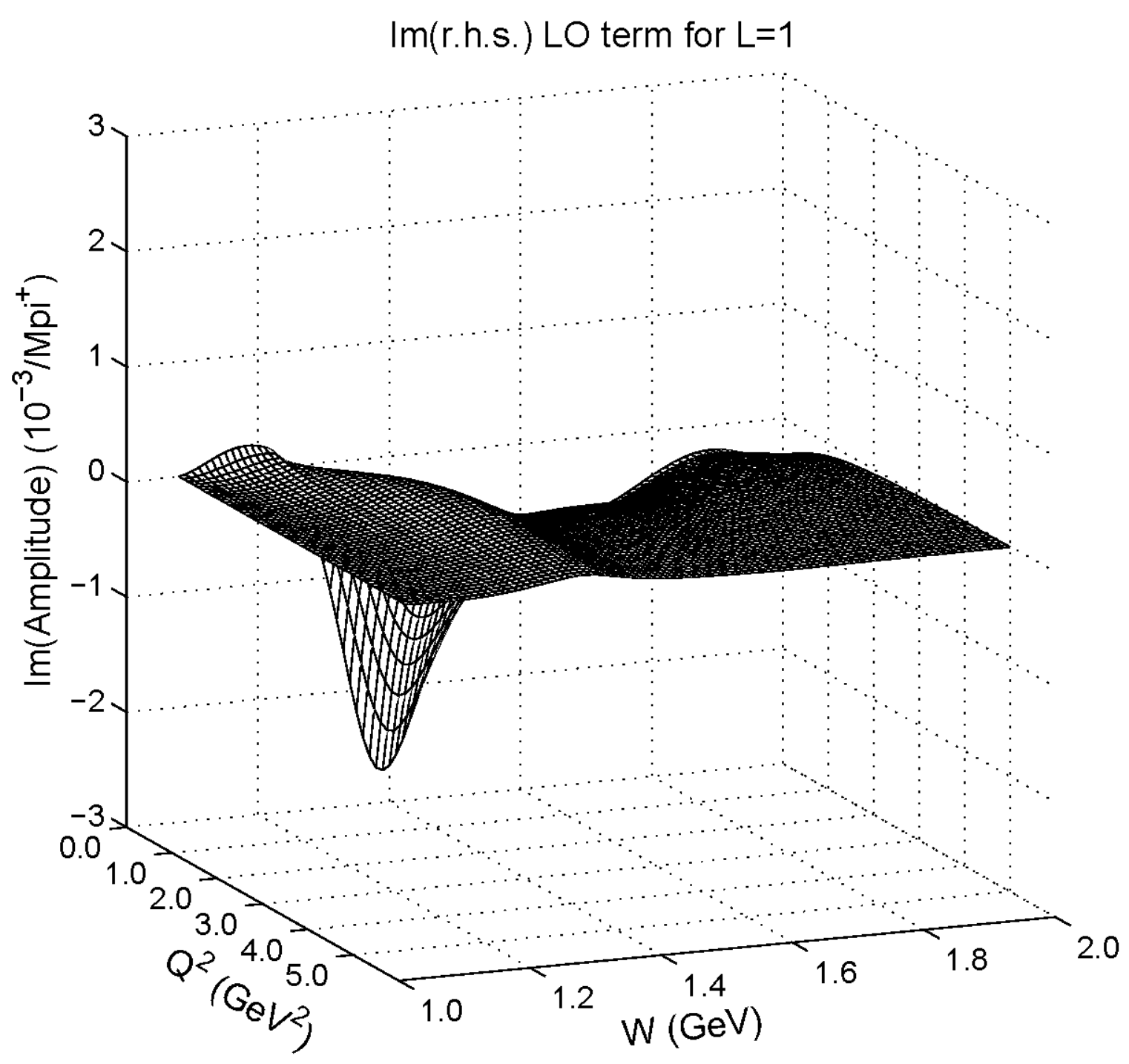}
\epsfxsize=0.48\textwidth\epsfbox{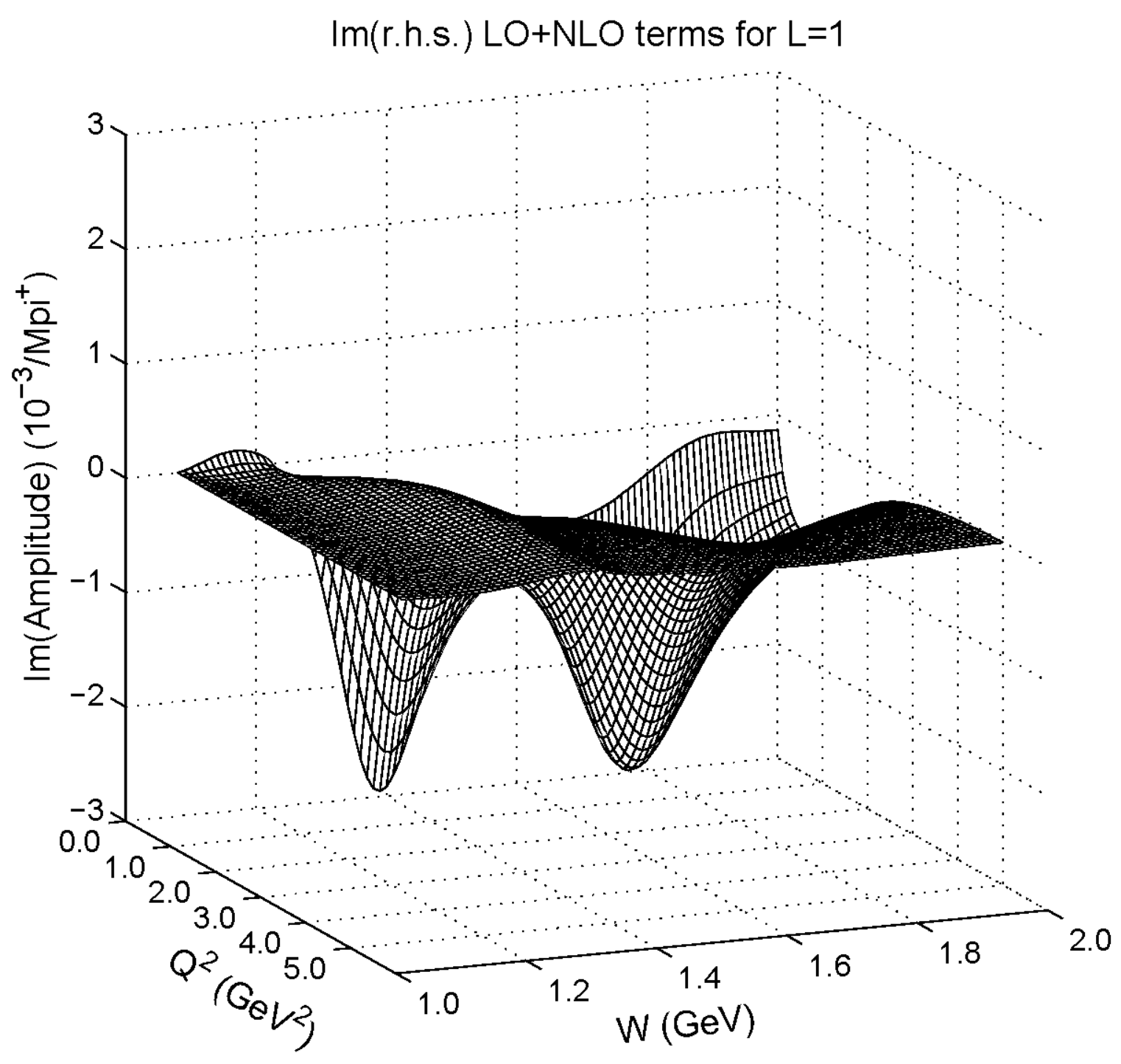}\\[1mm]
\epsfxsize=0.48\textwidth\epsfbox{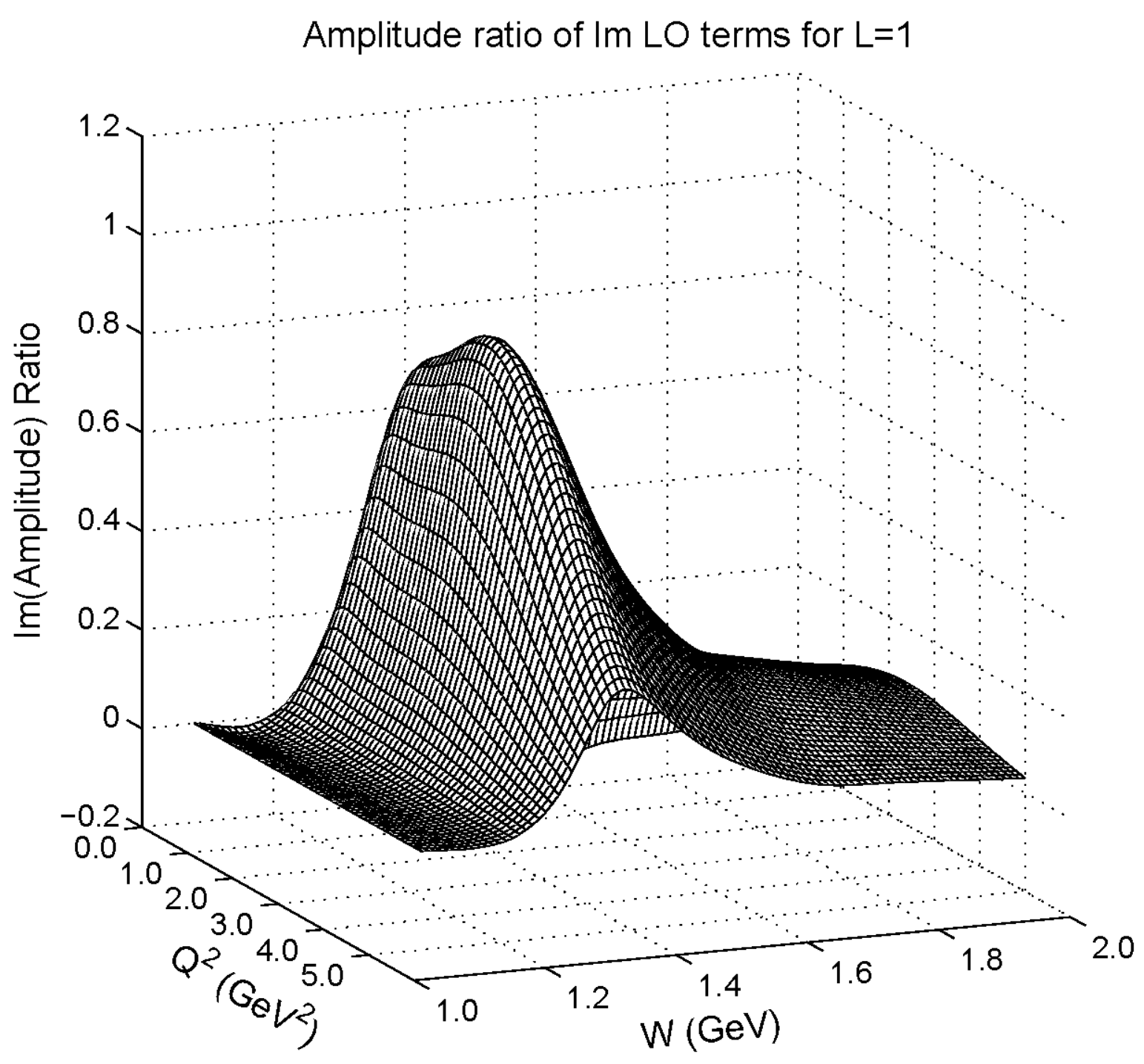}
\epsfxsize=0.48\textwidth\epsfbox{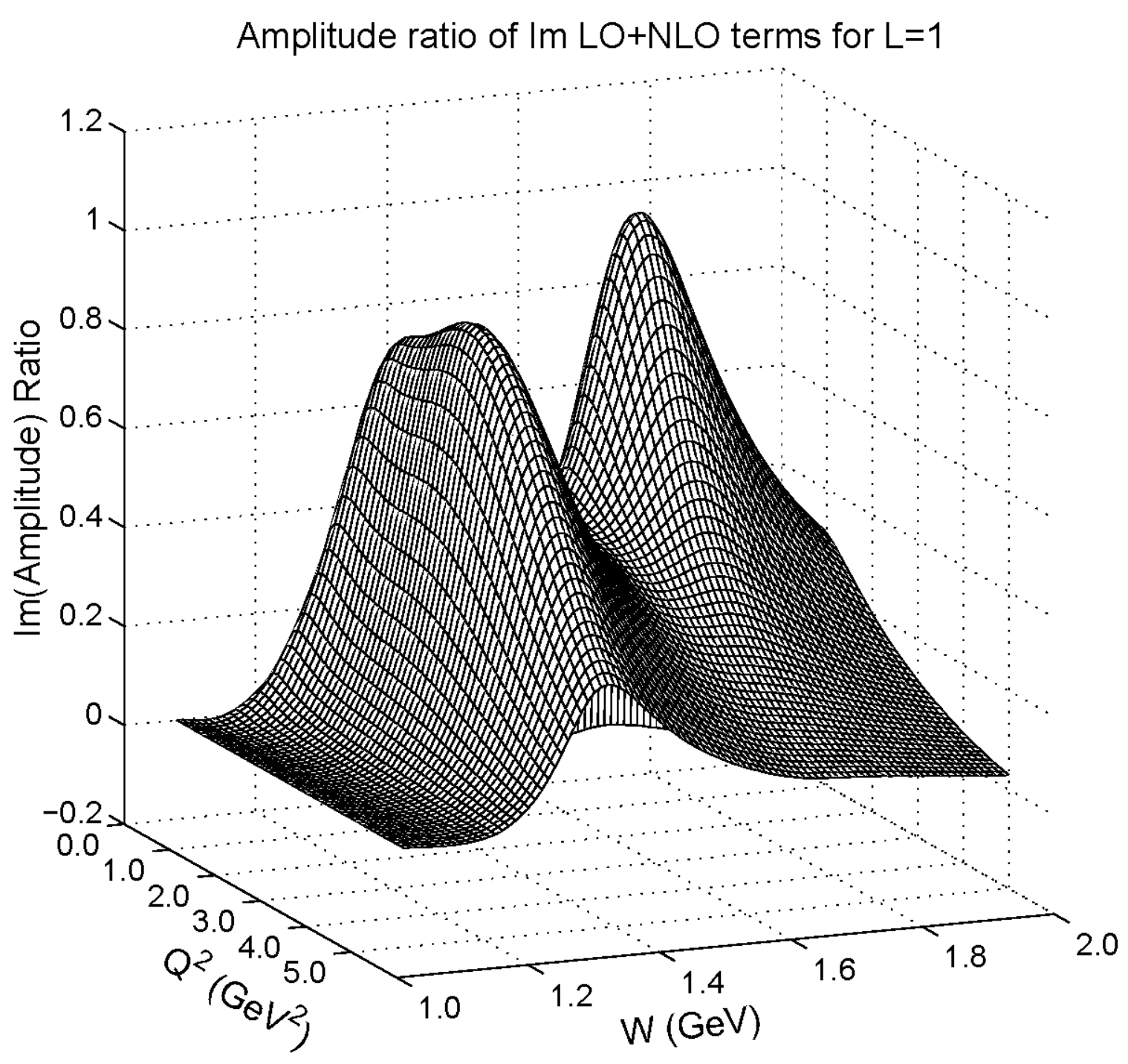}\\
\end{figure}
\begin{figure}[htp]
\epsfxsize=0.48\textwidth\epsfbox{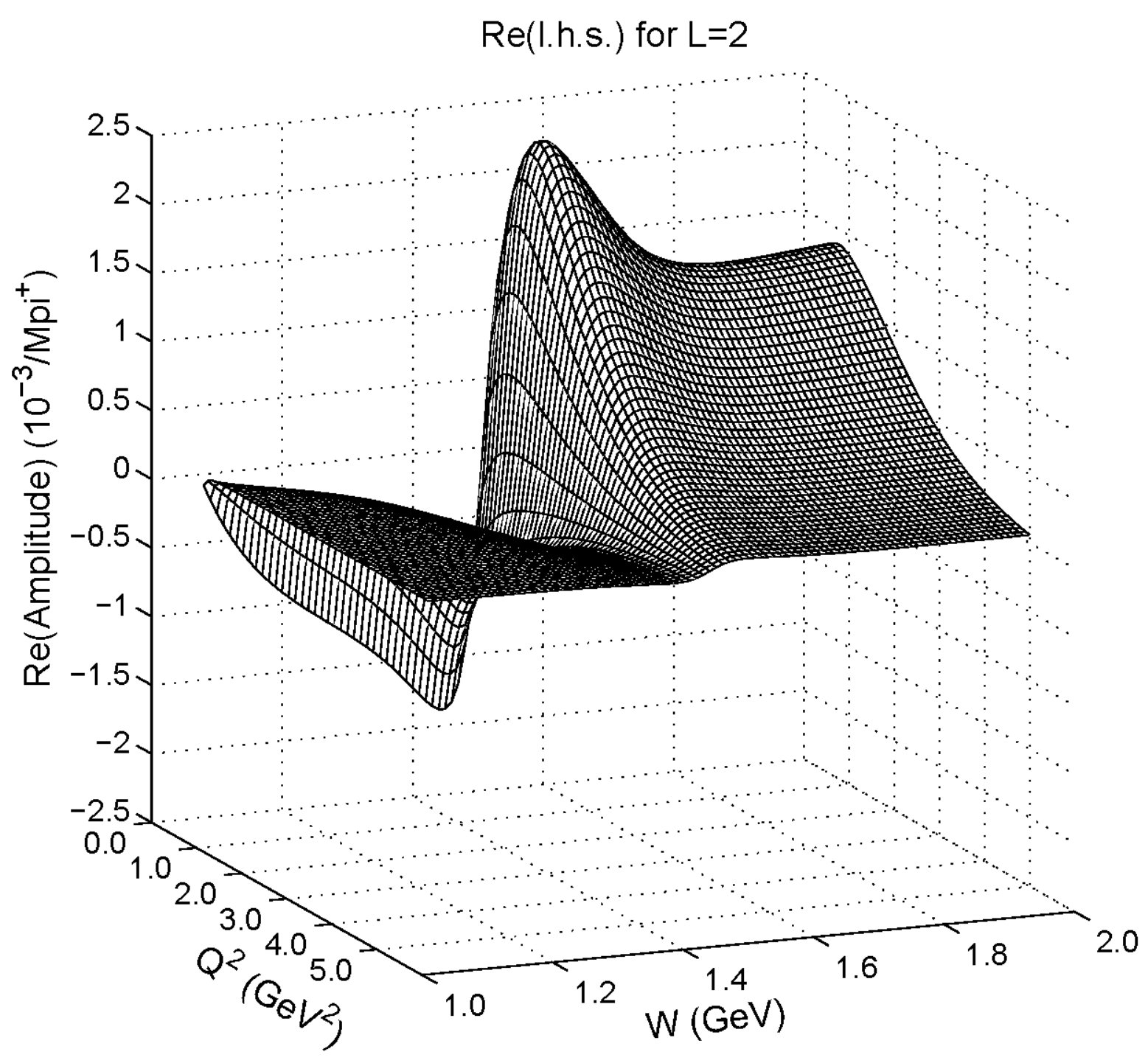}\\[1mm]
\epsfxsize=0.48\textwidth\epsfbox{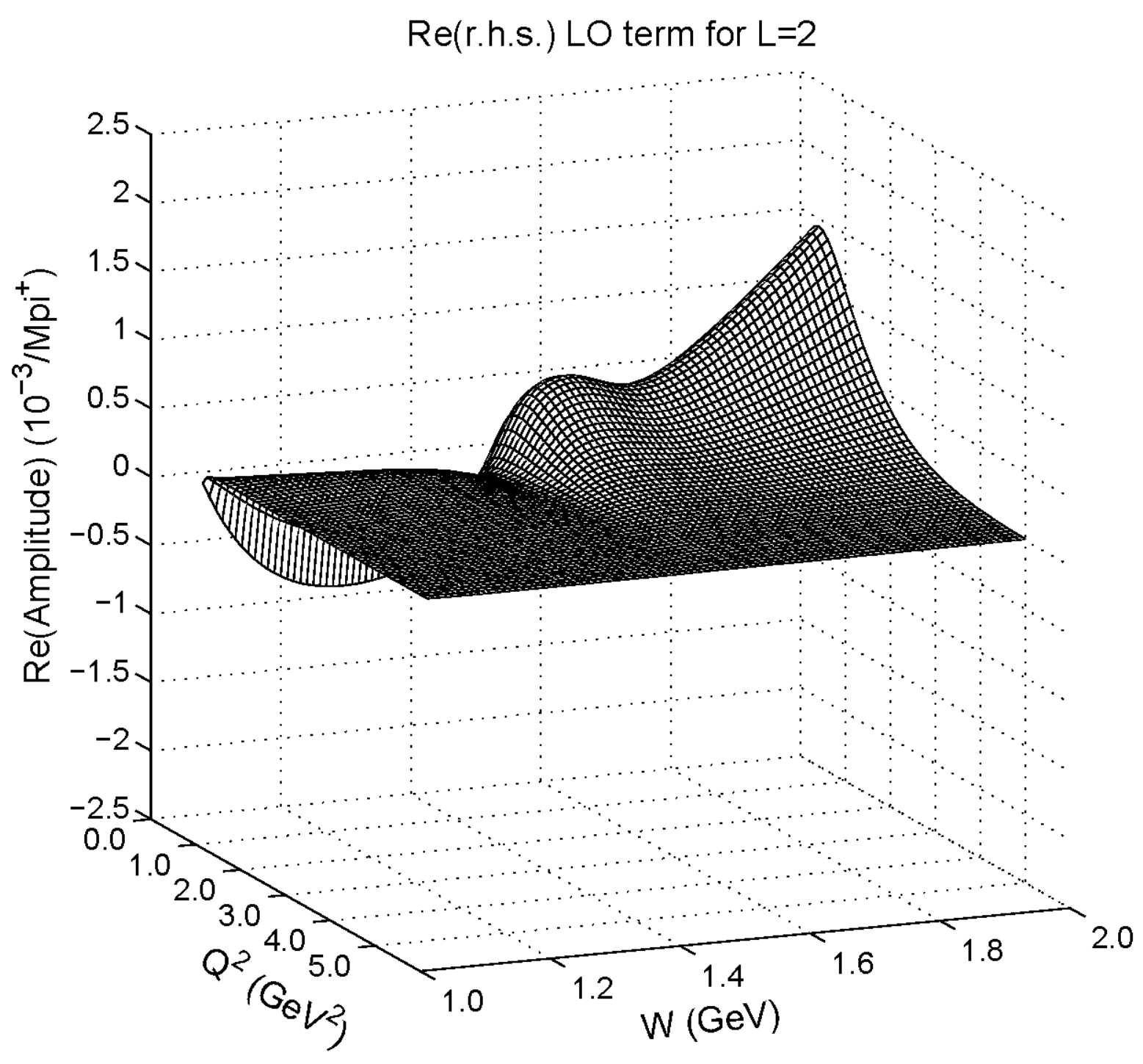}
\epsfxsize=0.48\textwidth\epsfbox{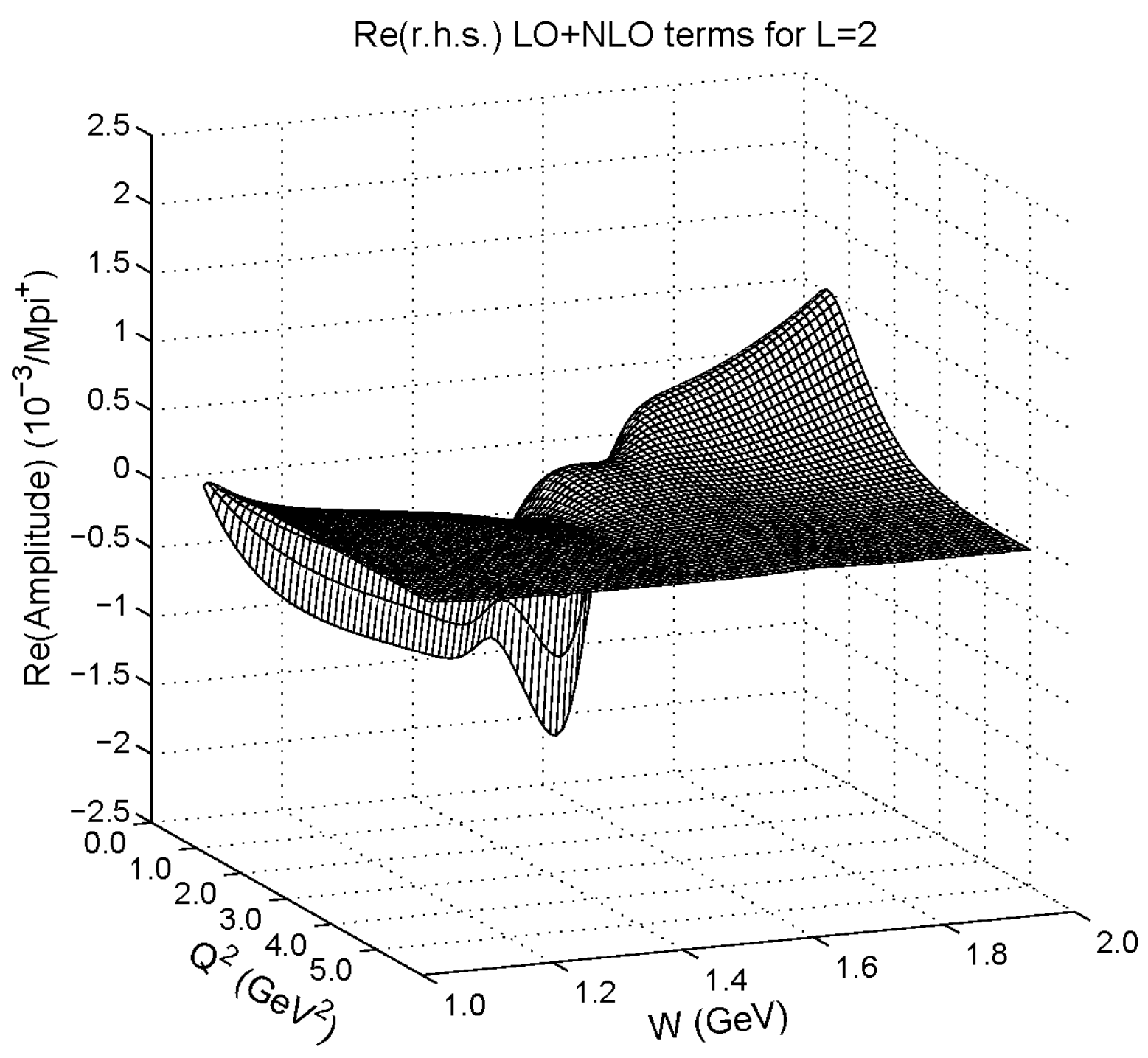}\\[1mm]
\epsfxsize=0.48\textwidth\epsfbox{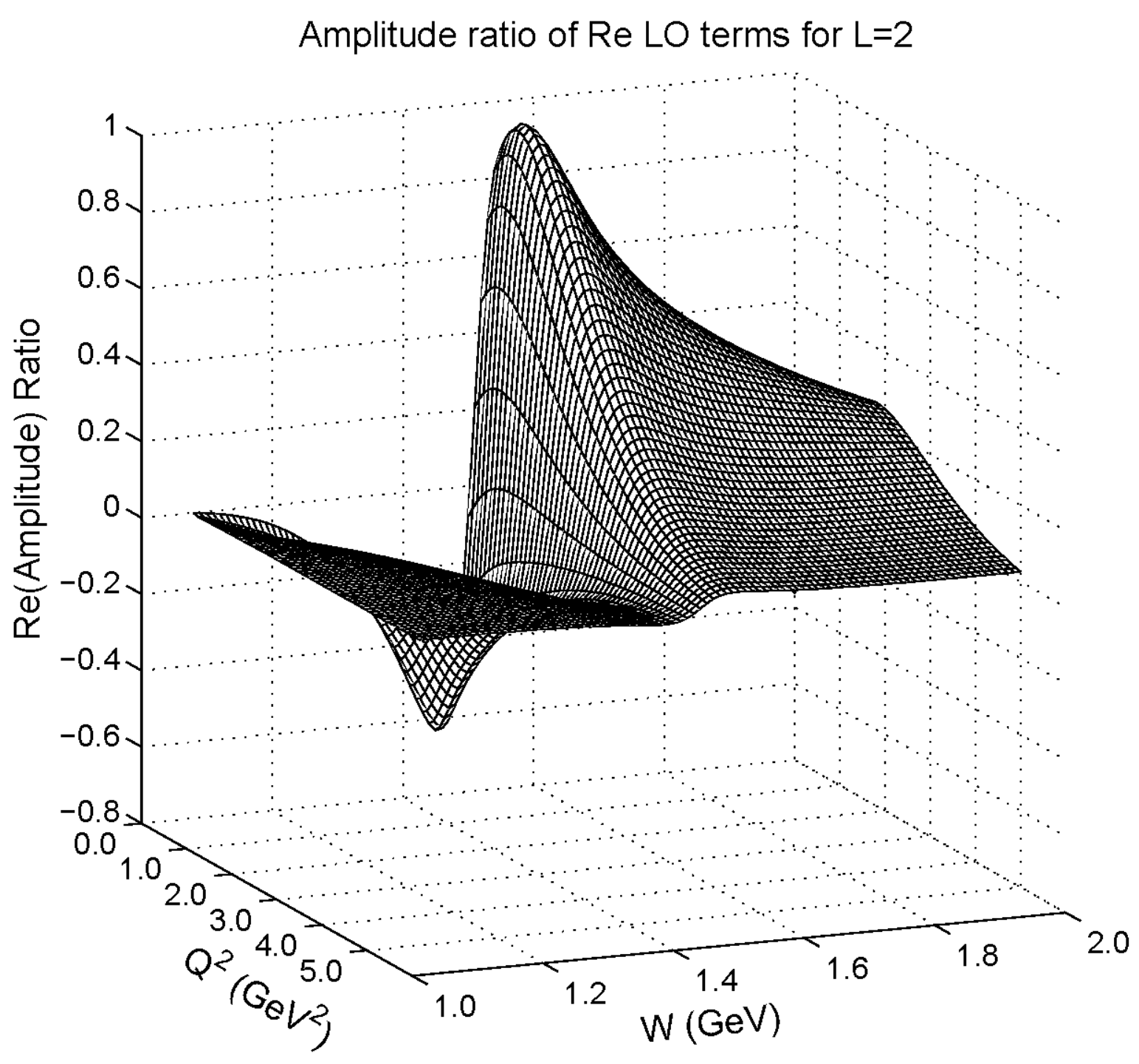}
\epsfxsize=0.48\textwidth\epsfbox{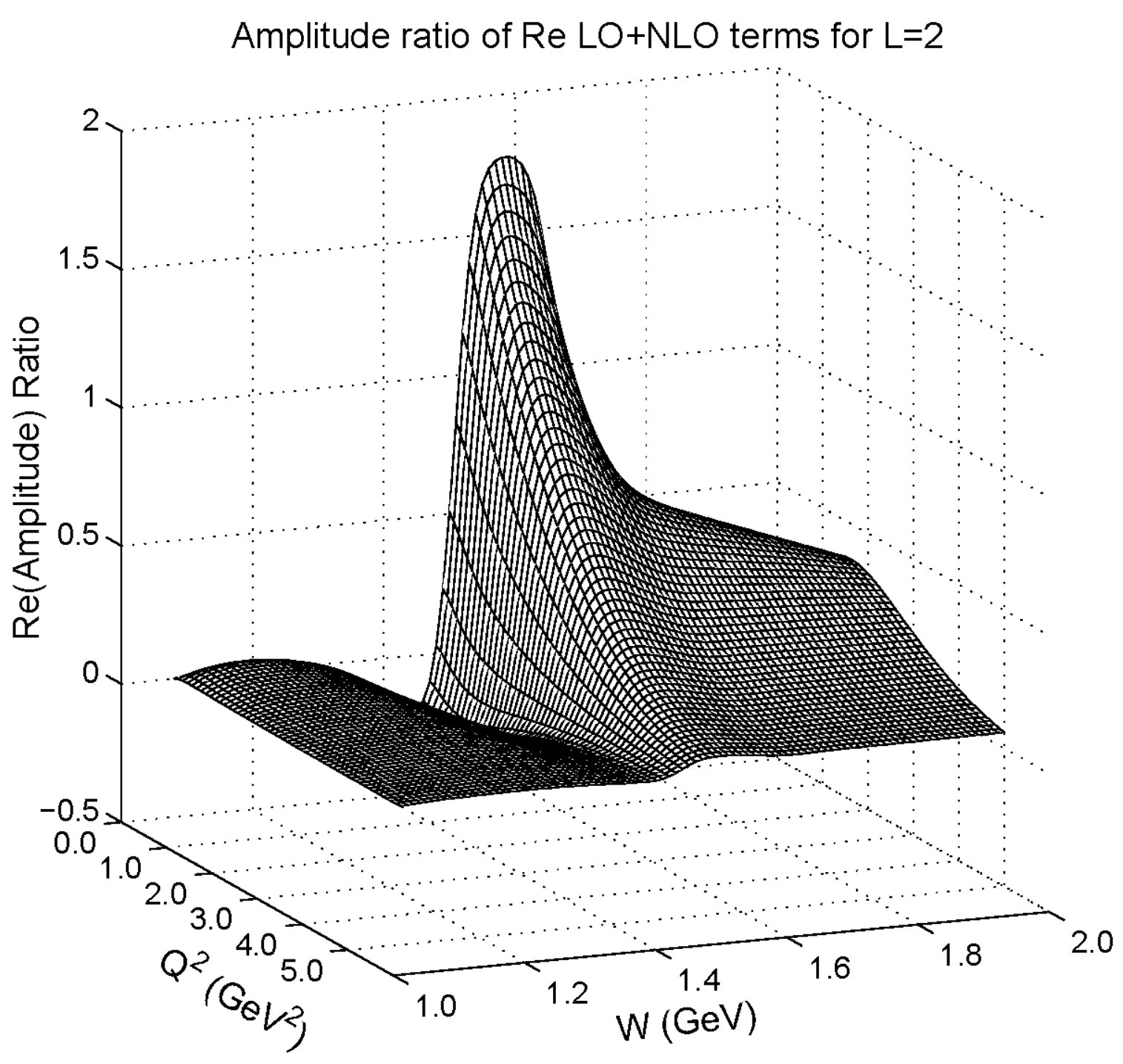}\\
\end{figure}
\begin{figure}[htp]
\epsfxsize=0.48\textwidth\epsfbox{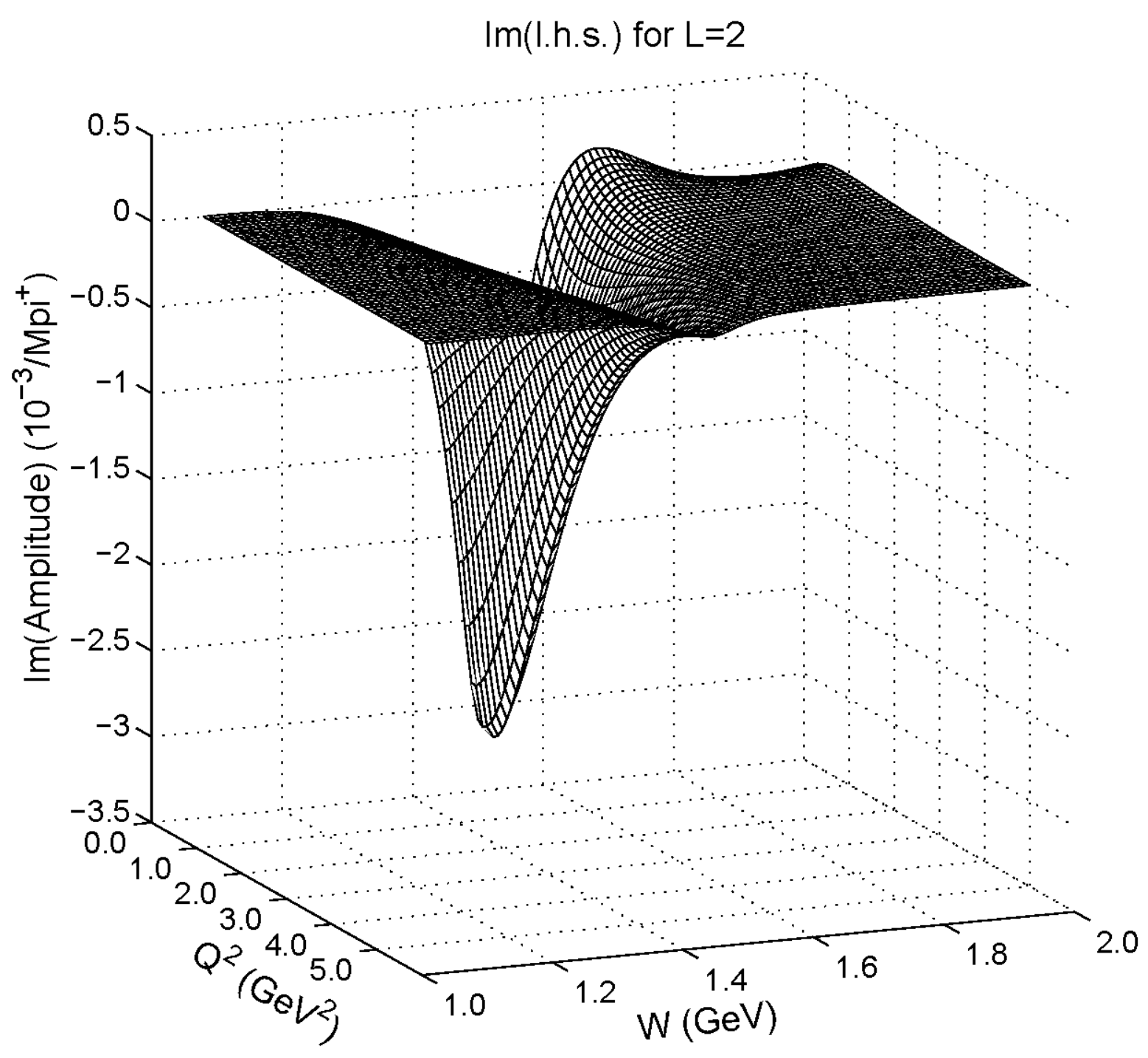}\\[1mm]
\epsfxsize=0.48\textwidth\epsfbox{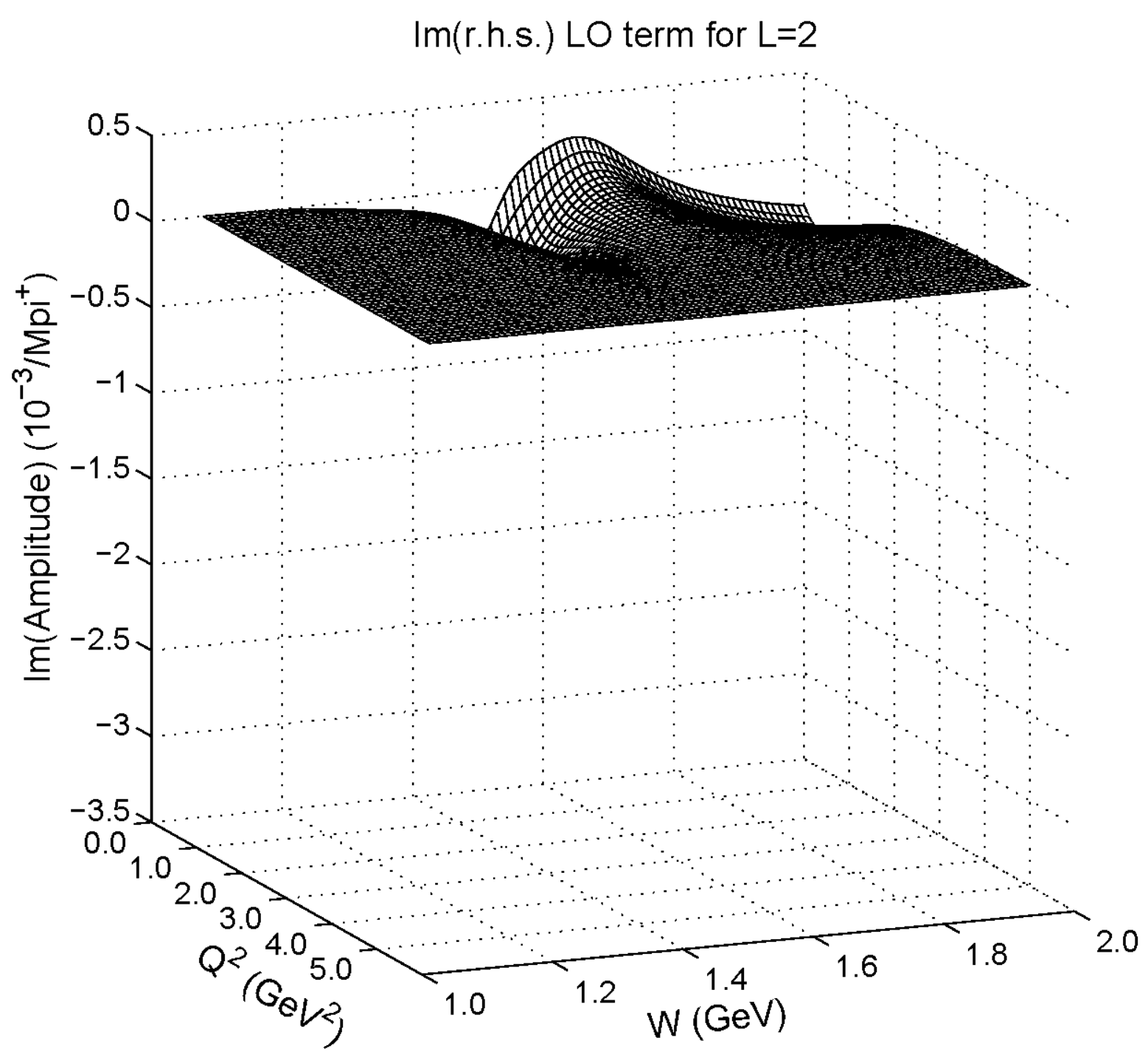}
\epsfxsize=0.48\textwidth\epsfbox{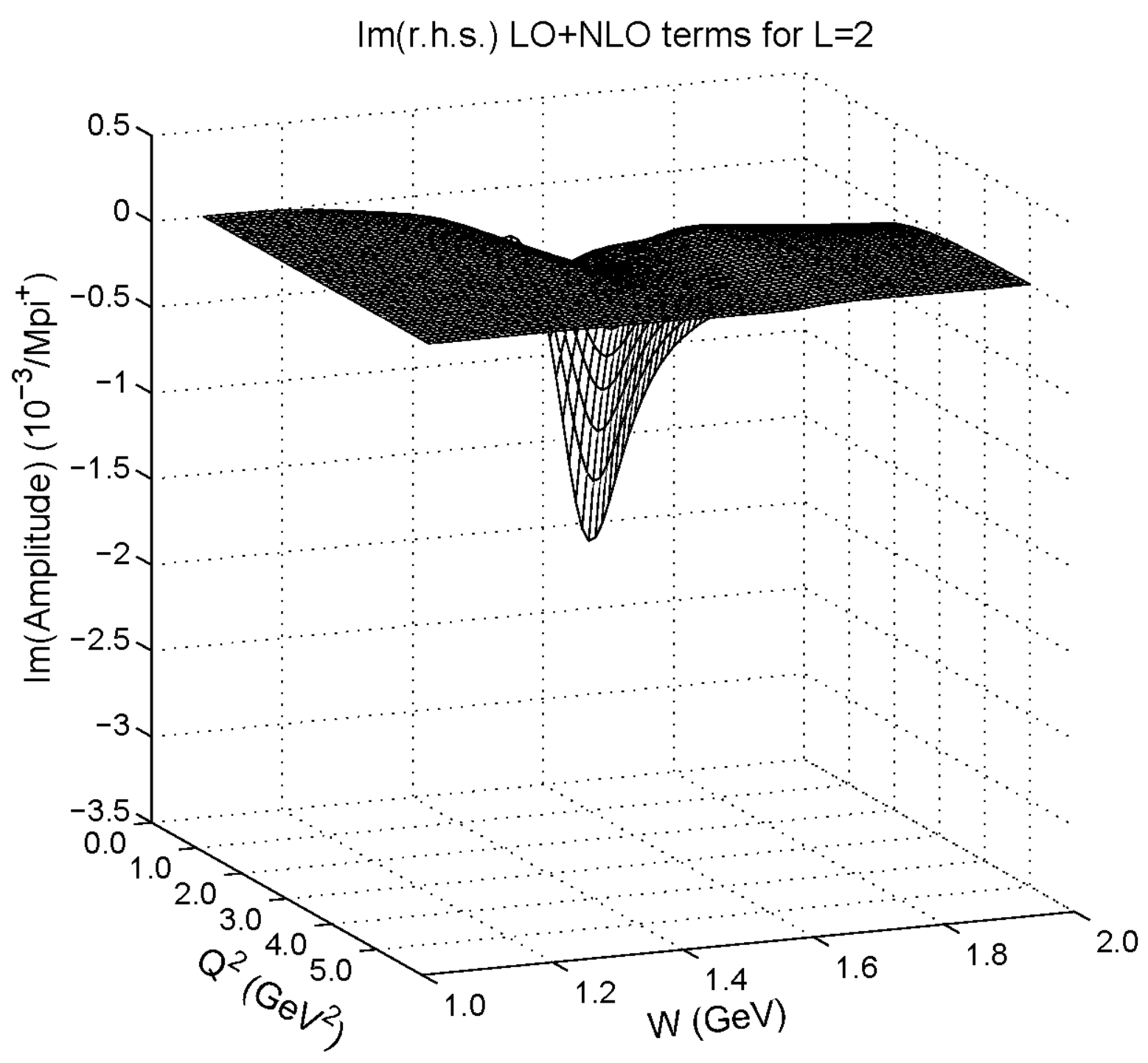}\\[1mm]
\epsfxsize=0.48\textwidth\epsfbox{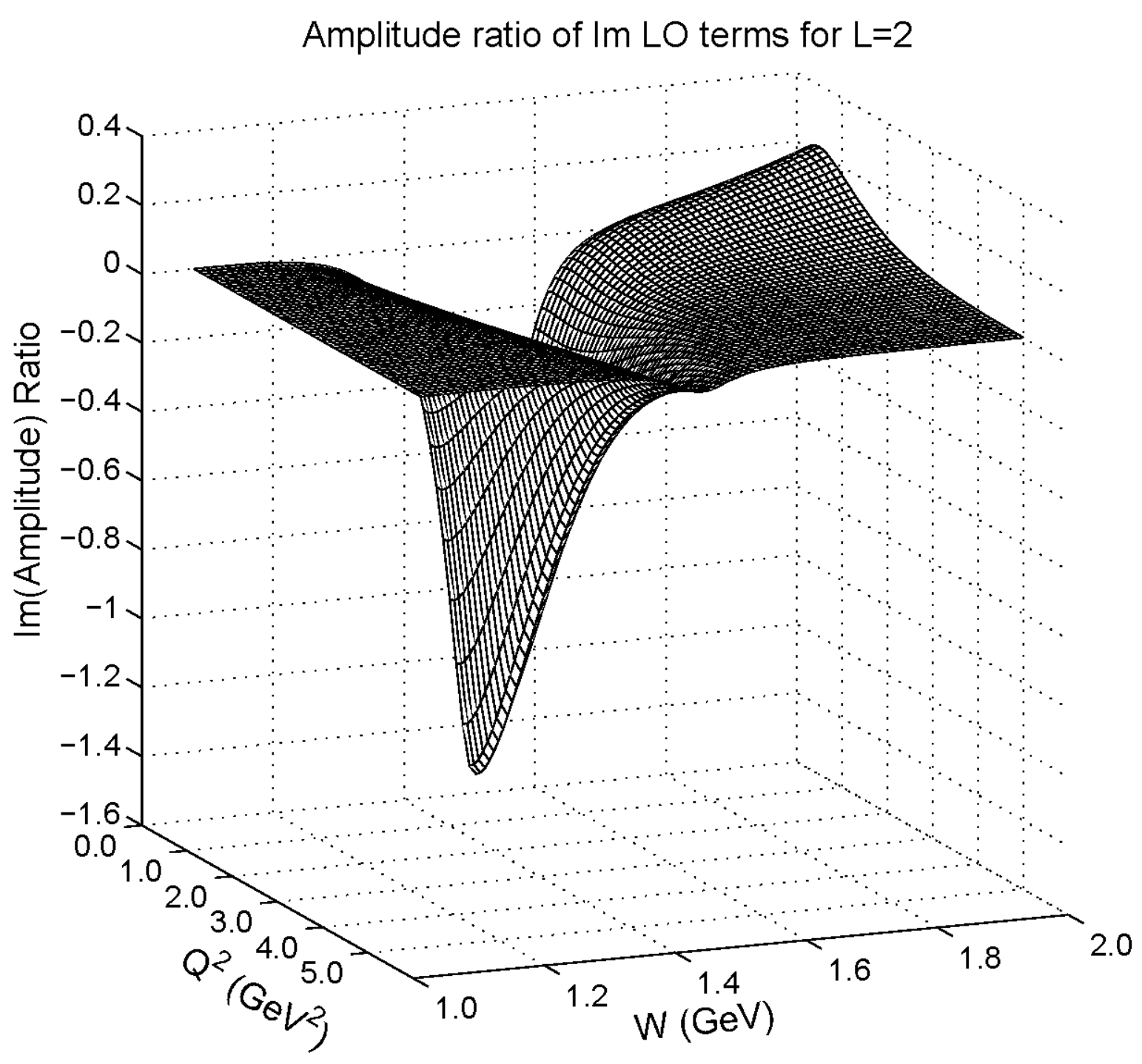}
\epsfxsize=0.48\textwidth\epsfbox{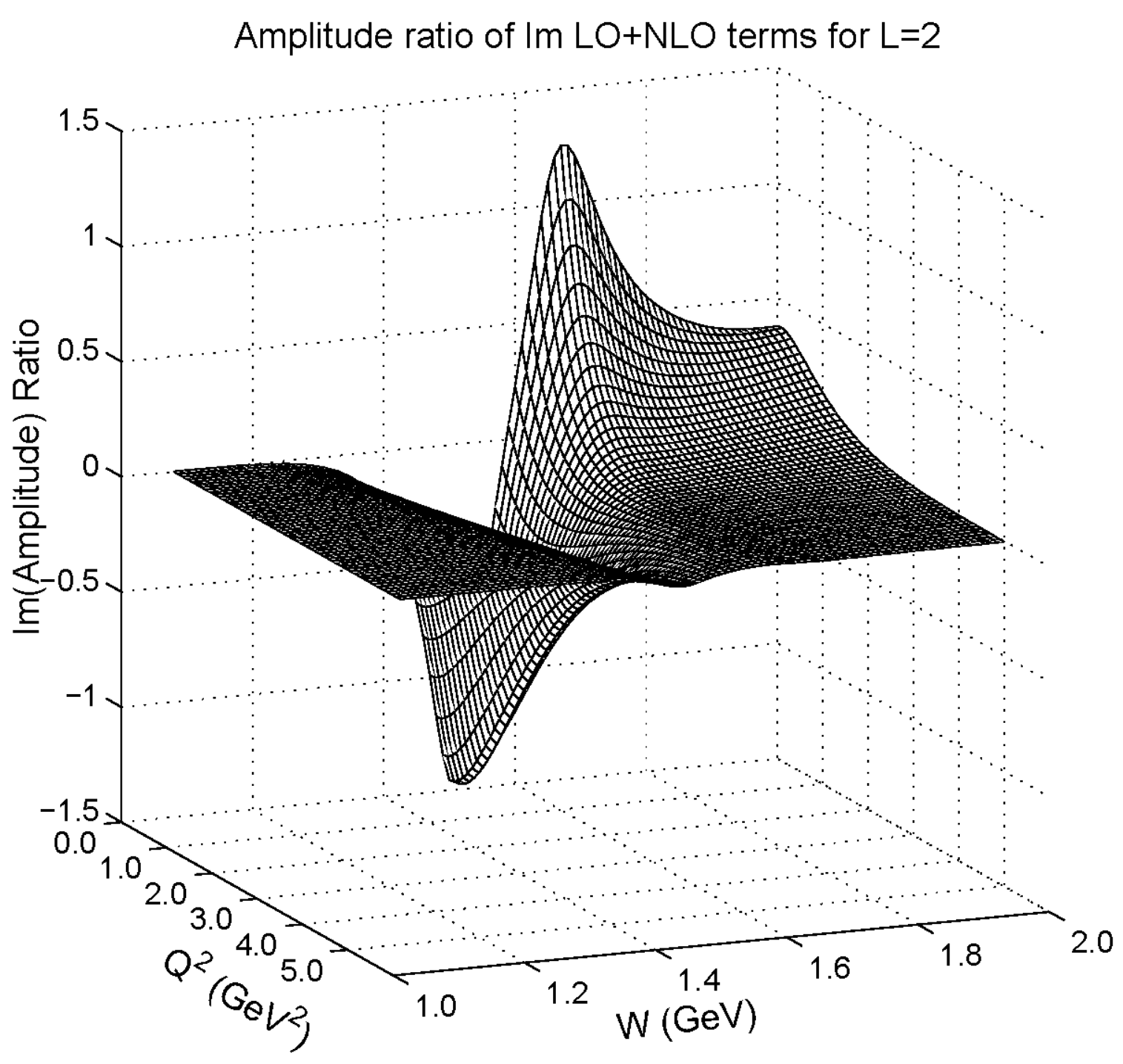}\\
\end{figure}
\begin{figure}[htp]
\epsfxsize=0.48\textwidth\epsfbox{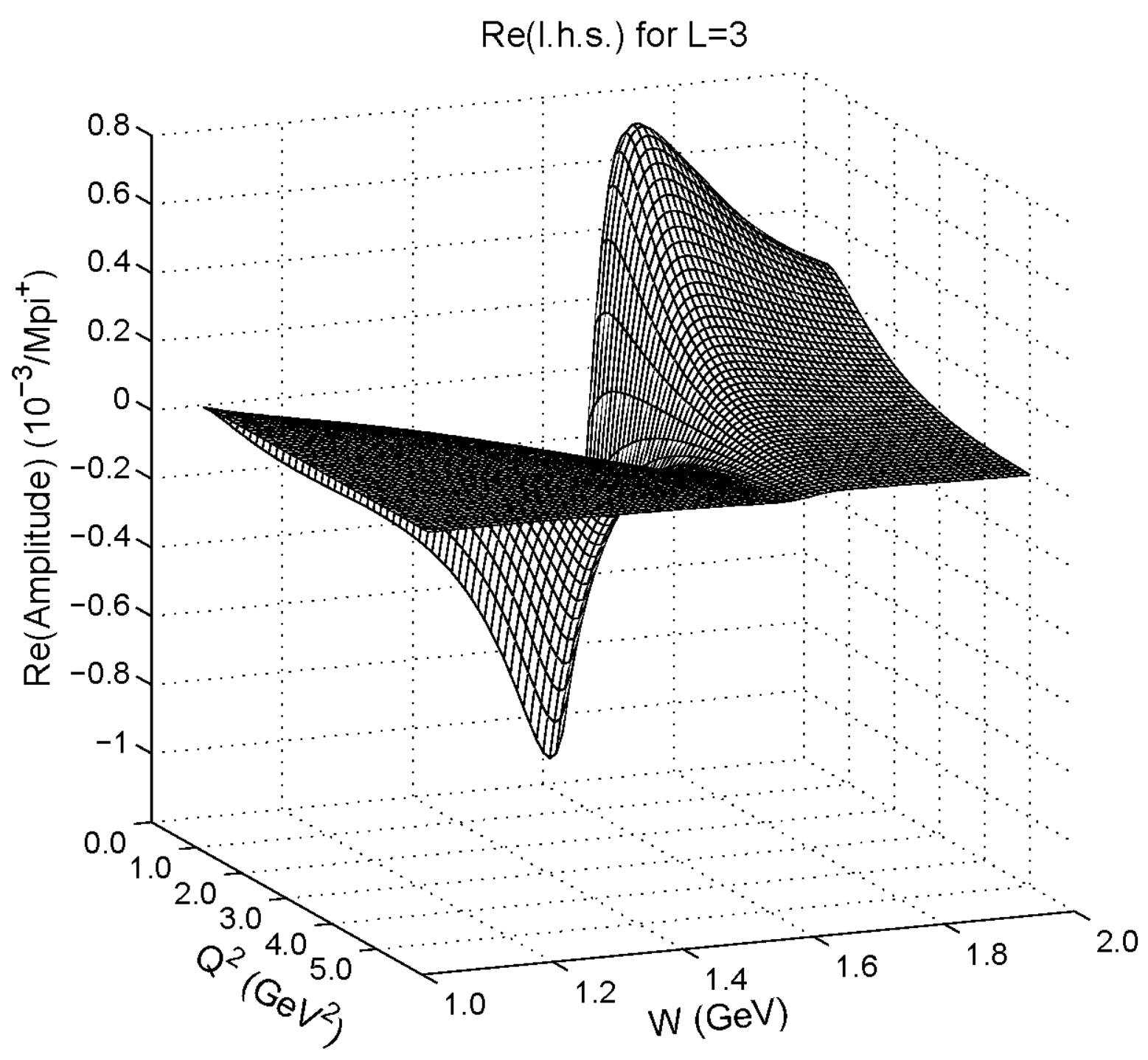}\\[1mm]
\epsfxsize=0.48\textwidth\epsfbox{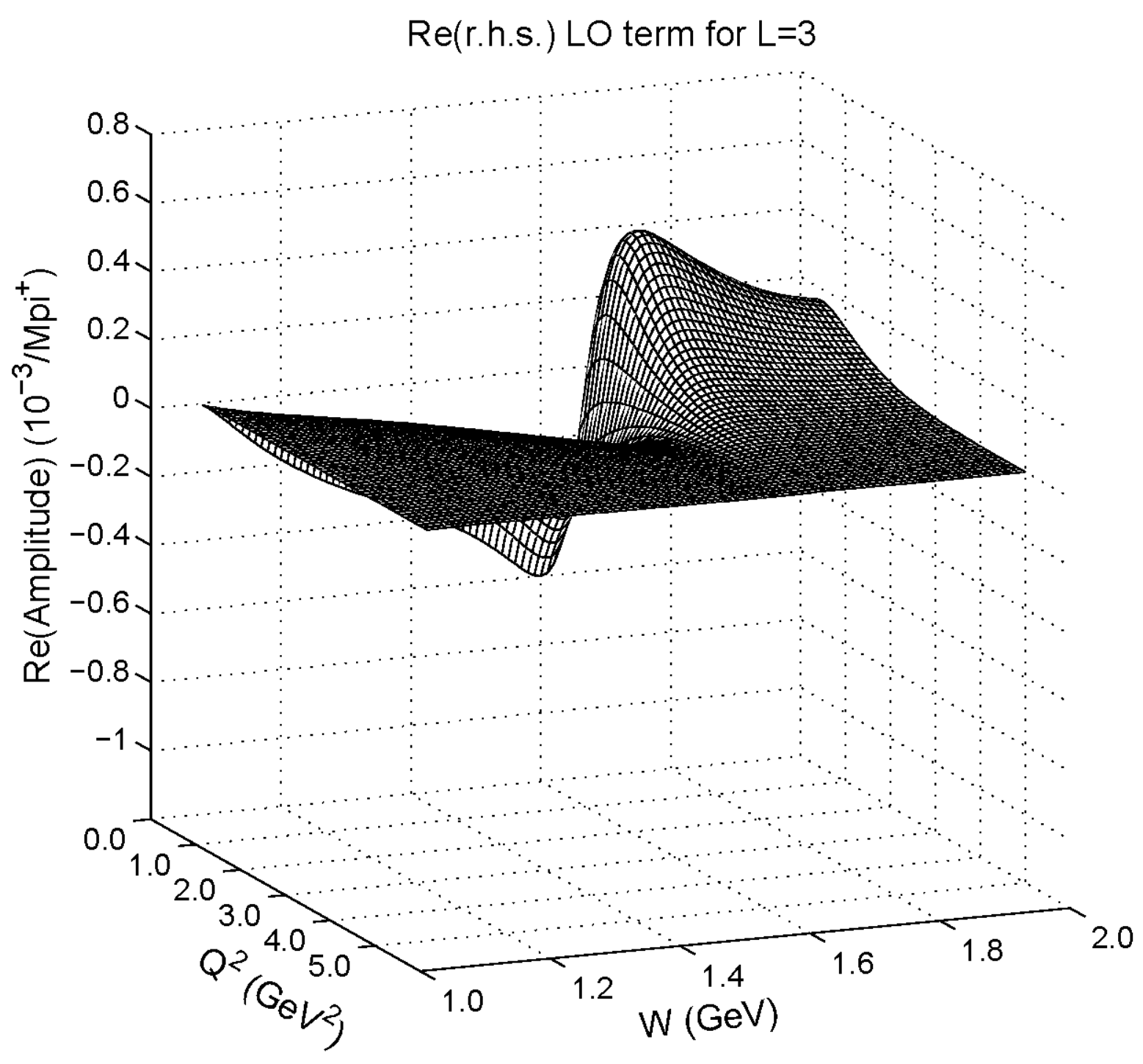}
\epsfxsize=0.48\textwidth\epsfbox{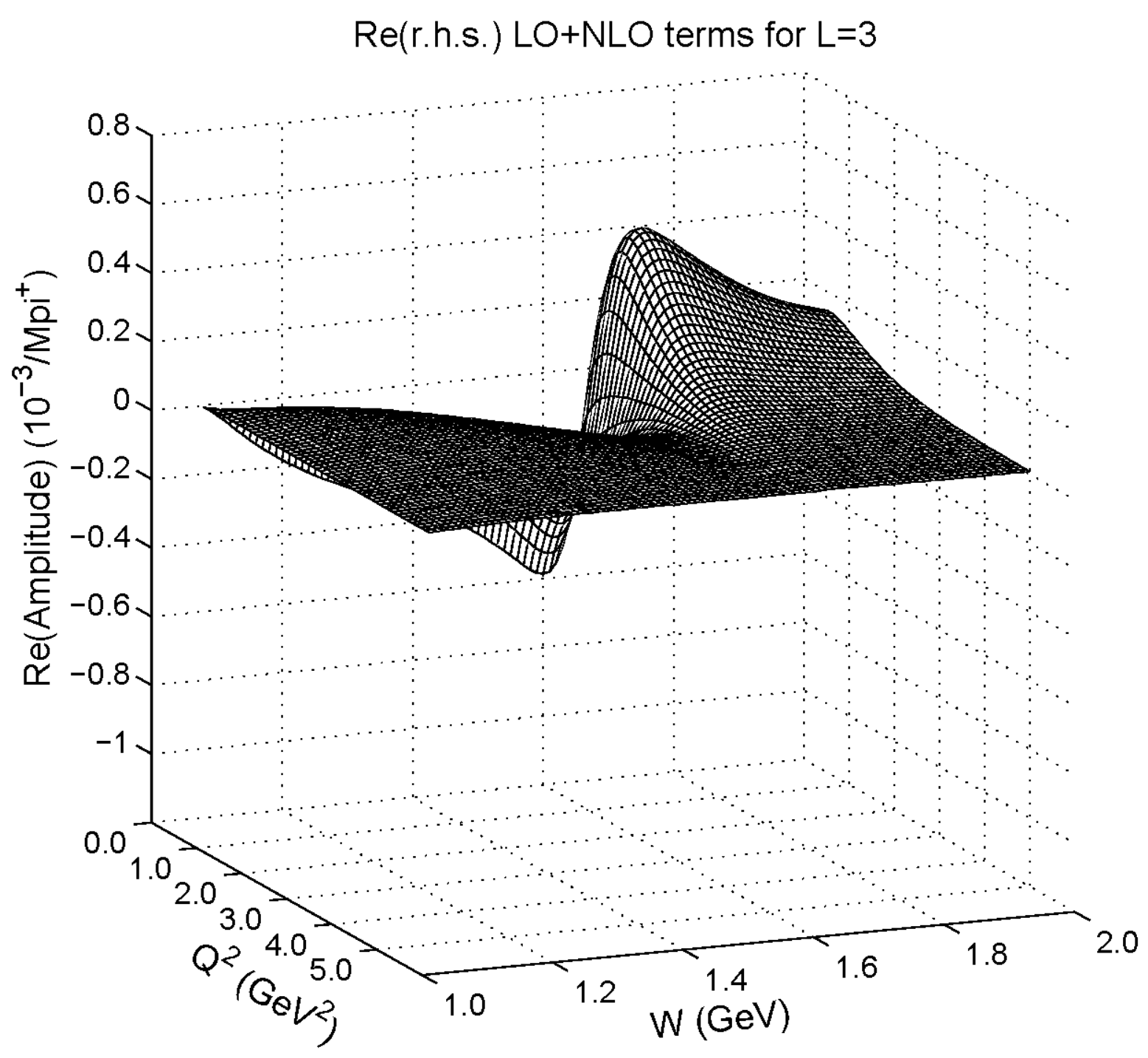}\\[1mm]
\epsfxsize=0.48\textwidth\epsfbox{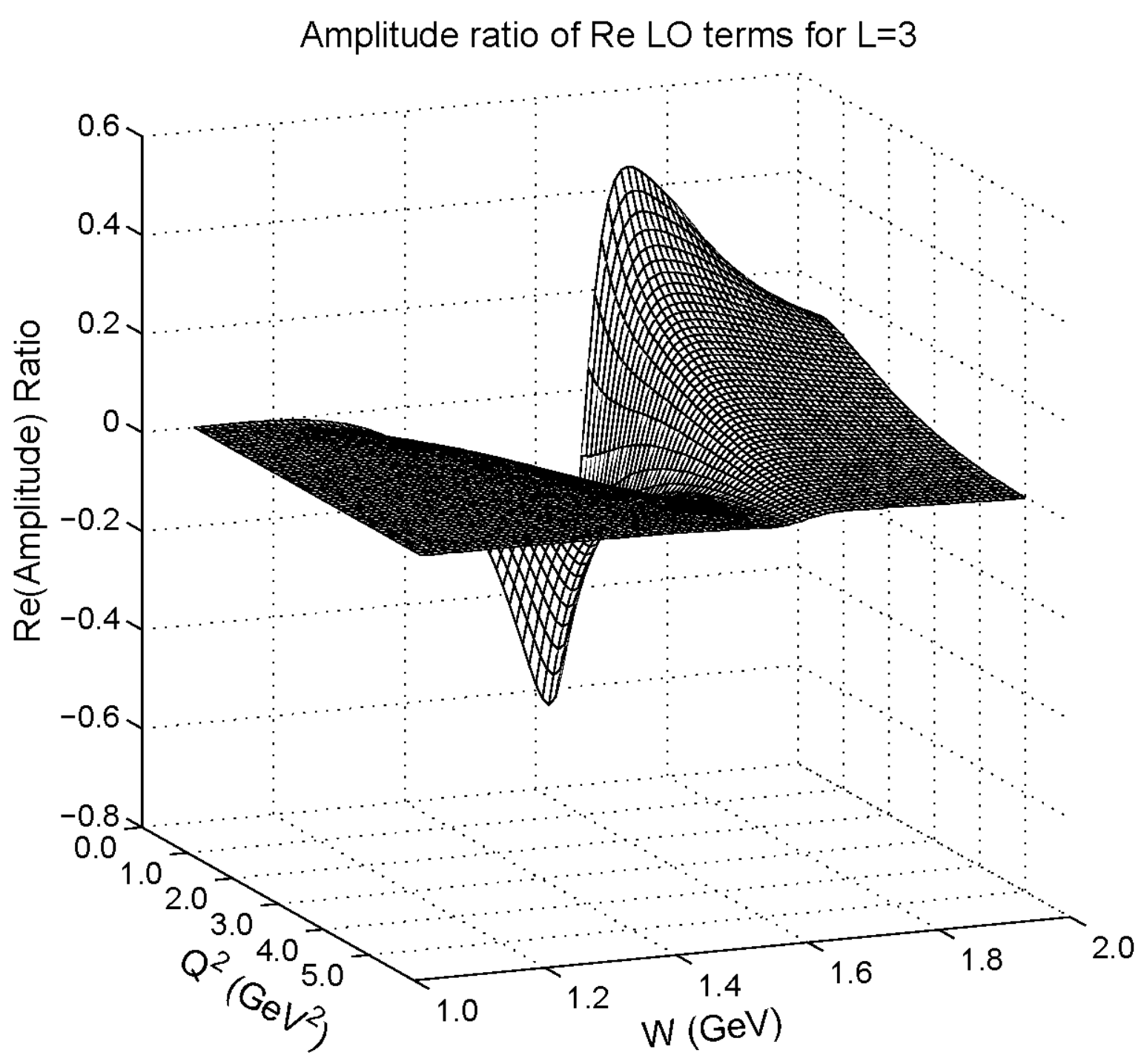}
\epsfxsize=0.48\textwidth\epsfbox{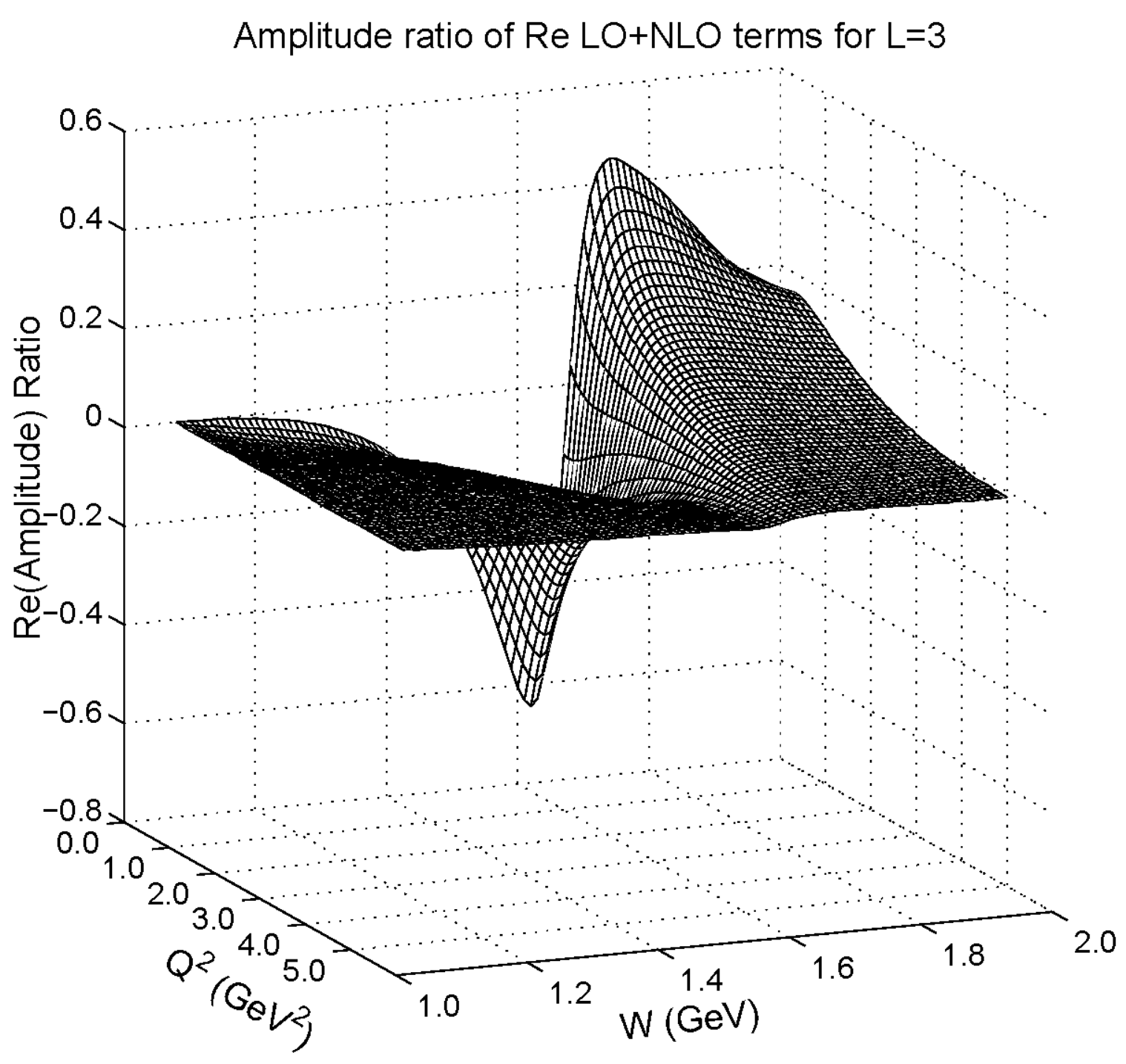}\\
\end{figure}
\begin{figure}[htp]
\epsfxsize=0.48\textwidth\epsfbox{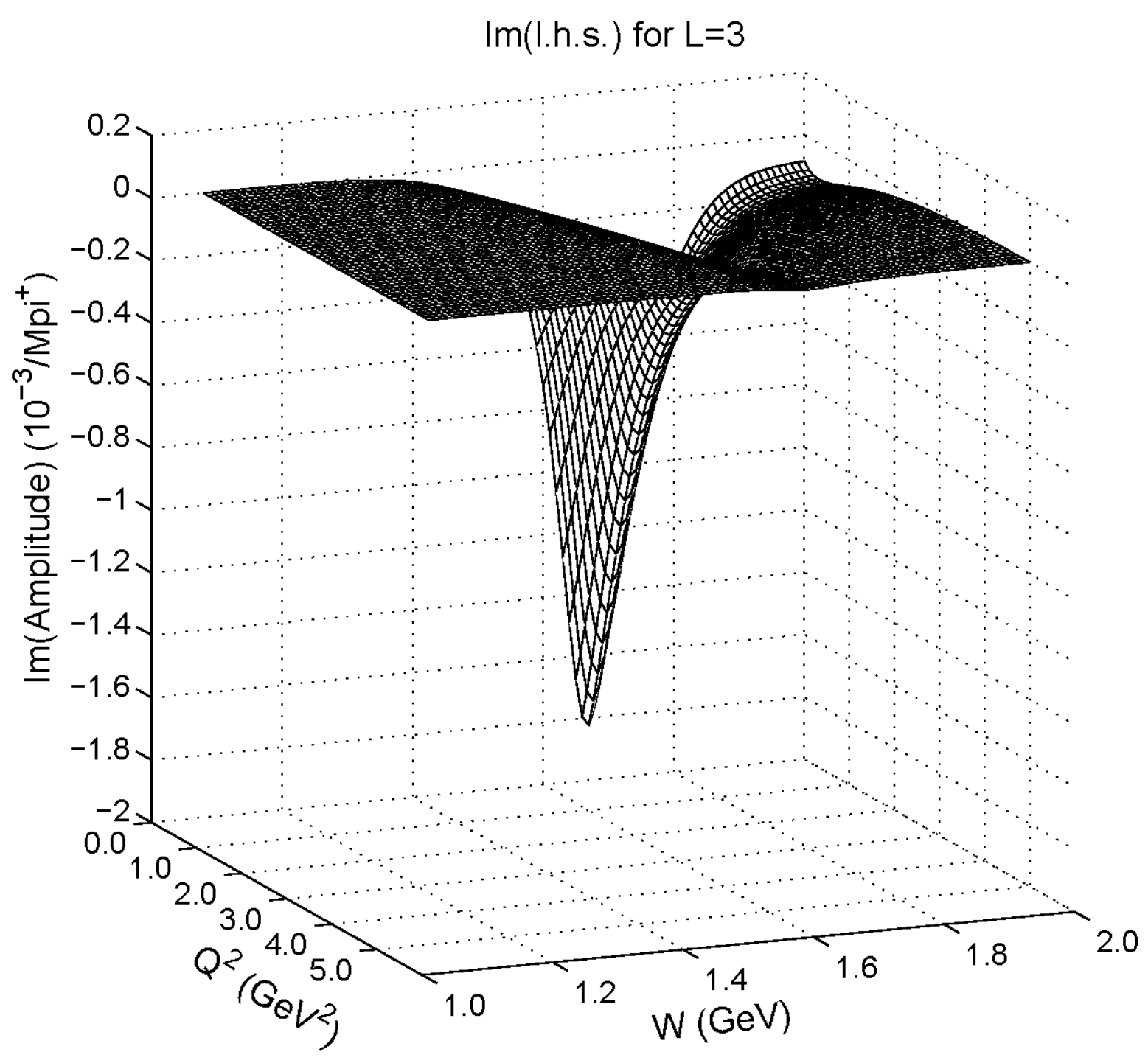}\\[1mm]
\epsfxsize=0.48\textwidth\epsfbox{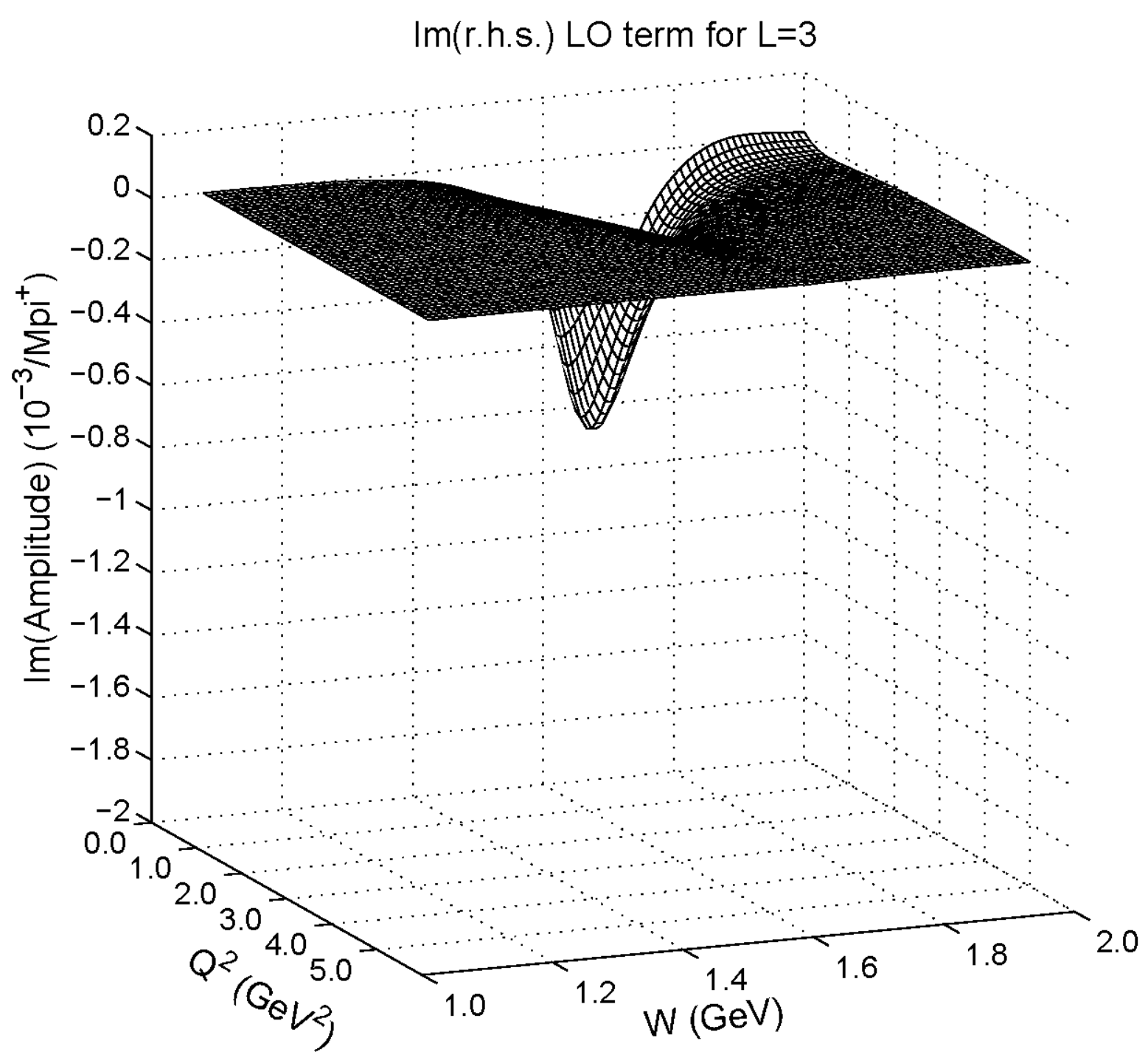}
\epsfxsize=0.48\textwidth\epsfbox{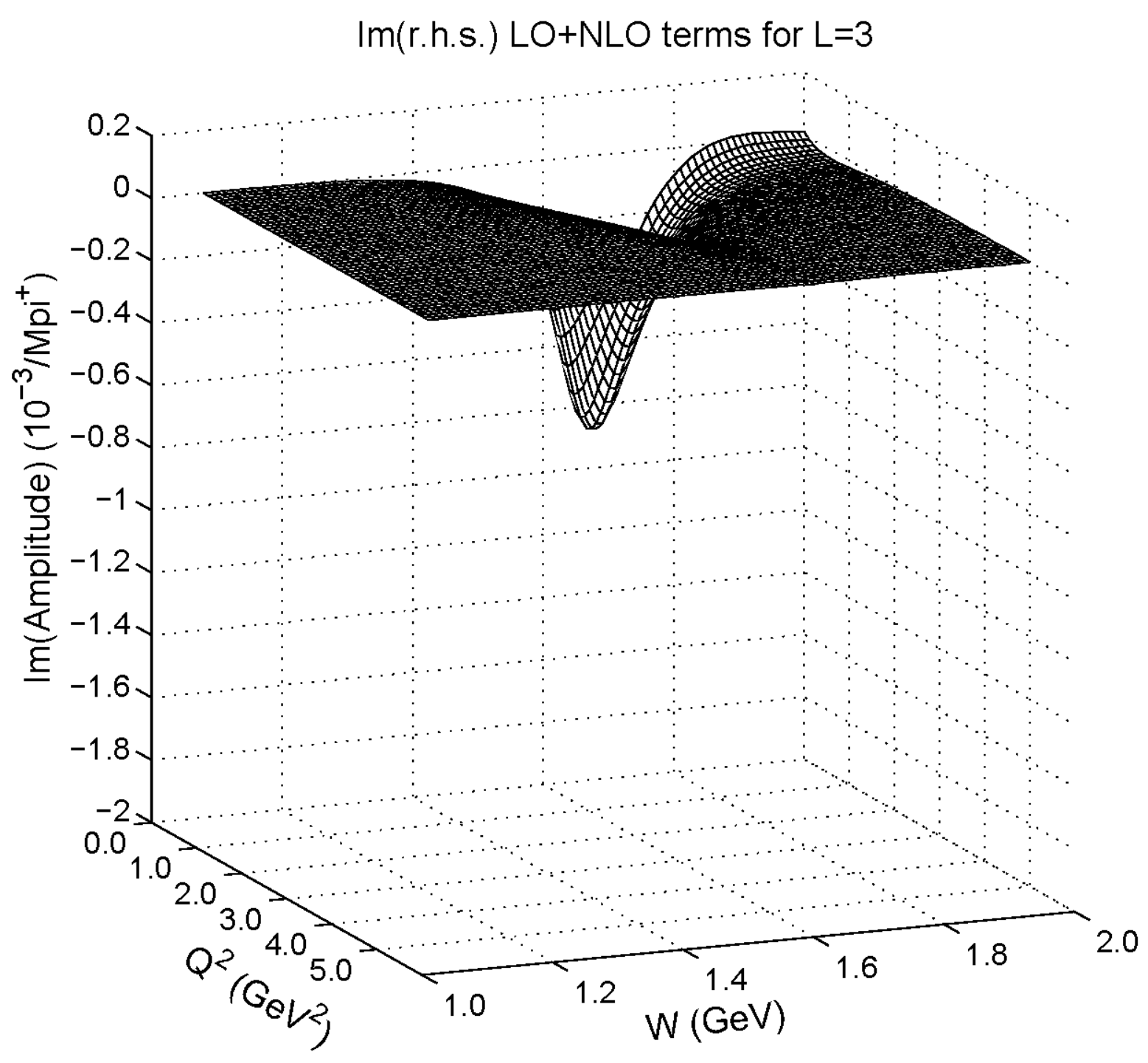}\\[1mm]
\epsfxsize=0.48\textwidth\epsfbox{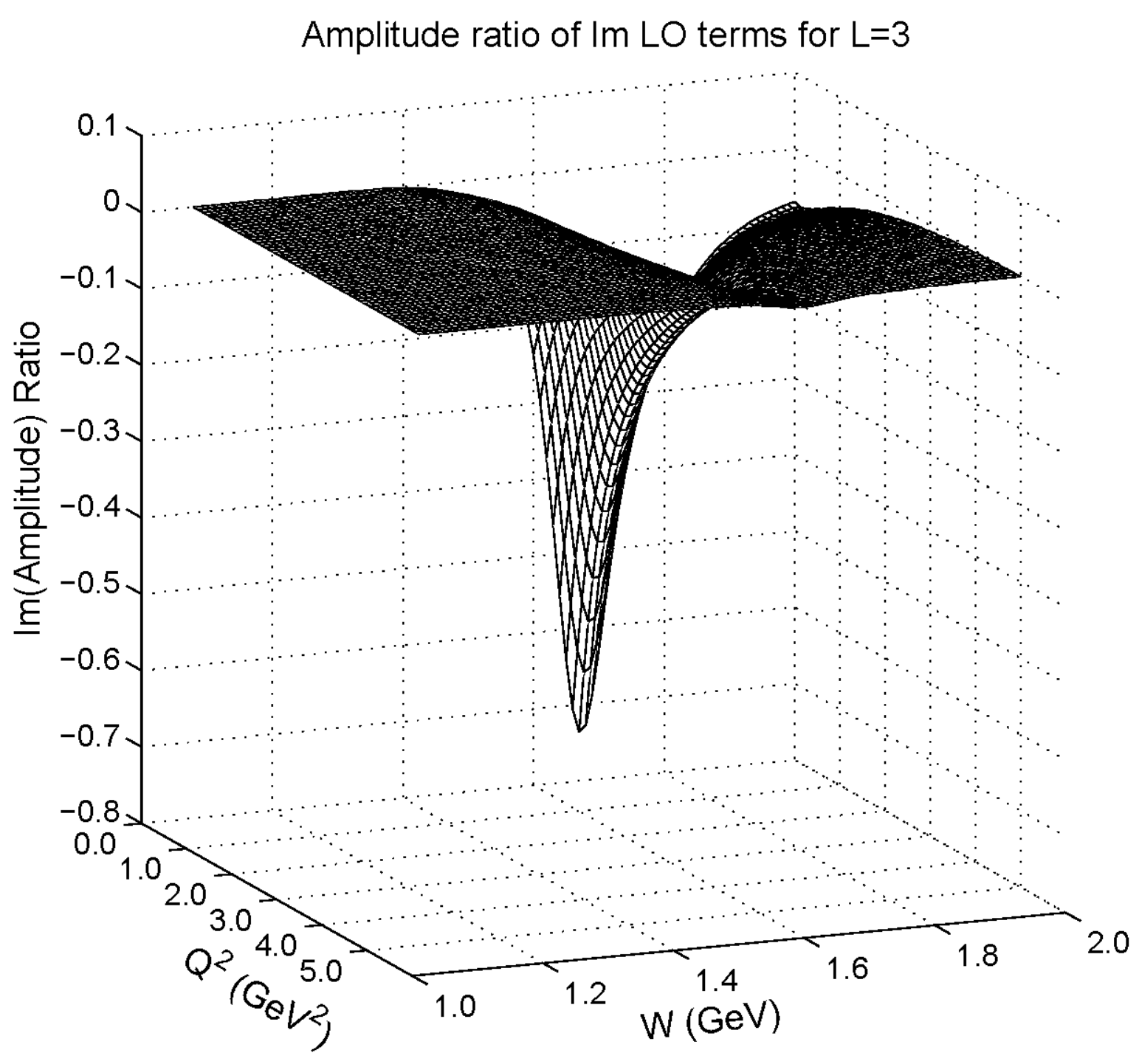}
\epsfxsize=0.48\textwidth\epsfbox{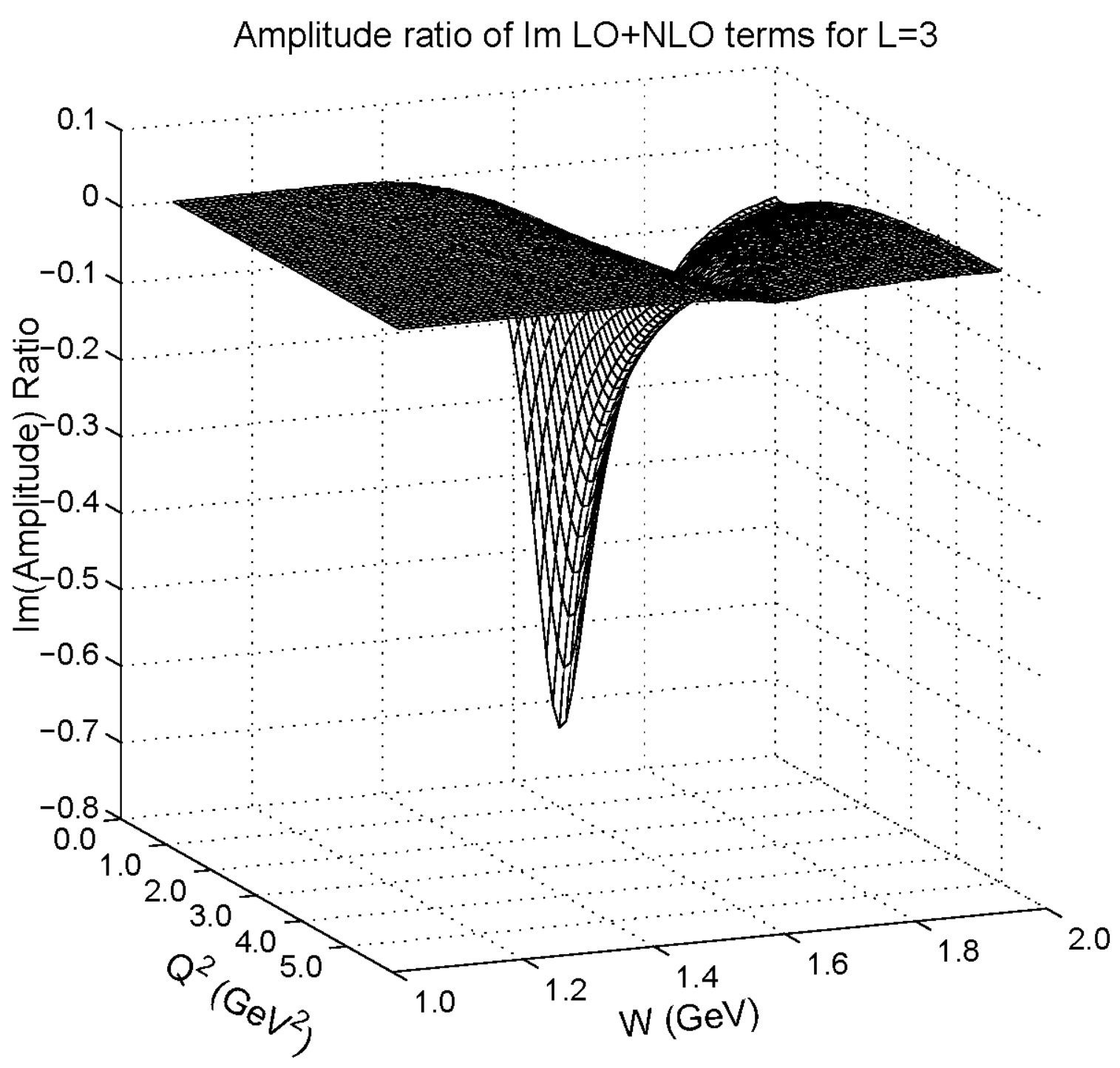}\\
\end{figure}
\begin{figure}[htp]
\epsfxsize=0.48\textwidth\epsfbox{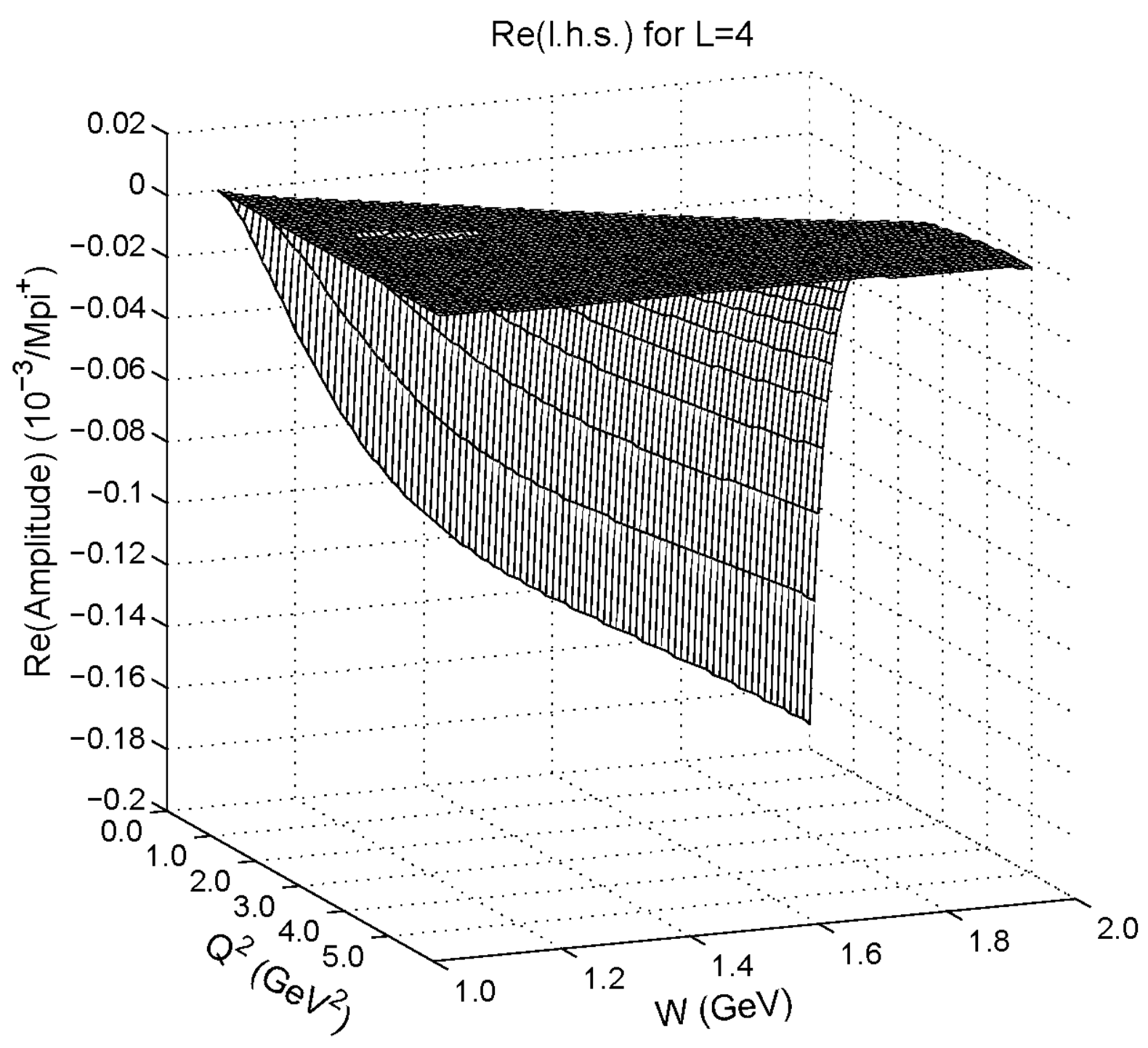}\\[1mm]
\epsfxsize=0.48\textwidth\epsfbox{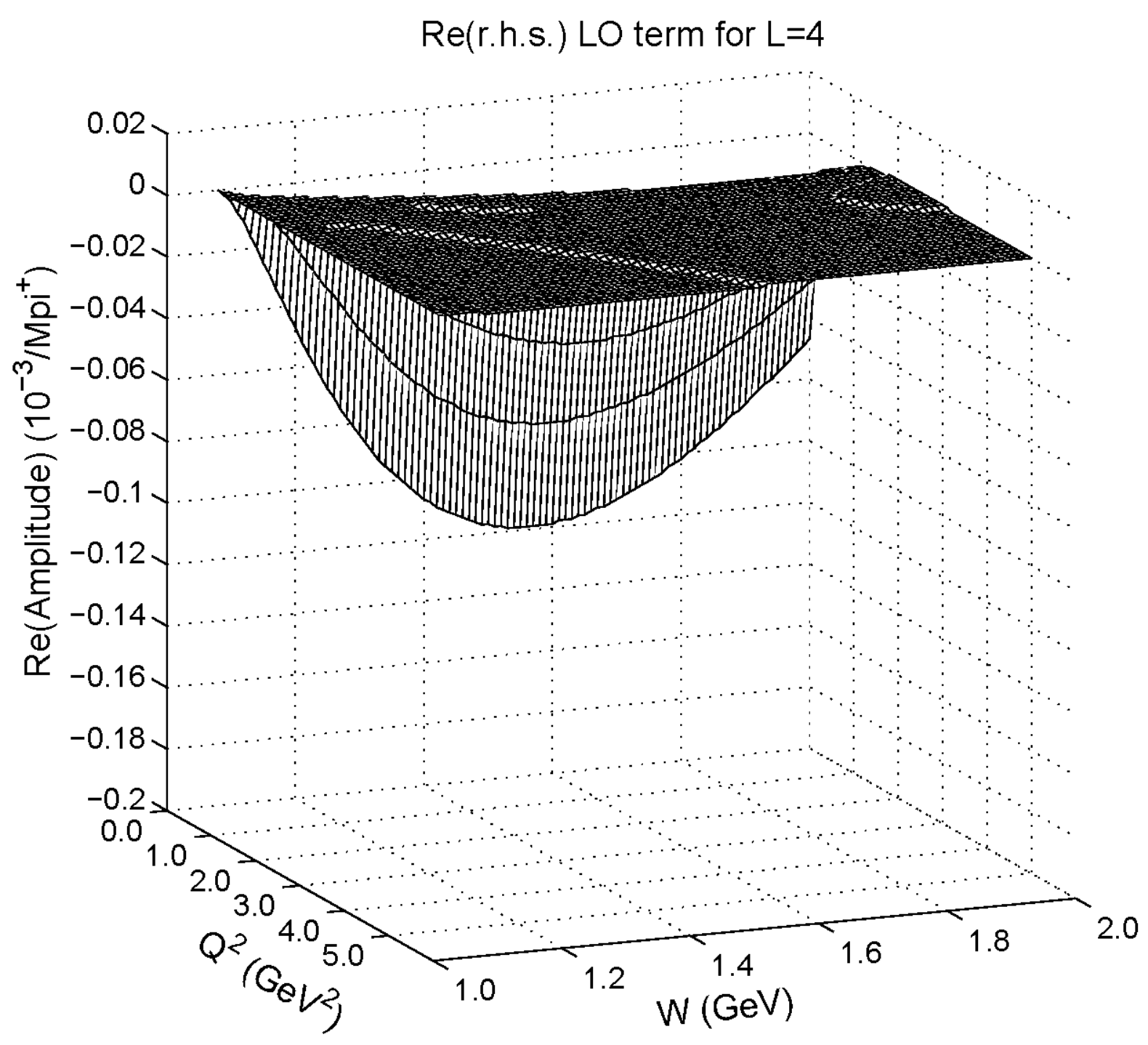}
\epsfxsize=0.48\textwidth\epsfbox{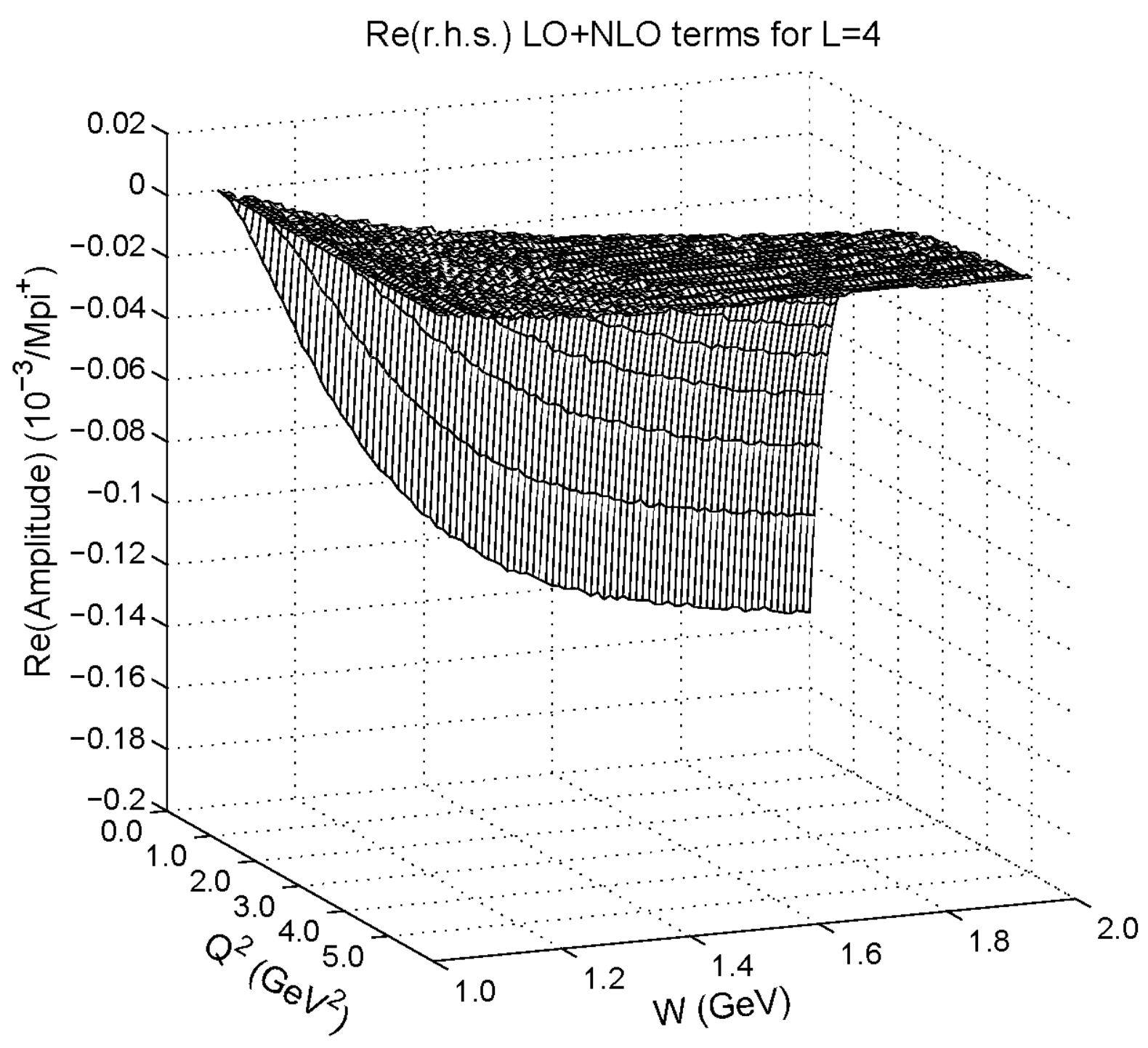}\\[1mm]
\epsfxsize=0.48\textwidth\epsfbox{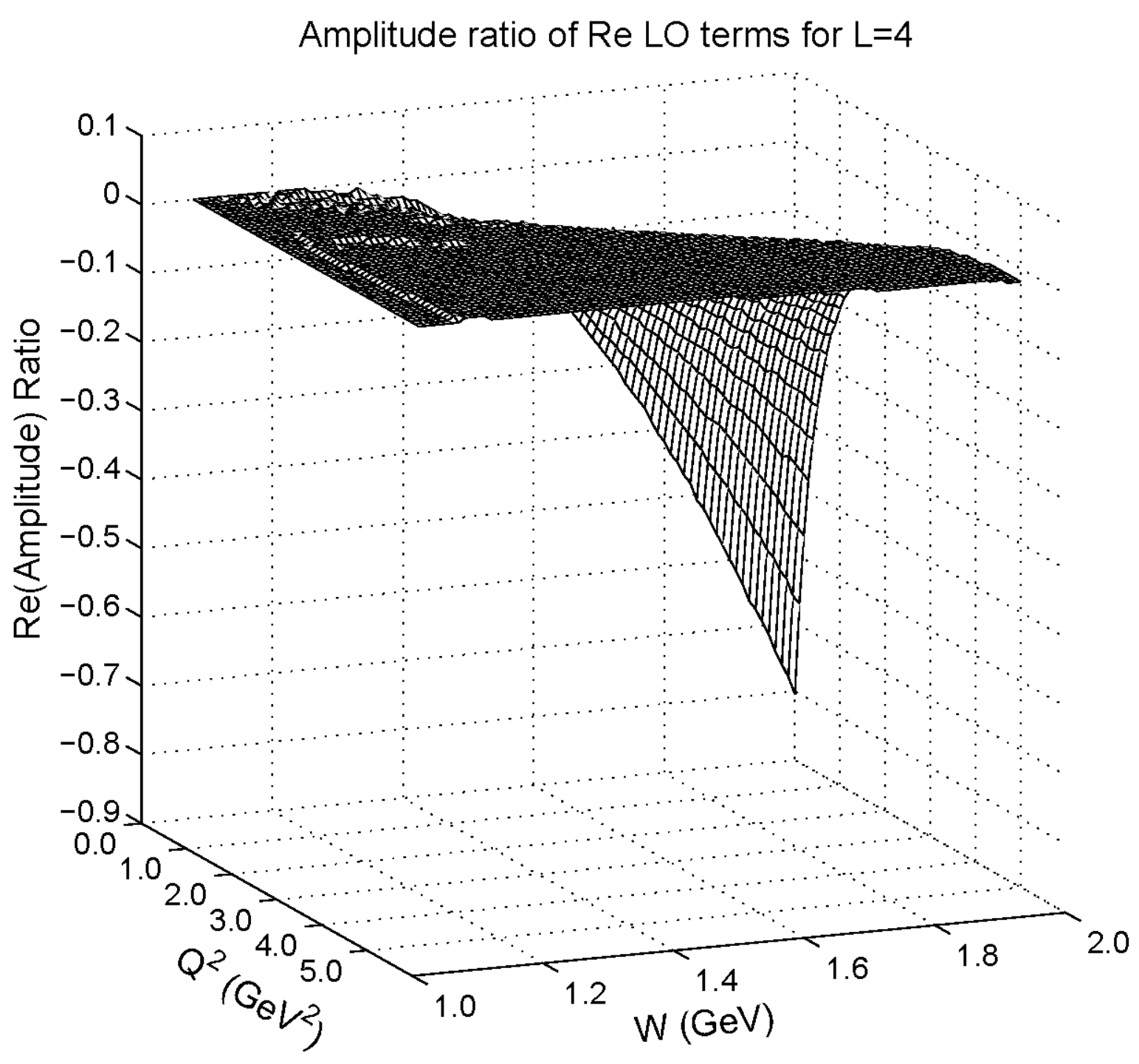}
\epsfxsize=0.48\textwidth\epsfbox{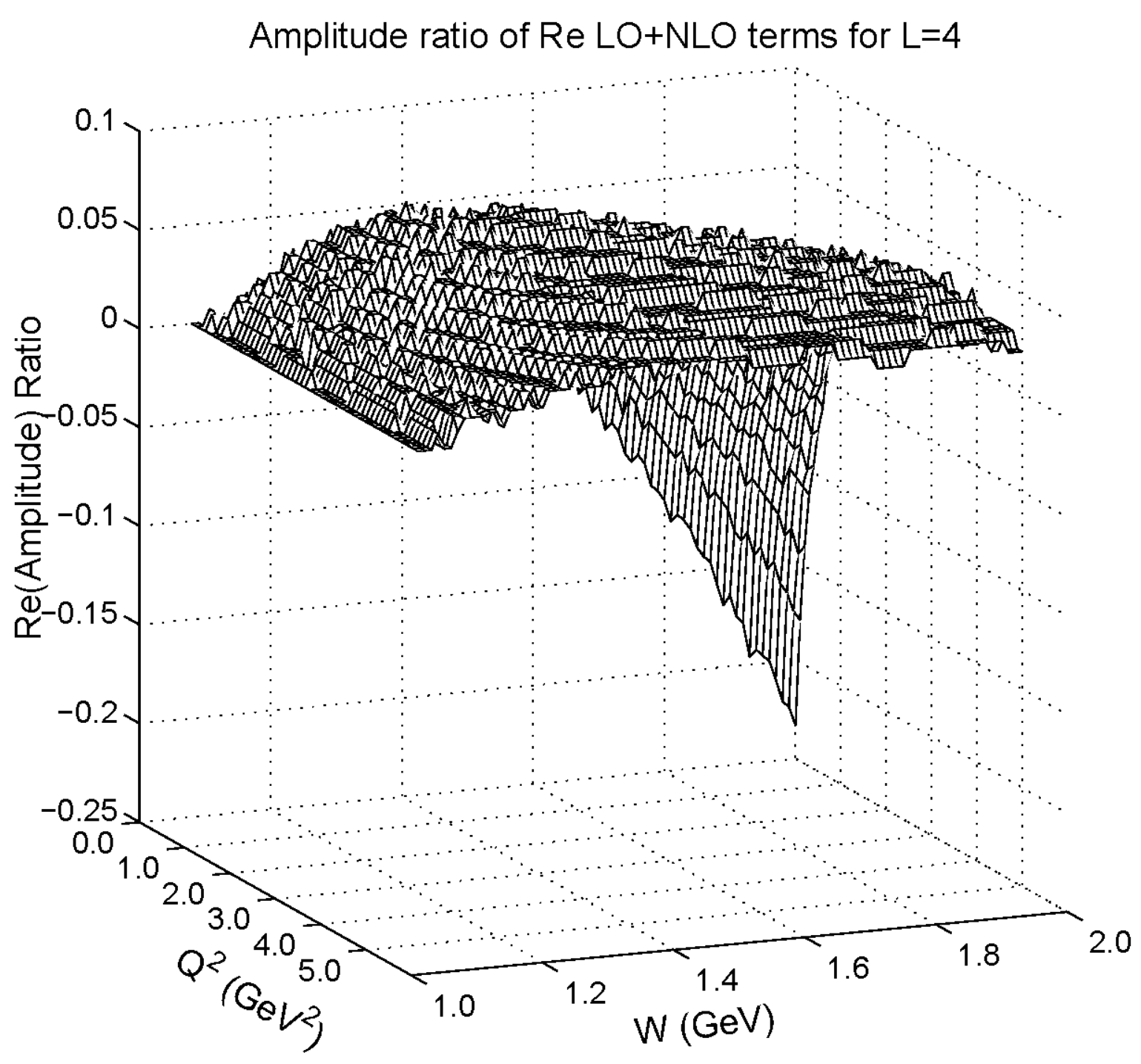}\\
\end{figure}
\begin{figure}[htp]
\epsfxsize=0.48\textwidth\epsfbox{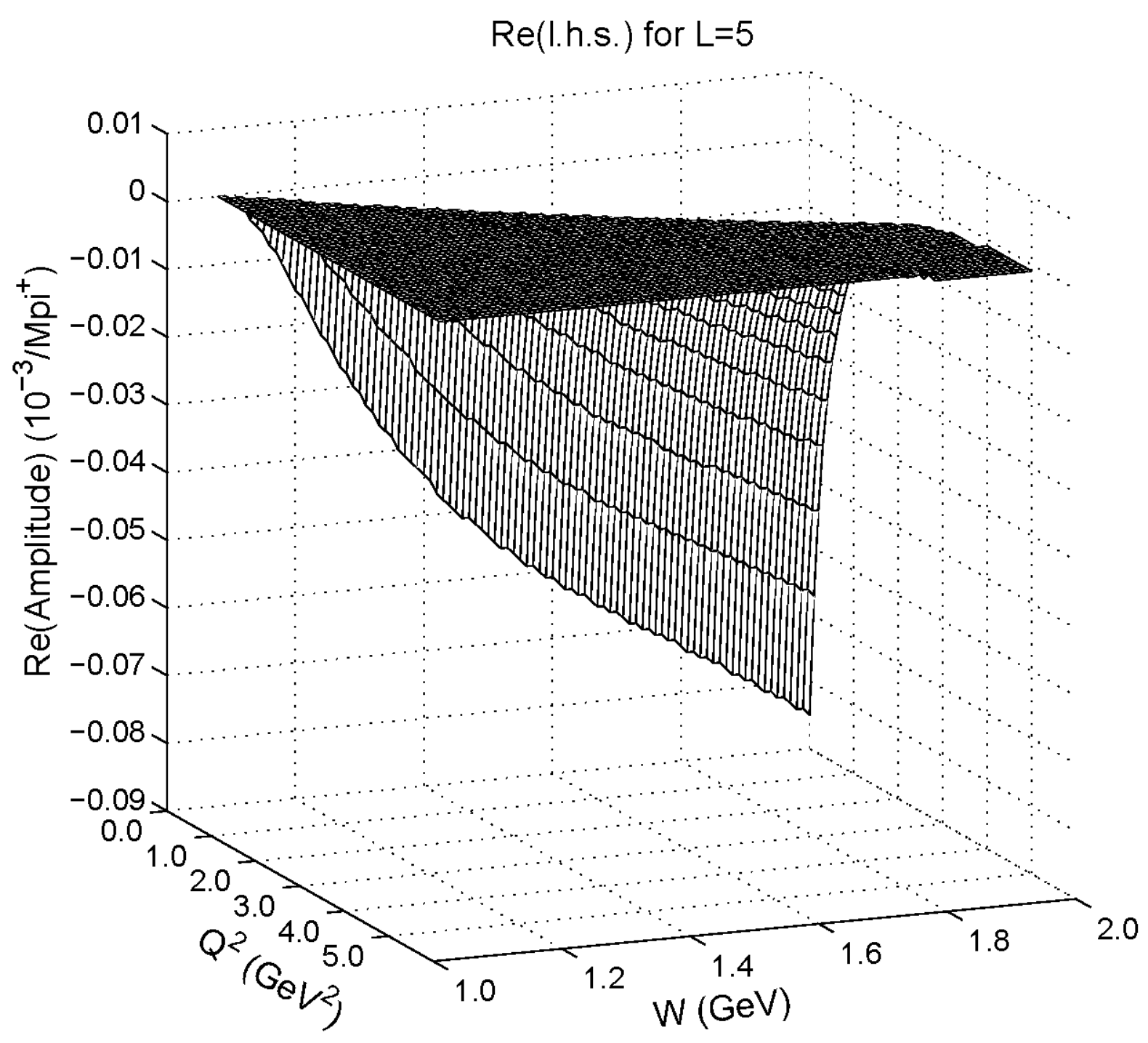}\\[1mm]
\epsfxsize=0.48\textwidth\epsfbox{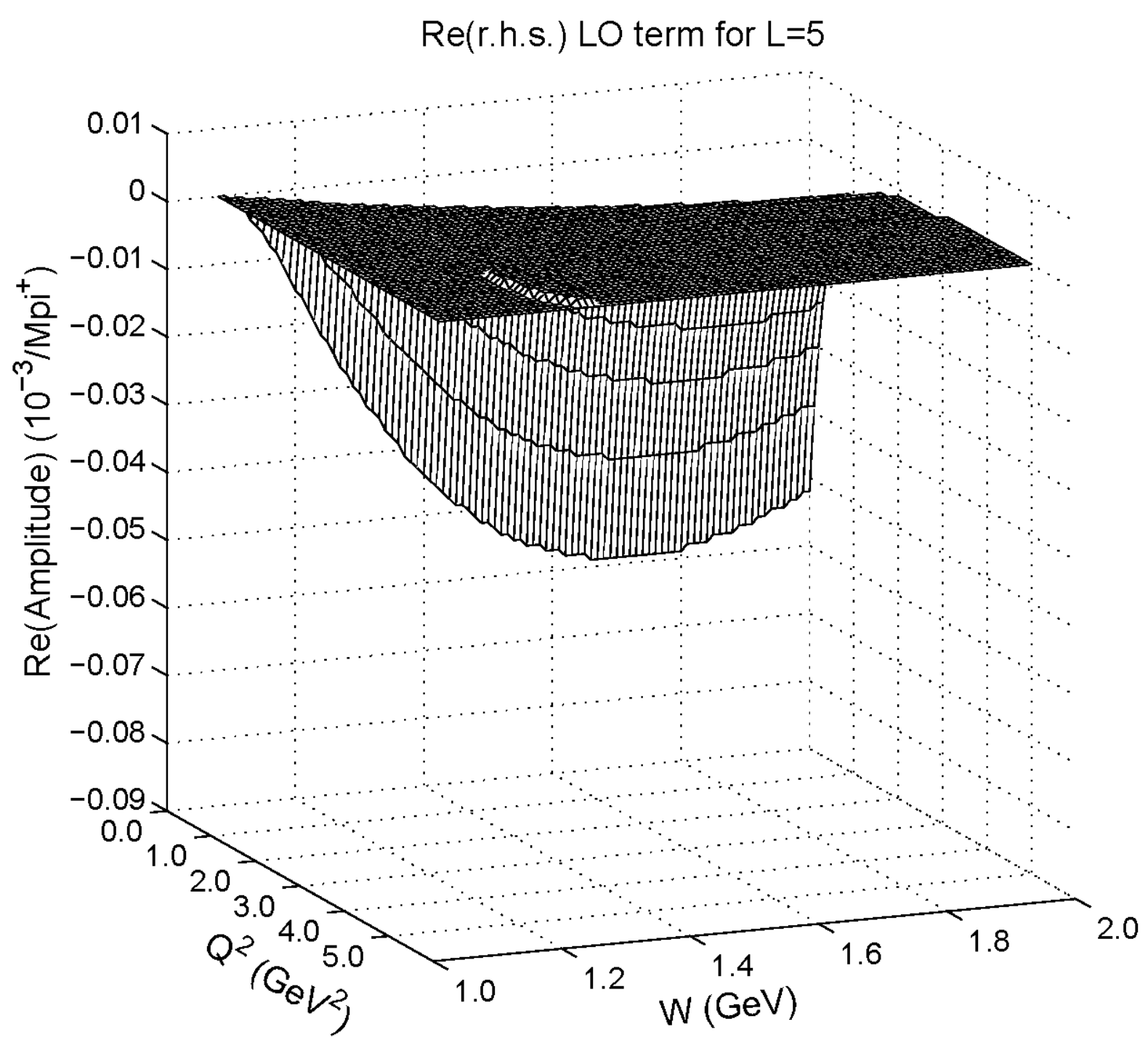}
\epsfxsize=0.48\textwidth\epsfbox{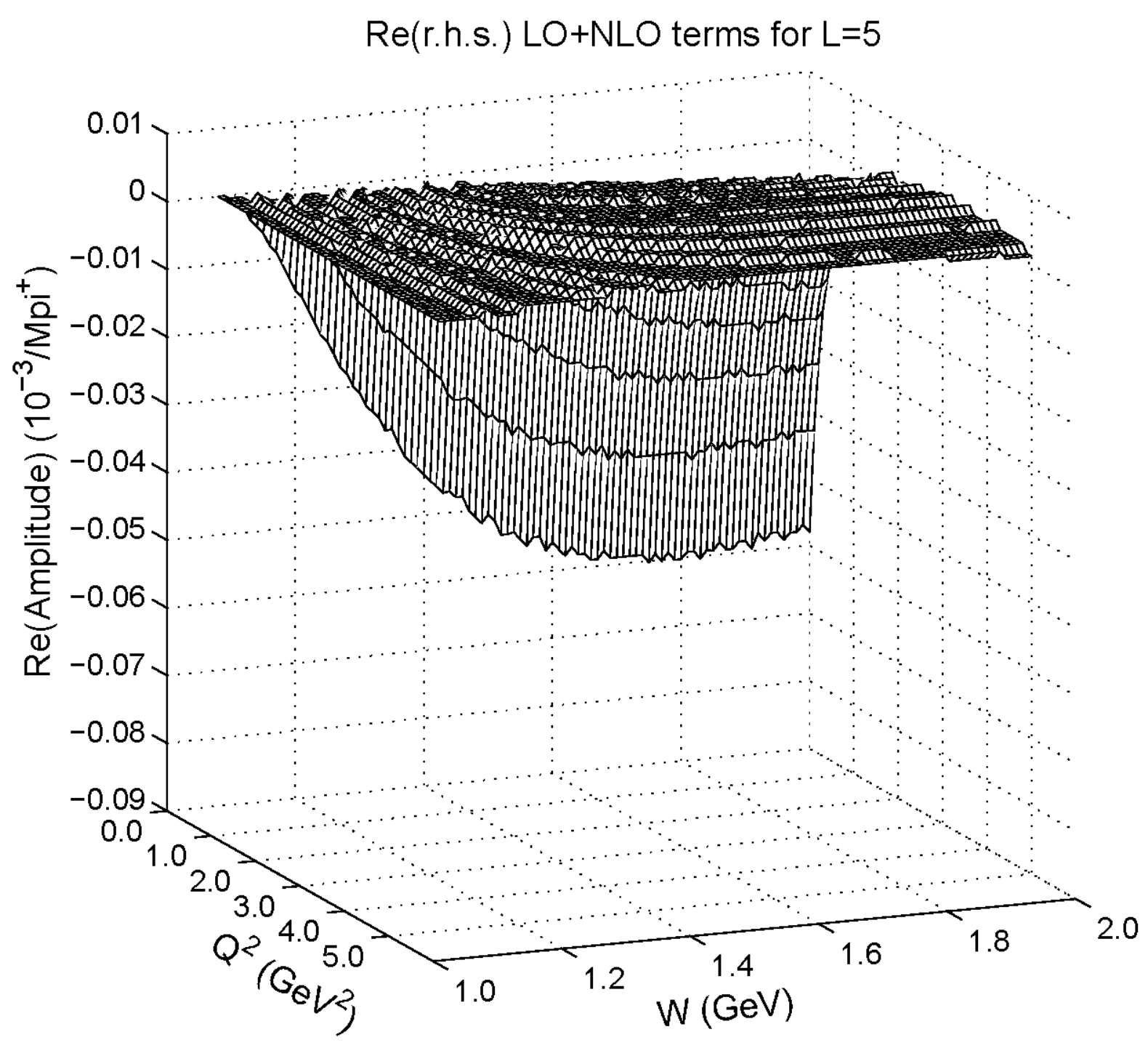}\\[1mm]
\epsfxsize=0.48\textwidth\epsfbox{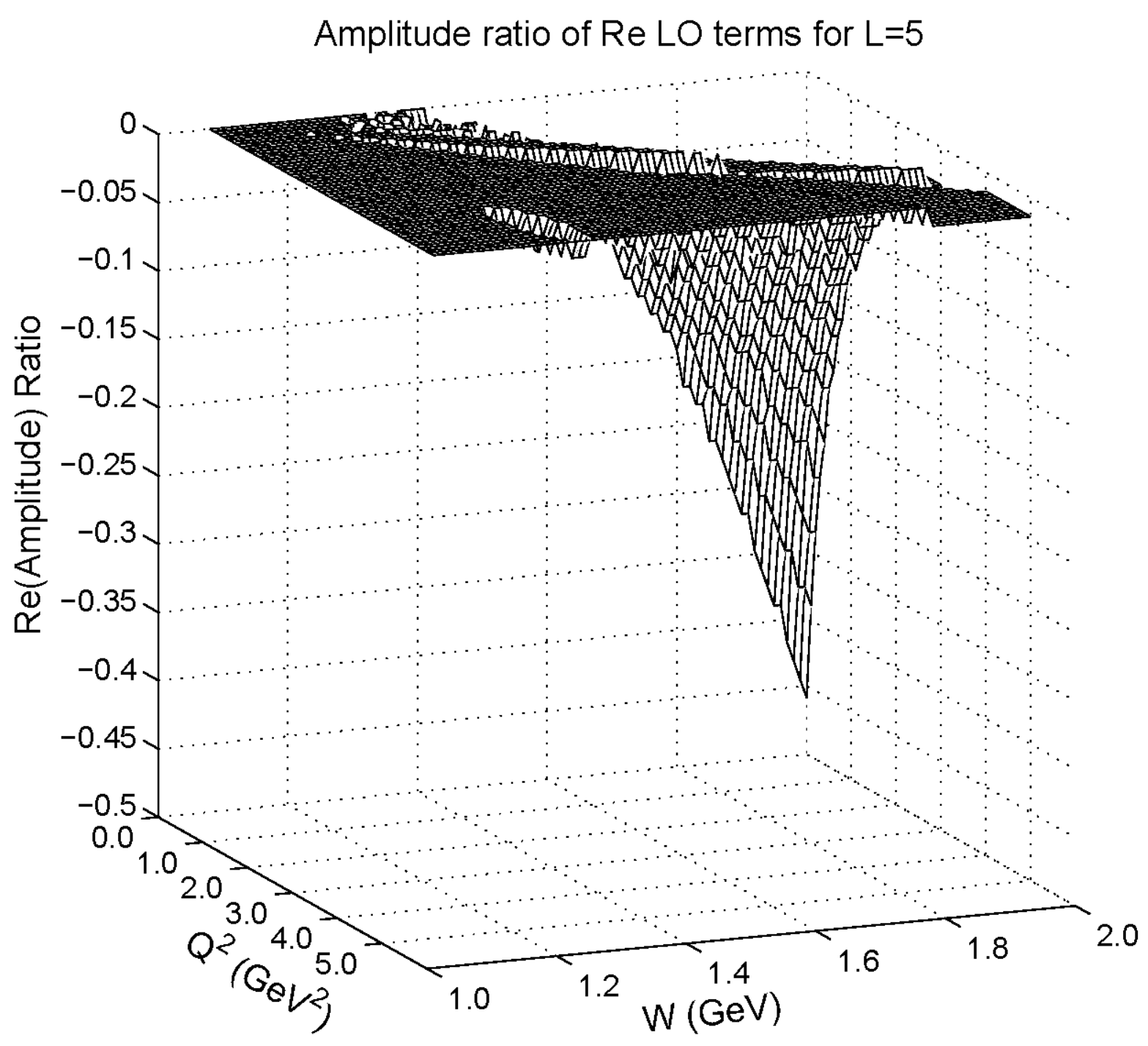}
\epsfxsize=0.48\textwidth\epsfbox{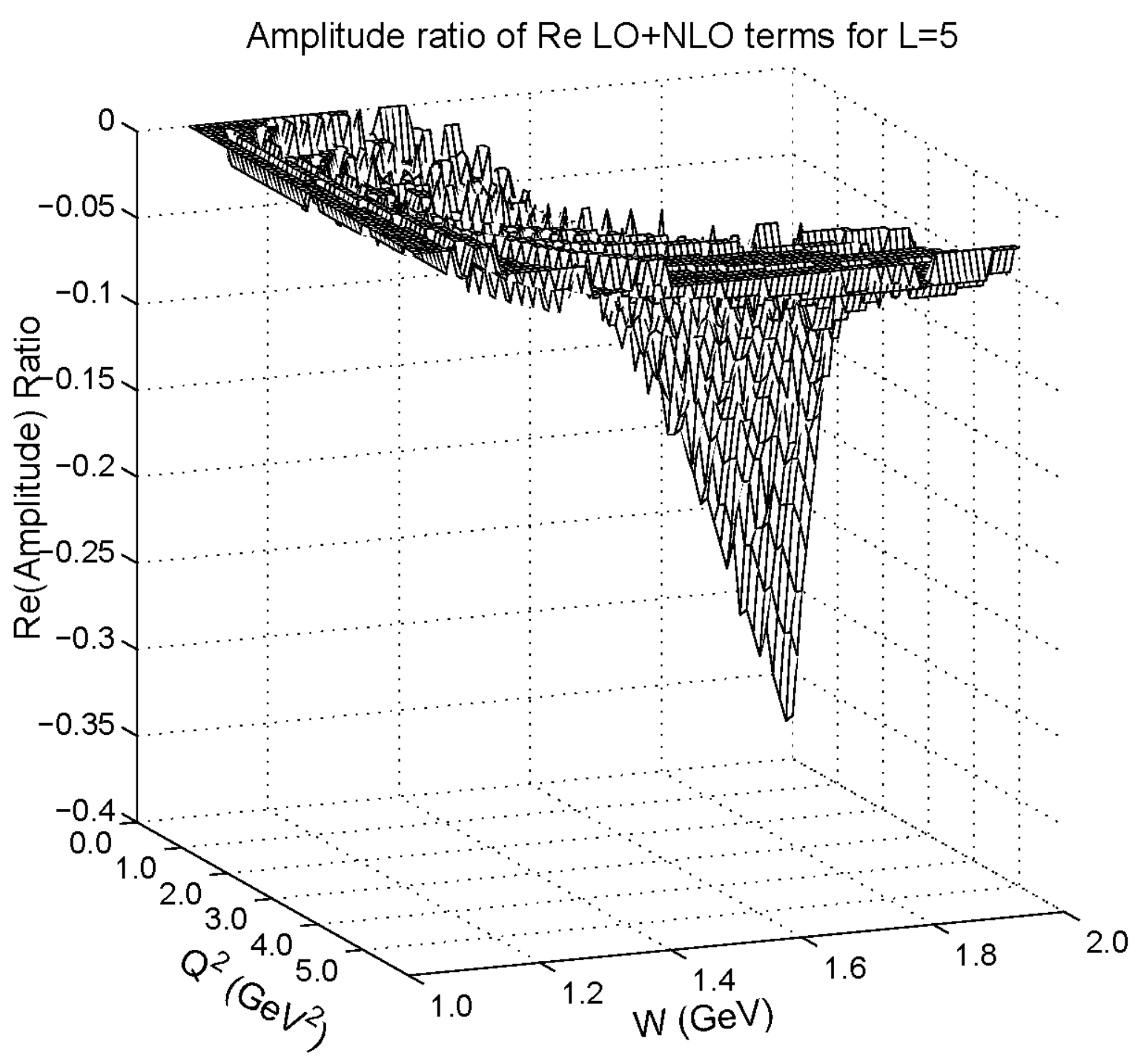}
\end{figure}
%
%
%

\section*{Acknowledgments}
This work was supported by the NSF under Grant No.\ PHY-0757394.

\end{document}